\documentclass[prd,nofootinbib,reprint,preprintnumbers,showkeys]{revtex4-2}
\usepackage{nlo-shower}

\usepackage{environ,graphicx}

\NewEnviron{prdfig}[2]{
  \BODY
}{}
\begin{document}

\title{Formulation of parton shower evolution beyond leading order for electron-positron annihilation}

\author{Zolt\'an Nagy}

\affiliation{
 Deutsches Elektronen-Synchrotron DESY, 
 Notkestr.\ 85, 22607 Hamburg, Germany
}

\email{Zoltan.Nagy@desy.de}

\author{Davison E.\ Soper}

\affiliation{
Institute for Fundamental Science,
University of Oregon,
Eugene, OR  97403-5203, USA
}

\email{soper@uoregon.edu}

\begin{abstract}
Parton shower evolution is given by a renormalization group equation that reflects the infrared behavior of perturbative QCD including color and spin and including quantum interference. With added approximations, such an evolution equation can provide the basis for a parton shower event generator. To date, splitting functions for parton shower evolution have been derived only to leading order in $\as$. We provide a framework that can provide splitting functions beyond leading order, and in particular at order $\as^2$. This requires a representation of Feynman graphs in which partons are off shell and are characterized by vector or spinor indices instead of on-shell spins. 
\end{abstract}

\date{9 May 2026}

\preprint{DESY-26-054}

\maketitle

\tableofcontents

\makeatletter
\let\toc@pre\relax
\let\toc@post\relax
\makeatother 
%--------------

%-------------------------------------------------

\section{Introduction}
\label{sec:introduction}

It is an unsolved problem to specify splitting functions for a parton shower in which the splitting functions are defined at order $\as^2$ or beyond. The splitting functions should be based on the soft and collinear singularities of quantum chromodynamics (QCD) \cite{NSAllOrder}. Thus what one needs is to translate the singularities of Feynman graphs into functions from which the parton splitting functions are constructed. This is not a trivial project beyond leading order in $\as$ because one has both real emissions and virtual exchanges and both soft and collinear singularities and combinations of these. Thus one seeks a method that constructs the needed singular functions directly from Feynman graphs. At order $\as^2$, we need splitting functions describing two real emissions, one real emission and one virtual loop, and two virtual loops. We describe here a general framework for constructing such functions for a parton shower for electron-positron annihilation to hadrons.\footnote{For lepton-hadron collisions and hadron-hadron collisions, one would need to include the factorization of parton distribution functions.} We also illustrate how this framework works for leading order splitting functions. We defer to later work the application of this framework at order $\as^2$.

There are many meanings in the literature for the name ``parton shower,'' just as there are many meanings for ``factorization.'' Having many meanings in circulation can lead to confusion. We start, therefore, with a specification of what we have in mind in this paper as a parton shower.

First of all, there are parton shower event generators. These began in the mid 1980s and before with work that included \textsc{Pythia} \cite{EarlyPythia}, \textsc{Herwig} \cite{EarlyHerwig}, \textsc{Isajet} \cite{EarlyIsajet} and work by Gottschalk \cite{Gottschalk}. Since then, they have become much more sophisticated, as reviewed in Ref.~\cite{SjostrandReview}. Parton shower event generators are very useful, even essential, for understanding events at electron-positron, electron-proton, and hadron-hadron colliders. A parton shower event generator is implemented in computer code that generates simulated events consisting of final state hadrons plus, sometimes, leptons and electro-weak vector bosons. The core of this event generation reflects perturbative QCD and generates partons (quarks and gluons). There is then a model of hadronization to turn the partons into hadrons. 

The parton shower part of the event generator, leaving aside hadronization, is often described using the structure of its computer code. However, we find it useful to use a more abstract and more mathematical description. The mathematical description \cite{NSI} uses linear algebra and the language of quantum statistical mechanics. We use states $\isket{\rho}$ that describe the quantum numbers of any number, $m = 2,3,\dots$, of partons. The space of such states is the {\em statistical space}. The quantum numbers for the partons consist, at a minimum, of their momenta and flavors: $\{p,f\}_m = \{p_1,f_1; p_2,f_2; \dots ; p_m,f_m\}$. Then there are states $\isket{\{p,f\}_m}$ in which there are $m$ partons such that parton 1 has momentum $p_1$ and flavor $f_1$ and, in general, parton $i$ has momentum $p_i$ and flavor $f_i$. These states form a basis for the statistical space. In this paper, we include also color and spin quantum numbers. 

The states evolve. The evolution can be thought of as the solution of a renormalization group equation, in which a scale parameter $\mu_\scS$ decreases from an initial high scale (say $1 \TeV$) to a much lower scale of a few GeV, at which the shower is turned off.\footnote{There is also a renormalization scale $\mu_\scR$. In the final splitting functions, it is helpful to set $\mu_\scR = \mu_\scS$.}\footnote{This description applies to first order showers with ordering variables related to transverse momentum or virtuality in a splitting. A first order angle ordered parton shower needs a somewhat different description.} (Then a hadronization model may be applied.) The shower scale $\mu_\scS$ is often related to a transverse momentum in parton splittings. As $\mu_\scS$ decreases, the number of partons either stays the same or increases. In a parton shower event generator, this evolution is quite literal. As the code runs, more and more partons are produced as partons split, with the scale $\mu_\scS$ that describes the splittings decreasing at each step.

In the mathematical description of this process, the state at a smaller value of the scale is related to the starting state at a hard scale $\mu_\scS = \mu_\scH$ by a linear operator $\cU(\mu_\scS,\mu_\scH)$:
\begin{equation}
\label{eq:showercU}
\sket{\rho(\mu_\scS)} = \cU(\mu_\scS,\mu_\scH) \sket{\rho(\mu_\scH)}
\;.
\end{equation}
The operator $\cU(\mu_\scS,\mu_\scH)$ obeys a renormalization group equation
\begin{equation}
\label{eq:showerevolution}
\mu_\scS \frac{d}{d\mu_\scS}\,\cU(\mu_\scS,\mu_\scH)
 = -\cS(\mu_\scS)\,\cU(\mu_\scS,\mu_\scH)
 \;.
\end{equation}
The solution of this can be written
\begin{equation}
\label{eq:Vexponential}
\cU(\mu_\scS,\mu_\scH)
=\mathbb{T} \exp\!\left(
\int_{\mu_\scS}^{\mu_\scH}\!\frac{d\mu}{\mu}\,\cS(\mu)
\right)
\;,
\end{equation}
where $\mathbb{T}$ denotes ordering according to the scale, with smaller $\mu$ to the left. The operator $\cS(\mu)$ is the shower splitting operator. It is typically evaluated at first order in perturbation theory, so that it is proportional to $\as$. At first order, it either creates a new parton or leaves the number of partons unchanged.

The description presented in the previous three paragraphs can, we think, describe the common structure of many forms of what is meant by a parton shower.

Now we offer a caveat. We define as a parton shower the solution, in principle, of the evolution equation (\ref{eq:showerevolution}). We distinguish this from a parton shower event generator, which requires that the evolution equation be implemented in computer code. This implementation may require approximations, as discussed in Sec.~\ref{sec:toeventgenerator} and later in Sec.~\ref{sec:conclusions}.

There is another caveat. At the end of the shower, one can apply an operator that measures an infrared safe observable $J$ with scale $\mu_\scJ$. Here infrared safety means that splittings below this scale do not affect the result beyond contributions suppressed by a power of the yet smaller scale divided by $\mu_\scJ$. An infrared safe observable is calculable in QCD with corrections for hadronization neglected. When $\mu_\scJ \ll \mu_\scH$, the value of the observable is typically affected by large logarithms, $\log(\mu_\scH/\mu_\scJ)$. One may wish to sum the contributions to the observable that contain the most powers of the logarithm for each power of $\as$. For analytical purposes, one typically considers a quantity $\cI_\scJ$ equal to the logarithm of an appropriate integral transform of the observable. Then one asks how accurately $\cI_\scJ$ is approximated. The accuracy of the summation is typically designated as leading logarithmic accuracy (LL), next-to-leading logarithmic accuracy (NLL), {\em etc}, This style of analysis is applied to particular sorts of infrared safe observables, typically those called {\em global event shapes}. 

One way to sum large logarithms like these is by simulating events using a parton shower of the form of Eqs.~(\ref{eq:showercU}), (\ref{eq:showerevolution}), and (\ref{eq:Vexponential}). Since the measurement of global event shapes provides a good test of QCD, the analysis of logarithmic accuracy is an important research problem. But it is not the subject of this paper.\footnote{We have, however, provided an analyses of the summation of logarithms in Refs.~\cite{NSThreshold, NSThresholdII, NSThrustSum}.} This paper is concerned with the application of a parton shower evolution equation to provide results for {\em any} infrared safe observable. For this purpose, we impose strict requirements on the nature of the statistical space and on the properties of the spitting operator $\cS(\mu)$. These strict requirements are not met by many formulations of Eqs.~(\ref{eq:showercU}), (\ref{eq:showerevolution}), and (\ref{eq:Vexponential}) that are useful for particular observables.

The variety of parton shower discussed in this paper may perhaps be called a {\em quantum parton shower}, borrowing from the title of Ref.~\cite{NSI}. For such a parton shower, we treat the operator $\cS(\mu)$ as having a perturbative expansion in powers of $\as$:
\begin{equation}
\label{eq:cSexpansion}
\cS(\mu) = \frac{\as}{2\pi}\,\cS^{(1)}(\mu) 
+ \left[\frac{\as}{2\pi}\right]^2\!\cS^{(2)}(\mu) + \cdots
\;.
\end{equation}
Then we examine the effect of using QCD perturbation theory applied to any starting state $\isket{\rho(\mu_\scH)}$ with high momentum partons whose momenta are not close to being collinear. After some number of splittings and virtual exchanges, we find infrared singularities, both in the form of factors $1/\epsilon$ from integrations over virtual loop momenta in $4 - 2\epsilon$ dimensions and in the form of singularities when subsets of produced partons become collinear with one another or soft ($p_i \to 0$). If the exponent in $\cU(\mu_\scS,\mu_\scH)$ is truncated at order $\as^N$, then we require that the small $\mu_\scS$ behavior of $\cU(\mu_\scS,\mu_\scH) \isket{\rho(\mu_\scH)}$ reflect these infrared singularities of QCD perturbation theory up to order $\as^N$. We specify this requirement more fully later in this paper, especially in Sec.~\ref{sec:IRsingularities}.

In order to make this work, we need to include color and spin as quantum numbers describing the partons. Furthermore, we need an amplitude (the ket amplitude) with $m$ partons in the final state and a conjugate amplitude (the bra amplitude) with $m$ partons in the final state. Both of these are given by a sum of Feynman diagrams. This means that we need basis states $\isket{\{p,f,c,c',s,s'\}_m}$ for the statistical space, with $\{c,s\}_m$ being the colors and spins for the ket amplitude and  $\{c',s'\}_m$ being the colors and spins for the bra amplitude. The splitting operator $\cS(\mu_\scS)$ is then an operator on the statistical space including colors and spins. The mathematical formulation of a quantum parton shower thus uses the language of quantum statistical mechanics, with $\isket{\{p,f,c,c',s,s'\}_m}$ representing basis states for the density matrix in color and spin space.

The approach to a parton shower using the quantum density matrix in  colors and spins, or just colors, is a topic of important recent research \cite{AngelesMartinez:2018cfz, Forshaw:2019ver, Forshaw:2020wrq, Forshaw:2020wrq, DeAngelis:2020rvq, Holguin:2020joq, Hoche:2020pxj, Platzer:2020lbr, Platzer:2022jny}. To emphasize that both the ket amplitude in color and spin space and the bra amplitude evolve, this approach is sometimes called amplitude evolution. There has also been extensive work using other methods aimed at adding order $\as^2$ corrections to the splitting functions in a parton shower \cite{JadachJHEP, JadachPolonica, HartgringNLO, SkandsNLO, HocheNLO1, HocheNLO2, HocheNLO3}.

%---------------------
\section{From the evolution equation to an event generator}
\label{sec:toeventgenerator}

Our aim in this paper is to define an operator based formalism that yields evolution equations for a parton shower in electron-positron annihilation. The splitting functions for this shower evolution have contributions proportional to $\as^1$, $\as^2$, or, in principle, any order $\as^n$. There is an approximation inherent in using a parton shower description: the scale $\mu$ of a shower splitting should be substantially smaller than the scale of the preceding splitting. This approximation can be improved by using more terms in the perturbative expansion of the splitting functions. Aside from this approximation, the evolution equation is exact: it includes full quantum color; it includes full quantum spin; it includes interference between the bra and ket amplitudes.

As noted in the Introduction, the evolution equation (\ref{eq:showerevolution}) for a quantum parton shower as analyzed in this paper has the same general form as the evolution equation for a parton shower event generator. However, it is typically not possible to implement the exact evolution equation as a parton shower event generator. Having the exact evolution equation is analogous to having the exact Schr\"odinger equation for the energy levels of a molecule. We need this starting point, but it does not produce numerical results without some approximations. Similarly, one will need approximations to produce a practical Monte Carlo event generator from the exact quantum shower evolution equations.

An example of the barriers to implementing Eq.~(\ref{eq:showerevolution}) for a quantum parton shower as a parton shower event generator concerns color. The most straightforward implementation of Eq.~(\ref{eq:showerevolution}) without color would involve the exponential of an operator that does not change the number of partons or their momenta or flavors and represents the inclusive probability for the partonic state to generate a splitting between two scales. With color, the operator in the exponential acts as a matrix on the color space. However, color space for many partons is too large to allow a numerical calculation of this exponential.\footnote{For example, the color space for twenty partons has something like $(20!)^2 \approx 10^{37}$ dimensions.}

Thus approximations are needed. The parton shower event generator can then be the solution of the evolution equation (\ref{eq:showerevolution}) using an approximate splitting operator $\cS_\mathrm{approx}$. One can then check how good the approximation is by calculating a cross section of interest using $\cS_\mathrm{approx}$ together with a finite number of contributions from $\Delta\cS = \cS - \cS_\mathrm{approx}$. The contributions proportional to $[\Delta\cS]^n$ should then be reasonably small. An example of this approach is available in Ref.~\cite{NSNewColor}.

In this paper, we define a formalism that can generate a parton shower evolution equation that fully accounts for the requirements of quantum mechanics. We also apply this to define first order splitting functions $\cS^{(1)}(\mu_\scS)$ in this formalism. We address the issues in implementing the resulting shower evolution approximately as a parton shower event generator only briefly, in Sec.~\ref{sec:conclusions}.

%----------------------------------------------------
\section{Singularities of QCD perturbation theory}
\label{sec:singularities}

The quantum parton shower formalism presented in this paper is based on the infrared (IR) singularities of QCD as seen in electron-positron annihilation. We consider $e^+e^- \to hadrons$. The analysis of the production of electroweak bosons or other particles that do not carry color is similar, but we do not address the changes needed to go beyond $e^+e^- \to hadrons$ in this paper. In addition, we treat all quarks as massless in order to keep the theory as simple as possible.

Consider a perturbative contribution to the QCD ket amplitude times the bra amplitude for the cross section to produce $m$ partons in electron-positron annihilation. These amplitudes include self-energy diagrams on final state parton lines with coefficients determined from the perturbative expansion of the Lehmann-Symanzik-Zimmermann (LSZ) factors on the final state parton lines described in Sec.~\ref{sec:LSZ}. 

Let $Q$ denote the total $e^+e^-$ momentum. The parton momenta are $\hat p_i$, $i \in \{1,\dots, \hat m\}$. For the Feynman diagrams considered, there can be poles $1/\epsilon$ in $d = 4-2\epsilon$ dimensions coming from virtual loop integrations and there can be singularities when some of the partons become collinear to one another and some partons become soft in the sense that their momenta approach zero. For the collinear singularities, there will be some number of directions in momentum space, labeled $j$, such that, for each of these directions, a subset $J(j)$ of the momenta become collinear with their total momentum,
\begin{equation}
\label{eq:momentumsum}
p_j = \sum_{i \in J(j)} \hat p_i
\;.
\end{equation}
That is, in the collinear limit, $\hat p_i = x_i p_j$ for $i \in J(j)$. There is also a set $S$ of soft final state partons, whose momenta approach zero in the limit. That is, in the soft limit, $\hat p_i = 0$ for $i \in S$. Thus in the singular limit, there are $m \le \hat m$ infinitely narrow jets with momenta $p_j$, $j \in \{1,\dots, m\}$.. 
 
When we speak of a singularity of a Feynman graph for a ket amplitude times a bra amplitude, it is important to note how strong a singularity is. We always have in mind integrating over a region near the singularity with the covariant measure given in Eq.~(\ref{eq:dp}) below and using integrations $d^d \ell$ for the virtual loops. We are interested in singularities that are strong enough to produce a logarithmically divergent integration if we integrate in $d = 4$ dimensions or a pole $1/\epsilon^n$ if we integrate in $4 - 2 \epsilon$ dimensions. We call this a {\em logarithmic singularity}. Weaker singularities are not significant in our analysis.

The relation of these singularities to the Feynman graphs is rather simple to understand if we work in a suitable gauge, such as the interpolating gauge of Doust \cite{Doust} and of Baulieu and Zwanziger \cite{BaulieuZwanziger}.  In Ref.~\cite{Gauge}, we have explored the properties of this gauge that make it useful for the description of parton showers. In interpolating gauge, gluons exchanged between partons with momenta in different directions generate only soft divergences. Furthermore, the interference between emitting a gluon from one parton in the ket amplitude and another parton in the bra amplitude generates only a soft divergence.

One can use Feynman gauge for the analysis of the infrared singularities, but then one must apply Ward identities to bring the singularity structure into its simplest form. This is a significant complication, so we use interpolating gauge in this paper. 

The gluon propagator in interpolating gauge is \cite{Gauge}:
\begin{equation}
\label{eq:gluonpropagator}
D^{\mu\nu}(q) = \frac{N^{\mu\nu}(q)}{q^2+\mi 0}
\;,
\end{equation}
where
\begin{equation}
  \label{eq:gluonnumerator}
  \begin{split}
    &N^{\mu\nu}(q) = - g^{\mu\nu} 
    + \frac{q^\mu\,\tilde{q}^\nu + \tilde{q}^\mu\,q^\nu}{q\cdot\tilde{q} 
    + \mi 0}
    \\
    & \quad
    - \left(1+\frac{1}{v^2}\right)\frac{q^\mu\,q^\nu}
    {q\cdot\tilde{q} + \mi 0}
    -\frac{\xi-1}{v^2}\,\frac{q^2\,q^\mu q^\nu}{(q\cdot \tilde q + \mi 0)^2}
    \;.
  \end{split}
\end{equation}
In $D^{\mu\nu}(q)$, $v$ is a parameter with $1 < v < \infty$. For instance, one might choose $v = 2$. There is another parameter $\xi$, which we usually choose to be $\xi = 1$. There is also a vector $n$, which we set to $n = Q/\sqrt{Q^2}$. The vector $\tilde q$ is related to $q$ in a frame in which $n = (1,\vec 0\,)$ by
\begin{equation}
\begin{split}
\tilde q^0 ={}& \frac{1}{v^2}\,q^0
\;,
\\
\tilde q^i  ={}& q^i\;,\hskip 1 cm i \in \{1,2,3\}
\;.
\end{split}
\end{equation}

The gauge propagator has two contributions \cite{Gauge},
\begin{equation}
\label{eq:TLdecomposition}
D^{\mu\nu}(q) = D^{\mu\nu}_\LT(q) + D^{\mu\nu}_\LL(q)
\;.
\end{equation}
The propagator for $T$ gluons is
\begin{equation}
\label{eq:DTexpansion}
D^{\mu\nu}_{\LT}(q) =  \frac{1}{q^2+\mi 0}
\sum_{s = \pm 1} \varepsilon^\mu(q,s)^*\,\varepsilon^\nu(q,s)
\;,
\end{equation}
where the polarization vectors are solutions of $q\cdot \varepsilon(q,s) = n \cdot \varepsilon(q,s) = 0$ with $\varepsilon^\star(q,s)\cdot \varepsilon(q,s') = - \delta_{s s'}$. Physical cross sections are constructed from the S matrix for quarks, antiquarks, and T gluons.

The propagator for L gluons is constructed from polarization vectors in the plane of $n$ and $q$. In a frame in which $n = (1,\vec 0\,)$, $D^{\mu\nu}_{\LL}(q)$ is
\begin{equation}
\begin{split}
\label{eq:DLexpansion}
D^{\mu\nu}_{\LL}(q) = {}& \frac{1}{q\cdot \tilde q +\mi 0}
\bigg\{
\frac{(q^\mu - q\cdot n\, n^\mu)(q^\nu - q\cdot n\, n^\nu)}
{v^2 \, \vec q^{\,2}}
\\& \quad
- n^\mu n^\nu
  -\frac{\xi-1}{v^2}\,\frac{q^\mu q^\nu}{q\cdot \tilde q + \mi 0}
\bigg\}
\;.
\end{split}
\end{equation}
Note that $D^{\mu\nu}_{\LL}(q)$ does not have a pole at $q^2 = 0$. Thus $D^{\mu\nu}_{\LL}(q)$ is not singular when $q$ is collinear with a lightlike momentum.

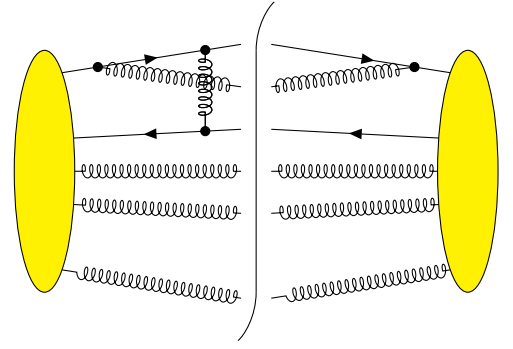
\begin{figure}
  \centering
  \begin{prdfig}{78e29c01223459ebb6c699866cffdb77}{singularity-example}
    \def\finalstatecut {
      % integral sign
      \draw (0.3,3.5) .. controls (0.2, 3.4) and (0,3) .. (0,2.5);
      \draw (0.0,2.5) -- (0,-2.5);
      \draw (-0.3,-3.5) .. controls (-0.2, -3.4) and (0,-3) .. (0,-2.5);
    }  
    
    \def\aampl#1 {
      \begin{scope}[#1]
        \draw [fill=yellow, name path=blob](0,0) ellipse (0.5 and 2); 
        
        % guidelines
        \coordinate(o) at(-10,0);
        \coordinate(o1) at(5,0);
        \coordinate(o2) at(0,0);
        
        % projection guidelines	
        \path [gray!30!white, name path=p] (3.25,3)--(3.25,-3);

        % guidelines of the outgoing particles
        \coordinate[rotate around={9:(o)}](a0) at(0,0);
        \coordinate[rotate around={3:(o)}](b0) at(0,0);
        \coordinate[rotate around={0:(o)}](c0) at(0,0);
        \coordinate[rotate around={-3:(o)}](d0) at(0,0);
        \coordinate[rotate around={-9:(o)}](e0) at(0,0);
        
        \coordinate[rotate around={9:(o)}](a00) at(o1);
        \coordinate[rotate around={3:(o)}](b00) at(o1);
        \coordinate[rotate around={0:(o)}](c00) at(o1);
        \coordinate[rotate around={-3:(o)}](d00) at(o1);
        \coordinate[rotate around={-9:(o)}](e00) at(o1);

        \path [gray!30!white, name path=g1] (a0)--(a00);
        \path [gray!30!white, name path=g2] (b0)--(b00);
        \path [gray!30!white, name path=g3] (c0)--(c00);
        \path [gray!30!white, name path=g4] (d0)--(d00);
        \path [gray!30!white, name path=g5] (e0)--(e00);

        % quark line
        \path[name intersections={of=g1 and blob, by={bi}}];
        \path[name intersections={of=g1 and p, by={b1i}}];

        \coordinate(a) at(bi);
        \coordinate(a1) at(b1i);
        
        % anti-quark line
        \path[name intersections={of=g2 and blob, by={bi}}];
        \path[name intersections={of=g2 and p, by={b1i}}];

        \coordinate(b) at(bi);
        \coordinate(b1) at(b1i);

        % real gluon emmision
        \coordinate [dot](r) at ($(a)!0.2!(a1)$){};
        \coordinate [](r1) at ($(a1)!0.5!(b1)$);

        % three other hard gluons   
        \path[name intersections={of=g3 and blob, by={bi}}];
        \path[name intersections={of=g3 and p, by={b1i}}];

        \coordinate(c) at(bi);
        \coordinate(c1) at(b1i);
        
        \path[name intersections={of=g4 and blob, by={bi}}];
        \path[name intersections={of=g4 and p, by={b1i}}];

        \coordinate(d) at(bi);
        \coordinate(d1) at(b1i);
        
        \path[name intersections={of=g5 and blob, by={bi}}];
        \path[name intersections={of=g5 and p, by={b1i}}];

        \coordinate(e) at(bi);
        \coordinate(e1) at(b1i);
      \end{scope}
    }

    \begin{tikzpicture}[scale=0.8, transform shape]
      \begin{feynman}[]
        %%%%%%%%%%%%%%%%%%%%%%%%%%%%%%%%%% 
        % Left hand side of the graph
        %%%%%%%%%%%%%%%%%%%%%%%%%%%%%%%%%% 
        \aampl{}
        % loop gluon	
        \coordinate[dot](l) at ($(a)!0.8!(a1)$){};	
        \coordinate(ll) at($(l) - (0,5)$);
        \coordinate[dot](l1) at(intersection of l--ll and b--b1);
        
        \draw [fermion] (a) -- (a1);
        \draw [anti fermion] (b) -- (b1);
        \draw [gluon] (l) -- (l1);
        \draw [gluon]  (r) -- (r1);
        \draw [gluon]  (c1) -- (c);
        \draw [gluon]  (d1) -- (d);
        \draw [gluon]  (e1) -- (e);

        %%%%%%%%%%%%%%%%%%%%%%%%%%%%%%%%%% 
        % Right hand side of the graph
        %%%%%%%%%%%%%%%%%%%%%%%%%%%%%%%%%% 
        \aampl{xscale=-1, xshift=-7cm}
        \draw [anti fermion] (a) -- (a1);
        \draw [fermion] (b) -- (b1);
        \draw [gluon] (l) -- (l1);
        \draw [gluon]  (r1) -- (r);
        \draw [gluon]  (c) -- (c1);
        \draw [gluon]  (d) -- (d1);
        \draw [gluon]  (e) -- (e1);	
        
        % integral sign
        \begin{scope}[yscale = 0.8, xshift=3.5cm]
          \finalstatecut
        \end{scope}
      \end{feynman}
    \end{tikzpicture}
  \end{prdfig}
  \caption{Example of singularities in QCD.
    \label{fig:d11exampleA}}
\end{figure}

As argued above, in interpolating gauge, the infrared singularities arise from graphs in which the $m$ infinitely narrow jets arise from $\hat m$ on-shell partons with momenta $p_i$ that emerge from a hard density-matrix element. This is illustrated in Fig.~\ref{fig:d11exampleA}, in which two of the $\hat m$ partons are grouped into one infinitely narrow jet and there is a $1/\epsilon$ factor that arises from the exchange of an infinitely soft gluon. 

To start with, the loop momentum $\ell$ of a virtual gluon that is part of a soft exchange in the ket amplitude flows through the hard subgraph for the amplitude or the conjugate amplitude. It is the integration over $\ell$ that produces a pole $1/\epsilon$.  Because $\ell$ is soft, we set $\ell \to 0$ inside the hard subgraph. If we expand the hard subgraph in powers of $\ell$, contributions with any power greater than zero do not produce a $1/\epsilon$ pole. 

The logarithmic infrared singularities have a simple flavor structure. The final state partons have flavors $\hat f_i$. A parton with $i\in S$ not only has $\hat p_i = 0$ in the singular limit but also must be a gluon, so that emitting this parton does not change the flavor of the emitting parton. Thus the flavor of jet $j$ is
\begin{equation}
\label{eq:flavorsum}
f_j = \sum_{i \in J(j)} \hat f_i
\;,
\end{equation}
with summation of flavors defined in a simple way as adding vectors in an $N_\Lf$ dimensional vector space, where $N_\Lf$ is the number of flavors and the gluon is represented by the zero vector. The same result is obtained using the Feynman rules for the diagram for the ket amplitude or the diagram for the bra amplitude.

%%%%%%%%%%%%%%%%%%%% FIGURE %%%%%%%%%%%%%%%%%%%%%%%%%%
%
% 
%%%%%%%%%%%%%%%%%%% END FIGURE %%%%%%%%%%%%%%%%%%%%%%%%

Thus the infrared singularities are represented by graphs like Fig.~\ref{fig:d11exampleA}, in which the yellow blobs represent hard subgraphs that produce $m$ on-shell partons with momenta $p_j$ and flavors $f_j$. In this example, $m = \hat m - 1$.

In each graph that exhibits an infrared singularity, there are $m$ partons emerging from the hard density matrix. In the singular limit, they have momenta $p_j$, with $p_j^2 = 0$, and flavors $f_j$. Notice that $p_j$ and $f_j$ are the same in the hard scattering ket amplitude and in the hard scattering bra amplitude since they are determined by Eqs.~(\ref{eq:momentumsum}) and (\ref{eq:flavorsum}).  The on-shell partons in the hard scattering ket amplitude have colors $\{c\}_m$ and spins $\{s\}_m$.  The on-shell partons in the hard scattering bra amplitude have colors $\{c'\}_m$ and spins $\{s'\}_m$. Thus the description of the infrared singularities requires colors $\{c,c'\}_m$, spins $\{s,s'\}_m$, momenta $\{p\}_m$ and flavors $\{f\}_m$. Separate momenta and flavors for the bra and ket amplitudes are not required.

%---------------------
\section{Infrared singularities in the statistical space}
\label{sec:IRsingularities}

In this section, we review the general formulation of a quantum parton shower given in Ref.~\cite{NSAllOrder} as it applies to electron-positron annihilation. In this formulation, the soft and collinear singularities of QCD are represented by an operator $\cD(\mu_\scS)$ that has a perturbative expansion in powers of $\as$. Then the parton shower evolution operator $\cU(\mu_\scS,\mu_\scH)$ is derived from $\cD(\mu_\scS)$ using Eq.~(\ref{eq:cUdef}) below. The splitting operator $\cS(\mu)$ is obtained from the derivative of $\cU(\mu_\scS,\mu_\scH)$ according to Eq.~(\ref{eq:showerevolution}).

The operator $\cD(\mu_\scS)$ acts on a linear vector space, the statistical space. We turn first to a description of this space.

%-----------------------------
\subsection{Statistical space}
\label{sec:statisticalspace}

As stated in Sec.~\ref{sec:introduction}, we view a parton shower using the language of quantum statistical mechanics. At any stage of the shower, specified by a scale parameter $\mu_\scS$, the state of the partons is described by a vector $\isket{\rho(\mu_\scS)}$ in a vector space that we refer to as the {\em statistical space}. This vector tells how many partons there are and what their quantum numbers are. In the earliest versions of parton showers \cite{EarlyPythia, EarlyHerwig, EarlyIsajet}, the quantum numbers consisted of just the momenta and flavors of the partons, specified by a list $\{p,f\}_m = \{p_1,f_1; p_2,f_2; \dots p_m,f_m\}$. Then $\isket{\rho(\mu_\scS)}$ was a linear combination of basis states $\isket{\{p,f\}_m}$. With just momenta and flavors, $\isbrax{\{p,f\}_m}\isket{\rho(\mu_\scS)}$ represents the probability that there are $m$ partons with momenta and flavors $\{p,f\}_m$. As we have seen in the previous section, we need separate colors and spins for the bra and ket amplitudes. Our description of colors and spins follows Ref.~\cite{NSI}.

Each gluon has a spin index $s$, which takes $2 - 2\epsilon$ values when we work in a $d = 4 - 2\epsilon$ dimensional space-time. These values correspond to the $2 - 2\epsilon$ choices for the polarization vector $\varepsilon(p,s)$ with $p\cdot\varepsilon(p,s) = n\cdot\varepsilon(p,s) = 0$, with the normalization $-\varepsilon(p,s)\cdot \varepsilon(p,s')^* = \delta_{s,s'}$.\footnote{When $\epsilon = 0$ we use helicities.} Similarly, each quark or antiquark has a spin index s that takes two values. Then we use spin basis states $\ket{\{s\}_m} = \ket{\{s_1,s_2,\dots,s_m\}}$ labeled by the spin indices. We take the spin basis states to be orthogonal and normalized:
\begin{equation}
\brax{\{s'\}_m}\ket{\{s\}_m} = \delta_{\{s\}_m}^{\{s'\}_m}
\;.
\end{equation}

Each gluon has a color index $c$, which takes $N_\Lc^2-1$ values. Here $N_\Lc$ is the number of colors, which we take to be possibly different from 3. Each quark or antiquark has a color index $c$, which takes $N_\Lc$ values. Instead of specifying the colors for the individual partons, it is useful to specify the color state of all of the $m$ partons together, restricting the joint state to be invariant under rotations in the color space. That is, the partons together are in a color singlet state. Then we use color basis states for $m$ partons that we designate as $\ket{\{c\}_m}$. There are many possible choices for color basis states. We prefer the trace basis \cite{tracebasis}, as specified in Sec.~7 of Ref.~\cite{NSI}. Other investigators prefer different choices of basis \cite{colorflowbasis, orthogonalbasis}. For the purposes of this paper, we do not need to choose a color basis as long as each state $\ket{\{c\}_m}$ is a color singlet.

We allow for the possibility that the color basis states $\iket{\{c\}_m}$ are not orthogonal and normalized, as is the case for the trace basis. Then we also define a dual basis $\iket{\{c\}_m}_\scD$ such that 
\begin{equation}
\dualL\brax{\{c'\}_m}\ket{\{c\}_m} = \delta_{\{c\}_m}^{\{c'\}_m}
\;,
\end{equation}
as discussed later at Eq.~(\ref{eq:colordualbasis}). For spin, we always choose an orthogonal and normalized basis, so that $\iket{\{s\}_m}_\scD = \iket{\{s\}_m}$.

To include color and spin, we use the language of quantum statistical mechanics. In this description, a pure state in the color and spin space is a quantum vector
\begin{equation}
\ket{\Psi} = \sum_{\{c,s\}_m} \psi(\{c,s\}_m)\ket{\{c,s\}_m}
\;.
\end{equation}
The associated probability for this state is
\begin{equation}
\begin{split}
\brax{\Psi}\ket{\Psi} ={}& \sum_{\{c,c',s,s'\}_m}
\psi^*(\{c',s'\}_m)\, \psi(\{c,s\}_m)
\\&\times
\brax{\{c'\}_m}\ket{\{c\}_m}
\brax{\{s'\}_m}\ket{\{s\}_m}
\;.
\end{split}
\end{equation}
A statistical mixture of pure states is described by a mixed state
\begin{equation}
\begin{split}
\label{eq:rhocolorspin}
\sket{\rho} ={}& 
\!\!\!\sum_{\{c,c',s,s'\}_m}\!\!\!
\rho(\{c,c',s,s'\}_m)\ket{\{c,s\}_m}
\bra{\{c',s'\}_m}
.
\end{split}
\end{equation}
The function $\rho$ here gives the matrix elements of the density matrix in spin and color. The probability for this mixed state is
\begin{equation}
\begin{split}
P ={}& \!\!
\sum_{\{c,c',s,s'\}_m}\!\!
\rho(\{c,c',s,s'\}_m)
\brax{\{c',s'\}_m}\ket{\{c,s\}_m}
\;.
\end{split}
\end{equation}

We define basis vectors $\sket{\{c,c',s,s'\}_m}$ in the color and spin part of the statistical space so that  $\sket{\rho}$  in Eq.~(\ref{eq:rhocolorspin}) is written as
\begin{equation}
\begin{split}
\label{eq:rhocolorspinII}
\sket{\rho} ={}& 
\!\!\!\sum_{\{c,c',s,s'\}_m}\!\!\!
\rho(\{c,c',s,s'\}_m)
\sket{\{c,c',s,s'\}_m}
\;.
\end{split}
\end{equation}

Including color and spin, a set of $m$ partons is produced starting with a set of fewer hard partons according to a Feynman graph for the ket amplitude. These partons have a color and spin state $\iket{\{c,s\}_{m}}$. This set of partons, with the same momenta and flavors $\{p,f\}_{m}$, is produced according to a possibly different Feynman graph for the bra amplitude, starting from $m$ partons in a possibly different color and spin state. These partons have a color and spin state $\ibra{\{c',s'\}_{m}}$. 

With this description, the partonic state $\isket{\rho(\mu_\scS)}$ is expanded in basis vectors according to
\begin{equation}
  \begin{split}
    \label{eq:stateexpansion}
    \sket{\rho(\mu_\scS)} =
    \sum_{m}
    \frac{1}{{m}!}
    \int{}& d\{p,f,c,c',s,s'\}_{m} 
    \\
    &\times
    \sket{\{p,f,c,c',s,s'\}_{m}}
    \\
    &\times
    \sbrax{\{p, f,c,c',s,s'\}_{m}}
    \sket{\rho(\mu_\scS)}
    \;.
  \end{split}
\end{equation}
A factor $1/{m}!$ is included because the partons are treated as identical. The integration measure is
\begin{equation}
\begin{split}
\label{eq:dpfccss}
\int\!d\{p,f,c,c',s,s'\}_{m} 
={}&  
\int\!d\{p\}_{m}
\sum_{\{f,c,c',s,s'\}_{m}}
\;,
\end{split}
\end{equation}
where
\begin{equation}
  \label{eq:dp}
  \begin{split}
    \int\!d\{p\}_{m} ={}& 
    \prod_{i=1}^{m}
    \left\{
      \int\!\frac{d^{4-2\epsilon}p_i}{(2\pi)^{4-2\epsilon}}\,
      (2\pi)\,\delta_+(p_i^2)\right\}
    \\&\times
    (2\pi)^{4-2\epsilon}\delta\!\left(\sum_i p_i - Q\right)
    \;.
  \end{split}
\end{equation}

The completeness relation for the basis states is
\begin{equation}
\begin{split}
\label{eq:completeness}
1 ={}& 
\sum_{m}
\frac{1}{{m}!}
\int\!d\{p,f,c,c',s,s'\}_{m}
\\&\times
\sket{\{p,f,c,c',s,s'\}_{m}}
\sbra{\{p,f,c,c',s,s'\}_{m}}
\;.
\end{split}
\end{equation}

At the end of the perturbative shower at scale $\mu_\Lf$, we have a state $\isket{\rho(\mu_\Lf)}$. This state is expanded as a linear combination of basis states as in Eq.~(\ref{eq:stateexpansion}). Assuming that we do not need to apply a hadronization model, we next apply an operator $\cO_\LJ$ that measures an observable on this state. That is
\begin{equation}
\begin{split}
\cO_\LJ &\sket{\{p,f,c,c',s,s'\}_{m}}
\\&
= J(\{p,f,c,c',s,s'\}_{m})
\sket{\{p,f,c,c',s,s'\}_{m}}
\;,
\end{split}
\end{equation}
where the eigenvalue $J$ is a numerical function that gives the value of the observable for partons with the quantum numbers specified. Typically, $J$ is a function only of the momenta of the partons, but we can allow for a more general dependence.\footnote{The partons described in the basis states $\{p,f,c,c',s,s'\}_{m}$ carry labels $i = 1,\dots,m$. However, we take the $m$ partons to be physically identical, distinguished only by their different quantum numbers. Thus for any allowed observable $\cO_\LJ$, we have $J(\{\tilde p, \tilde f, \tilde c, \tilde c', \tilde s, \tilde s'\}_{m}) = J(\{p,f,c,c',s,s'\}_{m})$ when $\tilde p_i = p_{\pi(i)}$ and likewise for the other quantum numbers for any permutation $\pi \in S_{m}$.}

Finally, we multiply by a statistical bra state $\isbra{1}$ defined by
\begin{equation}
  \label{eq:bra1}
  \begin{split}
    \sbrax{1}&\sket{\{p,f,c,c',s,s'\}_{m}}
    \\
    &\quad
    = 
    \brax{\{c'\}_{m}}\ket{\{c\}_{m}}
    \brax{\{s'\}_{m}}\ket{\{s\}_{m}}
    \,.
  \end{split}
\end{equation}
The quantum inner products give the probability for color and spin parts of the statistical basis states. This gives us
\begin{equation}
  \begin{split}
    \sbra{1}\cO_\LJ \sket{\{p,f,c,c',s,s'\}_{m}} 
    \hskip - 2.2 cm &
    \\
    ={}& 
    J(\{p,f,c,c',s,s'\}_{m})
    \\
    &\times\brax{\{c'\}_{m}}\ket{\{c\}_{m}}
    \brax{\{s'\}_{m}}\ket{\{s\}_{m}}
    \,.
  \end{split}
\end{equation}
% 

%------------------------------
\subsection{Infrared singular operator}
\label{sec:cD}

The representation of singularities of QCD in Fig.~\ref{fig:d11exampleA} is instructive, but not yet adapted for constructing a parton shower. The problem is that it depicts what happens only exactly at an infrared singularity. For instance, if $\hat p_k$ and $\hat p_l$ are the momenta of two final state partons, we can represent the singularity when $\hat p_k = \lambda_k p_j$ and $\hat p_l = (1-\lambda_k) p_j$, but it does not help away from this limit. What we need is an approximation that is built from the singular subdiagram like that in the example in Fig.~\ref{fig:d11exampleA} but that multiplies the singular factor by the hard subdiagram evaluated exactly at the singularity. This approximate version is illustrated in Fig.~\ref{fig:d11exampleB}.

%%%%%%%%%%%%%%%%%%%% FIGURE %%%%%%%%%%%%%%%%%%%%%%%%%%
% 
\begin{figure}
  \centering
  \begin{prdfig}{9251d71c68692d90875fbaaaa278d4a8}{singularity-example-factorized}
    \def\finalstatecut {
      % integral sign
      \draw (0.3,3.5) .. controls (0.2, 3.4) and (0,3) .. (0,2.5);
      \draw (0.0,2.5) -- (0,-2.5);
      \draw (-0.3,-3.5) .. controls (-0.2, -3.4) and (0,-3) .. (0,-2.5);
    }  

    \def\aampl#1 {
      \begin{scope}[#1]
        \draw [fill=yellow, name path=blob](0,0) ellipse (0.5 and 2); 
        
        % guidelines
        \coordinate(o) at(-10,0);
        \coordinate(o1) at(5,0);
        \coordinate(o2) at(0,0);
        
        % projection guidelines
        \coordinate(p0) at($(3.5,3)-(2.2,0)$);
        \coordinate(p1) at($(3.5,-3)-(2.2,0)$);
        \coordinate(p2) at($(3.5,3)-(2.45,0)$);
        \coordinate(p3) at($(3.5,-3)-(2.45,0)$);
        
        \path [red!30!white, name path=pr,draw] (p0)--(p1);
        \path [red!30!white, name path=pr1] (p2)--(p3);
        \path [red!30!white, name path=p] (3.25,3)--(3.25,-3);

        % guidelines of the outgoing particles
        \coordinate[rotate around={9:(o)}](a0) at(0,0);
        \coordinate[rotate around={3:(o)}](b0) at(0,0);
        \coordinate[rotate around={0:(o)}](c0) at(0,0);
        \coordinate[rotate around={-3:(o)}](d0) at(0,0);
        \coordinate[rotate around={-9:(o)}](e0) at(0,0);
        
        \coordinate[rotate around={9:(o)}](a00) at(o1);
        \coordinate[rotate around={3:(o)}](b00) at(o1);
        \coordinate[rotate around={0:(o)}](c00) at(o1);
        \coordinate[rotate around={-3:(o)}](d00) at(o1);
        \coordinate[rotate around={-9:(o)}](e00) at(o1);

        \path [gray!30!white, name path=g1] (a0)--(a00);
        \path [gray!30!white, name path=g2] (b0)--(b00);
        \path [gray!30!white, name path=g3] (c0)--(c00);
        \path [gray!30!white, name path=g4] (d0)--(d00);
        \path [gray!30!white, name path=g5] (e0)--(e00);

        % quark line
        \path[name intersections={of=g1 and blob, by={bi}}];
        \path[name intersections={of=g1 and p, by={b1i}}];
        \path[name intersections={of=g1 and pr, by={b2i}}];
        \path[name intersections={of=g1 and pr1, by={b3i}}];

        \coordinate [](a) at(bi) ;
        \coordinate [](a1) at(b1i);
        \coordinate [crossed dot](a2) at(b2i){};
        \coordinate [](a3) at(b3i);

        % anti-quark line
        \path[name intersections={of=g2 and blob, by={bi}}];
        \path[name intersections={of=g2 and p, by={b1i}}];
        \path[name intersections={of=g2 and pr, by={b2i}}];
        \path[name intersections={of=g2 and pr1, by={b3i}}];

        \coordinate [](b) at(bi);
        \coordinate [](b1) at(b1i);
        \coordinate [crossed dot](b2) at(b2i){};
        \coordinate [](b3) at(b3i);

        % real gluon emmision
        \coordinate [dot](r) at ($(a2)!0.2!(a1)$){};
        \coordinate [](r1) at ($(a1)!0.5!(b1)$);

        % three other hard gluons   
        \path[name intersections={of=g3 and blob, by={bi}}];
        \path[name intersections={of=g3 and p, by={b1i}}];
        \path[name intersections={of=g3 and pr, by={b2i}}];
        \path[name intersections={of=g3 and pr1, by={b3i}}];

        \coordinate [](c) at(bi);
        \coordinate [](c1) at(b1i);
        \coordinate [crossed dot](c2) at(b2i);
        \coordinate [](c3) at(b3i);

        \path[name intersections={of=g4 and blob, by={bi}}];
        \path[name intersections={of=g4 and p, by={b1i}}];
        \path[name intersections={of=g4 and pr, by={b2i}}];
        \path[name intersections={of=g4 and pr1, by={b3i}}];
        
        \coordinate [](d) at(bi);
        \coordinate [](d1) at(b1i);
        \coordinate [crossed dot](d2) at(b2i);
        \coordinate [](d3) at(b3i);

        \path[name intersections={of=g5 and blob, by={bi}}];
        \path[name intersections={of=g5 and p, by={b1i}}];
        \path[name intersections={of=g5 and pr, by={b2i}}];
        \path[name intersections={of=g5 and pr1, by={b3i}}];

        \coordinate [](e) at(bi);
        \coordinate [](e1) at(b1i);
        \coordinate [crossed dot](e2) at(b2i);
        \coordinate [](e3) at(b3i);

      \end{scope}
    }

    \begin{tikzpicture}[scale=0.9, transform shape]
      \begin{feynman}[]
        %%%%%%%%%%%%%%%%%%%%%%%%%%%%%%%%%%% 
        % Left hand side of the graph
        %%%%%%%%%%%%%%%%%%%%%%%%%%%%%%%%%%% 
        \aampl{}
        
        % loop gluon	
        \coordinate [dot](l) at ($(a)!0.8!(a1)$);	
        \coordinate(ll) at($(l) - (0,5)$);
        \coordinate[dot](l1) at(intersection of l--ll and b--b1);
        
        %	draw the diagram	
        \draw[fermion] (a) -- (a3);
        \draw[fermion] (a2) -- (a1);
        \draw[anti fermion] (b) -- (b3);
        \draw[anti fermion] (b2) -- (b1);
        \draw[gluon] (l) -- (l1);
        
        \draw[gluon] (r) --(r1);
        \draw[gluon] (c3) -- (c);
        \draw[gluon] (c1) -- (c2);
        \draw[gluon] (d3) -- (d);
        \draw[gluon] (d1) -- (d2);
        \draw[gluon] (e3) -- (e);
        \draw[gluon] (e1) -- (e2);

        % left position of the brace
        \coordinate(w0) at(p1);
        
        %%%%%%%%%%%%%%%%%%%%%%%%%%%%%%%%%% 
        % Right hand side of the graph
        %%%%%%%%%%%%%%%%%%%%%%%%%%%%%%%%%% 
        \aampl{xscale=-1, xshift=-7cm}
        
        %	draw the diagram	
        \draw[anti fermion] (a) -- (a3);
        \draw[anti fermion] (a2) -- (a1);
        \draw[fermion](b) -- (b3);
        \draw[ fermion](b2) -- (b1);
        
        \draw[gluon] (r1) --(r);
        \draw[gluon] (c) -- (c3);
        \draw[gluon] (c2) -- (c1);
        \draw[gluon] (d) -- (d3);
        \draw[gluon] (d2) -- (d1);
        \draw[gluon] (e) -- (e3);
        \draw[gluon] (e2) -- (e1);

        % right position of the brace
        \coordinate(w1) at(p1);

        \draw [decoration={brace}, decorate] (w1) -- (w0)
        node [pos=0.5, below] {${\cal D}^{(1,1)}(G)$};
        
        % integral sign
        \begin{scope}[yscale = 0.8, xshift=3.5cm]
          \finalstatecut
        \end{scope}
      \end{feynman}
    \end{tikzpicture}
  \end{prdfig}
  \caption{\label{fig:d11exampleB}
  Example of the contribution of a particular graph $G$ with one real emission and one virtual exchange to the singular operator ${\cal D}^{(1,1)}$. The translation from the graph to a contribution to ${\cal D}$ has several steps. This figure is a representation of Eq.~(\ref{eq:rhosingularity}) later in the paper.}
\end{figure}
% 
%%%%%%%%%%%%%%%%%%%%%%% END FIGURE %%%%%%%%%%%%%%%%%%%%%%%%%%%%%%%%%%%%%%%%%%%%%%%%%%
% 

Let us denote the momentum of parton $j$ as it enters the singular amplitude by $q_j$. The singular diagram conserves momentum exactly, so that $q_j$ is a linear combination of the final state momenta $\hat p_i$ and the momenta of exchanged virtual gluons. When the singular amplitude is evaluated close to, but not exactly at, the singular limit for the $\hat p_i$ and with soft exchanged gluons close to having zero momentum, the resulting momentum $q_j$ is close to but not exactly equal to $p_j$. 

We need one more ingredient. As we integrate over the splitting parameters, we get further from the singular limit. We introduce a scale $\mu_\scS$ that restricts how far the final state after splittings deviates from the final state at the singular point.  For instance, one could say that the transverse momentum in a nearly collinear splitting is restricted to $k_\perp^2 < \mu_\scS^2$. There is evidently some freedom in choosing the definition of the integration restrictions imposed by $\mu_\scS$. We propose what we think is a sensible set of choices in Sec.~\ref{sec:hardness} below. In the formalism of this paper, $\mu_\scS$ does not restrict integrations over virtual loops.

This construction will define a perturbative operator $\cD(\mu_\scS)$ that operates on the statistical space. This operator depends on the shower scale $\mu_\scS$. We construct $\cD(\mu_\scS)$ so that it reproduces the poles and logarithmic singularities of QCD order by order in perturbation theory. This requirement allows some freedom: we have to match singularities, but finite remainders remain unspecified. We arrange the definition so that the inclusive probability $\sbrax{1}\sket{\rho}$ associated with a state $\rho$ is not changed by applying $\cD(\mu_\scS)$:
\begin{equation}
\sbra{1}\cD(\mu_\scS) = \sbra{1}
\;.
\end{equation}

Acting on a statistical basis state $\isket{\{p,f,c,c',s,s'\}_m}$, $\cD(\mu_\scS)$  gives a linear combination of basis states $\isket{\{\hat p,\hat f,\hat c,\hat c',\hat s,\hat s'\}_{\hat m}}$ according to
\begin{equation}
  \begin{aligned}[c]
    \cD(\mu_\scS&)\sket{\{p,f,c,c',s,s'\}_m} 
    \\
    ={}& 
    \sum_{\hat m\ge m} \frac{1}{\hat m!}
    \int\!d\{\hat p,\hat f,\hat c,\hat c',\hat s,\hat s'\}_{\hat m}\,
    \sket{\{\hat p,\hat f,\hat c,\hat c',\hat s,\hat s'\}_{\hat m}}
    \\
    &
    \times 
    \sbra{\{\hat p,\hat f,\hat c,\hat c',\hat s,\hat s'\}_{\hat m}}
    \cD(\mu_\scS)
    \sket{\{p,f,c,c',s,s'\}_m}
    \;.
  \end{aligned}
\end{equation}
The operator $\cD(\mu_\scS)$ has a perturbative expansion
\begin{equation}
\begin{split}
\label{eq:cDexpansion}
\cD(\mu_\scS)
={}& 1 + \sum_{n=1}^\infty \left[\frac{\as(\mu_\scR)}{2\pi}\right]^n
\cD^{(n)}(\mu_\scS,\mu_\scR)
\;.
\end{split}
\end{equation}
Here $\mu_\scR$ is the renormalization scale that is used in the $\MSbar$ renormalization of virtual diagrams. We retain $\mu_\scR$ as a free parameter. Normally, we do not indicate the dependence of $\as(\mu_\scR)$ or the coefficients $\cD^{(n)}(\mu_\scS,\mu_\scR)$ on $\mu_\scR$.

At $n$th order, there are terms with $n_\scR$ real emissions and $n_\scV$ virtual loops:
\begin{equation}
  \begin{split}
    \cD^{(n)}(\mu_\scS) 
    =
    \mathop{\sum_{n_\scR=0}^n \sum_{n_\scV=0}^n}_{n_\scR + n_\scV = n}
    \cD^{(n_\scR, n_\scV)}(\mu_\scS)
    \;.
  \end{split}
\end{equation}

To define the shower operator $\cU(\mu_\scS,\mu_\scH)$, we also use the inverse operator to $\cD$, which is defined by its perturbative expansion:
\begin{equation}
\begin{split}
\label{eq:Dinverse}
\cD^{-1} ={}& 1
-  \left[\frac{\as}{2\pi}\right]\cD^{(1)}
- \left[\frac{\as}{2\pi}\right]^2\left\{
 \cD^{(2)} - \cD^{(1)}\cD^{(1)}\right\}
\\&
+ \cO(\as^3)
\;.
\end{split}
\end{equation}

The shower evolution operator is defined using $\cD(\mu^2)$ by \cite{NSAllOrder}
\begin{equation}
\begin{split}
\label{eq:cUdef}
\cU(\mu_\scS,\mu_\scH) ={}& 
\cD^{-1}(\mu_\scS)\cD(\mu_\scH)
\;.
\end{split}
\end{equation}
Then the splitting operator $\cS(\mu_\scS)$ is defined by Eq.~(\ref{eq:showerevolution}).

When we apply the perturbative operator $\cD^{(n_\scR,n_\scV)}(\mu_\scS)$ to a state $\sket{\{p,f,c,c',s,s'\}_m}$ with $m$ partons, we obtain states with $\hat m = m + n_\scR$ partons. We define $\cD^{(n_\scR,n_\scV)}(\mu_\scS)$ so that the corresponding matrix elements contain powers of $1/\epsilon$ and singularities when the $\hat m$ final state parton momenta $\hat p_j$ are grouped into $m$ infinitely narrow jets. These $1/\epsilon$ poles and collinear and soft singularities must match the corresponding poles and singularities of QCD, including their dependence on color and spin indices.

%----------------------------
\subsection{Momentum mapping}
\label{sec:momentummapping}

We need to define the momenta in the singular diagrams that contribute to $\cD(\mu_\scS)$. We start with momenta $p_j$ for $m$ partons. Then we emit partons, so that the there are $\hat m$ final state partons with momenta $\hat p_i$. We require that the parton shower evolution maintain momentum conservation:
\begin{equation}
\label{eq:momentumconservaton}
\sum_{i = 1}^{\hat m} \hat p_i = \sum_{j = 1}^m p_j
\;.
\end{equation}
A massless parton cannot split into more than one massless partons with non-zero momenta when the daughter partons are not exactly collinear. Additionally, when a soft gluon is emitted, momentum conservation cannot hold if the momenta of the original partons remain the same. Thus we need a momentum mapping that transfers some momentum between partons in order to retain momentum conservation.

Let $G$ label the combination of Feynman graphs used for the bra and ket amplitudes. Define splitting variables $\zeta_G$ that are adapted to the structure of the singular diagrams. The splitting variables $\zeta_{G}$ can include, for example, the momentum fraction $1-z$ and transverse momentum $k_\LT$ of an emitted gluon. Also define an integration measure $d\zeta_G$ over the splitting variables. We define the final state momenta $\hat p_j$ as functions of the $m$ initial momenta $p_i$ and splitting variables $\zeta_G$:
\begin{equation}
\label{eq:Rmap}
\{\hat p\}_{\hat m} = R(G;\zeta_{G},\{p\}_m)
\;.
\end{equation}
This momentum mapping function obeys momentum conservation, Eq.~(\ref{eq:momentumconservaton}). A familiar example of this that is sometimes used for a first order dipole shower is the momentum mapping of Catani and Seymour \cite{CataniSeymour}. 

We make different choices. When parton $l$ splits to partons $l$ and $m\!+\!1$ in both the bra and ket amplitudes, there is a singularity when $\hat p_{l}$ and $\hat p_{m\!+\!1}$ become collinear with $p_l$. Exactly at the singular point, the momenta of the remaining partons are unchanged: $\hat p_i = p_i$. When  $\hat p_{l}$ and $\hat p_{m\!+\!1}$ are not exactly collinear with $p_l$, we choose for $i \le m$, $i \ne l$,
\begin{equation}
\label{eq:pitohatpi0}
\hat p_i^\mu = \lambda\, \Lambda^\mu_{\ \nu} p_i^\nu
\;,
\end{equation}
where $\lambda$ is a scalar and $\Lambda^\mu_{\ \nu}$ is a Lorentz transformation, which are specified in Appendix \ref{sec:CollinearMomentumMapping}.

When one or more partons are emitted near a soft singularity, it is simplest to choose a momentum mapping that does not depend on which hard partons the soft partons were emitted from. We choose a momentum mapping given by a scaling and a Lorentz transformation for all of the partons $i$ with $i \le m$, as in Eq.~(\ref{eq:pitohatpi0}). The definition of $\lambda$ and $\Lambda^\mu_{\ \nu}$ is given in Appendix \ref{sec:SoftMomentumMapping}.

The definitions in Appendices \ref{sec:CollinearMomentumMapping} and \ref{sec:SoftMomentumMapping} provide a global momentum mapping, in which the needed momentum for momentum conservation is taken from all of the remaining partons in the event. This contrasts with a local momentum mapping, in which the needed momentum is taken just one parton, as in Ref.~\cite{CataniSeymour}. The effects of these two kinds of mapping were compared in Ref.~\cite{panscales}.

%------------------------------
\subsection{Splitting function}
\label{sec:splittingfunction}

Combining Eq.~(\ref{eq:showerevolution}) and Eq.~(\ref{eq:cUdef}) allows us to express the splitting operator $\cS(\mu_\scS)$ in terms of the derivative of $\cD(\mu_\scS)$:
\begin{equation}
\label{eq:cDevolution}
\mu_\scS \frac{d\cD(\mu_\scS)}{d\mu_\scS}
= \cD(\mu_\scS)\,\cS(\mu_\scS)
\;.
\end{equation}
This has the form of an evolution equation for $\cD(\mu_\scS)$. The solution of this with a boundary condition at $\mu_\scS = \mu_\Lf$ is
\begin{equation}
\begin{split}
\cD(\mu_\Lf)^{-1}\cD(\mu_\scS) ={}& 
\cU(\mu_\Lf,\mu_\scS)
\;,
\end{split}
\end{equation}
where
\begin{equation}
\begin{split}
\label{eq:cDsolution}
\cU(\mu_\Lf,\mu_\scS)
={}&
\mathbb{T}
\exp\left(\int_{\mu_\Lf}^{\mu_\scS}\!\frac{d\mu}{\mu}\,
\cS(\mu)
\right)
\;.
\end{split}
\end{equation}
Here $\mathbb{T}$ denotes ordering according to the scale, with larger $\mu^2$ to the right. If we expand Eq.~(\ref{eq:cDsolution}) in powers of $\cS$, we have
\begin{equation}
\begin{split}
\cU(\mu_\Lf,\mu_\scS) ={}& 
1
+ \int_{\mu_\Lf}^{\mu_\scS}\!\frac{d\mu}{\mu}\,\cS(\mu)
\\ &
+ \int_{\mu_\Lf}^{\mu_\scS}\!\frac{d\mu_1}{\mu_1}
\int_{\mu_\Lf}^{\mu_1}\!\frac{d\mu_2}{\mu_2}\
\cS(\mu_2)\,\cS(\mu_1)
\\ &
+ \int_{\mu_\Lf}^{\mu_\scS}\!\frac{d\mu_1}{\mu_1}
\int_{\mu_\Lf}^{\mu_1}\!\frac{d\mu_2}{\mu_2}
\int_{\mu_\Lf}^{\mu_2}\!\frac{d\mu_3}{\mu_3}
\\&\quad\times
\cS(\mu_3)\,\cS(\mu_2)\,\cS(\mu_1)
\\&+\cdots
\;.
\end{split}
\end{equation}

At first order, this is
\begin{equation}
\label{cD1fromS}
\cU^{(1)}(\mu_\Lf,\mu_\scS) = 
\int_{\mu_\Lf}^{\mu_\scS}\!\frac{d\mu}{\mu}\ \cS^{(1)}(\mu)
\;.
\end{equation}
At second order,
\begin{equation}
\begin{split}
\label{cD2fromS}
\cU^{(2)}&(\mu_\Lf,\mu_\scS) 
\\={}& 
\int_{\mu_\Lf}^{\mu_\scS}\!\frac{d\mu}{\mu}\ S^{(2)}(\mu^2)
\\&
+\int_{\mu_\Lf}^{\mu_\scS}\!\frac{d\mu_1}{\mu_1}
\int_{\mu_\Lf}^{\mu_1}\!\frac{d\mu_2}{\mu_2}\,
\cS^{(1)}(\mu_2)\,\cS^{(1)}(\mu_1)
\;.
\end{split}
\end{equation}
The second order infrared singular operator $\cD^{(2)}$ typically contains nested singularities. For instance, one can have one collinear splitting at a scale $\mu_1$ followed by a second collinear splitting at a much smaller scale $\mu_2$. This corresponds to a first order interaction that is generated by $\cS^{(1)}(\mu_1)$ followed by a second first order interaction that is generated by $\cS^{(1)}(\mu_2)$. Of course, two first order interactions will occur in a parton shower, which generates any number of first order splittings or virtual interactions contained in $\cS^{(1)}$. However, this product of two $\cS^{(1)}$ should not be part of the second order splitting operator $\cS^{(2)}$. This is the structure indicated in Eq.~(\ref{cD2fromS}).

%---------------------------------------------------------------
\subsection{Need for a bigger space}
\label{sec:moreisbetter}

In order to generate the operators $\cD(\mu_\scS)$ and the splitting operators $\cS(\mu)$ beyond leading order in $\as$, we need to use the Feynman diagrams for QCD, with the goal of extracting their associated logarithmic infrared singularities. There are many Feynman diagrams, and their infrared singularities are not always simple. Our method to attack this problem is to represent the Feynman diagrams as linear operators on a vector space. Then we can use linear algebra to simplify their structure.

Inside a Feynman diagram for the ket amplitude or the bra amplitude, partons do not have definite spins, and the most useful description of color may be different from what we use to describe color in the statistical space used for on-shell partons. It is especially significant that the momenta of parton lines inside a Feynman diagram are generally not on shell. Furthermore, the momenta and flavors inside a Feynman diagram for the ket amplitude are not the same as the momenta inside a Feynman diagram for the bra amplitude. Indeed, these two diagrams are typically not the same.

For these reasons, the statistical space described in the previous section is not adequate for the purpose at hand: we will need a bigger vector space. The vector space that we propose has a structure adapted to the Feynman diagrams. We devote much of this paper to defining this space. The construction used to build up Feynman diagrams from elementary operators also has significant structure. We also devote much of this paper to explaining this structure. We leave for later papers the construction of order $\as^2$ diagrams and the extraction of their infrared singularities. In this paper, we illustrate the formalism by applying it to order $\as^1$ diagrams. 

%---------------------------------------------------------------
\section{Quantum ket amplitude space}
\label{sec:quantumketspace0}

We need a precise language suitable for the description of Feynman diagrams for the ket quantum amplitude and for the conjugate bra quantum amplitude. 

%--------------------
\subsection{Vector space for the ket amplitude}
\label{sec:quantumketspace}

We consider a Feynman diagram for the amplitude for the emission of some number of partons starting with a hard state with $m_0$ partons. Inside the Feynman diagram, at an intermediate stage there are $m$ partons with labels $i \in \{1,\dots,m\}$. Parton $i$ has momentum $q_i$. These momenta are generally off shell. Each parton also has a flavor $f_i \in \{\Lg, \Lu, \bar \Lu, \Ld, \bar \Ld, \dots\}$. 

Each parton has a color index $a$ with $a \in \{1,\dots, N_\Lc^2 - 1\}$ if it is a gluon or $a \in \{1,\dots,N_\Lc\}$ if it is a quark or antiquark. The corresponding color basis states for parton $i$ are $\ket{a_i}$ with 
\begin{equation}
\brax{\bar a_i}\ket{a_i} = \delta_{\bar a_i,a_i}
\;.
\end{equation}
This contrasts with the description of color in the statistical space in Sec.~\ref{sec:statisticalspace}, where we allow a basis for the color space that is not orthogonal and normalized, as we describe later in Eqs.~(\ref{eq:colordualbasis}) and (\ref{eq:colorcompleteness}).

The parton also carries an index $r$ that specifies the component of the parton wave function that carries its Lorentz transformation properties: the $(1/2,0) \oplus (0,1/2)$ representation for a quark or antiquark, or the $(1/2,1/2)$ representation for a gluon. We now consider the $r$ index in some detail.

{\em Gluons.} Consider first the basis vectors for space-time vectors that represent gluons. Written in terms of the vector components, the inner product for vectors is the Lorentz invariant inner product
\begin{equation}
\label{eq:vectorinnerproduct}
\brax{A}\ket{B} = A^*_\nu B^\nu
= (A^*)^\mu g_{\mu\nu} B^\nu
\;.
\end{equation}
We define basis vectors $\iket{r}$ such that
\begin{equation}
\begin{split}
\ket{B} ={}& \sum_r \ket{r} B^{r}
\;.
\end{split}
\end{equation}
Then also
\begin{equation}
\bra{A} = \sum_r (A^*)^r\bra{r}
\;.
\end{equation}
Then
\begin{equation}
\brax{A}\ket{B} = \sum_{r,\bar r}
(A^*)^r\brax{r}\ket{\bar r} B^{\bar r}
\;.
\end{equation}
Thus the inner product of two basis vectors is the metric tensor
\begin{equation}
\label{eq:rrinnerproduct}
\brax{r}\ket{\bar r} = g_{r \bar r}
\;.
\end{equation}

It is also useful to define dual basis vectors $\iket{r}_\scD$ by
\begin{equation}
\ket{r}\dualR = g^{r \bar r }\ket{\bar r}
\;.
\end{equation}
Then
\begin{equation}
\dualL\brax{r}\ket{\bar r} = \delta_{\bar r}^r
\;.
\end{equation}
Using the dual basis vectors, we have the completeness relations
\begin{equation}
\label{eq:ketrcompleteness}
\sum_r \ket{r}\,\dualL\bra{r} = 1
\;.
\end{equation}

The components of vectors $\bra{A}$ and $\ket{B}$ can be obtained by taking inner products with the basis vectors:
\begin{equation}
\begin{split}
\dualL\brax{r}\ket{B} ={}& B^r
\;,
\\
\brax{A}\ket{r} = {}& A^*_r
\;.
\end{split}
\end{equation}

{\em Quarks.} For quark partons, we need Dirac spinors. If $A$ and $B$ are Dirac spinors, we use the inner product
\begin{equation}
\brax{A}\ket{B} = \bar A B
= A^*_\alpha (\gamma^0)_{\alpha\beta} B_\beta
\;.
\end{equation}
Here $\gamma^0$ plays the role of the metric tensor in the spinor space. We follow the usual convention that uses the bar in $\bar A$ with indices left implicit instead of displaying raised or lowered indices.

In order to have a notation that applies for both quarks and gluons, we choose a representation of the gamma matrices in which $\gamma^0$ is a real, symmetric matrix 
\begin{equation}
\gamma^0_{ij} = (\gamma^0_{ij})^* =\gamma^0_{ji}
\;.
\end{equation}
We also use $\gamma^0 \gamma^{\mu\,\dagger} \gamma^0 = \gamma^\mu$.

We define basis vectors $\ket{r}$ such that
\begin{equation}
\begin{split}
\ket{B} ={}& \sum_r \ket{r} B_r
\;.
\end{split}
\end{equation}
Then also
\begin{equation}
\begin{split}
\bra{A} ={}& \sum_r (A^*)_r\bra{r}
\;,
\end{split}
\end{equation}
so
\begin{equation}
\brax{A}\ket{B} = \sum_{r,\bar r}
A^*_r \brax{r}\ket{\bar r} B_{\bar r}
\;.
\end{equation}
Thus the inner product of two basis vectors is the spinor version of the metric tensor
\begin{equation}
\label{eq:rrinnerproductdirac}
\brax{r}\ket{\bar r} = \gamma^0_{r \bar r}
\;.
\end{equation}

We define dual basis vectors by
\begin{equation}
\ket{r}\dualR = \gamma^0_{r \bar r }\ket{\bar r}
\;.
\end{equation}
Then
\begin{equation}
\begin{split}
\dualL\brax{r}\ket{\bar r} ={}& \delta_{r \bar r}
\;.
\end{split}
\end{equation}
This gives us the completeness relation
\begin{equation}
\begin{split}
\label{eq:rcompleteness}
\sum_r \ket{r}\,\dualL\bra{r} ={}& 1
\;.
\end{split}
\end{equation}

The components of vectors $\bra{A}$ and $\ket{B}$ can be obtained by taking inner products with the basis vectors:
\begin{equation}
\begin{split}
\dualL\brax{r}\ket{B} ={}& B_r
\;,
\\
\brax{A}\ket{r} = {}& \bar A_r
\;.
\end{split}
\end{equation}

It may be helpful to note that there is another completeness relation: $\sum_r \iket{r}_\scD\,\ibra{r} = 1$. However, this relation is not as useful as Eq.~(\ref{eq:rcompleteness}) because $\ibrax{r}\iket{B}$ is $(\gamma^0 B)_r$ rather than $B_r$ and $\ibrax{A}\iket{r}_\scD$ is $(\bar A \gamma^0)_r$ rather than $\bar A_r$.

{\em Antiquarks.} The description of antiquarks is the same as for quarks, with the same basis vectors $\iket{r}$ and $\iket{r}_\scD$. The difference between quarks and antiquarks comes in the Feynman rules: an incoming antiquark is represented by a spinor $\bar V$ instead of $U$, an outgoing antiquark is represented by a spinor $V$ instead of $\bar U$, and the order of $\gamma$ matrices is reversed along an antiquark line compared to a quark line.

{\em Combined quantum numbers.} As outlined above, each parton is described by a set of quantum numbers $\{q_i,f_i,a_i,r_i\}$. In order to make our formulas more compact, we adopt the abbreviation
\begin{equation}
w_i = \{q_i,f_i,a_i,r_i\}
\end{equation}
for the quantum numbers that describe a parton in a Feynman diagram. Thus we use basis states $\iket{w_i}$ to describe the state of one parton. The dual basis states are ${}_\scD\!\ibra{w_i}$, where the dual basis vectors differ from the basis vectors $\ibra{w_i}$ only in their vector/spinor factor $\ibra{r_i}$. The inner product is
\begin{equation}
\begin{split}
\dualL\brax{\bar w_i}\ket{w_i}
={}& \dualL\brax{\bar q,\bar f,\bar a,\bar r}\ket{q,f,a,r}
\\
={}& (2\pi)^d \delta^d(\bar q - q)
\,
\delta_{\bar f,f} \delta_{\bar a,a} \delta_{\bar r,r}
\;.
\end{split}
\end{equation}

A set of $m$ partons is then described by a vector 
\begin{equation}
\begin{split}
\ket{\{w\}_m} 
={}& \ket{\{w_1,w_2,\cdots,w_m\}}
\\
={}& \ket{w_1}_1\otimes\ket{w_2}_2\cdots\otimes\ket{w_m}_m
\;. 
\end{split}
\end{equation}
Here the vector $\iket{w}_i$ in the $i$th position represents the quantum numbers $w$ of parton $i$. For any number of partons, the inner product of basis states is
\begin{equation}
\begin{split}
& \dualL\brax{\{\bar w\}_{\bar m}}
\ket{\{w\}_m}
\\ &\qquad
=\delta_{\bar m,m}
\prod_{i=1}^m
(2\pi)^d \delta^d(\bar q_i - q_i)\,
\delta_{\bar f_i,f_i} \delta_{\bar a_i,a_i} \delta_{\bar r_i,r_i}
\;.
\end{split}
\end{equation}

We abbreviate the integral needed for summing over basis states for $m$ partons as
\begin{equation}
\begin{split}
\label{eq:intqfar}
\int\!d\{w\}_m ={}& \prod_{i=1}^m 
\int\!d w_i 
\;,
\end{split}
\end{equation}
where
\begin{equation}
\begin{split}
\label{eq:intdwi}
\int\!dw_i ={}& 
\sum_{f_i}\sum_{c_i}\sum_{r_i}
\int \frac{d^d q_i}{(2\pi)^d}\,
\;.
\end{split}
\end{equation}
Note that we integrate over off-shell momenta $q_i$ in $d = 4 - 2\epsilon$ dimensions. We have a completeness relation
\begin{equation}
\label{eq:wcompletentness}
1 = 
\sum_m \int\!d\{w\}_m\,
\ket{\{w\}_m}\, \dualL\bra{\{w\}_m}
\;.
\end{equation}

The basis vectors $\ket{\{w\}_m}$ are not symmetric. In some cases we have to consider permutations of the single parton quantum numbers. In order to flip two partons we define a generic permutation operators $\bm{\pi}(l,k)$ by
\begin{equation}
\begin{split}
\bm{\pi}(l,k){}&\ket{\{w_1,\dots, w_l,\dots,w_k,\dots, w_{m}\}}
\\
={}& \ket{\{w_1,\dots,w_k,\dots,w_l,\dots, w_{m}\}}
\\
={}& \ket{w_1}_1\cdots\otimes\ket{w_k}_l\cdots\otimes\ket{w_l}_k\cdots\otimes\ket{w_m}_m
\;.
\end{split}
\end{equation}
Most often, we use two specific permutation operators, defined by
\begin{equation}
\begin{split}
\label{eq:PilandPi}
\bm{\Pi}_l\ket{\{w\}_{m\!+\!1}} ={}& \bm{\pi}(l,m+1)\ket{\{w\}_{m\!+\!1}}\;,
\\
\bm{\Pi}\ket{\{w\}_{m\!+\!2}} ={}& \bm{\pi}(m+1,m+2)\ket{\{w\}_{{m\!+\!2}}}\;.
\end{split}
\end{equation}
The operator $\bm{\Pi}_l$ exchanges parton $l$ with the last one, and $\bm{\Pi}$ exchanges the last two partons. 

We will use this vector space to represent and generate Feynman graphs. More specifically we are interested in Feynman graphs when $m$ hard lines are dressed with real and virtual radiation. We will define operators acting on the ket amplitude space with the aim that the usual Feynman graphs can be understood as matrix elements of these operators.  The simplest operator is the unit operator.  The simplest operator is the unit operator,
\begin{equation}
\dualL\bra{\{\bar{w}\}_{m}}\bm{1}\ket{\{w\}_m} =
\begin{prdfig}{60f4a778e3394ccd13958e154e7fb39c}{quantum-unit-operator}
\begin{tikzpicture}[baseline={(current bounding box.center)}]
  \begin{feynman}[]
    \vertex[empty dot,label={[left] $w_1$}] (v1) at (0,0) {};
    \coordinate [label={[right] $\bar w_1$}] (l1) at ($(v1) + (2cm,0)$);
    \vertex[empty dot,label={[left] $w_2$}] (v2) at ($(v1)-(0,0.5cm)$) {};
    \coordinate [label={[right] $\bar w_2$}] (l2) at ($(v2) + (2cm,0)$);
    \vertex[empty dot,label={[left] $w_m$}] (vm) at ($(v2)-(0,1.5cm)$) {};
    \coordinate [label={[right] $\bar w_m$}] (lm) at ($(vm) + (2cm,0)$);
    \diagram*{
      (v1)--[](l1);
      (v2)--[](l2);
      (vm)--[](lm);
    };
    % decorations
    \vertex [] at ($(v2) + (1cm,-0.375cm)$) {$\cdot$};
    \vertex [] at ($(v2) + (1cm,-0.750cm)$) {$\cdot$};
    \vertex [] at ($(v2) + (1cm,-1.125cm)$) {$\cdot$};    
  \end{feynman}
\end{tikzpicture}
\end{prdfig}
\;.
\end{equation}
This is trivial but helps to understand the concept.

%-----------
\subsection{Separation of harder and softer amplitudes}

In a parton shower, one separates harder interactions among partons from softer interactions. Put differently, one separates interactions that are less infrared singular from interactions that are more infrared singular. The first step of this separation was illustrated in Figs.~\ref{fig:d11exampleA} and \ref{fig:d11exampleB}. In Fig.~\ref{fig:d11exampleA}, partons with momenta $q_i$ for $i \in \{1,\dots, m\}$ emerge from a ket amplitude $H_\mathrm{ket}$ We take the diagrams in $H_\mathrm{ket}$ to be amputated on their external legs. We further suppose that the momenta $q_i$ are close to momenta $p_i$ that are exactly lightlike. The $p_i$ are not close to vanishing and not close to being collinear, so that all of the dot products $p_i \cdot p_j$ are larger than some scale $\mu_\mathrm{h}^2$. This implies that that the internal lines in $H_\mathrm{ket}$ are off shell by an amount of order at least $\mu_\mathrm{h}^2$. 

In Fig.~\ref{fig:d11exampleA}, there is also a bra amplitude $H_\mathrm{bra}$, from which $m$ partons with momenta $q_i'$ emerge. These momenta are also close to the lightlike momenta $p_i$, so that the internal lines of $H_\mathrm{bra}$ are also off shell by an amount of order at least $\mu_\mathrm{h}^2$. We will focus on the ket amplitude below. The analysis of the bra amplitude is analogous, with just a few differences that we will mention where needed.

In Fig.~\ref{fig:d11exampleA}, the partons coming from $H_\mathrm{ket}$ then participate in further interactions that can be represented by an amplitude $G_\mathrm{ket}$ composed of Feynman diagrams like the diagram illustrated in Fig.~\ref{fig:d11exampleA} that has one real parton emission and one virtual exchange. We suppose that the momentum of the emitted gluon is either very soft or else very nearly collinear to the momentum of the mother parton and that the momentum of the exchanged gluon is very soft. Thus the Feynman amplitude $G_\mathrm{ket}$ is close to being infrared singular. There are many possible Feynman diagrams for the amplitude. In general, we take the diagrams $G_\mathrm{ket}$ to produce $\hat m$ partons with on shell momenta $\hat p_j$ with $\hat m \ge m$ and to be amputated on their external legs and then multiplied by self-energy diagrams according of the LSZ prescription for the final state in an S-matrix element. Accompanying the bra amplitude, we can have a different Feynman diagram that also produces $\hat m$ partons with the same on shell momenta $\hat p_j$.

Thus in Fig.~\ref{fig:d11exampleA}, the parton lines from $H_\mathrm{ket}(\{q\}_m)$ enter a diagram $G_\mathrm{ket}(\{q\}_m, \{\hat p\}_{\hat m})$ in which partons are emitted and exchanged. The final state momenta $\hat p_i$ are on shell. The combined graph is then
\begin{equation}
\begin{split}
\label{eq:Gtotstart}
G_\mathrm{ket}^\mathrm{tot}
={}& \prod_{i=1}^m \left[ 
\int\!\frac{d^{4-2\epsilon} q_i}{(2\pi)^{4-2\epsilon}} \right]
\\&\times
G_\mathrm{ket}(\{\hat p\}_{\hat m},\{q\}_m)\,
H_\mathrm{ket}(\{q\}_m)
\;.
\end{split}
\end{equation}
When final state partons become collinear or soft and the momenta of exchanged partons approaches zero, the graph $G_\mathrm{ket}$ is infrared singular. In the infrared singular limit, the momenta $q_i$ approach the on-shell momenta $p_i$. 

In the infrared singular limit, we approximate $H_\mathrm{ket}(\{q\}_m)$ by expanding $H_\mathrm{ket}(\{q\}_m)$ about the point $\{q\}_m = \{p\}_m$, in powers of $q_i - p_i$. The lowest order contribution is $H_\mathrm{ket}(\{p\}_m)$, giving
\begin{equation}
\begin{split}
\label{eq:Gtotend}
G_\mathrm{ket}^\mathrm{tot}
\approx{}& \prod_{i=1}^m \left[ 
\int\!\frac{d^{4-2\epsilon} q_i}{(2\pi)^{4-2\epsilon}} \right]
\\&\times
G_\mathrm{ket}(\{\hat p\}_{\hat m},\{q\}_m)\,
H_\mathrm{ket}(\{p\}_m)
\;.
\end{split}
\end{equation}
This is the form indicated in Fig.~\ref{fig:d11exampleB}. We simply drop higher order contributions because the factors of $q_i - p_i$ remove the infrared singularity.\footnote{More precisely, the original logarithmic singularity that gave a pole $1/\epsilon$ after integrating over the region in splitting variables and exchanged momenta now does not give a pole.} Once we replace $H_\mathrm{ket}(\{q\}_m)$ by $H_\mathrm{ket}(\{p\}_m)$, the momenta $q_i$ now flow through a multiparton subgraph that does not depend on the $q_i$. That is, the points indicated by $\otimes$ symbols in the ket amplitude in Fig.~\ref{fig:d11exampleB} are effectively all at the same space-time location, so that the yellow blob in the ket amplitude and the yellow blob in the bra amplitude have been replaced by multiparton point vertices.

In Eq.~(\ref{eq:Gtotend}), an integration over $\{q\}_m$ remains. The momenta $q_i$ are determined using momentum conservation in the graph $G_\mathrm{ket}$ by the final state momenta $\hat p_j$ and by the momenta $\ell$ exchanged in virtual graphs like the graph illustrated in Fig.~\ref{fig:d11exampleB}. The momenta $\hat p_j$ can be restricted to a region such that the real emissions in $G_\mathrm{ket}$ are not far from being collinear or soft. This leaves the momenta $\ell$, which flow through the point vertex created by the approximation in Eq.~(\ref{eq:Gtotend}). These momenta can be restricted to be smaller than a scale $\mu^2$ characteristic of the hard subgraph (see Sec.~\ref{sec:Gamma-H-operators}). With these choices, the approximate equality Eq.~(\ref{eq:Gtotend}) can be roughly maintained. Then the infrared singularities of $G_\mathrm{ket}^\mathrm{tot}$ can be reproduced within the structure of Eq.~(\ref{eq:Gtotend}).

%--------------------------------------------------
\subsubsection*{Restrictions on types of graphs}
\label{sec:AllowedGraphTypes}

The analysis in this paper starts with the representation (\ref{eq:Gtotend}) of the complete amplitude as a hard factor $H_\mathrm{ket}$ and an amplitude $G_\mathrm{ket}$ that contains infrared singularities. It is the infrared singularities that are of interest. Accordingly, parts of $G_\mathrm{ket}$ that are not infrared singular should be associated with the hard subgraph $H_\mathrm{ket}$. In particular, $G_\mathrm{ket}$ should not contain subgraphs in which two or more of the initial $m$ partons combine to make fewer than $m$ partons.

The construction in this paper works at the level of cross sections, defined from the ket amplitude, the conjugate bra amplitude, and a momentum mapping algorithm. Thus we will later introduce further manipulations to factor contributions to the cross section that are not infrared singular into the hard subgraph $H_\mathrm{ket}$.

%--------------------------------------------------
\subsubsection*{Labeling}
\label{sec:Labeling}

We will try to keep the notation as simple as possible by adopting the convention that when a parton with label $l$ splits into two or more partons, one of the daughter partons is labeled $l$ and the other partons acquire new labels. For instance, if there were $\tilde m$ partons before the splitting and parton $l$ splits into two partons, the daughter partons are labelled $l$ and $\tilde m \!+\!1$. 

We will also need graphs in which parton $l$ exchanges a virtual parton with momentum $\ell$ with another parton, creating an infrared singularity from the integration range $\ell \to 0$. In this case, we retain the label $l$ for the parton after the exchange.

%--------------------------------------------------
\subsection{\label{sec:IandFstatesforamplitude}Initial and final states for the Feynman amplitude}

We will represent the Feynman graph $G_\mathrm{ket}(\{\hat p\}_{\hat m},\{q\}_m)$ using operators that act on the quantum ket amplitude space. For this purpose, we need vectors in the quantum ket amplitude space that specify the initial and final states of the partons. In this section, we specify these initial and final state vectors.

A closely analogous construction will apply to the quantum bra amplitude space, in which we represent Feynman graphs $G_\mathrm{bra}(\{\hat p\}_{\hat m},\{q\}_m)$. We state the differences needed for the quantum bra amplitude as needed.

%--------------------------------------------------
\subsubsection*{Initial state spinors and vectors}
\label{sec:InitialStateWaveFunction}

The parton lines in the quantum ket amplitude space evolve according to Feynman graphs with off-shell propagators (with some approximations). At the start of this evolution, we need spinor or polarization vector factors. 

For an incoming gluon with label $i$ in the ket amplitude, we insert a vector with components 
\begin{equation}
\label{eq:wavefctnINketg}
\dualL\brax{r_i}\ket{\chi_\Lg(p_i,s_i)} 
=
-\varepsilon^{r_i}(p_i,s_i)
\;.
\end{equation}
We use a minus sign with initial state polarization vectors as a convenient phase convention that accounts for the minus sign in the normalization $\varepsilon \cdot \varepsilon^* = -1$.

In Eq.~(\ref{eq:wavefctnINketg}), $(p_i,s_i)$ are the momentum and spin of this parton as it leaves the hard subgraph $H_\mathrm{ket}$ in Fig.~\ref{fig:d11exampleB}. It is, of course, an approximation to use an on-shell polarization vector, since the gluon momentum in the Feynman graph will typically be somewhat off shell.

For an incoming quark with label $i$ in the ket amplitude, we insert a spinor with components
\begin{equation}
\label{eq:wavefctnINketq}
\dualL\brax{r_i}\ket{\chi_q(p_i,s_i)}
=
\left[\frac{\s{n}}{2 p_i\cdot n}\, U(p_i,s_i)\right]_{r_i}
\;.
\end{equation}
The justification for choosing the factor $\s{n}/[2 p_i \cdot n]$ is given in Ref.~\cite{NSI} (with an equivalent form for this factor). Here we simply note the normalization
\begin{equation}
\frac{\overline U(p_i,s_i)\,\s{n}\, U(p_i,s_i)}{2 p_i \cdot n} = 1
\;.
\end{equation}

For an incoming antiquark in the ket amplitude, we insert a conjugate spinor with components
\begin{equation}
\label{eq:wavefctnINketqbar}
\dualL\brax{r_i}\ket{\chi_{\bar q}(p_i,s_i)}
=
\left[\overline V(p_i,s_i)\,\frac{\s{n}}{2 p_i\cdot n}\right]_{r_i}
\:.
\end{equation}

For incoming partons in the bra amplitude, let $(p_i,s'_i)$ be the momentum and spin of parton $i$  as it leaves the hard subgraph $H_\mathrm{bra}$ in Fig.~\ref{fig:d11exampleB}. We use initial spinor and polarization vector factors
\begin{equation}
\begin{split} 
\label{eq:wavefctnsINbra}
\brax{\chi_g(p_i,s'_i)}\ket{r'_i} ={}& -\varepsilon_{r'_i}(p_i, s'_i)^* 
\;,
\\
\brax{\chi_q(p_i,s'_i)}\ket{r'_i} ={}& 
\frac{\big[\,\overline U(p_i, s'_i)\s{n}\big]_{r'_i}}{2p_i \cdot n}
\;,
\\
\brax{\chi_{\bar q}(p_i,s'_i)}\ket{r'_i} ={}& 
\frac{\big[\s{n} {V}(p_i, s'_i)\big]_{r'_i}}{2p_i \cdot n}
\;.
\end{split}
\end{equation}
% 

%--------------------------------------------------
\subsubsection*{Initial state color}
\label{sec:InitialStateColor}

The partons in the hard subgraph were in a color state $\ket{\{c\}_{m}}$. In the component form used in the quantum ket amplitude space, this is
\begin{equation}
\sum_{\{a\}_m} \ket{\{a\}_m}\brax{\{a\}_m}\ket{\{c\}_{m}}
\;.
\end{equation}
We can use the trace basis for the colors in the statistical space, as in \textsc{Deductor} (see Ref.~\cite{NSI}). Then if $\ket{\{c\}_{m}}$ is a color string with a quark with label $1$ at one end, an antiquark with label $m$ at the other end, and $m-2$ gluons in between, we have
\begin{equation}
\begin{split}
\label{eq:tracebasisstates}
\brax{\{a\}_{m}}\ket{\{c\}_{m}}
={}& n(S)^{-1/2}\, [t^{a_2} t^{a_3} \cdots t^{a_{m\!-\!1}}]_{a_1 a_m}
\;,
\end{split}
\end{equation}
where $n(S) = N_\Lc C_\LF^{m-2}$ \cite{NSI}. It is significant that the state $\ket{\{c\}_{m}}$ is invariant under color rotations of all of the partons together. The trace basis is convenient, but one could use a different basis for the space of color singlet $m$-parton states.

If we use basis vectors $\ket{\{c\}_{m}}$ that are not orthogonal and normalized (as in the trace basis), then we define dual basis vectors\footnote{See Ref.~\cite{NSI}. As long as only a few emissions have occurred starting with a color state $\{c\}_{m}$, one can determine the final color state using some simple algebra without constructing the dual color basis vectors.} by
\begin{equation}
\label{eq:colordualbasis}
\dualL\brax{{\{c'\}_m}}\ket{{\{c\}_m}} = 
\delta^{\{c'\}_{m}}_{\{c\}_m}
\;.
\end{equation}
These appear in the completeness relations
\begin{equation}
\begin{split}
\label{eq:colorcompleteness}
1 ={}& \sum_{\{c\}_{m}}
\ket{\{c\}_{m}}\, \dualL\bra{\{c\}_{m}}
\;,
\\
1 ={}& \sum_{\{c'\}_{m}}
\ket{\{c'\}_{m}}\dualR\,
\bra{\{c'\}_{m}}
\;.
\end{split}
\end{equation}
%

%--------------------------------------------------
\subsubsection*{Initial state vector}
\label{sec:InitialStateKet}

Let the statistical space state of the partons coming from the hard scattering be $\sket{\{p,f,c,c',s,s'\}_m}$. The part of this that applies to the ket amplitude $H_\mathrm{ket}$ has quantum numbers $\{p,f,c,s\}_m$ with $m$ partons with lightlike momenta $p_i$, flavors $f_i$, spins $s_i$ and combined color state $\{c\}_m$. Then, following the notation of the preceding subsections, the initial state for the partons in the ket amplitude is
\begin{equation}
\begin{split}
\label{eq:InitialKetVector}
\ket{\chi(\{p,f,c,s\}_m)}\hskip - 1.8 cm {}&
\\ ={}& \int\!d\{\bar w\}_m\, \ket{\{\bar w\}_m}
\\ &\times
\brax{\{\bar a\}_{m}}\ket{\{c\}_{m}}
\prod_{i=1}^m \delta_{\bar f_i, f_i}\,
\dualL\brax{\bar r_i}\ket{\chi_{f_i}(p_i,s_i)}
\;.
\end{split}
\end{equation}
Recall that $\{w\}_m = \{q, f, a, r\}_m$. The flavors $f_i$ have been fixed and the amplitudes for the color indices $\bar a_i$ and the vector/spinor indices $r_i$ have been specified. There is an integration over the momenta $q_i$. These are completely free to start with, but are determined by momentum conservation from the final state momenta $\hat p_j$ and the momenta $\ell$ of exchanged partons.

In the bra amplitude, the initial state vector is
\begin{equation}
\begin{split}
\label{eq:InitialBraVector}
\bra{\chi(\{p,f,c',s'\}_m)}\hskip - 1.7 cm {}&
\\ ={}& \int\!d\{\bar w'\}_m\, \bra{\{\bar w'\}_m}
\\ &\times
\brax{\{c'\}_{m}}\ket{\{\bar a'\}_{m}}
\prod_{i=1}^m \delta_{\bar f_i, f_i}\,
\brax{\chi_{f_i}(p_i,s'_i)}\ket{\bar r'_i}
\;.
\end{split}
\end{equation}
%

%--------------------------------------------------
\subsubsection*{Final state vector}
\label{sec:FinalStateKet}

We also need a final state vector for the Feynman amplitude, corresponding to a state with quantum numbers $\{\hat p,\hat f,\hat c,\hat s\}_{\hat m}$. There are $\hat m$ partons with lightlike momenta $\hat p_i$, flavors $\hat f_i$, spins $\hat s_i$ and combined color state $\{\hat c\}_{\hat m}$.

For starting state $\iket{\chi(\{p,f,c,s\}_m)}$ in the quantum ket amplitude space, a given graph $G_\mathrm{ket}$ produces a state $\iket{\Psi}$,
\begin{equation}
\ket{\Psi} = \int\!d\{w\}_{\hat m}\ \ket{\{w\}_{\hat m}}\,
\dualL\brax{\{w\}_{\hat m}}\ket{\Psi}
\;.
\end{equation}
The labels $i$ on the parton lines here reflect the structure of the graph and the labels on the parton lines in the initial state $\iket{\chi(\{p,f,c,s\}_m)}$. We assign lightlike final state momenta $\hat p_j$ to the graph, determined by the combined ket and bra graphs $G_\mathrm{ket}$ and $G_\mathrm{bra}$ according to the momentum mapping from Sec.~\ref{sec:momentummapping}. Then to determine the amplitude for the final partons to have combined quantum numbers $\{\hat p,\hat f,\hat c,\hat s\}_{\hat m}$, we need to take the inner product $\ibrax{\psi(\{\hat p,\hat f,\hat c,\hat s\}_{\hat m})}\iket{\Psi}$ of $\iket{\Psi}$ with a state $\iket{\psi(\{\hat p,\hat f,\hat c,\hat s\}_{\hat m})}$ in the quantum ket amplitude space. 

The vector/spinor factors for outgoing partons in the ket amplitude are
\begin{equation}
\begin{split}
\label{eq:wavefctnsOUTketparts}
\brax{\psi_\Lg(\hat p_i,\hat s_i)}\ket{\hat r_i} 
={}& \varepsilon_{\hat r_i}(\hat p_i, \hat s_i)^*    
\;,
\\
\brax{\psi_q(\hat p_i,\hat s_i)}\ket{\hat r_i} ={}& 
\overline U_{\hat r_i}(\hat p_i, \hat s_i)
 \;,
\\
\brax{\psi_{\bar q}(\hat p_i,\hat s_i)}\ket{\hat r_i} ={}& 
{V}_{\hat r_i}(\hat p_i, \hat s_i)
\;.
\end{split}
\end{equation}
These are the standard factors for outgoing particles in the S-matrix.

The vector for outgoing particles in the ket amplitude space is then
\begin{equation}
\begin{split}
\label{eq:wavefctnOUTket}
\bra{\psi(\{\hat p,\hat f,\hat c,\hat s\}_{\hat m})}\hskip -1.5 cm {}&
\\={}& \int d\{\bar{w}\}_{\hat m}\, \dualL\bra{\{\bar{w}\}_{\hat m} }
\\ &\times 
\prod_{i=1}^{\hat m} \left\{(2\pi)^d \delta^d(\bar{q}_i - \hat p_i)\,
\delta_{\hat f_i\bar{f}_i}\right\}
\\ &\times 
\dualL\brax{\{\hat c\}_{\hat m}}\ket{\{\bar{a}\}_{\hat m}}
\prod_{i=1}^{\hat m} 
\brax{\psi_{\hat f_i}(\hat p_i, \hat s_i)}\ket{\bar{r}_i}
\;.
\end{split}
\end{equation}
Note that the momenta $q_i$ in $\ket{\Psi}$ are set to the lightlike momenta $\hat p_i$.

In the state for outgoing partons in the quantum bra amplitude, the partons have quantum numbers $\{\hat p,\hat f,\hat c',\hat s'\}_{\hat m}$. The vector/spinor factors for the outgoing partons are 
\begin{equation}
\begin{split}
\label{eq:wavefctnsOUTbraparts}
\dualL\brax{r'_i}\ket{\psi_\Lg(\hat p_i,\hat s'_i)} 
={}& \varepsilon^{\hat r'_i}(\hat p_i, \hat s'_i)
\;,
\\
\dualL\brax{\hat r'_i}\ket{\psi_q(\hat p_i,\hat s'_i)} 
={}& U_{\hat r'_i}(\hat p_i, \hat s'_i)
\;,
\\
\dualL\brax{\hat r'_i}\ket{\psi_{\bar q}(\hat p_i,\hat s'_i)} 
={}& \overline{V}_{\hat r'_i}(\hat p_i, \hat s'_i)
\;.
\end{split}
\end{equation}
The vector for outgoing particles in the Feynman bra amplitude is then
\begin{equation}
\begin{split}
\label{eq:wavefctnOUTbra}
\ket{\psi(\{\hat p, \hat f, \hat c', \hat s'\}_{\hat m})}\hskip - 1.5 cm {}&
\\={}& \int d\{\bar{w}\}_{\hat m}\, \ket{\{\bar{w}\}_{\hat m}}
\\&\times 
\prod_{i=1}^{\hat m} 
\left\{(2\pi)^d \delta^d(\bar{q}_i - \hat p_i)\,
\delta_{\hat f_i\bar{f}_i}\right\}
\\ &\times 
\brax{\{\bar{a}\}_{\hat m}}\ket{\{\hat c'\}_{\hat m}}\dualR\,
\prod_{i=1}^{\hat m} 
\dualL\brax{\bar{r}_i}
\ket{\psi_{\hat f_i}(\hat p_i, \hat s'_i)}
\;.
\end{split}
\end{equation}
%

%-------------------------------
%
\section{The Feynman amplitude}
\label{sec:FeynmanAmplitude}

In this section, we describe the general structure of the Feynman amplitudes $G_\mathrm{ket}$ that can exhibit infrared singularities and describe how these amplitudes can be implemented using operators on the quantum ket amplitude space.

%----------------------------
\subsection{The point vertex}
\label{sec:pointvertex}

The initial state vector for the Feynman ket amplitude, $\iket{\chi(\{p,f,c,s\}_m)}$ was given in  Eq.~(\ref{eq:InitialKetVector}). Its inner product with a ket amplitude basis state
\begin{equation}
\dualL\bra{\{\hat w\}_m} = \dualL\bra{\{\hat q,\hat f,\hat a,\hat r\}_m}
\end{equation}
is
\begin{equation}
\begin{split}
\label{eq:InitialKetMatrixElement}
\dualL\brax{\{\hat w\}_m} {}&\ket{\chi(\{p,f,c,s\}_m)} 
\\ ={}& 
\brax{\{\hat a\}_{m}}\ket{\{c\}_{m}}
\prod_{i=1}^m \delta_{\hat f_i, f_i}\,
\dualL\brax{\hat r_i}\ket{\chi_{f_i}(p_i,s_i)}
\;.
\end{split}
\end{equation}
We now seek to represent this matrix element as a simple Feynman graph, constructed as a time-ordered product of field operators.

Let $\phi(x;f_i, a_i,r_i)$ be a field operator appropriate for flavor $f_i$, with vector/spinor index $r_i$ and color index $a_i$:
\begin{equation}
\begin{split}
\phi(x;\Lg, a_i, r_i) ={}& 
\brax{\phi(x)}\ket{\Lg, a_i, r_i}
= A_{a_i r_i}(x)
\;,
\\
\phi(x;q, a_i, r_i) ={}& 
\brax{\phi(x)}\ket{q, a_i, r_i} 
= \overline\psi_{q,a_i r_i}(x)
\;,
\\
\phi(x;\bar q, a_i, r_i) ={}&  
\brax{\phi(x)}\ket{\bar q, a_i, r_i}
= \psi_{\bar q,a_i r_i}(x)
\;.
\end{split}
\end{equation}
Here $q$ denotes a quark flavor, $q \in \{ u, d, \dots\}$, and $\bar q$ denotes an antiquark flavor, $\bar q \in \{\bar u, \bar d, \dots\}$. We use these to form a composite operator that consists of a product of $m$ field operators at position $x$:
\begin{equation}
\begin{split}
\label{eq:cHdef}
\cH(x) ={}& 
\sum_{\{a,r\}_m}
\frac{1}{\prod_f N_{f}!}\,
\prod_{i=1}^m
\phi(x;f_i, a_i, r_i)
\\
&\times
\brax{\{a\}_{m}}\ket{\{c\}_{m}}
\prod_{i=1}^m 
\dualL\brax{r_i}\ket{\chi_{f_i}(p_i,s_i)}
\;.
\end{split}
\end{equation}
The operator $\cH(x)$ depends on the initial state quantum numbers $\{p,f,c,s\}_m$, although we do not display this dependence explicitly. We have divided by a statistical factor $\prod_f N_{f}!$, where $N_{f}$ is the number of fields of flavor $f \in \{ \Lg, u, \bar u, d, \bar d, \dots\}$ in $\cH(x)$. 

We also use single parton field operators
\begin{equation}
\begin{split}
\bar\phi(x_i;\Lg,\hat a_i,\hat r_i) ={}& 
\dualL\brax{\Lg,\hat a_i,\hat r_i}\ket{\bar\phi(x)}
= A_{\hat a_i}^{\hat r_i}(x)
\;,
\\
\bar\phi(x_i;q,\hat a_i,\hat r_i) ={}& 
\dualL\brax{q,\hat a_i,\hat r_i}\ket{\bar\phi(x)} 
= \psi_{q,\hat a_i,\hat r_i}(x_i)
\;,
\\
\bar\phi(x_i;\bar q,\hat a_i,\hat r_i) ={}&  
\dualL\brax{\bar q, \hat a_i,\hat r_i}\ket{\bar\phi(x)} 
= \overline\psi_{\bar q,\hat a_i,\hat r_i}(x_i)
\;.
\end{split}
\end{equation}
These are the conjugate field operators to the operators that were used in $\cH(x)$, with $\bar\psi$ replaced by $\psi$ and $\psi$ replaced by $\bar\psi$. In addition, the gluon field now has an upper vector index $r$ instead of a lower index.

Using these operators, with $\cH(x)$ at $x = 0$, we can construct a lowest order amputated graph with $m$ outgoing partons with momenta, flavors, colors, and vector/spinor indices $\{\hat q,\hat f,\hat a, \hat r\}_{m}$:
\begin{equation}
\begin{split}
\Gamma_\scH^{(0)} ={}&
\prod_{j=1}^m \int\!d^dx_j\,e^{\mi \hat q_j\cdot x_j}
\bra{0}
\mathbb{T}
\Big[
\cH(0)
\\& \times
\prod_{j=1}^m
\bar\phi(x_j; \hat f_j,\hat a_j,\hat r_j)
\Big]
\ket{0}_\mathrm{Born,1PI}
\;.
\end{split}
\end{equation}
Here we consider only connected graphs at order $\as^0$, amputated on their external legs, so that they are also one particle irreducible (1PI).

With the use of the definition (\ref{eq:cHdef}) of $\cH(x)$, this is
\begin{equation}
\begin{split}
\Gamma_\scH^{(0)} ={}&
\sum_{\{a,r\}_m}\,
\frac{1}{\prod_f N_{f}!}\,
\prod_{j=1}^m \int\!d^dx_j\,e^{\mi \hat q_j\cdot x_j}
\\&\times
\bra{0}
\mathbb{T}
\Big[
\prod_{i=1}^m
\phi(0; f_i, a_i, r_i)
\\& \times
\prod_{j=1}^{m} 
\bar\phi(x_j;\hat f_j,\hat a_j,\hat r_j)
\Big]
\ket{0}_\mathrm{Born, 1PI}
\\
&\times
\brax{\{a\}_{m}}\ket{\{c\}_{m}}
\prod_{i=1}^{m}
\dualL\brax{r_i}\ket{\chi_{f_i}(p_i,s_i)}
\;.
\end{split}
\end{equation}

We can now apply Wick's theorem to this. The number of fields of flavor $f$ in the final state must be the same as the number $N_f$ in $\cH(x)$. There are $N_f !$ ways to connect the initial fields of flavor $f$ to the final fields. All the ways of connecting the fields of the same flavor are equivalent. We simply label the final state lines according to the labels of the initial state lines in $\cH(0)$. This gives an expression with a product of single parton propagators,
\begin{equation}
  \begin{aligned}
    &\Gamma_\scH^{(0)} =
    \sum_{\{a,r\}_m}\,
    \prod_{j=1}^m \int\!d^dx_j\,e^{\mi \hat q_j\cdot x_j}\,
    \prod_{i=1}^m \delta_{\hat f_i, f_i}\,
    \\ 
    &\,\times
    \prod_{j=1}^m \!
    \bigg[
    \bra{0}
    \mathbb{T}
    \Big[
    \phi(0; f_j, a_i, r_i)
    \bar\phi(x_j; f_j,\hat a_j,\hat r_j)
    \Big]
    \ket{0}
    \bigg]_\mathrm{amp}
    \\ 
    &\,\times
    \brax{\{\bar a\}_{m}}\ket{\{c\}_{m}}\,
    \prod_{i=1}^{m}
    \dualL\brax{\bar r_i}\ket{\chi_{f_i}(p_i,s_i)}
    \;,
  \end{aligned}
\end{equation}
where ``amp'' instructs us to amputate the final state propagators. The single particle propagators are
\begin{equation}
\begin{split}
\int\!d^dx_j\,{}&e^{\mi \hat q_j\cdot x_j}
\bra{0}
\mathbb{T}
\big[
\phi(0;f_j, a_j, r_j)\,
\bar\phi(x_j; f_j,\hat a_j,\hat r_j)
\big]
\ket{0}
\\
={}& \mi\,D(\hat q_j;f_j)_{\hat r_j  r_j}
\delta_{\hat a_j  a_j}
\;.
\end{split}
\end{equation}
Thus $\Gamma_\scH^{(0)}$ is
\begin{equation}
\begin{split}
\Gamma_\scH^{(0)} ={}&
\sum_{\{a,r\}_m}\,
\bigg[
\prod_{j=1}^{m}
\mi\,D(\hat q_j;f_j)_{\hat r_j  r_j}
\delta_{\hat a_j  a_j}
\bigg]_\mathrm{amp}
\\ &\times
\brax{\{a\}_{m}}\ket{\{c\}_{m}}
\prod_{i=1}^m \delta_{\hat f_i, f_i}\,
\dualL\brax{r_i}\ket{\chi_{f_i}(p_i,s_i)}
\;.
\end{split}
\end{equation}
We amputate the propagators, giving,
\begin{equation}
\begin{split}
\Gamma_\scH^{(0)} ={}&
\brax{\{\hat a\}_{m}}\ket{\{c\}_{m}}
\prod_{i=1}^m \delta_{\hat f_i, f_i}\,
\dualL\brax{\hat r_i}\ket{\chi_{f_i}(p_i,s_i)}
\;.
\end{split}
\end{equation}
Comparing this to Eq.~(\ref{eq:InitialKetMatrixElement}), we see that ${}_\scD\!\ibrax{\{\hat w\}_m} \ket{\chi(\{p,f,c,s\}_m)}$ in the ket amplitude is created by Born Feynman graphs for the operator $\cH(0)$ in which a color singlet combination $m$ quark, antiquark and gluon field operators combine at a point $x = 0$. The field operators have color indices specified by $\brax{\{\hat a\}_{m}}\ket{\{c\}_{m}}$ and vector/spinor indices specified by ${}_\scD\!\ibrax{\hat r_i}\ket{\chi_{f_i}(p_i,s_i)}$.

%-----------------------------------------
\subsection{The complete Feynman diagram}
\label{sec:CompleteDiagram}

The Feynman diagrams $G_\mathrm{ket}$ begin with the point vertex created by the multiparton operator $\cH(0)$. The simplest diagrams are the diagrams $\Gamma_\scH$ that include the multiparton point vertex and are one particle irreducible. The graph for the amplitude $G_\mathrm{ket}$ in Fig.~\ref{fig:d11exampleB} is an example of such a graph. (To make this more evident visually, one should join all of the lines ending in an $\otimes$ symbol at a single point.) The graph for the bra diagram $G_\mathrm{bra}$ in Fig.~\ref{fig:d11exampleB} is one particle reducible because it can be decomposed into two graphs by cutting the parton line that emits the gluon.

We can define an operator $\bm \Gamma_\scH$ that creates the graphs $\Gamma_\scH$ beyond the Born graph:
\begin{equation}
\label{eq:Gamma-Gamma-H-expand}
\bm \Gamma_\scH = \bm 1 
+ \sum_{n_\scR = 0}^\infty \sum_{n_\scV = 1}^\infty
\left[\frac{\as}{2\pi}\right]^{n_\scR/2 + n_\scV}
\bm \Gamma_\scH^{(n_\scR, n_\scV)}
\;.
\end{equation}
The operators $\bm \Gamma_\scH^{(n_\scR, n_\scV)}$ attach real and virtual radiation to the $\cH(x)$ vertex in such a way that the operator $\bm \Gamma_\scH$ represents only 1PI graphs. We will discuss the $\bm \Gamma_\scH^{(n_\scR, n_\scV)}$ operators in detail in Sec.~\ref{sec:Gamma-H-operators}.

A general amputated Feynman graph $F_\mathrm{ket}$ that contains the vertex $\cH(0)$ consists of a 1PI graph $\Gamma_\scH$ that contains the multiparton vertex $\cH(0)$ convolved with a graph $W_\mathrm{ket}$ that contains parton emissions and QCD vertex and self-energy loop corrections. Using operators on the quantum ket amplitude space, this structure is
\begin{equation}
\label{eq:FisWGammaH}
\bm F = \bm W\, \bm \Gamma_\scH
\;.
\end{equation}
The operator $\bm W$ is built from standard 1PI irreducible subgraphs for loop corrections to QCD vertices and propagators together with parton emission vertices. We discuss the details of this in the next section.

The infrared sensitive operator $\cD$ for the shower is defined based on the graphs generated by $\bm F$ times self-energy graphs generated by an operator $\bm L$ that supplies the LSZ factors needed to create an S-matrix element from an amputated graph with on-shell outgoing partons. For graphs $F_\mathrm{ket}$ for the Feynman ket amplitude, there is a final state vector $\ibra{\psi(\{\hat p,\hat f,\hat c,\hat s\}_{\hat m})}$ in the quantum ket amplitude space specified by Eq.~(\ref{eq:wavefctnOUTket}). With this vector,
\begin{equation}
\begin{split}
\bra{\psi(\{\hat p,\hat f,\hat c,\hat s\}_{\hat m})}\bm L
\ket{\{w\}_{\hat m}}\hskip - 3 cm {}&
\\
={}&\brax{\psi(\{\hat p,\hat f,\hat c,\hat s\}_{\hat m})}
\ket{\{w\}_{\hat m}}
\prod_{l=1}^{\hat m} \sqrt{R(\hat p_l,\hat f_l)} 
\;,
\end{split}
\end{equation}
where $R(\hat p_l, \hat f_l)$ is the residue of the propagator pole at $\hat p_l^2 = 0$, which we have calculated at order $\as$ in Ref.~\cite{Gauge}. The complete matrix element for the quantum ket amplitude is then generated by the operator
\begin{equation}
\label{eq:GisLWGammaH}
\bm G = 
\bm L\,\bm F = \bm L\, \bm W\, \bm\Gamma_\scH
\;.
\end{equation}
%

%-------------------------------------
\subsection{Operators for 1PI graphs}

We need a tool set to build the  operator $\qW$ that was introduced in Eq.~\eqref{eq:FisWGammaH}. The graphs created by $\qW$ contain parton emissions and loop corrections to propagators and QCD vertices. We can build these graphs from standard 1PI graphs, but first we have to describe these graphs using our operator formalism. 

We define an operator $\qGamma_l^{(n_\scR, n_\scV)}$ that describes a 1PI graph in which parton $l$ initiates an interaction that contains $n_\scV$ virtual loops in which $n_\scR$ partons are emitted. Its non-zero matrix elements are
\begin{equation}
\begin{split}
\label{eq:Gamma-gen-matrix-element}
&\left[\aspi\right]^{n_\scR/2 + n_\scV} \dualL\bra{\{\bar w\}_{m+n_\scR}}
\qGamma^{(n_\scR,n_\scV)}_l\ket{\{w\}_{m}}
\\
&\qquad
=\prod_{\substack{i=1\\i\neq l}}^m
\left[(2\pi)^d\delta^d(\bar q_i - q_i)\,\delta_{\bar f_i,f_i} \delta_{\bar a_i,a_i}\delta_{\bar r_i,r_i}\right]
\\
&\qquad\quad
\times (2\pi)^d \delta^d\!\left(q_l -\sum_{k\in R} \bar q_k\right)
\\
&\qquad\quad
\times\Gamma^{(n_\scV)}\big(w_l; \{\bar{w}_i\;|\; i\in R\}\big) 
\;.
\end{split}
\end{equation}
Here parton $l$ splits into $n_\scR + 1$ partons with labels $l$ and $ m+1,...,m+n_\scR$. The label set $R$ is defined by 
\begin{equation}
R = \{l\}\cup \{m+1,...,m+n_\scR\}
\;.
\end{equation}
When $n_\scR = 0$ then $R=\{l\}$.

In the first line on the right hand side of Eq.~\eqref{eq:Gamma-gen-matrix-element} we have delta functions that say that the states of the non-interacting partons are unchanged. In the second line we have the momentum conservation delta function for the interacting partons. The function $\Gamma^{(n_\scV)}_l$ in the last line is defined by the 1PI Feynman graph with $n_\scR+1$ external legs at $n_\scV$ loop level: 
\begin{equation}
\begin{split}
    \Gamma^{(n_\scV)}_l\big(w_l;& \{\bar{w}_i\;|\; i\in R\} \big) = 
    \\
    &
    \begin{prdfig}{52f11f447d44e6bac3d6688355eb753e}{Gamma-gen1}
      \begin{tikzpicture}[baseline=(current bounding box.center)]
        \begin{feynman}[]
          \vertex[blob, fill=none, minimum size=1cm] (v1) 
          at (0,0) {1PI};
          \vertex[empty dot,label={[left] $w_l$}] (w1) 
          at ($(v1) + (180:1.5cm)$) {};
          \coordinate [label={[right] $\bar{w}_l$}] (u1) 
          at ($(v1) + (45:1.5cm)$);
          \coordinate [label={[right] $\bar{w}_{m\!+\!1}$}] (um1) 
          at ($(v1) + (15:1.5cm)$);
          \coordinate [label={[right] $\bar{w}_{m+n_\scR}$}] (umr) 
          at ($(v1) + (-45:1.5cm)$);
          \diagram*{
            (w1)--[](v1);
            (v1)--[](u1);
            (v1)--[](um1);
            (v1)--[](umr);
          };      
          % decorations
          \vertex [rotate=90] at ($0.5*(um1) + 0.5*(umr)$) {. . .};
        \end{feynman}
      \end{tikzpicture}
    \end{prdfig}
    \;\;. 
\end{split}
\end{equation}
On the left hand side of the diagram we have the incoming parton, $l$, which is connected to a vertex in the $\iket{\{w\}_m}$ state through a tree level propagator. This connection is represented by the small empty circle. On the
right hand side, we have the $n_\scR + 1$ outgoing partons (with no propagators). The function $\Gamma^{(n_\scV)}_l$ is fully symmetric in the $\{\bar{w}_i\;|\; i\in R\}$ variables. 

In the operator $\qGamma_l^{(0,n_\scV)}$ there is no emitted parton. This operator represents the  self-energy graph at order $n_\scV$ multiplied in the initial state by a tree level propagator. The full self energy (times a tree level propagator) is
\begin{equation}
\label{eq:selfenergy-operator}
\qSigma_l = \sum_{n_\scV = 1}^\infty\left[\aspi\right]^{n_\scV} 
\qGamma^{(0,n_\scV)}_l
\;.
\end{equation}
Then the full all-order propagator with a tree level propagator removed in the final state is created by the operator
\begin{equation}
\label{eq:qPl}
\qP_l = \bm1+\qSigma_l + \qSigma_l\qSigma_l
+ \qSigma_l\qSigma_l\qSigma_l +\cdots 
= \frac{\bm1}{\bm1-\qSigma_l}
\;.
\end{equation}

Now we can define an operator $\qW_l^{(n_\scR)}$ that describes parton emissions from line $l$. This operator starts with a full propagator for line $l$ followed by the emission of $n_\scR$ partons using a 1PI graph with any number of virtual loops. These operators can be multiplied sequentially, so that, for instance, $\qW_l^{(3)}\qW_l^{(2)}$ describes graphs with propagation on line $l$ followed by two parton emissions followed by propagation followed by three parton emissions. Thus we define
\begin{equation}
\begin{split}
\label{eq:F-operators}
\qW_l^{(1)} ={}&  \bigg(\qGamma^{(1,0)}_l 
+ \sum_{{n_\scV}=1}^\infty\left[\aspi\right]^{n_\scV}
\qGamma^{(1,{n_\scV})}_l\bigg)\frac{\bm1}{\bm1-\qSigma_l}\;,
\\
\qW_l^{(2)} ={}& \bigg(\qGamma^{(2,0)}_l 
+ \sum_{{n_\scV}=1}^\infty\left[\aspi\right]^{n_\scV}
\qGamma^{(2,{n_\scV})}_l\bigg)\frac{\bm1}{\bm1-\qSigma_l}\;,
\\
\qW_l^{(n_\scR)}={}& \sum_{{n_\scV}=1}^\infty\left[\aspi\right]^{n_\scV}
\qGamma^{(n_\scR,{n_\scV})}_l\frac{\bm1}{\bm1-\qSigma_l}
\;.
\end{split}
\end{equation}
For $n_\scR > 2$ there is no elementary vertex, so that the series expansion of $\qW_l^{(n_\scR)}$ starts with the $\as$ term. 

In the NLO shower we need only a handful of these partial operators to define the operator $\qW$. They are
\begin{equation}
\begin{split}
\label{eq:Fs-NLO-expansion}
\qSigma_l  ={}&  \aspi \,
\qGamma^{(0,{1})}_l + \cO(\as^2)
\;,
\\
\qW_l^{(1)} ={}& \qGamma^{(1,0)}_l 
\\&
+ \aspi\left(\qGamma^{(1,1)}_l 
+  \qGamma^{(1,0)}_l \qGamma^{(0,1)}_l\right) + \cO(\as^2)
\;,
\\
\qW_l^{(2)} ={}& \qGamma^{(2,0)}_l + \cO(\as)
\;.
\end{split}
\end{equation}

Now we have all the building blocks for the operator $\qW$. In this operator we include all possibilities when one or more partons split. We also include self energy corrections on the initial state line of the graph, which are part of the operators $\qW_{l}^{(n_\scR)}$. In $\qW$, we cannot have partons exchanged between different parton lines, say $l$ and $k$, because that would create a new 1PI subgraph that includes the vertex $\cH(0)$. Such a 1PI graph is generated by  $\bm\Gamma_\scH$ rather than $\qW$. 

For an order $\as^2$ shower, we need only contributions in which one or two partons are emitted. These are\footnote{With more than two emitted partons, more complicated combinations of the $\qW_l^{(n_\scR)}$ are needed.}
\begin{equation}
\begin{split}
\qW ={}& \bm{1} + \sum_{l=1}^m \sum_{n_\scR = 1}^2
 \left[\aspi\right]^{n_\scR/2}
 \qW_l^{(n_\scR)}
\\& 
+ \aspi \sum_{l=1}^m \sum_{k=1}^{m+1}
\left[ 1 + \delta_{kl}\, \bm \Pi\right]
\qW_k^{(1)}\qW_l^{(1)}
\\&
+ \cdots
\;.
\end{split}
\end{equation}
Here $\bm \Pi$, Eq.~(\ref{eq:PilandPi}), exchanges partons ${m\!+\!1}$ and ${m\!+\!2}$. After expanding this in $\as$ and using Eq.~\eqref{eq:Fs-NLO-expansion} we obtain
\begin{equation}
\begin{split}
\label{eq:G-for-NLO-shower}
\qW = \bm{1} &+ \left[\aspi\right]^{1/2}\sum_{l=1}^m  \qGamma^{(1,0)}_l
\\
&
+ \left[\aspi\right]^{3/2}\sum_{l=1}^m \left(\qGamma^{(1,1)}_l+ \qGamma^{(1,0)}_l \qGamma^{(0,1)}_l\right)
\\
&
+ \aspi\sum_{l=1}^m \qGamma^{(2,0)}_l
+ \aspi\sum_{\substack{l,k=1\\ k\neq l}}^m 
\qGamma^{(1,0)}_k\qGamma^{(1,0)}_l
\\
&
+ \aspi\sum_{l=1}^m \left(1+\bm{\Pi}\right)\qGamma^{(1,0)}_l\qGamma^{(1,0)}_l
\\
&+ \aspi\sum_{l=1}^m  \qGamma^{(1,0)}_{m\!+\!1}\qGamma^{(1,0)}_l
+ \cdots
\;.
\end{split}
\end{equation}
Here the first term is the trivial unit operator. The second term gives all the possible single real emission graphs. The third term describes all the graphs with one real emission and one virtual loop correction. We can have either a vertex correction in $\qGamma^{(1,1)}_l$ or a self energy on the initial state line, $\qGamma^{(1,0)}_l\qGamma^{(0,1)}_l$. In the fourth term there is the contribution to two parton emissions from the elementary four-point vertex, $\qGamma^{(2,0)}_l$.  The last three terms are the contributions to the double real emission graphs from two separate single parton emissions. In the fifth term we have all the graphs with two independent emitters, $l$ and $k$.  In the sixth term a single hard line emits two partons. The last term we have the graph in which parton $l$ emits parton $m\!+\!1$, which then emits parton $m\!+\!2$. In Fig.~\ref{fig:G-for-nlo-shower} we illustrated the non-zero and non-trivial matrix elements of the operator $\qW$ in Eq.~\eqref{eq:G-for-NLO-shower}. 

The omitted terms in Eq.~(\ref{eq:G-for-NLO-shower}) have a higher power of $\as$ than $\as^{3/2}$ for one parton emitted or a higher power of $\as$ than $\as^1$ for two partons emitted or have more than two partons emitted. When combined with the quantum bra amplitude, the terms retained are all that are required for an NLO shower.

%%%%%%%%%%%%%%%%%%%% FIGURE %%%%%%%%%%%%%%%%%%%%%%%%%%
% -------------------- Figure -----------------------------
\begin{figure*}[tb]
 \centering
 \begin{equation*}
   \begin{split}
     \dualL\bra{\{\bar{w}\}_{m\!+\!1}}{}&\bm{W}\ket{\{w\}_m}  
       \\
      = {}&
      \sqrt{\frac{\as}{2\pi}}\sum_{l=1}^m \Biigg[2cm]{\{}
      \begin{prdfig}{5a89aa90871a4f9c84556fa05961da4d}{G2-real-term}
      \begin{tikzpicture}[baseline={(current bounding box.center)}]
        \begin{feynman}[] 
          \vertex[empty dot,label={[above] $w_l$}] (v1) 
          at (0,0)  {};
          \vertex[dot] (v2) at ($(v1)+(1cm,0)$) {};
          \coordinate [label={[above] $\bar{w}_l$}] (i) 
          at ($(v2) + (35:1cm)$);
          \coordinate [label={[below] $\bar{w}_{m\!+\!1}$}] (j) 
          at ($(v2) + (-35:1cm)$);
          \diagram*{
            (v1)--[](v2);
            (v2)--[](i);
            (v2)--[](j);
          };      
        \end{feynman}
      \end{tikzpicture}
    \end{prdfig}
      + \aspi \Biigg[1.3cm][
      \begin{prdfig}{8c56094f45fee6412566bdfc8f3e7485}{G2-virtual-term1}
        \begin{tikzpicture}[baseline=(current bounding box.center)]
          \begin{feynman}[]
            \vertex[empty dot,label={[above] $w_l$}] (ol) at (0,0) {};
            \vertex[blob, pattern=none, fill=none, minimum size=0.65cm] (vl) 
            at ($(ol)+(1cm,0)$) {1PI};
            \coordinate [label={[above] $\bar{w}_l$}] (l) 
            at ($(vl) + (35:1cm)$);
            \coordinate [label={[below] $\bar{w}_{m\!+\!1}$}] (m1) 
            at ($(vl) + (-35:1cm)$);  
            \diagram*{
              (ol)--[](vl);
              (vl)--[](l);
              (vl)--[](m1);
            };      
          \end{feynman}
        \end{tikzpicture}
      \end{prdfig}
      +
      \begin{prdfig}{598c4b672421f743e32adcf2e94a656d}{G2-virtual-term2}
        \begin{tikzpicture}[baseline=(current bounding box.center)]
          \begin{feynman}[]
            \vertex[empty dot,label={[above] $w_l$}] (ol) at (0,0) {};
            \vertex[blob, pattern=none, fill=none, minimum size=0.65cm] (sl) 
            at ($(ol)+(0.7cm,0)$) {1PI};
            \vertex[dot] (vl) at ($(sl)+(0.7cm,0)$) {};
            \coordinate [label={[above] $\bar{w}_l$}] (l) 
            at ($(vl) + (35:1cm)$);
            \coordinate [label={[below] $\bar{w}_{m\!+\!1}$}] (m1) 
            at ($(vl) + (-35:1cm)$);  
            \diagram*{
              (ol)--(sl)--[](vl);
              (vl)--[](l);
              (vl)--[](m1);
            };      
          \end{feynman}
        \end{tikzpicture}
      \end{prdfig}
      \Biigg[1.3cm]] + \cO(\as^2) \Biigg[2cm]{\}}
      \\
      ~\\
      \dualL\bra{\{\bar{w}\}_{m\!+\!2}}{}&\bm{W}\ket{\{w\}_m} 
      \\
      = {}&
     \aspi\sum_{l=1}^m
     \begin{prdfig}{c7586c55c94490548a097f0471cd0925}{G2-real2-term2}
       \begin{tikzpicture}[baseline=(current bounding box.center)]
         \begin{feynman}[]
           \vertex[empty dot,label={[above] $w_l$}] (ol) at (0,0) {};
           \vertex[dot] (vl) at ($(ol)+(1cm,0)$) {};
           \coordinate[label={[above] $\bar{w}_l$}] (l) 
           at ($(vl) + (45:1cm)$);
           \coordinate[label={[below] $\bar{w}_{m\!+\!2}$}] (m1) 
           at ($(vl) + (-45:1cm)$);  
           \coordinate[label={[right] $\bar{w}_{m\!+\!1}$}] (m2) 
           at ($(vl) + (0:1cm)$);  
           \diagram*{
             (ol)--[](vl);
             (vl)--[](l);
             (vl)--[](m1);
             (vl)--[](m2);
           };      
         \end{feynman}
       \end{tikzpicture}
     \end{prdfig}
     +\quad \aspi\sum_{\substack{l,k=1\\k\neq l}}^m
     \begin{prdfig}{387879e987282dee691ffab83d0a7e17}{G2-real2-term1}
       \begin{tikzpicture}[baseline=(current bounding box.center)]
         \begin{feynman}[]
           \vertex[empty dot,label={[above] $w_l$}] (ol) 
           at (0,0) {};
           \vertex[dot] (vl) at ($(ol)+(1cm,0)$) {};
           \coordinate[label={[right] $\bar{w}_l$}] (l) 
           at ($(vl) + (35:1cm)$);
           \coordinate[label={[right] $\bar{w}_{m\!+\!1}$}] (m1) 
           at ($(vl) + (-35:1cm)$);  
           \vertex[empty dot,label={[above] $w_k$}] (ok) 
           at ($(ol)+(0,-1.5cm)$) {};
           \vertex[dot] (vk) at ($(ok)+(1cm,0)$) {};
           \coordinate[label={[right] $\bar{w}_k$}] (k) 
           at ($(vk) + (35:1cm)$);
           \coordinate[label={[right] $\bar{w}_{m\!+\!2}$}] (m2) 
           at ($(vk) + (-35:1cm)$);  
           \diagram*{
             (ol)--[](vl);
             (vl)--[](l);
             (vl)--[](m1);
             (ok)--[](vk);
             (vk)--[](k);
             (vk)--[](m2);
           };      
         \end{feynman}
       \end{tikzpicture}
     \end{prdfig}
     \\
     &
     + \aspi\sum_{l=1}^m\Biigg[2cm]\{
     \begin{prdfig}{ec98078e8b3c7a14941f5991b9051fb6}{G2-real2-term3}
       \begin{tikzpicture}[baseline=(current bounding box.center)]
         \begin{feynman}[]
           \vertex[empty dot,label={[above] $w_l$}] (ol) at (0,0) {};
           \vertex[dot] (v1) at ($(ol)+(1cm,0)$) {};
           \vertex[dot] (v2) at ($(v1)+(1cm,0)$) {};
           \coordinate[label={[above] $\bar{w}_l$}] (l) 
           at ($(v2) + (35:1cm)$);
           \coordinate [label={[below] $\bar{w}_{m\!+\!1}$}] (m1) 
           at ($(v1) + (-35:1cm)$);  
           \coordinate [label={[below] $\bar{w}_{m\!+\!2}$}] (m2) 
           at ($(v2) + (-35:1cm)$);  
           \diagram*{
             (ol)--[](v1)--(v2)--(l);
             (v2)--[](m2);
             (v1)--[](m1);
           };      
         \end{feynman}
       \end{tikzpicture}
     \end{prdfig}
     \!\! + \quad
     \begin{prdfig}{b6055ed9dc57d675e4ec5e6939fa9acd}{G2-real2-term4}
       \begin{tikzpicture}[baseline=(current bounding box.center)]
         \begin{feynman}[]
           \vertex[empty dot,label={[above] $w_l$}] (ol) at (0,0) {};
           \vertex[dot] (v1) at ($(ol)+(1cm,0)$) {};
           \vertex[dot] (v2) at ($(v1)+(1cm,0)$) {};
           \coordinate[label={[above] $\bar{w}_l$}] (l) 
           at ($(v2) + (35:1cm)$);
           \coordinate [label={[below] $\bar{w}_{m\!+\!2}$}] (m1) 
           at ($(v1) + (-35:1cm)$);  
           \coordinate [label={[below] $\bar{w}_{m\!+\!1}$}] (m2) 
           at ($(v2) + (-35:1cm)$);  
           \diagram*{
             (ol)--[](v1)--(v2)--(l);
             (v2)--[](m2);
             (v1)--[](m1);
           };      
         \end{feynman}
       \end{tikzpicture}
     \end{prdfig}
     \!\! + \quad
     \begin{prdfig}{6160c49e8ae1e95fdd1d43ebbd40435b}{G2-real2-term5}
      \begin{tikzpicture}[baseline=(current bounding box.center)]
        \begin{feynman}[]
          \vertex[empty dot,label={[above] $w_l$}] (ol) at (0,0) {};
          \vertex[dot] (v1) at ($(ol)+(1cm,0)$) {};
          \vertex[dot] (v2) at ($(v1)+(1cm,0)$) {};
          \coordinate [label={[above] $\bar{w}_{m\!+\!1}$}] (l) 
          at ($(v2) + (35:1cm)$);
          \coordinate [label={[above] $\bar{w}_{l}$}] (m1) 
          at ($(v1) + (35:1cm)$);  
          \coordinate [label={[below] $\bar{w}_{m\!+\!2}$}] (m2) 
          at ($(v2) + (-35:1cm)$);  
          \diagram*{
            (ol)--[](v1)--(v2)--(l);
            (v2)--[](m2);
            (v1)--[](m1);
          };      
        \end{feynman}
      \end{tikzpicture}
    \end{prdfig}
    \Biigg[2cm]\}
    \\
    & + \cO(\as^2)
  \end{split}
\end{equation*}
\caption{\label{fig:G-for-nlo-shower}
Illustration of the matrix elements of the operator $\qW$ that are needed for an NLO shower. The lines for partons that do not split are omitted.
}
\end{figure*}
% -------------------- Figure -----------------------------
%%%%%%%%%%%%%%%%%%% END FIGURE %%%%%%%%%%%%%%%%%%%%%%%%

\subsection{The propagator and LSZ factor}
\label{sec:LSZ}

Recall from Eq.~(\ref{eq:qPl}) that the propagator for parton $l$ with a tree level propagator on the final state removed is given by $\bm P_l = 1/(1 - \bm \Sigma)$ where $\qSigma$ is the self energy times a tree level propagator on the initial state side, which has the expansion
\begin{equation}
\qSigma_l = \sum_{n_\scV = 1}^\infty\left[\aspi\right]^{n_\scV} 
\qGamma^{(0,n_\scV)}_l
\;.
\end{equation}
Specializing Eq.~(\ref{eq:Gamma-gen-matrix-element}) to the case of a two point function, $\qGamma_l^{(0,n_\scV)}$ is
\begin{equation}
  \label{eq:Gammal-0n}
  \begin{split}
    \left[\frac{\as}{2\pi}\right]^{n_\scV}{}&
    \dualL\bra{\{\bar w\}_{m}}
    \qGamma^{(0,n_\scV)}_l\ket{\{w\}_{m}}
    \\
    &=
    \prod_{i\neq l}\left[
      (2\pi)^d \delta^d(\bar q_i - q_i)\,\delta_{\bar f_i,f_i}
      \delta_{\bar a_i,a_i}\delta_{\bar r_i,r_i}
    \right]
    \\
    &\quad
    \times (2\pi)^d \delta^d(\bar q_i - q_i)\,
    \Gamma^{(n_\scV)}\big(w_l; \{\bar w_l\}\big)
    \;.
  \end{split}
\end{equation}
The function $\Gamma^{(n_\scV)}(w_l; \{\bar w_l\})$ is the self energy times a tree level propagator at the $n_\scV$ loop level. In this paper we list this function only at 1-loop level but for the NLO shower we need it also at the 2-loop level. 

For a quark line, with $f_l \in \{u,d,\dots\}$, we have, including the $\MSbar$ renormalization counter term,
\begin{equation}
\begin{split}
    \Gamma^{(1)}\big(&w_l; \{\bar w_l\}\big)
    \\ ={}& 
      \begin{prdfig}{89798e82ce35ae8df75bcd2f9b7c2d49}{quark-selfenery}
        \begin{tikzpicture}[baseline=(v1.base)]
          \begin{feynman}[]
            \vertex[empty dot,label={[below=2mm] $a_l, r_l$}] (v1) 
            at (0,0) {};
            \vertex[dot] (v2) at ($(v1)+(1.5cm,0)$) {};
            \vertex[dot] (v3) at ($(v2)+(2cm,0)$) {};
            \coordinate [label={[below] $\bar{a}_l, \bar{r}_l$}] (i) 
            at ($(v3) + (1.5cm,0)$);
            \diagram*{
              (v1)--[fermion, rmomentum'={$q_l$}]
              (v2)--[fermion,rmomentum={$ q_l-\ell$}](v3);
              (v3)--[gluon, half right, momentum={[black] $\ell$}](v2);
              (v3)--[fermion, rmomentum'={$q_{l}$}](i);
            };
          \end{feynman}
        \end{tikzpicture}
      \end{prdfig}
      \\
      &+
      \begin{prdfig}{afec493e9e759851046e651e91124a23}{quark-selfenery-cnt}
        \begin{tikzpicture}[baseline=(v1.base)]
          \begin{feynman}[]
            \vertex[empty dot,label={[below=2mm] $a_l, r_l$}] (v1) 
            at (0,0) {};
            \vertex[crossed dot, label={[above]$(1)$}] (v2) 
            at ($(v1)+(1.5cm,0)$) {};
            \coordinate [label={[below] $\bar{a}_l, \bar{r}_l$}] (i) 
            at ($(v2) + (1.5cm,0)$);
            \diagram*{ 
              (v1)--[fermion, rmomentum'={$q_{l}$}]
              (v2)--[fermion, rmomentum'={$q_{l}$}](i);
            }; 
          \end{feynman}
        \end{tikzpicture}
      \end{prdfig}
      \\
      ={}& \delta_{\bar f_l,f_l}\delta_{a_l,\bar{a}_l}
      \frac{\big[\Sigma^{(1)}(q_l)\s{q}_l\big]_{\bar{r}_l,r_l}}
      {q_l^2+\mi0}
\;.
\end{split}
\end{equation}
The self-energy diagram shown is renormalized using $\MSbar$ renormalization. The counter term is indicated by the $\otimes$ vertex.

For an anti-quark line we have
\begin{equation}
  \Gamma^{(1)}\big(w_l; \{\bar w_l\}\big)
  = \delta_{\bar f_l,f_l}\delta_{a_l,\bar{a}_l}
  \frac{\big[\s{q}_l\Sigma^{(1)}(q_l)\big]_{r_l,\bar{r}_l}}{q_l^2+\mi0}
\;.
\end{equation}
Here the function $-\mi\Sigma^{(1)}(q_l)$ is the integrated one-loop quark self-energy. The rest of the expression is just the tree level propagator on the initial state side of the graph.  The self-energy function $\Sigma^{(1)}(q)$ is analyzed in Ref.~\cite{Gauge}. 

\begin{figure*}[htb]
  \centering
  \begin{equation*}
    \begin{split}
     \Gamma^{(1)}\big(w_l; \{\bar w_l\}\big) ={}&
      \begin{prdfig}{c9236c0d818b9b946783c3d45b574e62}{gluon-selfenergy-cnt}
        \begin{tikzpicture}[baseline=(a.base)]
          \begin{feynman}
            \vertex [crossed dot, label={[above]$(1)$}] (a) {};
            \vertex [empty dot, left=1.5cm of a, 
            label={[below=2mm] $a_l, r_l$}](c) {};
            \vertex [right=1.5cm of a, 
            label={[below=2mm] $\bar a_l, \bar r_l$}](e) {};  
            \diagram*{
              (c) --[gluon, rmomentum'={$q_l$}]
              (a) --[gluon, rmomentum'={$q_l$}](e);
            };
          \end{feynman}
        \end{tikzpicture}
      \end{prdfig}
      + 
      \begin{prdfig}{bae76a183b89d5c4ecb0d756b5f0e284}{gluon-selfenergy-term1}
        \begin{tikzpicture}[baseline=(a.base)]
          \begin{feynman}
            \vertex [dot] (a) {};
            \vertex [empty dot, left=1.3cm of a, 
            label={[below=2mm] $a_l, r_l$}](c) {};
            \vertex at ($(a) + (1.6cm, 0cm)$)[dot] (d) {};
            \vertex [right=1.3cm of d, 
            label={[below=2mm] $\bar a_l, \bar r_l$}](e) {};  
            \diagram*{
              (a) -- [gluon, half left, rmomentum'={[arrow shorten=0.7]
              $\ell$}](d);
              (d) -- [gluon, half left, momentum'={[arrow shorten=0.7]
              $q_l-\ell$}](a);
              (c) -- [gluon, rmomentum'={$q_l$}](a); 
              (d) -- [gluon, rmomentum'={$q_l$}](e);
            };
          \end{feynman}
        \end{tikzpicture}
      \end{prdfig}
      \\
      &
      +
      \begin{prdfig}{ac183798c7a83feb32065531939cfea3}{gluon-selfenergy-term2}
        \begin{tikzpicture}[baseline=(a.base)]
          \begin{feynman}
            \vertex [dot] (a) {};
            \vertex [empty dot, left=1.3cm of a, 
            label={[below=2mm] $a_l, r_l$}](c) {};
            \vertex at ($(a) + (1.6cm, 0cm)$)[dot] (d) {};
            \vertex [right=1.3cm of d, 
            label={[below=2mm] $\bar a_l, \bar r_l$}](e) {};  
            \diagram*{
              (a) -- [ghost, half left, rmomentum'={[arrow shorten=0.7]
              $\ell$}](d);
              (d) -- [ghost, half left, momentum'={[arrow shorten=0.7]
              $q_l-\ell$}](a);
              (c) -- [gluon, rmomentum'={$q_l$}](a); 
              (d) -- [gluon, rmomentum'={$q_l$}](e);
            };	
          \end{feynman}
        \end{tikzpicture}
      \end{prdfig}
      +
      \begin{prdfig}{08be71173b061a33d44e2f59ba908a4a}{gluon-selfenergy-term3}
        \begin{tikzpicture}[baseline=(a.base)]
          \begin{feynman}
            \vertex [dot] (a) {};
            \vertex [empty dot, left=1.3cm of a, label={[below=2mm] 
            $a_l, r_l$}](c) {};
            \vertex at ($(a) + (1.6cm, 0cm)$)[dot] (d) {};
            \vertex [right=1.3cm of d, label={[below=2mm] 
            $\bar a_l, \bar r_l$}](e) {};  
            \diagram*{
              (a) -- [fermion, half left, rmomentum'={[arrow shorten=0.7]
              $\ell$}](d);
              (d) -- [fermion, half left,  momentum'={[arrow shorten=0.7]
              $q_l-\ell$}](a);
              (c) -- [gluon, rmomentum'={$q_l$}](a);
              (d) -- [gluon, rmomentum'={$q_l$}](e);
            };	
          \end{feynman}
        \end{tikzpicture}
      \end{prdfig}
    \end{split}
  \end{equation*}
\caption{\label{fig:selfenergies}
The matrix element $\Gamma^{(1)}\big(w_l; \bar w_l\big)$ for the gluon self energy, including a propagator for the initial state parton. 
  }
\end{figure*}
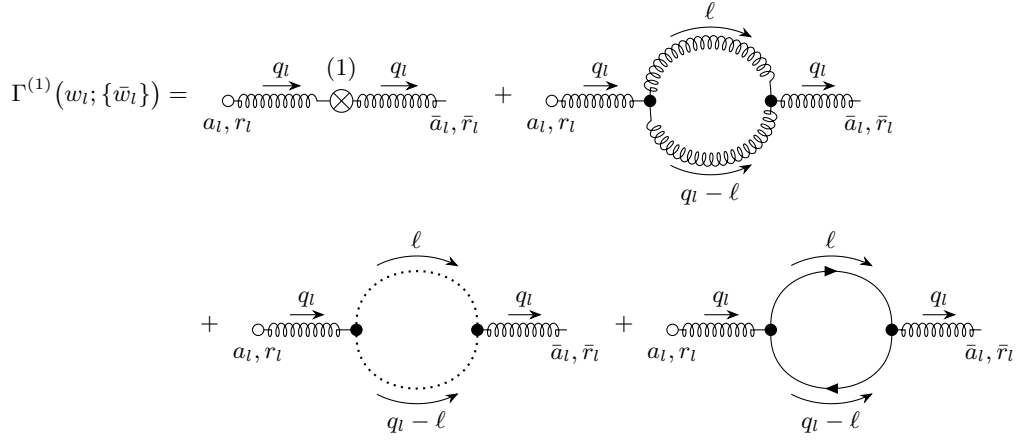

In the gluon case, we have an analogous expression,
\begin{equation}
\begin{split}
&\Gamma^{(1)}\big(w_l; \{\bar w_l\}\big)
\\
&\quad
= \delta_{\bar f_l,f_l}\delta_{a_l,\bar{a}_l}
\big[\Pi^{(1)}(q_l)\!\cdot\!D(q_l)\big]_{r_l,\bar{r}_l}
\;,
\end{split}
\end{equation}
where the function $-\mi\Pi^{(1)}_{\mu\nu}(q_l)$ is the integrated one-loop gluon self-energy graph and $D_{\mu\nu}(q_l)$ is the gluon propagator in interpolating gauge, defined in Eq.~\eqref{eq:gluonpropagator}. This is illustrated in Fig.~\eqref{fig:selfenergies}. The self-energy function $-\mi\Pi^{(1)}_{\mu\nu}(q)$ is analyzed and calculated in Ref.~\cite{Gauge}. 

%%%%%%%%%%%%%%%%%%%%%%% FIGURE %%%%%%%%%%%%%%%%%%%%%%
% -------------------- Figure -----------------------------
% -------------------- Figure -------------------
%%%%%%%%%%%%%%%%%%%%%%% END FIGURE %%%%%%%%%%%%%%%%%%%%%%

In order to form an S-matrix ket amplitude from a Green function amputated on its final state external legs, we need an LSZ operator $\bm L_l$ for each final state parton $l$. The operator $\qL_l$ inserts the square root of a propagator with a final tree-level propagator removed on line $l$. The operator $\qL_l$ acts on a final state parton and is multiplied by the final state wave function $\ibra{\psi(\{p,f,c,s\}_m)}$ defined in Eq.~(\ref{eq:wavefctnOUTket}). Taking this inner product sets the parton momentum $q_l$ to $p_l$, where $p_l^2 =0$. This gives the LSZ factor needed to make an S-matrix ket amplitude. 

We note from Ref.~\cite{Gauge} that the gluon propagator splits into a propagator for T gluons and a propagator for L gluons:
\begin{equation}
P_l = P_l^\LT + P_l^\LL
\;.
\end{equation}
The propagator $D^\LT_l(q_l) P_l^\LT(q_l)$ for T gluons has a pole at $q_l^2 = 0$, while the propagator for L gluons has a pole at $q_l\cdot\tilde q_l = 0$ but not at $q_l^2 = 0$.

For an L gluon with momentum $p_l$, the self energy $\Pi_\LL^{\mu\nu}(p_l)$ has terms proportional to $p_l^\mu p_l^\nu$, $n^\mu n^\nu$, and $p_l^\mu n^\nu + n^\mu p_l^\nu$. Thus $\varepsilon_\mu(p_l,s_l)\, \Pi_\LL^{\mu\nu}(p_l) = 0$, so that $\ibra{\psi(\{p,f,c,s\}_m)}P_l^\LL = 0$. Thus $P_l^\LL$ does not contribute to the LSZ factor.

The perturbative expansion of the operator $\bm L_l$ is defined by
\begin{equation}
\begin{split}
\label{eq:qLlpert}
\qL_l ={}& \sqrt{\qP_l}
\\
={}& \sqrt{\frac{1}{1 - \bm \Sigma}}
\\
={}& 1 + \aspi\, \frac{1}{2}\,\qSigma^{(1)}
\\&
+ \left[\aspi\right]^2
\left\{
\frac{1}{2}\,\qSigma^{(2)} + \frac{3}{8}\,\qSigma^{(1)}\qSigma^{(1)}
\right\}
\\&
+ \cO(\as^3)
\;.
\end{split}
\end{equation}

We use the matrix elements 
\begin{equation}
\begin{split}
&\bra{\psi(\{p,f,c,s\}_m)}\qL_l\ket{\{w\}_m}
\\
&\qquad\quad
=\brax{\psi(\{p,f,c,s\}_m)}\ket{\{w\}_m}\sqrt{R(p_l,f_l)}
\;,
\end{split}
\end{equation}
where $R(p_l,f_l)$ is the residue of the propagator pole at $p_l^2 =0$. In Ref.~\cite{Gauge}, we calculated their infrared singular part of $R(p,f)$ at first order, with the result 
\begin{equation}
\begin{aligned}[c]
\label{eq:Prop-residue}
&R(p,f) = 1-\aspi\frac{S_\epsilon}{\epsilon}\,\gamma_f
\\
&\quad
-\aspi\frac{S_\epsilon}{\epsilon}\, T^2_f
\left(
  \frac{1}{\epsilon}\left[\frac{\mur^2}{-\tdot{p}{p}}\right]^\epsilon
  +\frac{v-1}{v}
  -2\log\frac{2v}{1+v}
\right)
\\
&\quad
+  \cO(\epsilon^0)
\;.
\end{aligned}
\end{equation}
Here $T^2_f$ is the color charge, $T^2_q = T^2_{\bar q} = C_F$ and  $T^2_{\Lg} = C_A$, the constants $\gamma_f$ for gluons and quarks or antiquarks are
\begin{equation}
\begin{split}
\gamma_{\Lg} ={}& \frac{11}{6}\,C_A - \frac{2}{3}\,T_R n_f\;,
\\
\gamma_{q} ={}& \gamma_{\bar q} = \frac{3}{2}\, C_F
\;,
\end{split}
\end{equation}
and $S_\epsilon$ is the standard coefficient of $1/\epsilon$ for $\MSbar$ renormalization,
\begin{equation}
\label{eq:epsilonMSbar}
S_\epsilon = \frac{(4\pi)^\epsilon}{\Gamma(1-\epsilon)}
\;.
\end{equation}

The operators $\bm L_l$ multiply Green functions that depend on the gauge parameters $v$ and $n$. The dependence on these gauge parameters in the LSZ factors cancels the dependence of the Green functions, so that the S-matrix is gauge invariant \cite{Gauge}.

%-------------------------------------------------
\subsection{Vertex operators}
\label{sec:VertexOperators}

In Eq.~(\ref{eq:Gamma-gen-matrix-element}), we have defined vertex operators $\qGamma_l^{(n_\scR, n_\scV)}$ that describe 1PI graphs in which parton $l$ initiates an interaction that contains $n_\scV$ virtual loops in which $n_\scR$ partons are emitted. For a parton shower at LO or NLO, we need only $n_\scR  = 0$, 1, or 2. 

The operators for $n_\scR  = 0$ generate self-energy diagrams, which we described in the previous subsection.

The operator $\bm{\Gamma}^{(1,n_\scV)}_l$  describes the splitting of parton $l$ in an $m$ parton state $\ket{\{w\}_m}$ into two partons with labels $l$ and ${m\!+\!1}$. After the splitting, the partons have quantum numbers $\{\bar w\}_{m\!+\!1}=\{\bar q, \bar f, \bar a, \bar r\}_{m\!+\!1}$. Specializing Eq.~\eqref{eq:Gamma-gen-matrix-element}, the corresponding matrix element has the form,
\begin{equation}
\begin{split}
\label{eq:Gammal1n}
&\left[\frac{\as}{2\pi}\right]^{n_\scV +1/2}\,
\dualL\bra{\{\bar w\}_{m\!+\!1}}\qGamma^{(1,n_\scV)}_l\ket{\{w\}_{m}}
\\
&\qquad
= \prod_{i\ne l}\left[(2\pi)^d\delta^d(\bar q_i - q_i)\,\delta_{\bar f_i,f_i} \delta_{\bar a_i,a_i}\delta_{\bar r_i,r_i}\right]
\\
&\qquad\quad
\times (2\pi)^d \delta^d(\bar q_l + \bar q_{m\!+\!1} - q_l)\,
\\
&\qquad\quad
\times\Gamma^{(n_\scV)}\big(w_l; \{\bar w_l, \bar w_{m\!+\!1}\}\big)
\;.
\end{split}
\end{equation}
Here the function $\Gamma^{(n_\scV)}(w_l; \{\bar w_l, \bar w_{m\!+\!1}\})$ gives all the three point 1PI graphs with $n_\scV$ virtual loops. This function is symmetric in the variables $\bar w_l$ and  $\bar w_{m\!+\!1}$.

For a $1 \to 3$ splitting we use a splitting operator $\bm{\Gamma}^{(2,n_\scV)}_l$ that, acting on an $m$ parton state, creates two new partons with labels $m+1$ and $m+2$. Specializing Eq.~(\ref{eq:Gamma-gen-matrix-element}), we have
\begin{equation}
  \begin{split}
    \label{eq:Gammal2n}
    &\left[\frac{\as}{2\pi}\right]^{n_\scV +1}\,
    \dualL\bra{\{\bar w\}_{m\!+\!2}}
    \bm{\Gamma}^{(2,n_\scV)}_l\ket{\{w\}_{m}}
    \\
    &\qquad
    = \prod_{i\ne l}\left[(2\pi)^d\delta^d(\bar q_i - q_i)\,
    \delta_{\bar f_i,f_i} \delta_{\bar a_i,a_i}\delta_{\bar r_i,r_i}\right]
    \\
    &\qquad\quad
    \times (2\pi)^d \delta^d(\bar q_l + \bar q_{m\!+\!1} 
    + \bar q_{m\!+\!2} - q_l)
    \\
    &\qquad\quad
    \times\Gamma^{(n_\scV)}\big(w_l; \{\bar w_l, \bar w_{m\!+\!1}, 
    \bar w_{m\!+\!2}\}\big)
    \;.
  \end{split}
\end{equation}
The emission function $\Gamma^{(n_\scV)}(w_l; \{\bar w_l, \bar w_{m\!+\!1}, \bar w_{m\!+\!2}\})$ is the sum of all the 1PI graphs for the $1\to3$ processes with $n_\scV$ virtual loops. In the NLO shower we need this only at tree level, $n_\scV = 0$.

The function $\Gamma^{(0)}(w_l; \{\bar w_l, \bar w_{m\!+\!1}\})$ gives all of the tree level three-point 1PI graphs in Eq.~(\ref{eq:Gammal1n}). This function is symmetric in the variables $\bar w_l$ and  $\bar w_{m\!+\!1}$. To simplify the expression it is useful to break this symmetry by defining $\Gamma^{(0)}\big(w_l; \{\bar w_l, \bar w_{m\!+\!1}\}\big)$ to be the sum of two functions
\begin{equation}
 \begin{split}
\label{eq:V3sym}
&\Gamma^{(0)}\big(w_l; \{\bar w_l, \bar w_{m\!+\!1}\}\big)
\\
&\qquad
= V_3(w_l; \bar w_l, \bar w_{m\!+\!1}) + V_3(w_l; \bar w_{m\!+\!1}, \bar w_l)
\;.
\end{split}
\end{equation}
Using Eq.~(\ref{eq:V3sym}) in Eq.~(\ref{eq:Gammal1n}) then gives
\begin{equation}
\begin{split}
\label{eq:Gammal10}
&\left[\frac{\as}{2\pi}\right]^{1/2}\,
\dualL\bra{\{\bar w\}_{m\!+\!1}}\qGamma^{(1,0)}_l\ket{\{w\}_{m}}
\\
&\qquad
= \prod_{i\ne l}\left[(2\pi)^d\delta^d(\bar q_i - q_i)\,\delta_{\bar f_i,f_i} \delta_{\bar a_i,a_i}\delta_{\bar r_i,r_i}\right]
\\
&\qquad\quad
\times (2\pi)^d \delta^d(\bar q_l + \bar q_{m\!+\!1} - q_l)\,
\\
&\qquad\quad
\times
\big[
V_3(w_l; \bar w_l, \bar w_{m\!+\!1}) + V_3(w_l; \bar w_{m\!+\!1}, \bar w_l)
\big]
\;.
\end{split}
\end{equation}

For a $q \to q + \Lg$ splitting, the emission function $V_3(w_l; \bar w_l, \bar w_{m\!+\!1})$ is the three point vertex function with the requirement that parton ${m\!+\!1}$ is the gluon:
\begin{equation}
\begin{split}
\label{eq:Gammaqqg}
V_3(w_l;{}& \bar w_l, \bar w_{m\!+\!1}) = 
\begin{prdfig}{634e96916244b3837dcc52bc3f5c0282}{quark-gluon-vertex}
  \begin{tikzpicture}[baseline=(v1.base)]
    \begin{feynman}[]
      \vertex[empty dot,label={[below=2mm] $a_l, r_l$}] (v1) at (0,0) {};
      \vertex[dot] (v2) at ($(v1)+(2 cm,0)$) {};
      \coordinate [label={[right] $\bar {a}_l, \bar {r}_l$}] (i) 
      at ($(v2) + (55:1.5cm)$);
      \coordinate [label={[below] 
      $\bar {a}_{m\!+\!1},\bar {r}_{m\!+\!1}$}] (j) 
      at ($(v2) + (-55:1.5cm)$);
      % % 
      \diagram*{
        (v1)--[fermion,rmomentum'={$\bar q_l+ \bar q_{m\!+\!1}$}](v2);
        (v2)--[fermion,rmomentum={$\bar q_l$}](i);
        (v2)--[gluon,rmomentum={$\bar q_{m\!+\!1}$}](j);
      };
    \end{feynman}
  \end{tikzpicture}
\end{prdfig}
\\={}&
\mi \gs t^{\bar a_{m\!+\!1}}_{\bar a_l a_l}
\left[\gamma^{\bar r_{m\!+\!1}}\s{q}_l\right]_{\bar r_l r_l} \frac{\mi}{q_l^2 + \mi 0}
\\
&\times
\theta(f_l \in \{\Lu, \Ld,\dots\})\,\theta(\bar f_l = f_l)\,\theta(\bar f_{m\!+\!1} = \Lg)
\;.
\end{split}
\end{equation}
This function includes the $q \to q + g$ vertex and the propagator for the initial state quark with momentum $q_l = \bar q_l + \bar{q}_{m\!+\!1}$.

For a $\bar q \to \bar q + \Lg$ splitting, we have\footnote{The minus sign in the color factor here is often associated with the $\s{q}_l$ factor in the adjacent propagator.} 
\begin{equation}
  \label{eq:Gammabarqbarqg}
  \begin{split}
    V_3(w_l;{}& \bar w_l, \bar w_{m\!+\!1}) =
    \begin{prdfig}{de824c479cbae39d880b62925991d3dd}{anti-quark-gluon-vertex}
      \begin{tikzpicture}[baseline=(v1.base)]
        \begin{feynman}[]
          \vertex[empty dot,label={[below=2mm] $a_l, r_l$}] (v1) at (0,0) {};
          \vertex[dot] (v2) at ($(v1)+(2 cm,0)$) {};
          \coordinate [label={[right] $\bar {a}_l, \bar {r}_l$}] (i) 
          at ($(v2) + (55:1.5cm)$);
          \coordinate [label={[below] $\bar {a}_{m\!+\!1},
          \bar {r}_{m\!+\!1}$}] (j) at ($(v2) + (-55:1.5cm)$);
          % % 
          \diagram*{
            (v1)--[anti fermion,rmomentum'=
            {$\bar q_l+ \bar q_{m\!+\!1}$}](v2);
            (v2)--[anti fermion,rmomentum={$\bar q_l$}](i);
            (v2)--[gluon,rmomentum={$\bar q_{m\!+\!1}$}](j);
          };
        \end{feynman}
      \end{tikzpicture}
    \end{prdfig}
    \\={}&
    -\mi \gs t^{\bar a_{m\!+\!1}}_{a_l \bar a_l }\,
    \frac{\mi}{q_l^2 + \mi 0}\,
    \left[
      \s{q}_l \gamma^{\bar r_{m\!+\!1}}
    \right]_{r_l \bar r_l}
    \\&\times
    \theta(f_l \in \{\bar\Lu, \bar\Ld,\dots\})\,
    \theta(\bar f_l = f_l)\,
    \theta(\bar f_{m\!+\!1}\!=\! \Lg)
    \;.
  \end{split}
\end{equation}
Note that the order of the $\gamma$ matrices and the positions of the indices $\bar r_l$ and $r_l$ here are reversed compared to Eq.~(\ref{eq:Gammaqqg}). 

For a $\Lg \to  q + \bar q$ splitting, we define the emission function $V_3(w_l; \bar w_l, \bar w_{m\!+\!1})$ to be the three point vertex function with the requirement that parton ${m\!+\!1}$ is the antiquark:
\begin{equation}
  \label{eq:Gammagqbar}
  \begin{split}
    V_3(w_l&;\bar w_l, \bar w_{m\!+\!1}) = 
    \begin{prdfig}{13b0a568000f9250e9fe2c60f6a78a87}{gluon-quark-anti-quark-vertex}
      \begin{tikzpicture}[baseline=(v1.base)]
        \begin{feynman}[]
          \vertex[empty dot,label={[below=2mm] $a_l, r_l$}] (v1) 
          at (0,0) {};
          \vertex[dot] (v2) at ($(v1)+(2 cm,0)$) {};
          \coordinate [label={[right] $\bar {a}_l, \bar {r}_l$}] (i) 
          at ($(v2) + (55:1.5cm)$);
          \coordinate [label={[below] $\bar {a}_{m\!+\!1},
          \bar {r}_{m\!+\!1}$}] (j) at ($(v2) + (-55:1.5cm)$);
          % % 
          \diagram*{
            (v1)--[gluon,rmomentum'={$\bar q_l+ \bar q_{m\!+\!1}$}](v2);
            (v2)--[fermion,rmomentum={$\bar q_l$}](i);
            (v2)--[anti fermion,rmomentum={$\bar q_{m\!+\!1}$}](j);
          };
        \end{feynman}
      \end{tikzpicture}
    \end{prdfig}
    \\={}&
    \mi \gs t^{a_l}_{\bar a_l \bar a_{m\!+\!1}}
    \left[\gamma^\mu\right]_{\bar r_l \bar r_{m\!+\!1}}
     \frac{\mi N_{\mu r_l }(q_l)}{q_l^2 + \mi 0}
    \\&\times
    \theta(f_l \!=\!\Lg)\,
    \theta(\bar f_l \in \{\Lu, \Ld,\dots\})\,
    \theta(\bar f_{m\!+\!1} = - \bar f_l)
    \;,
  \end{split}
\end{equation}
where $N_{\mu\nu}(q)$ is the numerator of the gluon propagator, Eq.~(\ref{eq:gluonnumerator}). Here $- \bar f_l$ is the charge conjugate flavor to $\bar f_l$.  

For a $\Lg \to \Lg + \Lg$ splitting, we define the emission function as\footnote{Eq.~(\ref{eq:softggg}) is written analogously to Eq.~(\ref{eq:Gammaqqg}), using the color generator matrix for the $\bm 8$ representation of SU(3) $T^{\bar a_{m\!+\!1}}_{\bar a_l a_l} = \mi f_{a_l,\bar{a}_l,\bar{a}_{m\!+\!1}}$.} 
\begin{equation}
  \label{eq:softggg}
  \begin{aligned}[c]
   V_3(w_l&; \bar w_l,\bar w_{m\!+\!1}) =
    \begin{prdfig}{bdf17053fae02860b7895c61cbf8beed}{gluon-gluon-vertex-soft}
      \begin{tikzpicture}[baseline=(v1.base)]
        \begin{feynman}[]
          \vertex[empty dot,label={[below=2mm] $a_l, r_l$}] (v1) 
          at (0,0) {};
          \vertex[dot] (v2) at ($(v1)+(2cm,0)$) {};
          \coordinate [label={[right] $\bar{a}_l, \bar{r}_l$}] (i) 
          at ($(v2) + (55:1.5cm)$);
          \coordinate [label={[below] $\bar{a}_{m\!+\!1}, 
          \bar{r}_{m\!+\!1}$}] (j) at ($(v2) + (-55:1.5cm)$);
          \diagram*{
            (v1)--[gluon, rmomentum'={$\bar q_l+ \bar q_{m\!+\!1}$}](v2);
            (v2)--[gluon, rmomentum={$\bar q_l$}](i);
            (v2)--[{\EikHead}-, gluon, decoration={pre length=3mm},
            rmomentum={[black] $\bar q_{m\!+\!1}$},red](j);
          };
        \end{feynman}
      \end{tikzpicture}
    \end{prdfig}
    \\
    ={}&
    -\gs f_{a_l,\bar{a}_l,\bar{a}_{m\!+\!1}}\,
    \frac{\mi N_{\mu r_l }(q_l)}{q_l^2 + \mi 0}
    \\
    &
    \times \!
    \left[
      g^{\bar{r}_l\bar{r}_{m\!+\!1}}\,\bar{q}_{l}^{\mu}
      + g^{\mu\bar{r}_{m\!+\!1}}\,\bar{q}_l^{\bar{r}_l}
      -2g^{\mu\bar{r}_l}\,\bar{q}_l^{\bar{r}_{m\!+\!1}}
    \right]    
    \\
    &
    \times
    \theta(f_l \!=\! \bar f_l  \!=\!\bar f_{m\!+\!1}\!=\!\Lg)
    \;. 
  \end{aligned}
\end{equation}
This function $V_3(w_l; \bar w_l,\bar w_{m\!+\!1})$ is not the full gluon emission function. Rather, the sum of the partial emission function $V_3(w_l; \bar w_l,\bar w_{m\!+\!1})$ and the partial emission function with $\bar w_l \leftrightarrow \bar w_{m\!+\!1}$ is the full gluon emission function. It is this sum that appears in Eq.~\eqref{eq:Gammal10}. The two lines on the lower leg in the graphical representation here indicates the leg with index $m+1$ in $V_3(w_l; \bar w_l,\bar w_{m\!+\!1})$.

This decomposition of the $\Lg \to \Lg + \Lg$ emission function is useful for separating the singularities that appear when gluon $m\!+\!1$ is soft, $\bar{q}_{m\!+\!1} \to 0$, or gluon $l$ is soft, $\bar q_l \to 0$. The propagator $D_{\mu r_l}(q_l)$ contains a factor $1/(2\, \bar q_l\cdot \bar q_{m\!+\!1})$ when $\bar q_l^2 = \bar q_{m\!+\!1}^2 = 0$. This factor is singular when $\bar{q}_{l} \to 0$ and when $\bar{q}_{m\!+\!1} \to 0$. The partial emission function $V_3(w_l;  \bar w_l,\bar w_{m\!+\!1})$ contains a factor $\bar{q}_{l}$, which cancels the singularity for $\bar{q}_{l} \to 0$ but leaves the singularity for $\bar{q}_{m\!+\!1} \to 0$. The partial emission function with $l\leftrightarrow m\!+\!1$ has the opposite behavior. It is singular for $\bar{q}_{l} \to 0$ but not for $\bar{q}_{m\!+\!1} \to 0$.  

\begin{figure*}[htb]
  \centering
  \begin{equation*}
    \begin{split}
      \Gamma^{(1)}\big(w_l;{}& \{\bar{w}_l, \bar{w}_{m\!+\!1}\}\big) =
      \\&
      \begin{prdfig}{704919b26d154bf40fa1108254755720}{gluon-gluon-vertex-1loop-term0}
        \begin{tikzpicture}[baseline=(v1.base)]
          \begin{feynman}[]
            \vertex[empty dot,label={[below=2mm] $a_l, r_l$}] (v1) 
            at (0,0) {};
            \vertex[dot] (v2) at ($(v1)+(1.cm,0)$) {}; 
            \vertex[dot] (v3) at ($(v2)+(45:1.cm)$) {};
            \vertex[dot] (v4) at ($(v2)+(-45:1.cm)$) {};
            %%% % 
            \coordinate [label={[above] $\bar{a}_l, \bar{r}_l$}] (l) 
            at ($(v3) + (45:1.cm)$);
            \coordinate [label={[below] $\bar{a}_{m\!+\!1},
             \bar{r}_{m\!+\!1}$}] (m1) at ($(v4) + (-45:1.cm)$);
            \diagram*{
              (v1)--[rmomentum'={$q_l$}]
              (v2)--(v3)--[rmomentum={$\bar{q}_{l}$}](l);
              (v2)--(v4)--[rmomentum'={$\bar q_{m\!+\!1}$}](m1);
              (v3)--(v4);
            };
          \end{feynman}
        \end{tikzpicture}
      \end{prdfig}
      +\;
      \begin{prdfig}{1fc97cf53d6054e6c5b5a2ef8c708535}{gluon-gluon-vertex-1loop-cnt}
        \begin{tikzpicture}[baseline=(v1.base)]
          \begin{feynman}[]
            \vertex[empty dot,label={[below=2mm] $a_l, r_l$}] (v1) 
            at (0,0) {};
            \vertex[crossed dot, label={[above] $(1)$}] (v3) 
            at ($(v1)+(1cm,0)$) {};
            %%% 
            \coordinate [label={[above] $\bar{a}_l, \bar{r}_l$}] (l) 
            at ($(v3) + (45:1.5cm)$);
            \coordinate [label={[below] $\bar{a}_{m\!+\!1}, 
            \bar{r}_{m\!+\!1}$}] (m1) 
            at ($(v3) + (-45:1.5cm)$);
            % % 
            \diagram*{
              (v1)--[rmomentum'={$q_l$}](v3)--[rmomentum={$\bar{q}_{l}$}](l);
              (v3)--[rmomentum={$\bar q_{m\!+\!1}$}](m1);
            };
          \end{feynman}
        \end{tikzpicture}
      \end{prdfig}
      +\;
      \lBiigg[3.5cm]{[}
      \begin{prdfig}{cc11b7694e2e35b01d55b96c812a1710}{gluon-gluon-vertex-1loop-term1}
        \begin{tikzpicture}[baseline=(v1.base)]
          \begin{feynman}[]
            \vertex[empty dot,label={[below=2mm] $a_l, r_l$}] (v1) 
            at (0,0) {};
            \vertex[dot] (v2) at ($(v1)+(1.0cm,0)$) {}; 
            \vertex[dot] (v3) at ($(v2)+(45:1.cm)$) {};
            \vertex[dot] (v4) at ($(v2)+(-45:1.cm)$) {};
            %%% % 
            \coordinate [label={[above] $\bar{a}_l, \bar{r}_l$}] (l) 
            at ($(v3) + (45:1.cm)$);
            \coordinate [label={[below] $\bar{a}_{m\!+\!1}, 
            \bar{r}_{m\!+\!1}$}] (m1) 
            at ($(v4) + (-45:1.cm)$);
            \diagram*{
              (v1)--[gluon, rmomentum'={$q_l$}](v2)--[ghost, 
              with arrow=0.5](v3)
              --[gluon, rmomentum={$\bar{q}_{l}$}](l);
              (v2)--[ghost, with reversed arrow=0.5](v4)--[gluon, 
              rmomentum'={$\bar q_{m\!+\!1}$}](m1);
              (v3)--[ghost, with arrow=0.5](v4);
            };
          \end{feynman}
        \end{tikzpicture}
      \end{prdfig}
      +\;
      \begin{prdfig}{18c36aa3e039ce8e516a7601a6b7f1d0}{gluon-gluon-vertex-1loop-term2}
        \begin{tikzpicture}[baseline=(v1.base)]
          \begin{feynman}[]
            \vertex[empty dot,label={[below=2mm] $a_l, r_l$}] (v1) 
            at (0,0) {};
            \vertex[dot] (v2) at ($(v1)+(1.0cm,0)$) {};
            \vertex[dot] (v3) at ($(v2)+(45:1.cm)$) {};
            \vertex[dot] (v4) at ($(v2)+(-45:1.cm)$) {};
            %%% 
            \coordinate [label={[above] $\bar{a}_l, \bar{r}_l$}] (l) 
            at ($(v3) + (45:1.cm)$);
            \coordinate [label={[below] $\bar{a}_{m\!+\!1}, 
            \bar{r}_{m\!+\!1}$}] (m1) at ($(v4) + (-45:1.cm)$);
            % % 
            \diagram*{
              (v1)--[gluon, rmomentum'={$q_l$}]
              (v2)--[ghost, with reversed arrow=0.5](v3)
              --[gluon, rmomentum={$\bar{q}_{l}$}](l);
              (v2)--[ghost, with arrow=0.5](v4)
              --[gluon, rmomentum'={$\bar q_{m\!+\!1}$}](m1);
              (v3)--[ghost, with reversed arrow=0.5](v4);
            };
          \end{feynman}
        \end{tikzpicture}
      \end{prdfig}
      \\&
      +\;
      \begin{prdfig}{023f4e6f70eadb2ad30d88dc5990b97e}{gluon-gluon-vertex-1loop-term3}
        \begin{tikzpicture}[baseline=(v1.base)]
          \begin{feynman}[]
            \vertex[empty dot,label={[below=2mm] $a_l, r_l$}] (v1) 
            at (0,0) {};
            \vertex[dot] (v2) at ($(v1)+(1.cm,0)$) {};
            \vertex[dot] (v3) at ($(v2)+(1.2cm,0)$) {};
            %%% 
            \coordinate [label={[above] $\bar{a}_l, \bar{r}_l$}] (l) 
            at ($(v3) + (65:1.2cm)$);
            \coordinate [label={[below] $\bar{a}_{m\!+\!1}, 
            \bar{r}_{m\!+\!1}$}] (m1) 
            at ($(v3) + (-65:1.2cm)$);
            % % 
            \diagram*{
              (v1)--[gluon, rmomentum'={$q_l$}](v2)
              --[gluon, quarter left](v3)
              --[gluon, rmomentum={$\bar{q}_{l}$}](l);
              (v2)--[gluon, quarter right](v3)--
              [gluon, rmomentum={$\bar q_{m\!+\!1}$}](m1);
            };
          \end{feynman}
        \end{tikzpicture}
      \end{prdfig}
      +\;
      \begin{prdfig}{5b3f4cf5ef298e1ba24dd0c213a7a8e1}{gluon-gluon-vertex-1loop-term4}
        \begin{tikzpicture}[baseline=(v1.base)]
          \begin{feynman}[]
            \vertex[empty dot,label={[below=2mm] $a_l, r_l$}] (v1) 
            at (0,0) {};
            \vertex[dot] (v2) at ($(v1)+(1.cm,0)$) {};
            \vertex[dot] (v3) at ($(v2)+(55:1.2cm)$) {};
            %%% 
            \coordinate [label={[above] $\bar{a}_l, \bar{r}_l$}] (l) 
            at ($(v3) + (55:1.cm)$);
            \coordinate [label={[below] $\bar{a}_{m\!+\!1}, 
            \bar{r}_{m\!+\!1}$}] (m1) 
            at ($(v2) + (-45:1.5cm)$);
            % % 
            \diagram*{
              (v1)--[gluon, rmomentum'={$q_l$}](v2)
              --[gluon, quarter left](v3)
              --[gluon, rmomentum'={$\bar{q}_{l}$}](l);
              (v2)--[gluon, quarter right](v3);
              (v2)--[gluon, rmomentum'={$\bar q_{m\!+\!1}$}](m1);
            };
          \end{feynman}
        \end{tikzpicture}
      \end{prdfig}
      +\;
      \begin{prdfig}{23d00043ce3f28ba3946d660261217df}{gluon-gluon-vertex-1loop-term5}
        \begin{tikzpicture}[baseline=(origo.base)]
          \begin{feynman}[]
            \coordinate[] (origo) at (0,0);
            \vertex[empty dot,label={[below=2mm] $a_l, r_l$}] (v1) 
            at (0,0.5cm) {};
            \vertex[dot] (v2) at ($(v1)+(1.cm,0)$) {};
            \vertex[dot] (v3) at ($(v2)+(-55:1.2cm)$) {};
            %%% 
            \coordinate [label={[above] $\bar{a}_l, \bar{r}_l$}] (l) 
            at ($(v2) + (45:1.5cm)$);
            \coordinate [label={[below] $\bar{a}_{m\!+\!1}, 
            \bar{r}_{m\!+\!1}$}] (m1) 
            at ($(v3) + (-55:1cm)$);
            % % 
            \diagram*{
              (v1)--[gluon, rmomentum'={$q_l$}](v2)
              --[gluon, quarter left](v3);
              (v2)--[gluon, rmomentum'={$\bar{q}_{l}$}](l);
              (v2)--[gluon, quarter right](v3)--
              [gluon, rmomentum'={$\bar q_{m\!+\!1}$}](m1);
            };
          \end{feynman}
        \end{tikzpicture}
      \end{prdfig}
      \rBiigg[3.5cm]{]}\theta(f_l = \bar{f}_l = \bar{f}_{m\!+\!1}=\Lg)
     \end{split}
  \end{equation*}
  \caption{\label{fig:3g-1loop-vertex}
    The vertex function $\Gamma^{(1)}( w_l; \{\bar{w}_l, \bar{w}_{m\!+\!1}\})$ for the $\qGamma^{(1,1)}_{l}$ one-loop single emission operator, including a propagator for the initial parton. In the first graph, the internal lines in the loop can be quarks, antiquarks, or gluons. Graphs with a ghost loop are included separately.
  }
\end{figure*}
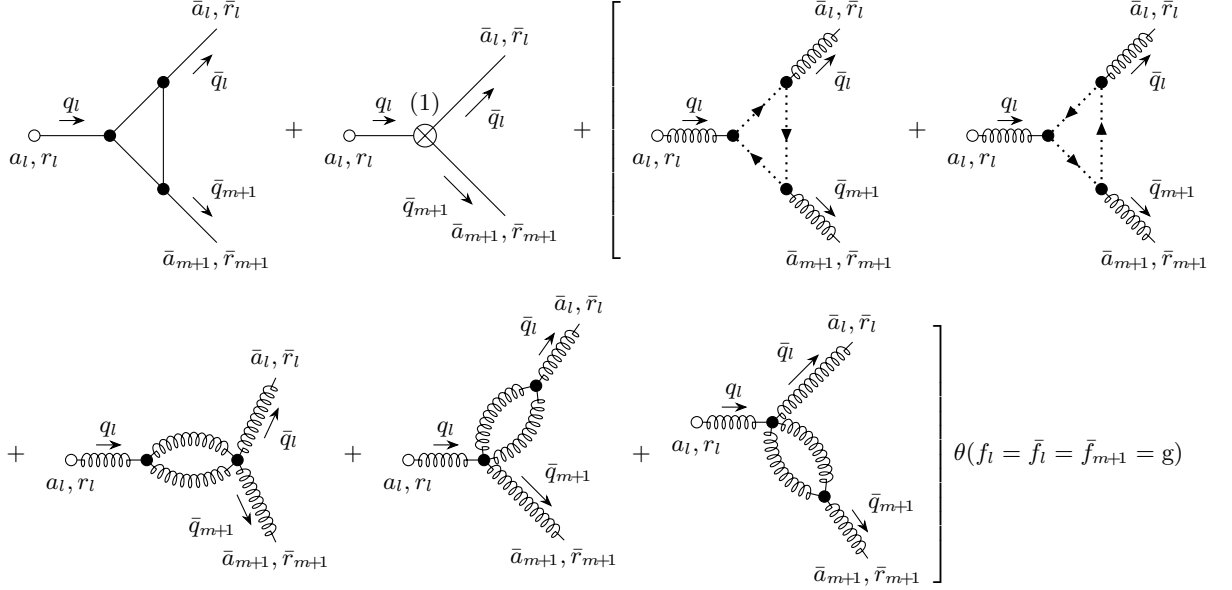

For the NLO shower, we also need the $1\to2$ splitting at one loop level. The function $\Gamma^{(1)}(w_l; \{\bar{w}_l, \bar{w}_{m\!+\!1}\})$ is rather complicated but we can define it by its Feynman graphs, which are given in Fig.~\ref{fig:3g-1loop-vertex}.
% 
% -------------------- Figure -------------------
% -------------------- Figure -------------------
% 

For the  $1 \to 3$ splittings created by a four point 1PI graph, we use the splitting operators $\qGamma^{(2,n_\scV)}$ as defined in Eq.~\eqref{eq:Gammal2n}. At tree level ($n_\scV=0$) the emission function $\Gamma^{(0)}(w_l; \{\bar w_l, \bar w_{m\!+\!1}, \bar w_{m\!+\!2}\})$ is the product of the propagator for the initial gluon times the four gluon vertex: 
\begin{equation}
  \begin{aligned}[c]
    &\Gamma^{(0)}\big(w_l; \{\bar w_l, \bar w_{m\!+\!1}, \bar w_{m\!+\!2}\}\big) 
    \\
    &\quad
    =\quad
    \begin{prdfig}{8cd1d757df7459d3fbf05e132981ece3}{4gluon-vertex}
      \begin{tikzpicture}[baseline=(v1.base)]
        \begin{feynman}[]
          \vertex[empty dot,label={[below=2mm] $a_l, r_l$}] (v1) at (0,0) {};
          \vertex[dot] (v2) at ($(v1)+(1.5cm,0)$) {};
          \coordinate [label={[right] $\bar{a}_l, \bar{r}_l$}] (i) 
          at ($(v2) + (35:1.75cm)$);
          \coordinate [label={[right] $\bar{a}_{m\!+\!1}, 
          \bar{r}_{m\!+\!1}$}] (j) 
          at ($(v2) + (0:1.75cm)$);
          \coordinate [label={[right] $\bar{a}_{m\!+\!2}, 
          \bar{r}_{m\!+\!2}$}] (k) 
          at ($(v2) + (-35:1.75cm)$);
          \diagram*{
            (v1)--[gluon, rmomentum'={$ q_l$}](v2);
            (v2)--[gluon, quarter left, rmomentum'={$\bar q_l$}](i);
            (v2)--[gluon, rmomentum'={$\bar q_{m\!+\!1}$}](j);
            (v2)--[gluon, quarter right, rmomentum={$\bar q_{m\!+\!2}$}](k);
          };
        \end{feynman}
      \end{tikzpicture}
    \end{prdfig}
    \\
    &\quad
    = \mi D_{r_l}^{\mu}(q_l)(-\mi\,\gs^2 )
    \\
    &\qquad
    \times
    \big\{
    f_{a_l \bar a_l e}f_{\bar a_{m\!+\!1} \bar a_{m\!+\!2} e}
    \left[ g_{\mu}^{\bar r_{m\!+\!1}} g^{\bar r_l \bar r_{m\!+\!2}}  
      - g_{\mu}^{\bar r_{m\!+\!2}} g^{\bar r_l \bar r_{m\!+\!1}} \right]
    \\
    &\qquad
    + f_{a_l \bar a_{m\!+\!1} e}f_{\bar a_l \bar a_{m\!+\!2} e}
    \left[ g_{\mu}^{\bar r_l} g^{\bar r_{m\!+\!1} \bar r_{m\!+\!2}}  
      - g_{\mu}^{\bar r_{m\!+\!2}} g^{\bar r_l \bar r_{m\!+\!1}} \right]
    \\
    &\qquad
    + f_{a_l \bar a_{m\!+\!2} e}f_{\bar a_l \bar a_{m\!+\!1} e}
    \left[ g_{\mu}^{\bar r_l} g^{\bar r_{m\!+\!1} \bar r_{m\!+\!2}}  
      - g_{\mu}^{\bar r_{m\!+\!1}} g^{\bar r_l \bar r_{m\!+\!2}} \right]\!
    \big\}
    \\
    &\qquad
    \times
    \theta(\bar f_l = \bar f_{m\!+\!1} = \bar f_{m\!+\!2} = f_l = \Lg)
    \;.
  \end{aligned}
\end{equation}
%

%-------------------------------------------------
\subsection{Exchange graphs}
\label{sec:ExchangeGraphs}

In Sec.~\ref{sec:CompleteDiagram} we have introduced the operator $\qGamma_\scH$ that creates 1PI graphs that include the point vertex $\cH(x)$ at which $m$ parton lines join. 

Many of the diagrams that contribute to $\bm \Gamma_\scH$ do not generate infrared singularities. In the end, the contributions from these graphs will not contribute to the infrared sensitive operator $\cD$ that generates the parton shower. For this reason, it is useful to define operators that generate graphs that {\em do} have infrared singularities. 

The simplest of these operators is $\qX_{lk}^{(0,1)}$, which corresponds to the graph in which a parton is exchanged between partons $l$ and $k$:
\begin{equation}
\begin{split}
\label{eq:qXlk01start}
\aspi\,
\dualL\bra{\{\bar w\}_m}
\qX_{lk}^{(0,1)}\ket{\{w\}_m}
\hskip - 2.5 cm {}&
\\
= {}& 
\mu_\scR^{2\epsilon}
\int\! \frac{d^d\ell}{(2\pi)^d}\,
X_{lk}^{(0,1)}\!\left(w_l, w_k; \bar{w}_l,\bar{w}_k; \ell\right)
\\&\times
\prod_{\substack{i=1\\ i\neq l,k}}^m
\left[
(2\pi)^d\delta(\bar{q}_i - q_i)\,
\delta_{\bar{f}_i, f_i}\,\delta_{\bar{a}_i, a_i}\,\delta_{\bar{r}_i, r_i}
\right]
\\&\times
(2\pi)^d \delta\!\left(q_l - \bar{q}_l - \ell \right)\,
(2\pi)^d \delta\!\left(q_k - \bar{q}_k + \ell \right)
\;.
\end{split}
\end{equation}

The function $X_{lk}^{(0,1)}\!\left(w_l, w_k; \bar{w}_l,\bar{w}_k; \ell\right)$ is given by the exchange graph 
 
\begin{equation}
  \begin{split}
  \label{eq:exchangegraph}
    &X_{lk}^{(0,1)}\big(w_l, w_k;\bar{w}_l,\bar{w}_k; \ell\big) 
    =
    \begin{prdfig}{559c55c946404313d7821c37361ab688}{parton-exchange-graph}
      \begin{tikzpicture}[baseline=(current bounding box.center)]
        \begin{feynman}[]
          \vertex[empty dot,label={[below=2mm] $w_l$}] (v1) at (0,0) {};
          \vertex[dot] (v2) at ($(v1)+(1.0cm,0)$) {};
          \vertex[empty dot,label={[above] $w_k$}] (v3) 
          at ($(v1)+(0,-1.5cm)$) {};
          \vertex[dot] (v4) at ($(v3)+(1.0cm,0)$) {};
          \coordinate[label={[below] $\bar{w}_l$}] (l) at ($(v2) + (1cm,0)$);
          \coordinate[label={[above] $\bar{w}_k$}] (k) at ($(v4) + (1cm,0)$);
          \diagram*{
            (v1)--[](v2)--[](l);
            (v3)--[](v4)--[](k);
            (v2)--[rmomentum'={[black] $\ell$}](v4);
          };
        \end{feynman}
      \end{tikzpicture}
    \end{prdfig}
    \;\;.
  \end{split}
\end{equation}
Using the Feynman rules, we see that this function is
\begin{equation}
  \begin{split}
    \label{eq:Exlk}
    X_{lk}^{(0,1)}\big({}& w_l, w_k;\bar{w}_l,\bar{w}_k; \ell\big) 
    \\
    ={}&\
    \sum_{f,a,r,r'}\frac{\mi\,N_{r' r}(\ell,f)}{\ell^2 + \mi 0}
    \\
    &\quad\times
    \theta(f = f_l - \bar f_l =  \bar f_k -  f_k)
    \\
    &\quad\times
    \big[V_3(w_l; \bar w_l, \bar w_\mathrm{x}) 
    + V_3(w_l; \bar w_\mathrm{x}, \bar w_l)\big]
    \\
    &\quad\times
    \big[V_3(w_k; \bar w_k, \bar w'_\mathrm{x})
    + V_3(w_k; \bar w'_\mathrm{x}, \bar w_k)\big]
    \;,
  \end{split}
\end{equation}
where the function $V_3$ is defined in Sec.~\ref{sec:VertexOperators}. The quantum numbers for the exchanged parton are $\bar w_\mathrm{x} = \{\ell,f,a,r\}$ and $\bar w'_\mathrm{x}= \{-\ell,-f,a,r'\}$ The function $N_{r' r}(\ell,f)$ is the numerator for the propagator of the exchanged parton, with vector/spinor indices $(r,r')$: 
\begin{equation}
  N_{r' r}(\ell,f) =
  \begin{cases}
    [\s{\ell}]_{r' r}  \quad & f\in\{u,d,\dots\}\;, \\[10pt]
    -[\s{\ell}]_{r r'} \quad & f\in\{\bar u,\bar d,\dots\}\;, \\[10pt]
    N_{r' r}(\ell)     \quad & f = \Lg\;,
  \end{cases}
\end{equation}
where $N_{\mu\nu}(\ell)$ for gluons is given in Eq.~(\ref{eq:gluonnumerator}).

In Eq.~(\ref{eq:Exlk}), a number of flavor choices are available. The flavor $f_l$ can be a quark flavor $q$, an antiquark flavor $\bar q$, or the gluon flavor $\Lg$. The same holds for the flavor $\bar f_l$ of parton $l$ after the interaction and for $f_k$, $\bar f_k$ and $f$. The flavor of the exchanged parton is determined by flavor conservation:
\begin{equation}
f = f_l - \bar f_l =  \bar f_k -  f_k 
\;.
\end{equation}

There is another exchange operator that we define for use in later papers. This operator gives a single exchange between two partons with a single emission. We define it by its non-zero matrix elements
\begin{equation}
  \begin{split}
    \label{eq:opE11-matrix}
    \left[\frac{\as}{2\pi}\right]^{3/2}
    \dualL\bra{\{\bar w\}_{m\!+\!1}}
    \qX^{(1,1)}_{lk}\ket{\{w\}_m}
    \hskip -4.0cm &
    \\
    = {}&
    \prod_{\substack{i=1\\ i\neq l,k}}^m
    \left[
      (2\pi)^d\delta(\bar{q}_i-q_i)\,
      \delta_{\hat{f}_i, f_i}\,\delta_{\bar{a}_i, a_i}\,
      \delta_{\bar{r}_i, r_i}
    \right]
    \\
    &\times
    (2\pi)^d \delta\big(q_k  + q_l - \bar{q}_l- 
    \bar{q}_k- \bar{q}_{m\!+\!1}\big)
    \\
    &\times
    X_{lk}^{(1,1)}(w_l, w_k; \bar{w}_{l},\bar{w}_{m\!+\!1},\bar{w}_k)
    \;.
  \end{split}
\end{equation}
The function $X_{lk}^{(1,1)}(w_l, w_k;\bar{w}_{l},\bar{w}_{m\!+\!1},\bar{w}_k)$ is given by the following graph:
\begin{equation}
  \begin{split}
  \label{eq:Xlk11}
    X_{lk}^{(1,1)}(w_l, w_k; \bar{w}_l,\bar{w}_k,{}&\bar{w}_{m\!+\!1}) 
     \\
     ={}&
    \begin{prdfig}{1544182867099e5f1b9d306e1f880561}{generic-exchange-with-emission-graph}
      \begin{tikzpicture}[baseline=(current bounding box.center)]
        \begin{feynman}[]
          \vertex[empty dot,label={[below=2mm] $w_l$}] (v1) at (0,0) {};
          \vertex[dot] (v2) at ($(v1)+(1.0cm,0)$) {};
          \vertex[empty dot,label={[above] $w_k$}] (v3) 
          at ($(v1)+(0,-1.5cm)$) {};
          \vertex[dot] (v4) at ($(v3)+(1.0cm,0)$) {};
          \coordinate[label={[above] $\bar{w}_l$}] (l) at ($(v2) + (1cm,0)$);
          \coordinate[label={[below] $\bar{w}_k$}] (k) at ($(v4) + (1cm,0)$);
          \vertex[dot] (v5) at ($(v2)+(0,-0.75cm)$) {};    
          \coordinate[label={[below] $\bar{w}_{m\!+\!1}$}] (m1) 
          at ($(v5) + (1cm,0)$);
          \diagram*{
            (v1)--(v2)--(l);
            (v3)--(v4)--(k);
            (v2)--(v5)--(v4);
            (v5)--(m1);
          };
        \end{feynman}
      \end{tikzpicture}
    \end{prdfig}
    \;.
  \end{split}	
\end{equation}
% 

%-------------------------------------------------
\subsection{1PI graphs that include the hard vertex}
\label{sec:Gamma-H-operators}

The operator $\qGamma_\scH$ that includes the point vertex $\cH(x)$ at which $m$ parton lines join has a perturbative expansion as in Eq.~(\ref{eq:Gamma-Gamma-H-expand}): 
\begin{equation}
\label{eq:Gamma-Gamma-H-expand-bis}
\qGamma_\scH = \bm 1 
+ \sum_{n_\scR = 0}^\infty \sum_{n_\scV = 1}^\infty
\left[\frac{\as}{2\pi}\right]^{n_\scR/2 + n_\scV}
\qGamma_\scH^{(n_\scR, n_\scV)}
\;.
\end{equation}
The lowest order contribution is $\qGamma_\scH^{(0,1)}$:
\begin{equation}
\label{eq:qGammaH01}
\qGamma_\scH^{(0,1)} = \sum_{l} \frac{1}{2}\sum_{k\ne l}
\qE^{(0,1)}_{lk}
+ \sum_l \qN^{(0,1)}_l
\;,
\end{equation}
where $\qE^{(0,1)}_{lk}$ corresponds to the exchange of a gluon between partons $l$ and $k$. There are Feynman graphs for the exchange of a quark also, but these graphs do not give infrared singularities, so we omit them. The operator $\qN^{(0,1)}_l$ is included in order to adjust the dependence of $\qGamma_\scH^{(0,1)}$ on the renormalization scale $\mur$. We will define it in Eq.~(\ref{eq:Kldef}) in the following section.

We base the definition of $\qE^{(0,1)}_{lk}$ on the exchange operator $\qX_{lk}^{(0,1)}$, restricted to gluon exchange. The initial state $\iket{\chi(\{p,f,c,s\}_m)}$ to which $\qX_{lk}^{(0,1)}$ is applied, given in Eq.~(\ref{eq:InitialKetVector}), includes a free integration over the momenta $q_i$ in the basis states $\iket{\{q,f,a,r\}_m}$:
\begin{equation}
\begin{split}
\label{eq:InitialKetVectorbis}
\ket{\chi(\{p,f,c,s\}_m)}\hskip - 1.5 cm {}&
\\ ={}& 
\left[\prod_{i=1}^m \int\! \frac{d^dq_i}{(2\pi)^d}\right]
\sum_{\{a,r\}_m}
\ket{\{q,f,a,r\}_m}
\\ &\times
\brax{\{a\}_{m}}\ket{\{c\}_{m}}
\prod_{i=1}^m
\dualL\brax{r_i}\ket{\chi_{f_i}(p_i,s_i)}
\;.
\end{split}
\end{equation}
The delta functions in Eq.~(\ref{eq:qXlk01start}) then fix the values of the $q_i$.  An integration over the exchanged momentum $\ell$ remains. That is,
\begin{equation}
\begin{split}
\label{eq:qXlk01A}
\aspi\,
\dualL\bra{\{\bar w\}_m}
\qX_{lk}^{(0,1)}\ket{\chi(\{p,f,c,s\}_m)}
\hskip - 4.5 cm {}&
\\
= {}& 
\sum_{\{a,r\}_m}\,
\brax{\{a\}_{m}}\ket{\{c\}_{m}}
\prod_{i=1}^m \dualL\brax{r_i}\ket{\chi_{f_i}(p_i,s_i)}
\\&\times
\prod_{\substack{i=1\\ i\neq l,k}}^m
\left[
\delta_{\bar{f}_i, f_i}\,\delta_{\bar{a}_i, a_i}\,\delta_{\bar{r}_i, r_i}
\right]
\\
&\times
\mu_\scR^{2\epsilon} 
\int\! \frac{d^d\ell}{(2\pi)^d}\,
X_{lk}^{(0,1)}\!\left(w_l, w_k; \bar{w}_l,\bar{w}_k;\ell\right)
\;.
\end{split}
\end{equation}
Now, after using the delta functions in $\qX_{lk}^{(0,1)}$, the momenta $q_i$ for spectator partons are given by $q_i = \bar q_i$. For the active partons, $q_l = \bar q_l + \ell$ and $q_k = \bar q_k - \ell$.

The integration over the momenta $q_i$ in Eq.~(\ref{eq:qXlk01A}) arose from an approximation. We began with Fig.~\ref{fig:d11exampleA}, represented schematically by Eq.~(\ref{eq:Gtotstart}). There is an integration over parton momenta $q_i$, but these momenta appear both in the graph $G_\mathrm{ket}(\{\hat p\}_{\hat m},\{q\}_m)$ and in the hard amplitude $H_\mathrm{ket}(\{q\}_m)$. In Eq.~(\ref{eq:Gtotend}), we then approximated $\{q\}_m$ in the hard amplitude by fixed momenta $\{p\}_m$. This approximation corresponds to Fig.~\ref{fig:d11exampleB}. This approximation leaves an integration over the momenta $\{q\}_m$. In the subgraph that we are considering for $\dualL\bra{\{\bar w\}_m} \qX_{lk}^{(0,1)}\ket{\{w\}_m}$, we should have $\bar q_l \approx p_l$ and $\bar q_k \approx p_k$, as long as later parton emissions are close to being soft or collinear. However, $q_l = \bar q_l + \ell$ and $q_k = \bar q_k - \ell$, so we are approximating the exchanged parton momentum $\ell$ as being soft.

We choose $\qE_{lk}^{(0,1)}$ to be $\qX_{lk}^{(0,1)}$, limited to gluon exchange, with approximations that are appropriate for the infrared limit of the integration over $\ell$:
\begin{equation}
\begin{split}
\label{eq:qGammalk01def}
\aspi\,
\dualL\bra{\{\bar w\}_m}
\qE_{lk}^{(0,1)}\ket{\chi(\{p,f,c,s\}_m)}
\hskip - 4.5 cm {}&
\\
= {}& 
\sum_{\{a,r\}_m}\,
\brax{\{a\}_{m}}\ket{\{c\}_{m}}
\prod_{i=1}^m 
\dualL\brax{r_i}\ket{\chi_{f_i}(p_i,s_i)}
\\&\times
\prod_{\substack{i=1\\ i\neq l,k}}^m
\left[
\delta_{\bar{f}_i, f_i}\,\delta_{\bar{a}_i, a_i}\,\delta_{\bar{r}_i, r_i}
\right]
\\
&\times
\mu_\scR^{2\epsilon} 
\int\! \frac{d^d\ell}{(2\pi)^d}\,
E_{lk}^{(0,1)}\!\left(w_l, w_k; \bar{w}_l,\bar{w}_k;\ell\right)
\;.
\end{split}
\end{equation}
We define $E_{lk}^{(0,1)}$ to be a modified version of $X_{lk}^{(0,1)}$ in Eq.~(\ref{eq:Exlk}). In the integration over $\ell$, there is a logarithmic singularity from $\ell \to 0$ only when the exchanged parton is a gluon, so we require that the exchanged parton be a gluon. Then in $X_{lk}^{(0,1)}$, there is a factor with a sum of two functions, $V_3(w_l; \bar w_l, \bar w_\mathrm{x}) + V_3(w_l; \bar w_\mathrm{x}, \bar w_l)$, where $\bar w_x = \{\ell,f,a,r\}$ represents the quantum numbers of the exchanged parton. The part of this that can give an infrared singularity when $\ell \to 0$ is the first term, $V_3(w_l; \bar w_l, \bar w_\mathrm{x})$, so we keep only this term. Similarly, there is a factor $V_3(w_k; \bar w_k, \bar w'_\mathrm{x}) + V_3(w_k; \bar w'_\mathrm{x}, \bar w_k)$, where the quantum numbers of the exchanged parton are $\bar w'_x = \{-\ell,-f,a,r'\}$. The part that can give an $\ell \to 0$ singularity is the first term, so we keep only $V_3(w_k; \bar w_k, \bar w'_\mathrm{x})$. 

In the case that parton $l$ is a quark, $V_3(w_l; \bar w_l, \bar w_\mathrm{x})$ is given in Eq.~(\ref{eq:Gammaqqg}). It contains a numerator factor $\s{q}_l = \s{\bar q}_l + \s{\ell}$. Since we are interested in the limit $\ell \to 0$, we replace this by $\s{\bar q}_l$. Thus we define an approximate version of $V_3$ for an incoming quark, $f_l \in \{\Lu, \Ld,\dots\}$, by
\begin{equation}
\begin{split}
\label{eq:GammaqqgIR}
\IR{V_3(w_l; \bar w_l, \bar w_x)}\hskip -2 cm {}&
\\={}&
\mi \gs t^{a}_{\bar a_l a_l}
\left[\gamma^{r}\s{\bar q}_l\right]_{\bar r_l r_l} 
\frac{\mi}{(\bar q_l + \ell)^2 + \mi 0}
\\
&\times
\theta(f_l \in \{\Lu, \Ld,\dots\})\,\theta(\bar f_l = f_l)\,\theta(f = \Lg)
\;.
\end{split}
\end{equation}
Note that we do not approximate $q_l^2 = (\bar q_l + \ell)^2$ in the denominator by $\bar q_l^2$. Although this would be a good approximation for $\ell$ near 0, we are integrating over $\ell$ and this could lead to an ultraviolet divergence for large $\ell$.

When parton $k$ is a quark, we use an analogous approximate version of $V_3(w_k; \bar w_k, \bar w'_\mathrm{x})$:
\begin{equation}
\begin{split}
\label{eq:GammaqqgkIR}
\IR{V_3(w_k; \bar w_k, \bar w'_x)}\hskip -2 cm {}&
\\={}&
\mi \gs t^{a}_{\bar a_k a_k}
\left[\gamma^{r'}\s{\bar q}_k\right]_{\bar r_k r_k} 
\frac{\mi}{(\bar q_k - \ell)^2 + \mi 0}
\\
&\times
\theta(f_k \in \{\Lu, \Ld,\dots\})\,\theta(\bar f_k = f_k)\,\theta(f = \Lg)
\;.
\end{split}
\end{equation}

In the case that parton $l$ is an antiquark, $V_3(w_l; \bar w_l, \bar w_\mathrm{x})$ is given in Eq.~(\ref{eq:Gammabarqbarqg}). We define an approximate version of this by
\begin{equation}
  \label{eq:GammabarqbarqgIR}
  \begin{split}
    \IR{V_3(w_l; \bar w_l, \bar w_x)} \hskip -2 cm {}&
    \\={}&
    -\mi \gs t^{a}_{a_l \bar a_l }\,
    \frac{\mi}{(\bar q_l + \ell)^2 + \mi 0}\,
    \left[
      \s{\bar q}_l \gamma^{r}
    \right]_{r_l \bar r_l}
    \\&\times
    \theta(f_l \in \{\bar\Lu, \bar\Ld,\dots\})\,
    \theta(\bar f_l = f_l)\,
    \theta(f = \Lg)
    \;.
  \end{split}
\end{equation}
When parton $k$ is an antiquark, we use an analogous approximate version of $V_3(w_k; \bar w_k, \bar w'_\mathrm{x})$:
\begin{equation}
\begin{split}
\label{eq:GammabarqbarqgkIR}
   \IR{V_3(w_k; \bar w_k, \bar w'_x)} \hskip -2 cm {}&
    \\={}&
    -\mi \gs t^{a}_{a_k \bar a_k }\,
    \frac{\mi}{(\bar q_k - \ell)^2 + \mi 0}\,
    \left[
      \s{\bar q}_k \gamma^{r'}
    \right]_{r_k \bar r_k}
    \\&\times
    \theta(f_k \in \{\bar\Lu, \bar\Ld,\dots\})\,
    \theta(\bar f_k = f_k)\,
    \theta(f = \Lg)
    \;.
\end{split}
\end{equation}

When parton $l$ is a gluon and the exchanged parton is a gluon, $V_3(w_l; \bar w_l, \bar w_\mathrm{x})$ is given in Eq.~(\ref{eq:softggg}). There is a propagator for the gluon with momentum $q_l = \bar q_l + \ell$. In the numerator of this propagator, we set $\ell \to 0$:
\begin{equation}
  \label{eq:softgggl}
  \begin{aligned}[c]
   \IR{V_3(w_l; \bar w_l,\bar w_x)} \hskip -1.5 cm {}&
    \\
    ={}&
    -\gs f_{a_l,\bar{a}_l,a}\,
    \frac{\mi N_{\mu r_l }(\bar q_l)}
    {(\bar q_l + \ell)^2 + \mi 0}
    \\
    &
    \times \!
    \left[
      g^{\bar{r}_l r}\,\bar{q}_{l}^{\mu}
      + g^{\mu r}\,\bar{q}_l^{\bar{r}_l}
      -2g^{\mu\bar{r}_l}\,\bar{q}_l^r
    \right]    
    \\
    &
    \times
    \theta(f_l = \bar f_l = f = \Lg)
    \;. 
  \end{aligned}
\end{equation}
Similarly when parton $k$ is a gluon we take
\begin{equation}
  \label{eq:softgggk}
  \begin{aligned}[c]
   \IR{V_3(w_k; \bar w_k,\bar w'_x)} \hskip -2 cm {}&
    \\
    ={}&
    -\gs f_{a_k,\bar{a}_k,a}\,
    \frac{\mi N_{\mu r_k }(\bar q_k)}
    {(\bar q_k - \ell)^2 + \mi 0}
    \\
    &
    \times \!
    \left[
      g^{\bar{r}_k r'}\,\bar{q}_{k}^{\mu}
      + g^{\mu r'}\,\bar{q}_k^{\bar{r}_k}
      -2g^{\mu\bar{r}_k}\,\bar{q}_k^{r'}
    \right]    
    \\
    &
    \times
    \theta(f_k = \bar f_k = f = \Lg)
    \;. 
  \end{aligned}
\end{equation}

Now $E_{lk}^{(0,1)}$ is $X_{lk}^{(0,1)}$ from Eq.~(\ref{eq:Exlk}) with these approximations that select the infrared singular part of $X_{lk}^{(0,1)}$:
\begin{equation}
  \begin{split}
    \label{eq:EGammalk}
    E_{lk}^{(0,1)}\big({}& w_l, w_k;\bar{w}_l,\bar{w}_k; \ell\big) 
    \\
    ={}&\
    \sum_{f,a,r,r'}\frac{\mi\,N_{r' r}(\ell,f)}{\ell^2 + \mi 0}
    \\
    &\quad\times
    \theta(f = \Lg)\, \theta(f_l = \bar f_l)\,
    \theta(\bar f_k =  f_k)
    \\
    &\quad\times
    \IR{V_3(w_l; \bar w_l, \bar w_\mathrm{x})}\,
    \IR{V_3(w_k; \bar w_k, \bar w'_\mathrm{x})}
    \;,
  \end{split}
\end{equation}

With these definitions, $\qE_{lk}^{(0,1)}$ can be pictured as
%---------------

% 
\begin{equation}
  \begin{split}
  \aspi\,
    \bra{\{\bar{w}\}_{m}}\textcolor{red}{\qE_{lk}^{(0,1)}}
    \ket{\chi(\{p,f,c,s\}_m}
    \hskip -3.5cm &
    \\
    &=  \IR[lr]{
    \begin{prdfig}{0b7638745416b7f75a4f90708d397ed0}{hard-vertex-exchange}
      \begin{tikzpicture}[baseline=(current bounding box.center)]
        \begin{feynman}[]
          \vertex[blob, fill=black, minimum size=0.55cm] (v1) at (0,0) {};
          \vertex[empty dot,label={[above] $u_1$}] (w1) 
          at ($(v1) + (135:1.15cm)$) {};
          \vertex[empty dot,label={[below=1mm] $u_m$}] (wm) 
          at ($(v1) + (-135:1.15cm)$) {};
          \vertex[empty dot,label={[left] $u_l$}] (wl) 
          at ($(v1) + (155:1.15cm)$) {};
          \vertex[empty dot,label={[left] $u_k$}] (wk) 
          at ($(v1) + (-155:1.15cm)$) {};
          \coordinate[] (ow1) at ($(v1) + (145:0.4cm)$);
          \coordinate[] (owm) at ($(v1) + (-145:0.4cm)$);
          \coordinate[] (owl) at ($(v1) + (165:0.4cm)$);
          \coordinate[] (owk) at ($(v1) + (-165:0.4cm)$);
          \coordinate [label={[right] $\bar{w}_1$}] (u1) 
          at ($(v1) + (45:1.5cm)$);
          \coordinate [label={[right] $\bar{w}_{m}$}] (um) 
          at ($(v1) + (-45:1.5cm)$);
          \vertex [dot,red] (vl) at ($(v1) + (20:1.1cm)$) {};
          \vertex [dot,red] (vk) at ($(v1) + (-20:1.1cm)$) {};         
          \coordinate [label={[right] $\bar{w}_l$}] (ul) 
          at ($(v1) + (20:1.5cm)$);
          \coordinate [label={[right] $\bar{w}_k$}] (uk) 
          at ($(v1) + (-20:1.5cm)$);
          \draw[{Latex[length=1mm]}-] (0.55cm,-0.05) arc (-155:155:0.15);
          \diagram*{
            (w1)--[](ow1);
            (wm)--[](owm);
            (wl)--[](owl);
            (wk)--[](owk);
            (v1)--[](u1);
            (v1)--[](um);
            (v1)--[red](vl)--[red](ul);
            (v1)--[red](vk)--[red](uk);
            (vl)--[red,gluon](vk);
          };      
          % decorations
          \vertex [rotate=90] at ($0.5*(wl) + 0.5*(wk)$) {. . .};
          \vertex [rotate=90] at ($0.5*(ul) + 0.5*(uk)$) {. . .};
        \end{feynman}
      \end{tikzpicture}
    \end{prdfig}
    }
    \;\;,
  \end{split}
\end{equation}
where $u_i = \{p_i, f_i, c_i, s_i\}$.

For a shower with $\as^2$ splitting functions, we also need the contributions of the operator $\qGamma_\scH$ that represent 1PI graphs with one real emission and one virtual loop. These are obtained by summing contributions in which the virtual exchange is between partons $l$ and $k$, with the emission of a parton labelled $m+1$:
\begin{equation}
\qGamma_\scH^{(1,1)} = \sum_{l} \frac{1}{2}\sum_{k\ne l}
( 1 + \qPi_l + \qPi_k)
\qE^{(1,1)}_{lk}
\;.
\end{equation}
Here  $\qPi_l$ exchanges parton $l$ with parton $m+1$ and $\qPi_k$ exchanges parton $k$ with parton $m+1$, as defined in Eq.~(\ref{eq:PilandPi}). The operator $\qE^{(1,1)}_{lk}$ is defined by its Feynman graph,
\begin{equation}
  \label{eq:qGamma11H}
  \begin{split}
    \left[\aspi\right]^{3/2}{}&
    \bra{\{\bar{w}\}_{m}}
    \textcolor{red}{\qE_{lk}^{(1,1)}}\ket{\chi(\{p,c,s\}_m)}  
    \\
    &\quad= 
    \begin{prdfig}{f3ca3d8d37df5b908f02fd474d352a2d}{hard-vertex-exchg-to-3g}
      \begin{tikzpicture}[baseline=(current bounding box.center)]
        \begin{feynman}[]
          \vertex[blob, fill=black, minimum size=0.55cm] (v1) at (0,0) {};
          \vertex[empty dot,label={[above] $u_1$}] (w1) 
          at ($(v1) + (135:1.15cm)$) {};
          \vertex[empty dot,label={[below=1mm] $u_m$}] (wm) 
          at ($(v1) + (-135:1.15cm)$) {};
          \vertex[empty dot,label={[above] $u_l$}] (wl) 
          at ($(v1) + (160:1.5cm)$) {};
          \vertex[empty dot,label={[below=1mm] $u_k$}] (wk) 
          at ($(v1) + (-160:1.5cm)$) {};
          \coordinate[] (ow1) at ($(v1) + (145:0.4cm)$);
          \coordinate[] (owm) at ($(v1) + (-145:0.4cm)$);
          \coordinate[] (owl) at ($(v1) + (165:0.4cm)$);
          \coordinate[] (owk) at ($(v1) + (-165:0.4cm)$);
          \coordinate [label={[right] $\bar{w}_1$}] (u1) 
          at ($(v1) + (55:1.5cm)$);
          \coordinate [label={[right] $\bar{w}_{m}$}] (um) 
          at ($(v1) + (-55:1.5cm)$);
          \vertex [dot,red] (vl) at ($(v1) + (30:1.25cm)$) {};
          \vertex [dot,red] (vk) at ($(v1) + (-30:1.25cm)$) {};         
          \vertex [dot,red] (vm) at ($(v1) + (0:1.25cm)$) {};         
          \coordinate [label={[right] $\bar{w}_l$}] (ul) 
          at ($(v1) + (30:2cm)$);
          \coordinate [label={[right] $\bar{w}_k$}] (uk) 
          at ($(v1) + (-30:2cm)$);
          \coordinate [label={[right] $\bar{w}_{m\!+\!1}$}] (um1) 
          at ($(v1) + (0:2cm)$);
          \draw[{Latex[length=1mm]}-] (0.6cm,-0.05) arc (-155:155:0.2);
          \diagram*{
            (w1)--[](ow1);
            (wm)--[](owm);
            (wl)--[](owl);
            (wk)--[](owk);
            (v1)--[](u1);
            (v1)--[](um);
            (v1)--[red](vl)--[red](ul);
            (v1)--[red](vk)--[red](uk);
            (vl)--[red](vm)--[red](vk);
            (vm)--[red](um1);
          };      
          % decorations
          \vertex [rotate=90] at ($0.5*(wl) + 0.5*(wk)$) {. . .};
        \end{feynman}
      \end{tikzpicture}
    \end{prdfig}
    \;\;, 
  \end{split}
\end{equation}
with the restriction that one or both of the exchanged partons must be a gluon. We omit the graphs in which both of the exchanged partons are quarks or antiquarks because these graphs do not have infrared singularities. The integration over the loop momentum $\ell$ in $\qE_{lk}^{(1,1)}$ is finite in the ultraviolet region in which $\ell$ is much larger than the external momenta $\bar q_l$, $\bar q_k$ and $\bar q_{m\!+\!1}$. Thus $\qE_{lk}^{(1,1)}$ is the same as $\qX_{lk}^{(1,1)}$, Eq.~(\ref{eq:Xlk11}), except for the restriction that one or both of the exchanged partons must be a gluon.

%-------------------------------------------------
\subsection{Renormalization scale dependence}
\label{sec:1PIrenormalizationscale}

Consider a renormalized Green function $G$ with $m$ external legs that is amputated on each external leg. The dependence of $G$ on the renormalization scale $\mur$ is
\begin{equation}
\label{eq:muRdependence}
\mur\,\frac{dG}{d\mur} = \sum_l \gamma_l \cdot G
\;.
\end{equation}
Here $\gamma_l$ is the anomalous dimension for the field strength renormalization of the field on line $l$, $\psi(x)$ for a quark or antiquark line and $A^{\mu}(x)$ for a gluon line. The dot in Eq.~(\ref{eq:muRdependence}) indicates contraction of vector/spinor indices, as in Eq.~(\ref{eq:gammaldefdetailed}) below. For $\psi(x)$, the dependence on spinor indices is trivial: $Z_\psi$ and $\gamma_\psi$ are proportional to unit matrices on the spinor indices. However, in interpolating gauge, $Z_A$ and $\gamma_A$ are non-trivial tensors in the vector indices. 

We can consider $Z_\psi$, $Z_A$, $\gamma_\psi$, and $\gamma_A$ to be operators on the vector/spinor part of the ket amplitude space. Then the anomalous dimensions are defined by
\begin{equation}
\begin{split}
\label{eq:gammaldef}
\mur\,\frac{d Z_\psi}{d\mur} ={}& 
2 \gamma_\psi\,Z_\psi
\;,
\\
\mur\,\frac{dZ_A^{\mu\nu}}{d\mur} ={}& 
2 \gamma_A\,Z_A
\;,
\end{split}
\end{equation}
where $Z_\psi$ and $Z_A$ are the field strength renormalization constants in interpolating gauge, which are given at first order in Ref.~\cite{Gauge}.\footnote{In Eq.~(158) of Ref.~\cite{Gauge}, the coupling renormalization relation was given as $g_\LB = Z_g g$. With the more standard conventions of Ref.~\cite{JCCfoundations}, this relation is $g_\LB = \mu^\epsilon Z_g g$.} 
Including the vector/spinor indices, this is
\begin{equation}
\begin{split}
\label{eq:gammaldefdetailed}
\mur\,\frac{d\, \dualL\bra[]{\bar r} Z_\psi \ket[]{r}}{d\mur} ={}& 
2 \sum_{r'}\dualL\bra[]{\bar r}\gamma_\psi\ket[]{r'}\,
\dualL\bra{r'}Z_\psi \ket{r}
\;,
\\
\mur\,\frac{d\, \dualL\bra[]{\bar r} Z_A \ket[]{r}}{d\mur} ={}& 
2 \sum_{r'}\dualL\bra[]{\bar r}\gamma_A\ket[]{r'}\,
\dualL\bra[]{r'}Z_A \ket{r}
\;.
\end{split}
\end{equation}

The anomalous dimensions are \cite{Gauge}
\begin{equation}
\begin{split}
\label{eq:gammalresult}
\dualL\bra[]{\bar r}\gamma_\psi\ket[]{r} ={}&  
\frac{1}{2}\,
\frac{\as}{2\pi}\, S_\epsilon 
\left\{
\frac{(v-1)^2}{v(v+1)} + 
\frac{\xi}{v}
\right\} C_\LF\,
\delta_{\bar r r}
\\&
+\cO(\as^2)
\;,
\\
\dualL\bra[]{\bar r}\gamma_A\ket[]{r} ={}& 
- \frac{1}{2}\,\frac{\as}{2\pi}\, S_\epsilon 
\left\{
c_\LA g^{\bar r}_r + \tilde c_\LA h^{\bar r}_r
\right\} 
\\&
+\cO(\as^2)
\;.
\end{split}
\end{equation}
In $\gamma_A$, the constants $c_\LA$ and $\tilde c_\LA$ are
\begin{equation}
\begin{split}
c_\LA ={}& \left[
\frac{22 v^3 + 35 v^2 + 20 v - 1}{6 v (v+1)^2}
- \frac{\xi}{2 v}
\right] C_\LA
\\&
- \frac{4}{3}\, T_\LR n_\Lf
\;,
\\
\tilde c_\LA ={}& - \frac{4 v (2v+1)}{3 (v+1)^2}\,C_\LA
\;.
\end{split}
\end{equation}

Our Green functions represented in $\qGamma_\scH$ have $m$ external quark, antiquark, and gluon legs and are amputated on those legs. However, if we were to leave out the interactions $\qN_l^{(0,1)}$ in Eq.~(\ref{eq:qGammaH01}), they would lack the $\mur$ dependence of Eq.~(\ref{eq:muRdependence}). We introduce the needed $\mur$ dependence by hand by including interactions $\qN_l^{(0,1)}$. We define these interactions by 
\begin{equation}
\begin{split}
\label{eq:qKlk01}
\aspi\,
\dualL\bra{\{\bar w\}_m}
\qN_{l}^{(0,1)}\ket{\chi(\{p,f,c,s\}_m)}
\hskip - 4.5 cm {}&
\\
= {}& 
\sum_{\{a,r\}_m}\,
\brax{\{a\}_{m}}\ket{\{c\}_{m}}
\prod_{i=1}^m
\dualL\brax{r_i}\ket{\chi_{f_i}(p_i,s_i)}
\\&\times
\prod_{\substack{i=1\\ i\neq l}}^m
\left[
\delta_{\bar{f}_i, f_i}\,\delta_{\bar{a}_i, a_i}\,\delta_{\bar{r}_i, r_i}
\right]
\\&\times
\delta_{\bar{f}_l, f_l}\,\delta_{\bar{a}_l, a_l}
N_{l}(f_l,\mur;\bar{r}_l, r_l)
\;.
\end{split}
\end{equation}
This operator has no momentum dependence and is proportional to the unit matrix in flavors, colors, and vector/spinor indices for spectator partons, those with indices $i \ne l$. For parton $l$, $\qN_{l}^{(0,1)}$ is proportional to the unit matrix in flavor and color. When parton $l$ is a quark or antiquark, $\qN_{l}^{(0,1)}$ is also proportional to the unit matrix in the spinor indices.  When parton $l$ is a gluon, $\qN_{l}^{(0,1)}$ has a nontrivial dependence on the vector indices. We define the constants $N_{l}(f_l,\mur;\bar{r}_l, r_l)$ by
\begin{equation}
\label{eq:Kldef}
N_{l}(f_l,\mur;\bar{r}_l, r_l) =
\log(\mur/\mu_{l})\,
\frac{\as(\mur)}{2\pi}
\dualL\bra[]{\bar{r}_l}\gamma_l^{(1)}\ket[]{r_l}
.
\end{equation}
Here $[\as/(2\pi)]\,\gamma_l^{(1)}$ is the contribution to $\gamma_l$ in Eq.~(\ref{eq:gammalresult}) that is proportional to the first power of $\as$ and $\mu_l$ is a fixed scale parameter. One could choose, for instance, $\mu_l^2 = \min_k(2 p_l\cdot p_k)$. This definition gives us the desired dependence on $\mur$:
\begin{equation}
\mur\,\frac{d}{d\mur}\,N_{l}(f_l,\mur,\bar{r}_l, r_l)
= \frac{\as(\mur)}{2\pi}\,
\dualL\bra[]{\bar{r}_l}\gamma_l^{(1)}\ket[]{r_l}
\;.
\end{equation}
%

%---------------------------------
\subsection{Conjugate Feynman amplitude}
\label{sec:FeynmanBraAmplitude}

In the ket Feynman amplitude, we can start with a vector $\ket{\chi}$ and apply operators $\qGamma$ to it. In the corresponding construction in the bra Feynman amplitude, we start with a vector $\ibra{\chi}$ and apply operators $\bm \Gamma^\dagger$ to it. Using the completeness relation (\ref{eq:wcompletentness}), this is
\begin{equation}
\begin{split}
\bra{\chi}\bm\Gamma_1^\dagger{}& \bm\Gamma_2^\dagger \cdots 
\\={}&
\sum_{m, \bar m}
\int\! d\{w'\}_{m} \int\! d\{\bar w'\}_{\bar m} 
\\&\times
\brax{\chi}\ket{\{w'\}_m}
\dualL\bra{\{w'\}_m}
\bm\Gamma_1^\dagger \bm\Gamma_2^\dagger \cdots
\ket{\{\bar w'\}_{\bar m}}
\\&\times
\dualL\bra{\{\bar w'\}_{\bar m}}
\;.
\end{split}
\end{equation}
Thus we need $\dualL\bra{\{w'\}_m}\bm\Gamma^\dagger \ket{\{\bar{w}'\}_{\bar{m}}}$.  As an example, for $\bm{\Gamma}_l^{(1,0)\dagger}$ in Eq.~(\ref{eq:Gammal1n}), we have 
\begin{equation}
\begin{split}
\label{eq:Gammal1ndagger}
&\left[\frac{\as}{2\pi}\right]^{1/2}\,
\dualL\bra{\{w'\}_{m}}\qGamma^{(1,0)\dagger}_l
\ket{\{\bar w'\}_{m\!+\!1}}
\\
&\qquad
= \prod_{i\ne l}\left[(2\pi)^d\delta^d(\bar q'_i - q'_i)\,\delta_{\bar f'_i,f'_i} 
\delta_{\bar a'_i,a'_i}\delta_{\bar r'_i,r'_i}\right]
\\
&\qquad\quad
\times (2\pi)^d \delta^d(\bar q'_l + \bar q'_{m\!+\!1} - q'_l)\,
\\
&\qquad\quad
\times 
\overline\Gamma^{(0)}\big(w'_l; \{\bar w'_l, \bar w'_{m\!+\!1}\}\big)
\;.
\end{split}
\end{equation}
The functions $\overline\Gamma^{(0)}(w'_l; \{\bar w'_l, \bar w'_{m\!+\!1}\})$ are quite simply related to the functions $\Gamma^{(0)}(w_l; \{\bar w_l, \bar w_{m\!+\!1}\})$ used in the ket amplitude. The spinor factors are the Dirac adjoints of the corresponding factors in $\Gamma(w_l; \{\bar w_l, \bar w_{m\!+\!1}\})$. Two significant differences are the replacement of $1/(q_l^2 + \mi 0)$ in the mother parton propagator by $1/(q_l^{\prime 2} - \mi 0)$ and the reversal of the order of $\gamma$ matrices.

%---------------------------------
\section{Quantum evolution space}
\label{sec:QuantumEvolutionSpace}

The operator $\cD(\mu_\scS)$ acts on a vector space that includes both bra and ket amplitudes, the statistical space described in Sec.~\ref{sec:statisticalspace}. Now we need a similar vector space that combines the evolution of Feynman diagrams for the ket amplitude and for the conjugate bra amplitude.

%--------------------
\subsection{Vector space for the ket$\times$bra amplitudes}
\label{sec:quantumbraketspace}

In order to work with the operators $\bm G(G_\mathrm{ket})$ and $\bm G^\dagger(G_\mathrm{bra})$ together, we use the tensor product of the vector space for the quantum ket amplitude and the vector space for the quantum bra amplitude. We call this the quantum evolution space. Typically, we need only basis vectors with the same number of partons in the amplitude and the conjugate amplitude. These are
\begin{equation}
\begin{split}
\sket{\{w,w'\}_{m}}
={}&\ket{\{w\}_{m}}\,  \dualL\bra{\{w'\}_{m}}
\;.
\end{split}
\end{equation}

Letting an operator $\bm G(G_\mathrm{ket})$ act on the ket vectors and an operator $\bm G^\dagger (G_\mathrm{bra})$ act on the bra vectors gives a product operator $\bm G(G_\mathrm{ket}) \otimes {\bm G}^\dagger(G_\mathrm{bra})$ defined by
\begin{equation}
\begin{split}
\label{eq:graphoperators}
\big[\bm G(G_\mathrm{ket}) \otimes {}&\bm G^\dagger(G_\mathrm{bra})
\big]
\sket{\{w,w'\}_{m}}
\\={}&  \bm G(G_\mathrm{ket}) \ket{\{w\}_{m}}\,
\dualL\bra{\{w'\}_{m}}
\bm G^\dagger(G_\mathrm{bra})
\;.
\end{split}
\end{equation}

It is useful to introduce names for operators on this space. As a convention, we name operators in the quantum evolution space with hatted boldface letters. For instance, we define elementary operators $\eC^{(1,0)}_l(\mu)$, $\eX^{(1,0)}_{lk}(\mu)$, $\eX_{lk}^{(0,1)}$ and $\eE_{lk}^{(0,1)}$ in this section.

We have already defined many elementary operators in the quantum space of Feynman amplitudes, such as $\bm{\Gamma}^{(1,0)}_l$. With them we can  immediately define related elementary operators in the quantum evolution space. 

%---------------------------------
\subsection{Hardness measure}
\label{sec:hardness}

We will define vertex operators in the vector space of Feynman amplitudes to build Feynman graphs. We would like to associate to every vertex with real radiation a scale that characterizes the hardness of the splitting. This scale should measure the distance from the infrared singular surfaces. 

To do this we define a hardness measure $h(q)$ that tests the hardness of a line carrying momentum $q$ in the graph. This function has to be infrared sensitive,
\begin{equation}
h(q) \to 0\quad \text{as}\quad q^2\to 0
\;.
\end{equation}
The simplest hardness function would be simply the virtuality,
\begin{equation}
\label{eq:q2-measure}
h(q) = q^2
\;.
\end{equation}
Another sensible choice is the virtuality based choice made in \textsc{Deductor}:
\begin{equation}
\begin{split}
\label{eq:Lambdasqcoll}
h(q) ={}& 
\frac{q^2}{2 q\cdot Q}\,Q^2
\;.
\end{split}
\end{equation}
This function is denoted as $\Lambda^{\!2}(q_l)$ and explained in Ref.~\cite{ShowerTime}.

Both functional forms for $h(q)$ are monotonic in a sequence of real parton emissions. Consider a first order splitting of parton $l$ with momentum $q_l$ followed by subsequent splittings of the daughter partons, which then have momenta $\bar q_i$ and $\bar q_j$. Then
\begin{equation}
q_l = \bar q_i + \bar q_j
\;.
\end{equation}
Assuming that there are only real parton emissions, we have
\begin{equation}
\begin{split}
\bar q_i^2 \ge{}& 0\;,\qquad \bar q_i \cdot Q > 0
\;,
\\
\bar q_j^2 \ge{}& 0\;,\qquad \bar q_j \cdot Q > 0
\;.
\end{split}
\end{equation}
Under these conditions, the function $h(q)$ defined by Eq.~(\ref{eq:Lambdasqcoll}) obeys
\begin{equation}
\begin{split}
\label{eq:Hinequality}
h(\bar q_i + \bar q_j) \ge{}& h(\bar q_i)
\;,
\\
h(\bar q_i + \bar q_j) \ge{}& h(\bar q_j)
\;.
\end{split}
\end{equation}
We present the proof of this in Appendix \ref{sec:hardnessinequality}. These relations hold trivially if $h(q) = q^2$.

Hardness measures based on transverse momentum are often used in parton shower algorithms. We do not consider these measures in this paper.

We measure the hardness of a splitting using the momenta of a set $A \subset \{1,\dots, m\}$ of partons after the splitting in the ket amplitude and a set $B \subset \{1,\dots, m\}$ of partons after the splitting in the bra amplitude. We define an operator $\eh(A,B;\mu)$ on the quantum evolution space by
\begin{equation}
\begin{split}
\label{eq:ehABdef}
\eh(A,&B;\mu)\sket{\{w, w'\}_{m}}
\\
={}& 
2\mu\,
\delta\bigg(\mu^2 - 
\max\!\bigg[h\bigg(\sum_{i\in A}{q}_i\bigg),
h\bigg(\sum_{j\in B}{q}'_j\bigg)\bigg]
\bigg)
\\&\times
\sket{\{w, w'\}_{m}}
\;.
\end{split}
\end{equation}

In most cases, we use two specializations of this operator. For single collinear emission in which parton $l$ splits into partons $l$ and $m+1$ in both the ket and bra amplitudes, we use
\begin{equation}
\begin{split}
\label{eq:coll-measure}
\eh_{l}(\mu)&\sket{\{w, w'\}_{m + 1}}
 \\={}&  \eh\big(\{l,m+1\},\{l,m+1\}; \mu\big)
 \sket{\{w, w'\}_{m + 1}}
\;.
\end{split}
\end{equation}
For the single soft emission in which parton $l$ splits into partons $l$ and $m+1$ in the ket amplitude and parton $k$ splits into partons $k$ and $m+1$ in the bra amplitude, we use
\begin{equation}
\begin{split}
\label{eq:soft-measure}
\eh_{lk}(\mu)&\sket{\{w, w'\}_{m + 1}}
 \\={}&  \eh\big(\{l,m+1\},\{k,m+1\}; \mu\big)
 \sket{\{w, w'\}_{m + 1}}
\;.
\end{split}
\end{equation}
%

%--------------------
\subsection{Operators on the quantum evolution space}
\label{sec:quantumbraketoperators}

We have seen how to express Feynman diagrams using operators on the ket amplitude space and operators on the bra amplitude space. Now we need to put these together.

For operators describing real emissions, we also introduce a new scale $\mu$ with the aid of the operator $\hat{\bm{h}}(A, B; \mu)$, which measures the hardness of the real emissions. In a later section, Sec.~\ref{sec:HardSoftExtended}, we will define the shower algorithm via a jet observable in which the scale $\mu$ is the jet resolution scale. 

For operators that describe virtual graphs, we do not define a new scale since the jet observable that we will use in the shower algorithm is sensitive only to real parton emissions, not virtual diagrams.

%---------------------------------
\subsubsection*{Tree level irreducible operators}

We define an operator $\eC^{(1,0)}_l(\mu)$ that describes the splitting of parton $l$ out of $m$ partons to partons $l$ and $m\!+\!1$ in both the ket amplitude and the bra amplitude: 
\begin{equation}
  \label{eq:hatCl}
  \begin{split}
    \frac{1}{\mu}\,
    \eC^{(1,0)}_l&(\mu)\sket{\{w,w'\}_{m}}
    \\= {}&  
    \eh_l(\mu)\,
    \frac{1}{2}
    \left[\bm\Gamma^{(1,0)}_l
      \otimes \bm\Gamma^{(1,0)\dagger}_l\right]
    \sket{\{w,w'\}_{m}}
    \;.
  \end{split}
\end{equation}
We include a factor $1/2$ to account for the symmetry between the two emitted partons and a factor $1/\mu$ so that the natural integration measure for $\eC^{(1,0)}_l(\mu)$ is $d\mu/\mu$. The scale operator $\eh_l(\mu)$ provides a delta function that restricts the allowed final states according to the given value of $\mu$.

Similarly we can represent simple interference diagrams that appear when parton $m + 1$ is emitted by parton $l$ in the ket amplitude and by parton  $k$ in the bra amplitude:
\begin{equation}
  \label{eq:eXlk}
  \begin{split}
    \frac{1}{\mu}\,
    \eX^{(1,0)}_{lk}&(\mu) \sket{\{w,w'\}_{m}}
    \\= {}&  
    \eh_{lk}(\mu)\,
    \left[\bm\Gamma^{(1,0)}_l
      \otimes \bm\Gamma^{(1,0)\dagger}_k\right]
    \sket{\{w,w'\}_{m}}
    \;.
  \end{split}
\end{equation}
% 

%---------------------
\subsubsection*{One loop irreducible operators}

We can use the exchange operators $\bm X_{lk}^{(0,1)}$ and $\bm E_{lk}^{(0,1)}$ defined in Eqs.~ (\ref{eq:qXlk01A}) and (\ref{eq:qGammalk01def}) to define operators that describes an exchange between two partons, $l$ and $k$:
\begin{equation}
\begin{split}
\label{eq:eXlk01eGammalk01}
\eX_{lk}^{(0,1)} ={}& 
\bm X_{lk}^{(0,1)}\otimes\bm{1}
+ \bm{1}\otimes\bm X_{lk}^{(0,1)\dagger}
\;,
\\
\hat {\bm E}_{lk}^{(0,1)} ={}& 
\bm E_{lk}^{(0,1)}\otimes\bm{1}
+ \bm{1}\otimes\bm E_{lk}^{(0,1)\dagger}
\;.
\end{split}
\end{equation}
The exchange graphs appear either on the ket or the bra side. 

Here we do not introduce a scale $\mu$ based on the momenta of emitted partons because no partons are emitted. However, it is important that the operators depend on $\mur$. When the exchanged parton is a gluon, the operator has a soft singularity that arises from the integration over the momentum of the exchanged gluon, leading to a contribution proportional to $\mur^{2\epsilon}/\epsilon$.

%-------------------------------------------------
\section{Extended statistical space}
\label{sec:extendedspace}

With the quantum evolution space, we can describe parton splittings and virtual graphs starting from $m$ partons with momenta $q_i$ in the ket amplitude and $q'_i$ in the bra amplitude. The parton momenta in these amplitudes are generally off shell. We now need to extend this to a larger space that can describe the final state after these splittings in which the parton momenta are on shell.

%---------------------------------
\subsection{Jet momentum space}
\label{sec:JetMomentumSpace}

We need one more factor in the space that describes Feynman diagrams with  parton emissions starting from $m$ partons. We need this extra factor in order to describe the momentum mapping that gives on-shell momenta for the partons associated with a graph $G = (G_\mathrm{ket},G_\mathrm{bra})$.

As described in Sec.~\ref{sec:singularities}, in the graph $G$, we can think of the on-shell final state partons as subjets of larger jets. Let there be $n_\scJ$ initial jets with momenta $\{p\}_{n_\scJ} = \{p_1,\dots,p_{n_\scJ}\}$. The initial jets are idealized as having $p_j^2 = 0$. Let there be $\hat n_\scJ$ final state subjets with momenta $\{\hat p\}_{\hat n_\scJ}$ with $\hat p_i^2 = 0$. For each initial jet $j$, a subset $J(j)$ of the final state subjets are part of jet $j$. Each final state subjet is a part of one of the initial jets. There is an infrared singularity in $G$ when the momenta $\hat p_i$ for $i \in J(j)$ are collinear with $p_j$ or are soft and the momenta $\ell_i$ of any exchanged gluons vanish.

When the $\hat p_i$ for $i \in J(j)$ are not zero or exactly collinear with  momentum $p_j$, then we need a momentum mapping from the $\{p\}_{n_\scJ}$ to the $\{\hat p\}_{\hat n_\scJ}$, as described in Sec.~\ref{sec:momentummapping}. We need to include information about the $\{p\}_{n_\scJ}$ in the vector space. Then we can include the operators that describe the momentum mapping along with the operators that describe the graph $G$ according to the Feynman rules.

As in Eq.~(\ref{eq:Rmap}), we relate the final jet momenta $\{\hat p\}_{\hat n_\scJ}$ to the initial jet momenta $\{p\}_{n_\scJ}$ and splitting parameters $\zeta_{G}$ by a momentum mapping function
\begin{equation}
\label{eq:Rmapbis}
\{\hat p\}_{\hat n_\scJ} = R(G;\zeta_{G},\{p\}_{n_\scJ})
\end{equation}
that obeys
\begin{equation}
\sum_{i=1}^{\hat n_\scJ} \hat p_i = 
\sum_{i=1}^{n_\Lj} p_i
\;.
\end{equation}
The momentum mapping also includes a Jacobian factor in the integration measure $d\zeta_G$ for integration over the splitting parameters so that
\begin{equation}
\label{eq:dzetaproperty}
d\{\hat p\}_{\hat n_\scJ} = d\zeta_{G}\ d\{p\}_{n_\scJ}
\;.
\end{equation}

The momentum mapping becomes trivial as the momenta $\{\hat p\}_{\hat n_\Lj}$ approach the infrared singularities of the graph $G$. Beyond that, there is some freedom in choosing the function $R$. The functions $R$ used in this paper are defined in Appendices \ref{sec:CollinearMomentumMapping} and \ref{sec:SoftMomentumMapping}. Often the momentum mapping for emitting two partons will be constructed as a product of mappings from $m$ partons to $m+1$ partons and then to $m+2$ partons.

In order to incorporate the momentum mapping function, we define a vector space, the jet momentum space, with basis vectors $\sket{\{p\}_{n_\scJ}}$ that specify the momenta of an arbitrary number $n_\scJ$ of jets. The completeness relation for these basis vectors is
\begin{equation}
\label{eq:completenessp}
1 = \sum_{{n_\scJ} =1}^\infty \frac{1}{{n_\scJ} !}
\int\! d\{p\}_{n_\scJ}\
\sket{\{p\}_{n_\scJ}}\sbra{\{p\}_{n_\scJ}}
\;,
\end{equation}
using the integration measure $d\{p\}_{n_\scJ}$ from Eq.~(\ref{eq:dp}).
 
We specify the jet momentum mapping with the use of an operator $\bm R(G)$ that acts on the space of jet momenta spanned by the basis vectors $\sket{\{p\}_{n_\scJ}}$. This operator is defined using the function $R(G;\zeta_{G},\{p\}_{n_\scJ})$ by 
\begin{equation}
\label{eq:bmRdef}
\bm R(G) \sket{\{p\}_{n_\scJ}} 
= \int\!d\zeta_G\ \sket{\{\hat p\}_{\hat n_\scJ}}
\;,
\end{equation}
where $\{\hat p\}_{\hat n_\scJ}$ is given in Eq.~(\ref{eq:Rmapbis}) and $\int\!d\zeta_G$ indicates integration over the splitting parameters, with the property (\ref{eq:dzetaproperty}). This operator is explained in more detail in Appendices \ref{sec:CollinearMomentumMapping}, \ref{sec:SoftMomentumMapping}, and \ref{sec:MomentumMappingOperator}.

\subsection{Extended statistical space}
\label{sec:ExtendedStatisticalSpace}

We supplement the vectors $\sket{\{w,w'\}_{m}}$ in the quantum evolution space with the jet basis vectors $\sket{\{p\}_{n_\scJ}}$. This gives us a direct product space, the extended statistical space, with basis vectors
\begin{equation}
\begin{split}
\label{eq:extendedstatspacevectors}
\sket{\{w,w'\}_{m};\{p\}_{n_\scJ}}
={}& \sket{\{w,w'\}_{m}}
\otimes \sket{\{p\}_{n_\scJ}}
\;.
\end{split}
\end{equation}

We define bra vectors in the extended statistical space so that the completeness relation is
\begin{equation}
\begin{split}
\label{eq:extendedcompleteness}
1 ={}&  \sum_{m,n_\scJ}
\frac{1}{n_\scJ !}
\!\int\!d\{w\}_m  \int\! d\{w'\}_m\!
\int\!d\{p\}_{n_\scJ}
\\&\times
\sket{\{w,w'\}_{m};\{p\}_{n_\scJ}}
\sbra{\{w,w'\}_{m};\{p\}_{n_\scJ}}
\;.
\end{split}
\end{equation}
In our analysis, we mostly use terms with $n_\scJ = m$. The integration measures are given by Eqs.~(\ref{eq:intqfar}) and (\ref{eq:dp}).

%--------------------------------------------------
\subsection{Spectator parton evolution}
\label{sec:SpectatorEvolution}

We have defined operators on the quantum evolution space that generate the elementary parts of Feynman graphs. We now need to extend these operators to the extended statistical space by associating momentum mapping operators with them. We use calligraphic letters with hats, for example $\xC$, to name these operators.

The extended statistical space includes a factor to describe the jet momenta $\{p\}_m$. With this factor, each splitting operator acting on the quantum evolution space can be accompanied by a momentum mapping operator acting on the jet momentum space. For example, to extend the interference emission operator $\eX^{(1,0)}_{lk}(\mu)$, Eq.~(\ref{eq:eXlk}), to the extended statistical space, we add the single emission soft momentum mapping operator $\bm R_{\rm soft}^{(1)}$, Eq.~(\ref{eq:Rsoftn}). The simplest definition would be
\begin{equation}
\label{eq:xCl}
\xX^{(1,0)}_{lk}(\mu)_\mathrm{simple} = \eX^{(1,0)}_{lk}(\mu)
\otimes \bm R_{\rm soft}^{(1)}
\;.
\end{equation}
With the inclusion of a momentum mapping operator like $\bm R_{\rm soft}^{(1)}$, at each step of Feynman graph evolution with one or more partons emitted, there is a corresponding step of momentum mapping. The jet momenta $\{p\}_m$ are mapped to new momenta $\{\hat p\}_{\hat m}$. 

With this formulation, we start with a quantum ket amplitude $\ket{\chi(\{p,f,c,s\}_m)}$  times a quantum bra amplitude $\bra{\chi(\{p,f,c',s'\}_m)}$, as given in Eq.~(\ref{eq:InitialKetVector}) and (\ref{eq:InitialBraVector}). We start with an initial vector $\sket{\{p\}_m}$ in the jet momentum space. This gives a vector in the extended statistical space. We then have evolution in the extended statistical space according to a graph $G$. In the jet momentum space, this gives a new state $\sket{\{\hat p\}_{\hat m}}$. We then multiply the evolved  quantum ket amplitude by a final state amplitude $\ibra{\psi(\{\hat p, \hat f, \hat c, \hat s\}_{\hat m})}$, Eq.~(\ref{eq:wavefctnOUTket}). We also multiply the evolved  quantum bra amplitude by a final state amplitude $\iket{\psi(\{\hat p, \hat f, \hat c', \hat s'\}_{\hat m})}$, Eq.~(\ref{eq:wavefctnOUTbra}). This procedure will be defined precisely in Sec.~\ref{sec:MappingS} below.

Many of the partons in the ket amplitude are spectator partons, for which there is no interaction. If we were to use just Eq.~(\ref{eq:xCl}), we would have for each spectator parton $i$ in the ket amplitude a factor
\begin{equation}
\label{eq:spinoverlapi}
\brax{\psi_{f_i}(\hat p_i, \hat s_i)}
\ket{\chi_{f_i}(p_i, s_i)}
\;.
\end{equation}
There would be a similar factor for each spectator parton in the bra amplitude. If $\hat p_i$ were equal to $p_i$, the factor in Eq.~(\ref{eq:spinoverlapi}) would be just $\delta_{\hat s_i, s_i}$ (using the identity (\ref{eq:PSPHspins})). However, the global momentum mappings used in this paper modify the momenta of spectator partons. This implies that if we were to use just Eq.~(\ref{eq:xCl}), the momentum mapping would have the side effect of modifying the spins of spectator partons.

In order to keep the spin of a spectator parton from changing when its momentum changes from $p_i$ to $\hat p_i$, we apply an operator $\sum_s \iket{\chi_{f_i}(\hat{p}_i, s)}\,\ibra{\psi_{f}(p_i, s)}$ acting on the ket amplitude space.  Then 
\begin{equation}
\begin{split}
\sum_s \ket{\chi_{f_i}(\hat{p}_i, s)}{}&
    \brax{\psi_{f_i}(p_i, s)}
    \ket{\chi_{f_i}(p_i, s_i)}
\\ ={}&
\ket{\chi_{f_i}(\hat{p}_i, s_i)}
\;.
\end{split}
\end{equation}
With this change, the factor in Eq.~(\ref{eq:spinoverlapi}) for each spectator parton becomes $\ibrax{\psi_{f_i}(\hat p_i, \hat s_i)}\iket{\chi_{f_i}(\hat p_i, s_i)} = \delta_{\hat s_i, s_i}$. We also use the corresponding operator acting on the bra amplitude space.

We update the vector/spinor factors for all of the spectator partons
both in the bra and ket amplitudes by transforming these factors using an operator $\eT$ in the quantum evolution space that has the effect of supplying the corresponding transformation for the spectator partons.
Let $A$ and $B$ be two sets of indices $i \in \{1,\dots,m\}$. The sets $A$
and $B$ are the sets of indices of partons that we will not consider to be spectator partons in the ket and bra state respectively.  The definition of $\eT$ for a momentum mapping $\{p\}_m \to \{\hat p\}_{m + n_\scR}$ with $n_\scR$ real emissions is  
\begin{equation}
  \begin{aligned}[c]
    \label{eq:Mketbraoperator}
    \sbra{\{\hat w, \hat w'\}_{m}}
    \eT(A, B;\{\hat p\}_{m+n_\scR},\{p\}_m)
    \sket{\{w, w'\}_{m}}
    \hskip -6.0 cm {}&
    \\
    ={}& 
    \prod_{i = 1}^m \Big\{
    \delta_{\hat f_i,f_i} \delta_{\hat f'_i,f'_i}
    \delta_{\hat a_i, a_i}\delta_{\hat a'_i, a'_i}
    \\
    &\qquad\times
    (2\pi)^d \delta(\hat q_i - q_i)\,
    (2\pi)^d \delta(\hat q'_i - q'_i)
    \Big\}
    \\
    &\times
    \prod_{i\in A} \delta_{\hat r_i, r_i}
    \prod_{i\in B} \delta_{\hat r'_i, r'_i}
    \\
    &\times
    \prod_{i\notin A} \bigg\{
    \sum_s \dualL\brax{\hat{r}_i} \ket{\chi_{f_i}(\hat{p}_i, s)}\,
    \brax{\psi_{f_i}(p_i, s)}\ket{r_i}
    \bigg\}
    \\
    &\times
    \prod_{i\notin B}\bigg\{
    \sum_s \dualL\brax{r'_i} \ket{\psi_{f_i}(p_i, s)}\,
    \brax{\chi_{f_i}(\hat{p}_i, s)}\ket{\hat{r}'_i}
    \bigg\}
    \;.
  \end{aligned}
\end{equation}

Since the operator $\eT$ acting on the quantum evolution space is closely associated with the momentum mapping, it is useful to combine it with the momentum mapping operator to define a momentum mapping operator $\xP$ in the extended statistical space. 

At second order level, we often need two kinds of mappings $\xP$ on the extended statistical space, each with $n_\scR =$ 1 or 2. The first is a collinear mapping $\xP_{\mathrm{coll}}^{(n_\scR)}(l)$ that accompanies emission  of one or two partons from parton $l$ in the ket amplitude and in the bra amplitude. The second is a soft mapping $\xP^{(n_\scR)}_{\mathrm{soft}}$ that accompanies the emission of one or two partons from parton $l$ in the ket amplitude and from parton $k$ in the bra amplitude.

For the collinear mappings we define $\xP_{\mathrm{coll}}^{(n_\scR)}(l)$ by its non-zero matrix elements as
\begin{equation}
  \begin{aligned}[c]
    \label{eq:Rcoll-ext}
    \sbra{\{\hat w, \hat w'\}_{m},\{\hat p\}_{m + n_\scR}}
    \xP_{\mathrm{coll}}^{(n_\scR)}(l)
    \sket{\{w, w'\}_{m},\{p\}_{m}} \hskip - 6.5 cm {}&
    \\
    ={}&
    \sbra{\{\hat w, \hat w'\}_{m}}
    \eT \big(\{l\}, \{l\}; \{\hat p\}_{m + n_\scR},\{p\}_m\big)
    \sket{\{w, w'\}_{m}}
    \\
    &\times
    \sbra{\{\hat p\}_{m + n_\scR}}
    \bm R_\mathrm{coll}^{(n_\scR)}(l)\sket{\{p\}_m}
    \;.
  \end{aligned}
\end{equation}
For the soft mapping we define $\xP^{(n_\scR)}_{\mathrm{soft}}$ as
\begin{equation}
  \begin{aligned}[c]
    \label{eq:Rsoft-ext}
    \sbra{\{\hat w, \hat w'\}_{m},\{\hat p\}_{m + n_\scR}}
    \xP^{(n_\scR)}_{\mathrm{soft}}
    \sket{\{w, w'\}_{m},\{p\}_{m}} \hskip - 5.8 cm {}&
    \\
    ={}&
    \sbra{\{\hat w, \hat w'\}_{m}}
    \eT \big(\{\},\{\}; \{\hat p\}_{m + n_\scR},\{p\}_m\big)
    \sket{\{w, w'\}_{m}}
    \\
    &\times
    \sbra{\{\hat p\}_{m + n_\scR}}
    \bm R_\mathrm{soft}^{(n_\scR)}
    \sket{\{p\}_m}
    \;.
  \end{aligned}
\end{equation}
Here $\{\}$ denotes the empty set. With this choice, the vector/spinor correction is applied to all of the $m$ partons. Note that in the jet momentum space the number of jets increases by $n_\scR$, while in the quantum evolution space the number of the partons remains the same. This momentum mapping operator will always be accompanied by a splitting operator in which the number of partons in the quantum evolution space increases by $n_\scR$ while the number of jets remains the same. Then the net effect is to increase both the number of partons and the number of jets by $n_\scR$.

% -----------------------------------------------------
\subsection{Operators for generating graphs}
\label{sec:GeneratingGraphs}

We are now prepared to extend operators on the quantum evolution space that generate the elementary parts of Feynman graphs to the extended statistical space. We discuss here only first order irreducible operators. At second and higher perturbative orders, there are new operators that cannot be reduced to products of the first order operators ordered in their scales from harder to softer. We anticipate analyzing the second order operators in future work.

\subsubsection*{Tree level operators}

To extend the collinear emission operator $\hat{\bm C}^{(1,0)}_l(\mu)$, Eq.~(\ref{eq:hatCl}), we define
\begin{equation}
\label{eq:xC10}
\xC^{(1,0)}_l(\mu) 
= \big[\eC^{(1,0)}_l(\mu)\otimes \bm{1}_\Lp\big]\, 
\xP_{\mathrm{coll}}^{(1)}(l)
\;.
\end{equation}
We start with the collinear momentum mapping operator $\xP_{\mathrm{coll}}^{(1)}(l)$, Eq.~(\ref{eq:Rcoll-ext}), with a single emission, including  adjustment to the vector/spinor amplitudes of the spectator partons. Next we have the actual collinear splitting operator trivially extended into the extended statistical space, using a unit operator $\bm{1}_\Lp$ on the jet momentum space.   

The operator $\eX^{(1,0)}_{lk}(\mu)$, Eq.~(\ref{eq:eXlk}), describes emission of a parton from parton $l$ in the ket amplitude and from parton $k$ in the bra amplitude. To extend this to the extended statistical space, we define
\begin{equation}
\label{eq:hatXlk10}
\xX_{lk}^{(1,0)}(\mu) = 
\big[\eX_{lk}^{(1,0)}(\mu)\otimes \bm{1}_\Lp\big]\, 
\xP_{\mathrm{soft}}^{(1)}
\;.
\end{equation}
This starts with the soft momentum mapping operator $\xP_{\mathrm{soft}}^{(1)}$, Eq.~(\ref{eq:Rsoft-ext}), with a single emission, including  adjustment to the vector/spinor amplitudes of the spectator partons. This is followed by the operator $\eX_{lk}^{(1,0)}(\mu)$, trivially extended to the extended statistical space.

\subsubsection*{One loop operators}

The extension of the one loop operators $\eX^{(0,1)}_{lk}$ and $\hat{\bm E}^{(0,1)}_{lk}$ in Eq.~(\ref{eq:eXlk01eGammalk01}) to the extended statistical space is very simple since the needed momentum mapping operator is simply the unit operator. We define
\begin{equation}
\begin{split}
\label{eq:xXlk01eGammalk01}
\xX^{(0,1)}_{lk} ={}& \eX^{(0,1)}_{lk} \otimes \bm{1}_\Lp 
\;,
\\
\xE^{(0,1)}_{lk} ={}& \hat{\bm E}^{(0,1)}_{lk} \otimes \bm{1}_\Lp 
\;.
\end{split}
\end{equation}

\subsubsection*{LSZ operators}

The LSZ operators $\bm L_l$ act on the space for the Feynman ket amplitudes. LSZ operators $\bm L_k^\dagger$ acting on the space for the Feynman bra amplitudes are defined in the same way. Then we can define operators $\hat{\bm L}$ that act on the complete quantum evolution space and include all of the final state parton lines by
\begin{equation}
\label{eq:hatbmL}
\hat{\bm L} = \prod_{l,k} \bm L_l\otimes \bm L_k^\dagger
\;.
\end{equation}
This operator has a trivial extension to the extended statistical space,
\begin{equation}
\label{hatcL}
\hat{\cL} = 
\hat{\bm L} \otimes \bm{1}_\Lp
\;,
\end{equation}
where we have only the trivial unit transformation in the jet momentum space.

At first order, this is
\begin{equation}
\begin{split}
\label{hatcL1}
\hat{\cL}^{(1)} ={}&
\sum_l \hat{\cL}_l^{(1)}
\;,
\end{split}
\end{equation}
with
\begin{equation}
\label{eq:LSZcL1}
\hat{\cL}_l^{(1)} = 
[\bm L^{(1)}_l\otimes \bm{1}]\otimes \bm{1}_\Lp
+ [\bm{1} \otimes \bm L^{(1) \dagger}_l]\otimes \bm{1}_\Lp
\;.
\end{equation}
%

%---------------------------------
\subsubsection*{Combined operators}
\label{sec:CombinedOps}

A complete graph $G = (G_\mathrm{ket}, G_\mathrm{bra})$ is generated by a corresponding operator $\hat {\bm G}(G)$ that acts on the quantum evolution space, as in Eq.~(\ref{eq:graphoperators}). If we combine this with the associated momentum mapping, we call the associated operator on the extended statistical space $\xG(G)$. We can generate complete graphs using operators $\xG(G)$ that are constructed as products of elementary operators such as $\xC^{(1,0)}_l(\mu)$, $\xX_{lk}^{(1,0)}(\mu)$, $\xE^{(0,1)}_{lk}$, and $\hat{\cL}_l^{(1)}$ given above.

%---------------------------------
\subsection{Mapping to the extended statistical space}
\label{sec:MappingH}

We next need to define how to transform from the statistical space, with its on-shell partons, to the extended statistical space. We map a basis vector $\sket{\{p,f,c,c',s,s'\}_m}$ in the statistical space to a vector $\mathbb{P}_\scH\sket{\{p,f,c,c',s,s'\}_m}$ in the extended statistical space. The subscript H on the operator $\mathbb{P}_\scH$ reminds us that this is the mapping to the extended statistical space of the partons from the hard subgraph. The mapping is defined by
\begin{equation}
\begin{split}
\label{eq:cPH}
\mathbb{P}_\scH \sket{\{p,f,c,c',s,s'\}_m} \hskip - 2.5 cm &
\\={}& 
\int\! d\{q,a,r\}_{m}
\int\! d\{q',a',r'\}_{m}
\\&\times 
\sket{\{q,f,a,r, q',f,a',r'\}_{m};\{p\}_{m}}
\\&\times
\dualL\brax{\{q,f,a,r\}_m} \ket{\chi(\{p,f,c,s\}_m)}
\\&\times
\brax{\chi(\{p,f,c',s'\}_m)}\ket{\{q',f,a',r'\}_m}
\;.
\end{split}
\end{equation}
Here $\iket{\chi(\{p,f,c,s\}_m)}$ is the initial state vector for the Feynman amplitude, Eq.~(\ref{eq:InitialKetVector}), and $\ibra{\chi(\{p,f,c',s'\}_m)}$ is the initial state vector for the conjugate Feynman amplitude, Eq.~(\ref{eq:InitialBraVector}). 

This definition requires some explanation. The number of partons in the extended statistical space and the list of momenta $\{p\}_m$ have been set to $m$ and $\{p\}_m$ given by the statistical space basis vector. The flavors $\{f\}_m$ and $\{f'\}_m$ for the incoming parton lines have been set to the flavors $\{f\}_m$ in the statistical basis vector. 

There are integrations over the momenta $\{q\}_m$ and $\{q'\}_m$. These momenta will be determined by momentum conservation as linear combinations of the momenta $\{\hat p\}_{\hat m}$ at the end of the evolution in the extended statistical space. Then, close to an infrared singularity when the $\{\hat p\}_{\hat m}$ are close to forming $m$ massless jets and momenta $\ell$ exchanged by virtual partons are close to zero, the $\{q\}_m$ will be close to the original lightlike, positive energy, momenta $\{p\}_m$.

There are sums over the color indices $\{a\}_m$ and $\{a'\}_m$ and vector/spinor indices $\{r\}_m$ and $\{r'\}_m$. The dependence on these indices is given by the inner products of the vectors $\iket{\chi(\{p,f,c,s\}_m)}$ and $\ibra{\chi(\{p,f,c',s'\}_m)}$ with the basis states ${}_\scD\!\ibra{\{q,f,a,r\}_m}$ and $\iket{\{q',f,a',r'\}_m}$ .

%-------------------
\subsection{Mapping to the statistical space}
\label{sec:MappingS}

Starting from the vector $\mathbb{P}_\scH\isket{\{p,f,c,c',s,s'\}_{m}}$ with $m$ partons in the extended statistical space, evolution according to graphs $G$ creates a vector $\isket{\Phi}$ with typically more partons, which have new quantum numbers. The new partons are emitted according to the Feynman graphs and the jet momenta $\{p\}_m$ are updated to $\{\hat p\}_{\hat m}$ according to the associated momentum mappings.

The evolved state can be expanded in basis vectors using the completeness relation (\ref{eq:extendedcompleteness}),
\begin{equation}
  \begin{split}
    \sket{\Phi} ={}&  
    \frac{1}{m!}\,\int\!d\{w\}_{m} \int\! d\{w'\}_{m} \int\!d\{p\}_{m}
    \\
    &\times
    \sket{\{w,w'\}_{m};\{p\}_{m}}
    \\
    &\times
    g_\Phi(\{w,w'\}_{m};\{p\}_{m})
    \;,
  \end{split}
\end{equation}
where the function $g_\Phi$ is
\begin{equation}
  g_\Phi = \sbrax{\{w,w'\}_{m};\{p\}_{m}}\sket{\Phi}\;.
\end{equation}

We need to define how to transform this vector in the extended statistical space, with its off-shell partons, to the statistical space with on-shell partons. We map a basis vector $\isket{\{w,w'\}_{m}; \{p\}_{m}}$ in the extended statistical space to a vector $\mathbb{P}_\scS\isket{w,w'\}_{m}; \{p\}_{m}}$ in the statistical space. The subscript S on the operator $\mathbb{P}_\scS$, with S for {\em soft}, reminds us that subsets of the $m$ partons are close to infrared singularities. The definition is  
\begin{equation}
  \label{eq:cPSmod}
  \begin{split} 
    \mathbb{P}_\scS\sket{\{q,f,a,r, q',f',a',r'\}_{m};\{p\}_{m}} 
    \hskip - 4 cm &
    \\
    ={}&
    \sum_{\{\tilde{f}, c, c', s, s'\}_{m}}
    \sket{\{p, \tilde{f}, c, c', s, s'\}_{m}}
    \\
    &\times
    \brax{\psi(\{p, \tilde{f}, c, s\}_{m})}\ket{\{q, f, a,r\}_{m}}
    \\
    &\times
    \dualL\brax{\{q', f', a',r'\}_{m}}
    \ket{\psi(\{p, \tilde{f}, c', s'\}_{m})}
    \;.
  \end{split}
\end{equation}
Here $\ibra{\psi(\{p, \tilde f, c, s\}_{m})}$ is the final state vector for the Feynman amplitude, Eq.~(\ref{eq:wavefctnOUTket}), and $\ket{\psi(\{p, \tilde f, c', s'\}_{m})}$ is the final state vector for the conjugate Feynman amplitude, Eq.~(\ref{eq:wavefctnOUTbra}).  

In Eq.~(\ref{eq:cPSmod}), the number of partons in the statistical space is the same as in the extended statistical space. The momenta in the statistical space vector are $\{p\}_{m}$ as specified in the jet momentum part of the extended statistical space basis vector. The momenta $p_i$ are determined by the splitting variables and by momentum mapping used for the emissions.

There are sums over the color states $\{c\}_{m}$ and $\{c'\}_{m}$ and spins $\{s\}_{m}$ and $\{s'\}_{m}$. The dependence on these is given by the inner products of the vectors $\ibra{\psi(\{p, \tilde f, c, s\}_{m})}$ and $\iket{\psi(\{p, \tilde f, c', s'\}_{m})}$ with the basis states $\iket{\{q, f, a, r\}_{m}}$ and ${}_\scD\!\ibra{\{q', f', a',r'\}_{m}}$. 

These inner products in Eq.~(\ref{eq:cPSmod}) contain delta functions that set $q_i$ to $p_i$ and $q'_i$ to $p_i$. For each splitting in the evolution, there were integrations over the momenta of the daughter partons, restricted by momentum conservation. With the delta functions that set $q_i = q'_i = p_i$, all of the other momenta in the Feynman graphs are fixed.

The flavor of parton $i$ in the statistical space is $\tilde f_i$. The inner products in Eq.~(\ref{eq:cPSmod}) contain delta functions that set $f_i$ to $\tilde f_i$ and $f'_i$ to $\tilde f_i$. Thus $f_i$ must be the same as $f'_i$

We note that $\mathbb{P}_\scS \mathbb{P}_\scH$ is the identity operator on the statistical space:
\begin{equation}
\label{eq:PSPH}
\mathbb{P}_\scS \mathbb{P}_\scH = 1
\;.
\end{equation}
This follows with the use of the identities
\begin{equation}
  \label{eq:PSPHspins}
  \begin{split}
    \sum_{r_i} \brax{\psi_{f_i}(p_i,\tilde s_i)}\ket{r_i}\, 
    \dualL\brax{r_i}\ket{\chi_{f_i}(p_i,s_i)}
    ={}& \delta_{\tilde s_i s_i}
    \;,
    \\
    \sum_{r'_i}\brax{\chi_{f_i}(p_i,s'_i)}\ket{r'_i}\,
    \dualL\brax{r'_i}\ket{\psi_{f_i}(p_i,\tilde s'_i)}
    ={}& \delta_{\tilde s'_i s'_i}
    \;.
  \end{split}
\end{equation}

To verify this for the ket amplitude, we write it out for the three possible flavor choices:
\begin{equation}
  \begin{split}
    -\varepsilon(p_i,s)^*\cdot\varepsilon(p_i,\tilde s) ={}&  
    \delta_{\tilde s s}\;,
    \\
    \overline U(p_i,\tilde s) \frac{\s{n}}{2p_i\cdot n}U(p_i,s) 
    ={}& \delta_{\tilde s s} \;,
    \\
    \overline V(p_i,s) \frac{\s{n}}{2p_i\cdot n}V(p_i,\tilde s) 
    ={}& \delta_{\tilde s s}
    \;.
  \end{split}
\end{equation}
For the bra amplitude, the proof is analogous.

The operator $\mathbb{P}_\scH \mathbb{P}_\scS$ is {\em not} the identity operator on the extended statistical space. It is, however, a projection operator:
\begin{equation}
  \label{eq:PHPS}
  (\mathbb{P}_\scH\,\mathbb{P}_\scS)\, 
  (\mathbb{P}_\scH\,\mathbb{P}_\scS) = 
  \mathbb{P}_\scH\,\mathbb{P}_\scS \;.
\end{equation}
% 

%-------------------------------------------------
\subsection{Infrared limit of interference}
\label{sec:interferenceIR}

In Eq.~(\ref{eq:hatXlk10}), we have defined an operator $\xX^{(1,0)}_{lk}(\mu)$ on the extended statistical space. This operator represents the interference between the emission of parton $m+1$ from parton $l$ in the ket amplitude and the emission of parton $m+1$ from parton $k$ in the bra amplitude. The operator $\xX^{(1,0)}_{lk}(\mu)$ creates a logarithmic infrared singularity in the soft limit $\bar q_{m\!+\!1} \to 0$ in $\PS \xX^{(1,0)}_{lk}(\mu) \PH$, in which the final state momenta $\bar q_l$, $\bar q_k$, and $\bar q_{m\!+\!1}$ are on shell.  However, there is a simplification available because not all of the contributions to $\xX_{lk}$ produce this logarithmic infrared singularity. We will define the infrared singular part of $\xX_{lk}^{(1,0)}(\mu)$, which we call $\iIR{\xX_{lk}^{(1,0)}(\mu)}$. Then
\begin{equation}
\omIR{\xX_{lk}^{(1,0)}(\mu)} = \xX_{lk}^{(1,0)}(\mu) 
- \IR{\xX_{lk}^{(1,0)}(\mu)}
\end{equation}
does {\em not} produce a logarithmic infrared singularity. The operator $\iIR{\xX_{lk}^{(1,0)}(\mu)}$ will occur frequently, so we give it a name:
\begin{equation}
  \label{eq:xElk10def}
  \xE_{lk}^{(1,0)}(\mu) = \IR{\xX_{lk}^{(1,0)}(\mu)}\;.
\end{equation}

To define $\iIR{\xX_{lk}^{(1,0)}(\mu)}$,  we first approximate Eq.~(\ref{eq:hatXlk10}) by approximating $\eX_{lk}^{(1,0)}(\mu)$:
\begin{equation}
\label{eq:hatElk10IR}
\IR{\xX_{lk}^{(1,0)}(\mu)} = 
\big[\IR{\eX_{lk}^{(1,0)}(\mu)}\otimes 
\bm{1}_\Lp\big]\, \xP_{\mathrm{soft}}^{(1)}
\;.
\end{equation}
We define $\iIR{\eX_{lk}^{(1,0)}(\mu)}$ using an approximated version of Eq.~(\ref{eq:eXlk}):
\begin{equation}
\label{eq:hatElkIR}
\begin{split}
\frac{1}{\mu}\,
\IR{\eX^{(1,0)}_{lk}(\mu)} \sket{\{w,w'\}_{m}}\hskip - 3.5 cm {}&
\\= {}&  
\eh_{lk}(\mu)\,
\left[\IR{\bm\Gamma^{(1,0)}_l}
\otimes {\IR{\bm\Gamma^{(1,0)\dagger}_k}}\right]
\sket{\{w,w'\}_{m}}
\;.
\end{split}
\end{equation}
The operator $\bm\Gamma^{(1,0)}_l$ that acts on the ket state has the form given in Eq.~(\ref{eq:Gammal10}):
\begin{equation}
\begin{split}
\label{eq:Gammal10bis}
&\left[\frac{\as}{2\pi}\right]^{1/2}\,
\dualL\bra{\{\bar w\}_{m\!+\!1}}\qGamma^{(1,0)}_l\ket{\{w\}_{m}}
\\
&\qquad
= \prod_{i\ne l}\left[(2\pi)^d\delta^d(\bar q_i - q_i)\,\delta_{\bar f_i,f_i} \delta_{\bar a_i,a_i}\delta_{\bar r_i,r_i}\right]
\\
&\qquad\quad
\times (2\pi)^d \delta^d(\bar q_l + \bar q_{m\!+\!1} - q_l)\,
\\
&\qquad\quad
\times
\big[
V_3(w_l; \bar w_l, \bar w_{m\!+\!1}) + V_3(w_l; \bar w_{m\!+\!1}, \bar w_l)
\big]
\;.
\end{split}
\end{equation}
Here the function $V_3(w_l; \bar w_l, \bar w_{m\!+\!1})$ is given for each flavor choice by Eqs.~(\ref{eq:Gammaqqg}), (\ref{eq:Gammabarqbarqg}), (\ref{eq:Gammagqbar}), and (\ref{eq:softggg}). The operator $\bm\Gamma^{(1,0)\dagger}_k$ is defined analogously.

To define $\iIR{\bm\Gamma^{(1,0)}_l}$, we note first of all that an infrared singularity appears only when parton $m+1$ is a gluon. Thus a $\Lg \to q + \bar q$ splitting is not infrared singular. In the case of a $q \to q + \Lg$ splitting or a $\bar q \to \bar q + \Lg$ splitting, only the contribution from $\hat f_{m\!+\!1} = \Lg$ is infrared singular. This contribution is provided by $V_3(w_l; \bar w_l, \bar w_{m\!+\!1})$ and not by $V_3(w_l; \bar w_{m\!+\!1}, \bar w_l)$. For a $\Lg \to \Lg + \Lg$ splitting, $V_3(w_l; \bar w_l, \bar w_{m\!+\!1})$ can provide an infrared singularity but $V_3(w_l; \bar w_{m\!+\!1}, \bar w_l)$ contains a factor $\bar q_{m\!+\!1}$ that removes the singularity for $\bar q_{m\!+\!1} \to 0$. Thus we can define the infrared singular part of $\qGamma^{(1,0)}_l$ by keeping $V_3(w_l; \bar w_l, \bar w_{m\!+\!1})$ in Eq.~(\ref{eq:Gammal10}) and eliminating $V_3(w_l; \bar w_{m\!+\!1}, \bar w_l)$ and also eliminating the flavor choice $\Lg \to q + \bar q$ entirely. Thus we define
\begin{equation}
\begin{split}
\label{eq:Gammal10BIR}
&\left[\frac{\as}{2\pi}\right]^{1/2}\,
\dualL\bra{\{\bar w\}_{m\!+\!1}}
\IR{\qGamma^{(1,0)}_l}\ket{\{w\}_{m}}
\\
&\qquad
= \prod_{i\ne l}\left[(2\pi)^d\delta^d(\bar q_i - q_i)\,
\delta_{\bar f_i,f_i} \delta_{\bar a_i,a_i}\delta_{\bar r_i,r_i}\right]
\\
&\qquad\quad
\times (2\pi)^d \delta^d(\bar q_l + \bar q_{m\!+\!1} - q_l)\,
\\
&\qquad\quad
\times
\IR{V_3(w_l; \bar w_l, \bar w_{m\!+\!1})}\,
\theta(\hat f_{m\!+\!1} = \Lg)
\;.
\end{split}
\end{equation}
The same change is applied to $\qGamma^{(1,0)\dagger}_k$. 

We now need a definition for $\iIR{V_3(w_l; \bar w_l, \bar w_{m\!+\!1})}$. This depends on the flavor choice. 

For a  $q \to q + \Lg$ splitting, one more approximation is needed. In Eq.~(\ref{eq:Gammaqqg}) for $V_3(w_l; \bar w_l, \bar w_{m\!+\!1})$, there is a factor $\s{q}_l = \s{\bar q_l} + \s{\bar q}_{m\!+\!1}$. In the infrared limit, $\bar q_{m\!+\!1} \to 0$. Thus we set $\bar q_{m\!+\!1}$ to zero in this factor in $V_3(w_l; \bar w_l, \bar w_{m\!+\!1})$, giving
\begin{equation}
\begin{split}
\label{eq:Gammaqqgapprox}
\IR{V_3(  w_l; \bar w_l, \bar w_{m\!+\!1})} \hskip - 2.3 cm{}&
\\ ={}&
\mi \gs t^{\bar a_{m\!+\!1}}_{\bar a_l a_l}
\left[\gamma^{\bar r_{m\!+\!1}}\s{\bar q}_l\right]_{\bar r_l r_l} \frac{\mi}{q_l^2 + \mi 0}
\\
&\times
\theta(f_l \in \{\Lu, \Ld,\dots\})\,\theta(\bar f_l = f_l)\,\theta(\bar f_{m\!+\!1} = \Lg)
\;.
\end{split}
\end{equation}

For a $\bar q \to \bar q + \Lg$ splitting, an analogous derivation gives
\begin{equation}
\begin{split}
\label{eq:Gammaqbarqbargapprox}
\IR{V_3(  w_l; \bar w_l, \bar w_{m\!+\!1})} \hskip - 2.3 cm{}&
\\ ={}&
-\mi \gs t^{\bar a_{m\!+\!1}}_{a_l \bar a_l }\,
\frac{\mi}{q_l^2 + \mi 0}\,
\left[
  \s{\bar q}_l \gamma^{\bar r_{m\!+\!1}}
\right]_{r_l \bar r_l}
\\
&\times
\theta(f_l \in \{\bar\Lu, \bar\Ld,\dots\})\,\theta(\bar f_l = f_l)\,\theta(\bar f_{m\!+\!1} = \Lg)
\;.
\end{split}
\end{equation}

For a $\Lg \to \Lg + \Lg$ splitting, we use Eq.~(\ref{eq:softggg}) with $N_{\mu r_l }(q_l) = N_{\mu r_l }(\bar q_l + \bar q_{m\!+\!1})$ replaced by $N_{\mu r_l }(\bar q_l)$:
\begin{equation}
\begin{split}
\label{eq:Gammaggggapprox}
\IR{V_3(  w_l; \bar w_l, \bar w_{m\!+\!1})} \hskip - 2.3 cm{}&
\\ ={}&
-\gs f_{a_l,\bar{a}_l,\bar{a}_{m\!+\!1}}\,
    \frac{\mi N_{\mu r_l }(\bar q_l)}{q_l^2 + \mi 0}
    \\
    &
    \times \!
    \left[
      g^{\bar{r}_l\bar{r}_{m\!+\!1}}\,\bar{q}_{l}^{\mu}
      + g^{\mu\bar{r}_{m\!+\!1}}\,\bar{q}_l^{\bar{r}_l}
      -2g^{\mu\bar{r}_l}\,\bar{q}_l^{\bar{r}_{m\!+\!1}}
    \right]
\\
&\times
\theta(f_l = \bar f_l = \bar f_{m\!+\!1} = \Lg)
\;.
\end{split}
\end{equation}

We make the analogous changes in $\qGamma^{(1,0)\dagger}_k$. This defines  $\iIR{\xX_{lk}^{(1,0)}(\mu)} = \xE_{lk}^{(1,0)}(\mu)$ for each flavor choice. 

%-------------------------------------------------
\section{Structure of graphs}
\label{sec:structure}

We now have a vector space, the extended statistical space, with enough structure to describe Feynman graphs for the ket amplitude and the bra amplitude and, in addition, enough structure to describe the momentum mapping that can take us from on-shell momenta $\{p\}_m$ for $m$ partons to on-shell momenta $\{\hat p\}_{\hat m}$ for $\hat m$ partons, with $\hat m \ge m$.

Let $\xG(G)$ be the operator on the extended statistical space that generates the graph $G = (G_\mathrm{ket},G_\mathrm{bra})$ and let $\xG$ be the sum over graphs of $\xG(G)$:
\begin{equation}
\label{eq:xGexpandedinG}
\xG = \sum_G \xG(G)
\;.
\end{equation}
The initial state to which we apply $\xG$ is a linear combination of basis states $\isket{\{w, w'\}_{m},\{p\}_{m}}$, obtained by applying the operator $\PH$ to an $m$ parton state $\isket{\{p,f,c,c',s,s'\}_m}$ in the statistical space. See Eq.~(\ref{eq:cPH}). We assume that no subsets of the starting momenta $p_i$ are close to being collinear or soft. The operator $\xG$ can be expanded in perturbation theory and then expanded according to how many real emissions and virtual loops it contains: 
\begin{equation}
\begin{split}
\label{eq:hatGexpansion}
\xG ={}& 1 + \frac{\as}{2\pi}\,\xG^{(1)} 
+ \left[\frac{\as}{2\pi}\right]^2\,\xG^{(2)} + \cdots
\\
={}& 1 + \frac{\as}{2\pi}\,\left[\xG^{(1,0)} + \xG^{(0,1)} \right]
\\&
+ \left[\frac{\as}{2\pi}\right]^2\,\left[\xG^{(2,0)} + \xG^{(1,1)}
+ \xG^{(0,2)} \right] 
+ \cdots
\end{split}
\end{equation}
Here $\xG^{(n_\scR, n_\scV)}$ has $n_\scR$ real emissions and $n_\scV$ virtual loops. The graphs $G$ then correspond to individual contributions to $\xG^{(n_\scR, n_\scV)}$.

We have operators that generate elementary parton splittings and exchanges. These operators provide factors in the Feynman graphs $G$. Using these operators, we can create operators $\hat \cG(G)$ that create complete graphs. These include LSZ factors discussed in Secs.~\ref{sec:LSZ} and \ref{sec:GeneratingGraphs}, so that they define an effective cross section. The graphs $G$ are renormalized with $\MSbar$ renormalization at a scale $\mu_\scR$. They thus include counter terms that remove ultraviolet divergences that come from virtual loops. We do not indicate the dependence of $G$ and the corresponding operators on $\mu_\scR$ in this and the following section.

The initial state in the extended statistical space is obtained by starting with a basis vector $\isket{\{p,f,c,c',s,s'\}_m}$ in the statistical space. Then $\mathbb{P}_\scH \isket{\{p,f,c,c',s,s'\}_m}$ gives us the initial state in the extended statistical space to which the operator $\xG(G)$ is applied. The operator $\xG(G)$, acting on a state $\isket{\{w, w'\}_{m},\{p\}_{m}}$, produces a linear combination of basis states $\isket{\{\hat w, \hat w'\}_{\hat m},\{\hat p\}_{\hat m}}$ in the extended statistical space. The parton momenta $\{\hat q, \hat q'\}_{\hat m}$ in $\{\hat w, \hat w'\}_{\hat m}$ are, at this point, undefined.

The operator $\xG(G)$ includes a momentum mapping. For this, we start with parton momenta $\{p\}_m$ in the initial state $\isket{\{w, w'\}_{m},\{p\}_{m}}$. The $p_i$ are lightlike momenta with $\sum p_i = Q$. Then the final momenta in $\{\hat p\}_{\hat m}$ in $\isket{\{\hat w, \hat w'\}_{\hat m},\{\hat p\}_{\hat m}}$ are $\{\hat p\}_{\hat m} = R(G;\zeta_G,\{p\}_m)$. Typically, the factors that contribute to $\xG(G)$ each contribute to the overall momentum mapping. 

Next, the operator $\mathbb{P}_\scS$ returns us to the statistical space, giving us
\begin{equation}
\label{eq:hatcGmatrixelement}
\sbra{\{\hat p,\hat f,\hat c,\hat c',\hat s,\hat s'\}_{\hat m}}
\PS\,\xG\, \PH 
\sket{\{p,f,c,c',s,s'\}_{m}}
\;.
\end{equation}
Here $\isket{\{\hat p,\hat f,\hat c,\hat c',\hat s,\hat s'\}_{\hat m}}$ has $\hat m$ partons with on-shell momenta $\{\hat p\}_{\hat m}$ that were determined by the momentum mapping in $\xG(G)$. The momenta $\{\hat q\}_{\hat m}$ and $\{\hat q'\}_{\hat m}$, which were undefined, are now are set to $\{\hat q\}_{\hat m} = \{\hat q'\}_{\hat m} = \{\hat p\}_{\hat m}$. It is only when the final state momenta are put on shell that the infrared singularities of interest to us appear. 

With this final state, the initial momenta $\{q,q'\}_m$ are linear combinations of the final momenta $\{\hat p\}_{\hat m}$ and the momenta $\ell$ of any exchanged gluons. The momenta $\ell$ do not enter that hard subgraph in Fig.~\ref{fig:d11exampleA} but rather flow through the effective multiparton vertices in Fig.~\ref{fig:d11exampleB}.

The operator $\PS$ sets the flavors $\{\hat f\}_{\hat m}$ in the bra amplitude equal to the flavors $\{\hat f'\}_{\hat m}$ in the ket amplitude. The operator $\PS$ also provides an appropriate mapping for colors and spins. 

In the following section, we introduce a factorization of this complete $\xG$ into a hard factor, with an effective lower cutoff on the scale $\mu$ of splittings, $\mu_\scS < \mu$, and a soft factor, with an effective upper cutoff $\mu < \mu_\scS$.

%--------------------------------------------------
\section{Hard-soft factorization in the extended statistical space}
\label{sec:HardSoftExtended}

The matrix element (\ref{eq:hatcGmatrixelement}) produced by $\xG(G)$ is singular when the momenta $\{\hat p\}_{\hat m}$ form $m$ infinitely narrow jets with momenta $\{p\}_m$. Of course, there are also singularities when subsets of the $\{\hat p\}_{\hat m}$ form more than $m$ infinitely narrow jets. Although $\xG(G)$ contains the infrared singularities that we seek to organize, it does not isolate these singularities. To isolate infrared singularities, we factor $\xG$ into an infrared singular factor times a nonsingular factor. Since we are interested in using parton shower evolution to reproduce the infrared singularities of QCD, we need only the singular factor.

We write $\xG$ in the form
\begin{equation}
\label{eq:hatcAdef2}
\xG = \xA\,\xK
\;,
\end{equation}
where $\xA$ is an infrared singular factor and $\xK$ is an operator on the extended statistical space that we are free to define as long as it has two properties. First, the perturbative expansion of $\xK$ begins with 1 at order $\as^0$. Second, $\xK$ is free of infrared singularities.  Eq.~(\ref{eq:hatcAdef2}) implies
\begin{equation}
\label{eq:hatcAdef1}
\xA = \xG\,\xK^{-1}
\;.
\end{equation}

The aim here is choose $\xK$ so as to make the singular factor $\xA$ as simple as possible. We can write Eq.~(\ref{eq:hatcAdef1}) as
\begin{equation}
\begin{split}
\xA ={}& 
\left(
1 + \frac{\as}{2\pi}\,\xG^{(1)} + \cdots \right)
\left(
1 - \frac{\as}{2\pi}\,\xK^{(1)} + \cdots \right)
\\ ={}&
1 + \frac{\as}{2\pi}\left(\xG^{(1)}
- \xK^{(1)}  \right) + \cdots
\;.
\end{split}
\end{equation}
The operator $\xG^{(1)}$ has contributions that we can label with an index $i$ from several sorts of real or virtual first order graphs,
\begin{equation}
\xG^{(1)} =  \sum_i \xG_{i}^{(1)} 
\;.
\end{equation}
Now we can choose corresponding contributions $\xK_i^{(1)}$ to remove unwanted nonsingular terms from $\xG_i^{(1)}$. 

We have encountered an example of this in Sec.~\ref{sec:interferenceIR}, in which $i$ denotes first order graphs representing interference between parton emission from parton $l$ and parton $k$. In this example
\begin{equation}
\begin{split}
\xG_{lk}^{(1,0)} ={}& 
\int_0^{\infty}\!\frac{d\mu}{\mu}\,
\xX_{lk}^{(1,0)}(\mu)
\;,
\\
\xA_{lk}^{(1,0)} ={}& 
\int_0^{\infty}\!\frac{d\mu}{\mu}\,
\IR{\xX_{lk}^{(1,0)}(\mu)}
\;,
\\
\xK_{lk}^{(1,0)} ={}& 
\int_0^{\infty}\!\frac{d\mu}{\mu}\,
\omIR{\xX_{lk}^{(1,0)}(\mu)}
\;.
\end{split}
\end{equation}
At higher orders, similar considerations apply.

The net result of these manipulations is 
\begin{equation}
  \begin{split}
    \label{eq:xGfactoriztion}
    \xG ={}& \xA\,
    \xK
    \;.
  \end{split}
\end{equation}
The operator $\xK$ is infrared finite. Thus the infrared singularities of $\xG$ are the same as the infrared singularities of $\xA$. For this reason, parton shower evolution can be based on $\xA$ alone.

It is useful to define $\xA$ so that it is independent of the renormalization scale $\mur$:
\begin{equation}
\label{eq:dxAdmur}
\mur\,\frac{d \xA(\mur)}{d\mur} = 0
\end{equation}
We normally do not display the functional dependence of operators on $\mur$, but here we do display $\mur$ as an argument of $\xA$. The dependence on $\mur$ includes the dependence of $\as(\mur)$ on $\mur$. With Eq.~(\ref{eq:dxAdmur}), we can adjust $\mur$ as needed to improve the usefulness of shower evolution. Eq.~(\ref{eq:dxAdmur}) should hold order by order in perturbation theory up to the order at which $\xA$ is defined. This is straightforward at order $\as^1$. We will address this issue at order $\as^2$ in future work.

We can now go further to separate singular and nonsingular contributions. Consider a graph $G = (G_\mathrm{ket},G_\mathrm{bra})$ with $\hat m$ final state partons and $m$ initial partons. The graph is singular when the momenta $\{\hat p\}_{\hat m}$ form $m$ infinitely narrow jets with momenta $\{p\}_m$. Of course, there are also singularities when subsets of the $\{\hat p\}_{\hat m}$ form more than $m$ infinitely narrow jets. Specifically, $G$ can be singular at points at which the emitted partons are grouped into jet subsets $J(j)$, $j \in \{1,\dots, n_J\}$ and a soft subset $S$. At these singular points, emitted parton momenta for $i \in S$ vanish and emitted parton momenta for $i \in J(j)$ are collinear. If we integrate the emitted parton momenta near the singularities using dimensional regularization, then the integration produces poles $1/\epsilon^n$. Additional poles appear from the virtual diagrams. 

To isolate the singularities, we use the hardness measure $h$ described in Sec.~ \ref{sec:hardness} to classify the final states according to how many of the splittings in $G$ were resolvable into separate jets at a scale $\mu_\scS$. Let $N_J(\mu_\scS)$ be the number of new jets created by the splittings in $\xA(G)$ that are resolvable at scale $\mu_\scS$, as determined by a jet algorithm to be defined below. We distinguish between the two cases $N_J(\mu_\scS) = 0$ and $N_J(\mu_\scS) \ge 1$. We write
\begin{equation}
\begin{split}
\label{eq:hatcAjetdecomposition}
\xA(G) ={}& \xA_{0}(G,\mu_\scS) + \xA_{\ge 1}(G,\mu_\scS)
\;,
\end{split}
\end{equation}
with
\begin{equation}
\begin{split}
\xA_{0}(G,\mu_\scS) ={}& \mathbb{J}_0(\mu_\scS)\star \xA(G)
\;,
\\
\xA_{i,\ge 1}(G,\mu_\scS) ={}&  
\mathbb{J}_{\ge 1}(\mu_\scS)\star \xA(G)
\;.
\end{split}
\end{equation}
Here we have used jet counting mappings $\mathbb{J}_0(\mu_\scS)$ and $\mathbb{J}_{\ge 1}(\mu_\scS)$ that are linear mappings on the space of operators $\xA$:
\begin{equation}
\begin{split}
\mathbb{J}_0(\mu_\scS)&\star \left(
\alpha\, \xA_\alpha
+\beta\, \xA_\beta 
\right) 
\\={}& 
\alpha\, \mathbb{J}_0(\mu_\scS)\star \xA_\alpha
+ \beta\, \mathbb{J}_0(\mu_\scS)\star \xA_\beta
\;,
\\
\mathbb{J}_{\ge 1}(\mu_\scS)&\star 
\left(\alpha\, \xA_\alpha
+\beta\, \xA_\beta \right) 
\\={}& 
\alpha\, \mathbb{J}_{\ge 1}(\mu_\scS)\star \xA_\alpha
+ \beta\, \mathbb{J}_{\ge 1}(\mu_\scS)\star \xA_\beta
\;.
\end{split}
\end{equation}
Acting on a contribution $\xA(G)$ that corresponds to an individual graph, $\mathbb{J}_0(\mu_\scS)\star \xA(G)$ is $\xA(G)$ with theta functions inserted into the integrations over splitting variables for each parton emission such that  the emissions produce no new resolvable jets. Similarly, $\mathbb{J}_{\ge 1}(\mu_\scS)\star \xA(G)$ produces at least one new jet that is resolvable at scale $\mu_\scS$.

Summing over graphs $G$, this gives the decomposition
\begin{equation}
\label{eq:hatcGsimpledecomposition}
\xA = \xA_0(\mu_\scS) + \xA_{\ge 1}(\mu_\scS)
\;.
\end{equation}
In the first term, $\xA_0(\mu_\scS)$ produces no extra jets that are resolvable at scale $\mu_\scS$. In the second term, $\xA_{\ge 1}(\mu_\scS)$ produces at least one extra jet resolvable at scale $\mu_\scS$.

We now define the jet counting mappings $\mathbb{J}_0(\mu_\scS)$ and $\mathbb{J}_{\ge 1}(\mu_\scS)$. The number of new, resolvable jets is determined by the graph $G$ and by the relation between the initial momenta $\{p\}_m$ and the final momenta $\{\hat p\}_{\hat m}$, as fixed by the momentum mapping associated with $G$. Graphs with no real emissions have $\hat m = m$ and $\{\hat p\}_{m} = \{p\}_{m}$. Therefore, these graphs contribute only to $\xA_0(\mu_\scS)$. It is graphs with real emissions that can contribute to $\xA_{\ge 1}(\mu_\scS)$.

We combine jets starting from the final state produced by the splittings in $\hat \cA(G)$ and working toward the starting hard state using the hardness measure $h$. The number $N_J(\mu_\scS)$ of new jets resolvable at scale $\mu_\scS$ satisfies either $N_J(\mu_\scS) = 0$ or $N_J(\mu_\scS)\ge 1$. We define the conditions under which $N_J(\mu_\scS) \ge 1$. 

{\em Combining two partons.} Consider combining two partons with labels $i$ and $j$. For $i,j$ to be resolvable jets, we require, first, that
\begin{equation}
h(\hat p_i + \hat p_j) > \mu_\scS^2
\;.
\end{equation}
In addition, we require one or the other of two conditions. We ask that there be some parton index $a$ such that $G_\mathrm{ket}$ has a splitting $a \to i + j$. Or we ask that there be some index $b$ such that $G_\mathrm{bra}$ has a splitting $b \to i + j$. If one of these conditions holds in addition to $h(\hat p_i + \hat p_j) > \mu_\scS^2$, then $N_J(\mu_s) \ge 1$. 

{\em Combining three partons.} Consider combining three partons with labels $i$, $j$, and $k$. For $i,j,k$ to contain new jets, we require, first, that
\begin{equation}
h(\hat p_i + \hat p_j + \hat p_k) > \mu_\scS^2
\;.
\end{equation}
In addition, we require one or the other of two conditions. We ask that there be some index $a$ such that $G_\mathrm{ket}$ has a splitting $a \to i + j + k$. This could be a sequence of elementary splittings or a single elementary splitting. Or we ask that there be some index $b$ such that $G_\mathrm{bra}$ has a splitting $b \to i + j + k$. If one of these conditions holds in addition to $h(\hat p_i + \hat p_j + \hat p_k) > \mu_\scS$, then $J(\mu_\scS) \ge 1$. 

If we work at higher than order $\as^2$, then we consider combining more than three final state partons in the same way. 

Notice that this algorithm combines partons into jets according to the hardness measure $h$ and also according to the shower history corresponding to a particular operator $\xA(G)$. We could, for instance, encounter a parton pair $(i,j)$ for which $h(\hat p_i + \hat p_j)$ is the smallest of the hardness measures for pairs of partons, but for which partons $i$ and $j$ are not the daughter partons in any splitting in the graph $G$.  Such accidental combinations are ignored, even though these jets would be combined in an ordinary jet algorithm of the sort that is applied to experimental data.\footnote{When there are many jets produced by the parton shower, the chance that the pair $(i,j)$ with the smallest $h(\hat p_i + \hat p_j)$ corresponds to the last splitting is typically quite small. This makes it impracticable to use just a simple jet algorithm that does not take the splitting history into account.} 

The operator
\begin{equation}
\begin{split}
\xA_0(\mu_\scS) ={}& 1 + \sum_{n=1}^\infty 
\left[\frac{\as}{2\pi}\right]^n
\sum_{n_\scR = 0}^n
\xA^{(n_\scR,n-n_\scR)}_0(\mu_\scS)
\end{split}
\end{equation}
is of special interest. Acting on a state $\sket{\{w, w'\}_{m_0},\{p\}_{m_0}}$ with $m_0 \ge m$, $\xA^{(n_\scR, n - n_\scR)}_0(\mu_\scS)$ creates $n_\scR$ extra partons that are unresolvable from the previous partons at scale $\mu_\scS$ and contains $n_\scV = n - n_\scR$ virtual loops. Evidently, $\xA^{(n_\scR,n - n_\scR)}_0(\mu_\scS)$ has infrared singularities from final state partons becoming collinear or soft and may also have infrared poles $1/\epsilon$ from integrating over loop momenta.

The operator $\xA(\mu_\scS)$ is related to another operator
\begin{equation}
\begin{split}
\xA_\scH(\mu_\scS)
={}& 1 + \xA^{-1}_0(\mu_\scS)
\xA_{\ge 1}(\mu_\scS)
\;.
\end{split}
\end{equation}
Up to second order, this is
\begin{equation}
\begin{split}
\xA_\scH^{(1)}(\mu_\scS) ={}& \xA_{\ge 1}^{(1)}(\mu_\scS)
\;,
\\
\xA_\scH^{(2)}(\mu_\scS) ={}& \xA_{\ge 1}^{(2)}(\mu_\scS)
- \xA_{0}^{(1)}(\mu_\scS)\,\xA_{\ge 1}^{(1)}(\mu_\scS)
\;.
\end{split}
\end{equation}
We now sketch an argument that $\xA_\scH(\mu_\scS)$ is free of infrared singularities. 

At first order, $\xA_{\ge 1}^{(1)}(\mu_\scS)$ produces the emission of a parton. Integrating over the splitting variables with the restriction that the splitting is resolvable at scale $\mu_\scS$ does not produce an infrared singularity. 

At second order, $\xA_{\ge 1}^{(2)}(\mu_\scS)$ could describe two emissions. According to Eq.~(\ref{eq:Hinequality}), the first emission must be the hardest. It is thus resolvable at scale $\mu_\scS$. The second, softer, emission could be unresolvable. Thus $\xA_{\ge 1}^{(2)}(\mu_\scS)$ could create an infrared singularity from the integration over the splitting variables of the second emission. However, just in the limit that produces the infrared singularity, there is a contribution to $\xA_{0}^{(1)}(\mu_\scS)\,\xA_{\ge 1}^{(1)}(\mu_\scS)$ that gives two emissions, the first hard and the second containing exactly the same infrared singularity. Because of the minus sign in this contribution, the singularities cancel. 

Similarly, $\xA_{\ge 1}^{(2)}(\mu_\scS)$ could describe a parton emission that is resolvable at scale $\mu_\scS$ followed by a virtual exchange graph or a virtual self-energy graph. The integration over the loop momentum in the virtual graph can then produce an infrared divergence. However, there is a contribution to $\xA_{0}^{(1)}(\mu_\scS)\,\xA_{\ge 1}^{(1)}(\mu_\scS)$ that gives a matching resolvable emission and then a matching infrared divergent virtual graph. Again, the singularities cancel.

At higher perturbative orders, $\xA_{\ge 1}(\mu_\scS)$ has at least one parton emission that is resolvable at scale $\mu_\scS$. This can be followed by softer emissions or by virtual graphs that produce no resolvable jets but do produce infrared singularities. However, the same infrared singularities will be cancelled by singularities in $\xA_{0}^{-1}(\mu_\scS)$.

We will examine in future papers how this cancellation works at second order in $\as$.

We conclude that $\xA(\mu_\scS)$ can be decomposed into a product
\begin{equation}
\label{eq:xAhardsoft}
\xA(\mu_\scS) = \xA_0(\mu_\scS)
\left\{  1 + \xA^{-1}_0(\mu_\scS)
\xA_{\ge 1}(\mu_\scS) 
\right\}
\;.
\end{equation}
The first factor, $\xA_0(\mu_\scS)$, generates infrared singularities. The second factor has no infrared singularities. The operator $\cD(\mus)$ first introduced in Sec.~\ref{sec:cD} is designed to reflect the infrared singularities of QCD. For this reason, we will base the definition of $\cD(\mus)$ on $\xA_0(\mus)$.

%--------------------------------------------------
\section{Hard-soft factorization in the statistical space}
\label{sec:HardSoftStatSpace}

Let us approximate the cross section for an infrared safe observable represented by an operator $\cO_\LJ$ in the statistical space,  starting with a statistical state $\isket{\{p,f,c,c',s,s'\}_m}$ as
\begin{equation}
\label{eq:sigmaJ}
\sigma^{(m)}_\scJ = \sbra{1}\cO_\LJ\,\cA\sket{\{p,f,c,c',s,s'\}_m}
\;.
\end{equation}
Here 
\begin{equation}
\cA = \PS \xA \PH
\;,
\end{equation}
where $\xA$ was defined in Eq.~(\ref{eq:xGfactoriztion}) from $\hat \cG$. Using
\begin{equation}
\cG = \PS \xG \PH
\;,
\end{equation}
the relation between $\cA$ and $\cG$ is
\begin{equation}
\label{eq:cGfromcA}
\cG = \cA\,\cK
\;,
\end{equation}
where 
\begin{equation}
\begin{split}
\cK ={}& 
[\PS\xA\PH]^{-1}\,\PS[\xA\xK]\PH 
\;.
\end{split}
\end{equation}
The operator $\xK$ is free of infrared singularities and poles. Thus the infrared behavior of $\PS[\xA\xK]\PH$ matches the infrared behavior of $\PS\xA\PH$. We conclude that $\cK$ is free of infrared singularities and poles. Thus $\cA$ is $\cG$ with an infrared finite part removed.

In Eq.~(\ref{eq:sigmaJ}), the basis state represents a hard state with $m$ partons in which none of the partons are nearly collinear with other partons or soft. We can think of these partons as jets, labelled with momenta, flavors, colors, and spins.\footnote{Each jet has a single flavor $f$. However, color and spin are treated in the description used in the statistical space, Sec.~\ref{sec:statisticalspace}, using labels $\{c,c',s,s'\}_{m_0}$.} 

We separate $\cA$ into two terms according to whether or not it produces resolvable jets at scale $\mu_\scS$:
\begin{equation}
\cA = \cA_0(\mu_\scS) + \cA_{\ge 1}(\mu_\scS)
\;,
\end{equation}
where
\begin{equation}
\begin{split}
\cA_0(\mu_\scS) ={}& \PS \xA_0(\mu_\scS) \PH
\;,
\\
\cA_{\ge 1}(\mu_\scS) ={}& \PS \xA_{\ge 1}(\mu_\scS) \PH
\;.
\end{split}
\end{equation}
Then we can write $\cA$ in a form analogous to Eq.~(\ref{eq:xAhardsoft}) in the extended statistical space:
\begin{equation}
\cA = \cA_0(\mu_\scS)
\left\{ 1 + \cA_0^{-1}(\mu_\scS)\cA_{\ge 1}(\mu_\scS)
\right\}
\;.
\end{equation}
As in Eq.~(\ref{eq:xAhardsoft}), we conclude that the operator in braces here is free of infrared singularities and poles because the infrared behavior of $\PS\xA_0\PH$ matches the infrared behavior the full operator of $\PS\xA\PH$.

%----------------
\subsection{Infrared renormalization via probability conservation}
\label{sec:probabilityconservation}

The zero jet operator has a perturbative expansion
\begin{equation}
\begin{split}
\cA_0(\mu_\scS) ={}&
1 + \frac{\as}{2\pi}\,\cA_0^{(1)}(\mu_\scS) 
\\&
+ \left[\frac{\as}{2\pi}\right]^2 \cA_0^{(2)}(\mu_\scS) + \cdots
\;.
\end{split}
\end{equation}
Although each operator $\cA_0^{(n)}(\mu_\scS)$ is infrared singular, when we measure the contribution to the total probability coming from this operator using $\sbra{1}\cA_0^{(n)}(\mu_\scS)$, we know that $\sbra{1}\cA_0^{(n)}(\mu_\scS)$ must be infrared finite because of cancellations between real emission graphs and virtual graphs. Thus, as in Ref.~\cite{NSAllOrder}. we can define another operator, which, using the ``$\star$'' notation of Sec.~\ref{sec:HardSoftExtended}, we denote by $\mathbb{P}\star \cA_0^{(n)}(\mu_\scS)$. This operator has no infrared singularities, does not create new partons, does not change the momenta and flavors of partons, and obeys
\begin{equation}
\label{eq:Popstart}
\sbra{1}\cA_0^{(n)}(\mu_\scS) = \sbra{1}
\big[\mathbb{P}\star \cA_0^{(n)}(\mu_\scS)\big]
\;.
\end{equation}
If partons did not have colors and spins, $\big[\mathbb{P} \star \cA_0^{(n)}\big] \isket{\{p,f\}_{m_0}}$ would simply be $\isket{\{p,f\}_{m_0}}$ times an eigenvalue obtained by integrating over the splitting variables for the unresolved splittings in $\cA_0^{(n)}(\mu_\scS)$. With colors and spins, the operator $\mathbb{P}\star \cA_0^{(n)}(\mu_\scS)$ does not change the momenta and flavors of partons but modifies the colors and spins of the partons in a statistical state on which it acts. 

In what follows, we will adopt a more compact notation, from Ref.~\cite{NSAllOrder}, for $\mathbb{P}\star \cA_0$:
\begin{equation}
\label{eq:Pop}
\mathbb{P}\star \cA_0^{(n)}(\mu_\scS) = \Pop{\cA_0^{(n)}(\mu_\scS)}
\;.
\end{equation}
With this notation, Eq.~(\ref{eq:Popstart}) is
\begin{equation}
\label{eq:Popstartmod}
\sbra{1}\cA_0^{(n)}(\mu_\scS) = \sbra{1}\,\Pop{\cA_0^{(n)}(\mu_\scS)}
\;.
\end{equation}

We will discuss details of the definition of $\iPop{\cdots}$ further in Sec.~ \ref{sec:ProbabililtyConservation1st}. For now, it suffices to outline the intended physical idea. Let $\cA_0^{(n)}(\mu_\scS)$ be applied to a statistical space state $\sket{\{p,f,c,c',s,s'\}_{m_0}}$ with $m_0$ partons. We can consider the $m_0$ partons to be $m_0$ jets if we extend the idea of a jet to be a QCD object that carries flavor, color, and spin in addition to momentum. The resulting state,
\begin{equation*}
\cA_0^{(n)}(\mu_\scS)\sket{\{p,f,c,c',s,s'\}_{m_0}}
\;,
\end{equation*}
is a linear combination of states with $\{m_0, m_0 + 1, \dots, m_0 + n\}$ partons. Thus the $m_0$ jets now have substantial internal structure. However, this structure cannot be resolved at scale $\mu_\scS$. Using the operator $\Pop{\cA_0^{(n)}(\mu_\scS)}$, we replace this state with a state 
\begin{equation*}
\Pop{\cA_0^{(n)}(\mu_\scS)}\sket{\{p,f,c,c',s,s'\}_{m_0}}
\end{equation*}
that has $m_0$ jets with the same momenta and flavors $\{p,f\}_{m_0}$ as before but with the colors and spins rearranged so that the total probabilities of the two states are the same
\begin{equation}
\begin{split}
\sbra{1} \cA_0^{(n)}&(\mu_\scS)\sket{\{p,f,c,c',s,s'\}_{m_0}}
\\={}&
\sbra{1}\Pop{\cA_0^{(n)}(\mu_\scS)}\sket{\{p,f,c,c',s,s'\}_{m_0}}
\;.
\end{split}
\end{equation}
Thus the effect of the unresolved interactions at scales smaller than $\mu_\scS$ is accounted for by replacing the original $m_0$ jets by the same number of what we can call infrared renormalized jets. We can think of this as being similar to ultraviolet renormalization.

To make use of this, define an infrared nonsingular operator $\cI(\mu_\scS)$ by
\begin{equation}
\cI(\mu_\scS) = \Pop{\cA_0(\mu_\scS)}
\left\{ 1 + \cA_0^{-1}(\mu_\scS)\cA_{\ge 1}(\mu_\scS)\right\}
\;.
\end{equation}
Then define an operator $\cD(\mu_\scS)$ that relates $\cI(\mu_\scS)$ and $\cA$ by
\begin{equation}
\label{eq:cDcIiscA}
\cD(\mu_\scS)\,\cI(\mu_\scS) = \cA
\;.
\end{equation}
We can write this as
\begin{equation}
\begin{split}
\cD(\mu_\scS)\Pop{\cA_0(\mu_\scS)}{}&
\left\{ 1 + \cA_0^{-1}(\mu_\scS)\cA_{\ge 1}(\mu_\scS)\right\}
\\={}& \cA_0(\mu_\scS)
\left\{ 1 + \cA_0^{-1}(\mu_\scS)\cA_{\ge 1}(\mu_\scS)\right\}
\;.
\end{split}
\end{equation}
This implies
\begin{equation}
\cD(\mu_\scS)\Pop{\cA_0(\mu_\scS)} = \cA_0(\mu_\scS)
\;,
\end{equation}
so that
\begin{equation}
\label{eq:cDfromcA0}
\cD(\mu_\scS) = \cA_0(\mu_\scS)\Pop{\cA_0(\mu_\scS)}^{-1}
\;.
\end{equation}
We learn two things from this representation. First, $\cD(\mu_\scS)$ creates parton splittings, but all of these splittings are unresolvable at scale $\mu_\scS$. Second,
\begin{equation}
\begin{split}
\sbra{1}\cD(\mu_\scS) ={}& \sbra{1}\cA_0(\mu_\scS)\Pop{\cA_0(\mu_\scS)}^{-1}
\\={}&  \sbra{1}\Pop{\cA_0(\mu_\scS)}\Pop{\cA_0(\mu_\scS)}^{-1}
\;,
\end{split}
\end{equation}
so
\begin{equation}
\label{eq:1cDis1}
\sbra{1}\cD(\mu_\scS) = \sbra{1}
\;.
\end{equation}

Eq.~(\ref{eq:cDcIiscA}) gives us an interesting result when we insert it into Eq.~(\ref{eq:sigmaJ}):
\begin{equation}
\sigma^{(m)}_\scJ = \sbra{1}\cO_\LJ
\cD(\mu_\scS)\cI(\mu_\scS)
\sket{\{p,f,c,c',s,s'\}_m}
\;.
\end{equation}
Suppose that $\cO_\LJ$ is infrared safe at a scale $\mu_\scJ$ that is larger than $\mu_\scS$. Since $\cO_\LJ$ does not resolve the splittings in $\cD(\mu_\scS)$, the operators $\cO_\LJ$  and  $\cD(\mu_\scS)$ commute:
\begin{equation}
\sigma^{(m)}_\scJ = \sbra{1}\cD(\mu_\scS) \cO_\LJ
\cI(\mu_\scS)
\sket{\{p,f,c,c',s,s'\}_m}
\;.
\end{equation}
Then Eq.~(\ref{eq:1cDis1}) gives us
\begin{equation}
\label{eq:sigmaJresult}
\sigma^{(m)}_\scJ = \sbra{1} \cO_\LJ
\cI(\mu_\scS)
\sket{\{p,f,c,c',s,s'\}_m}
\;.
\end{equation}

In Eq.~(\ref{eq:sigmaJ}) the cross section $\sigma^{(m)}_\scJ$ is written using what might be called bare jets, so that we need to work in $4 - 2\epsilon$ dimensions to express the result for $\sigma^{(m)}_\scJ$. When $\mu_\scJ > \mu_\scS$, we can use Eq.~(\ref{eq:sigmaJresult}), in which $\sigma^{(m)}_\scJ$ is written using what we can call infrared renormalized jets, created by the operator $\cI(\mu_\scS)$. This equation can be applied in four dimensions.

We define a shower generator $\cS(\mu)$ by
\begin{equation}
\label{eq:generatorofcD}
\mu_\scS \frac{d}{d\mu_\scS}\,\cD(\mu_\scS) = \cD(\mu_\scS)\,\cS(\mu_\scS)
\end{equation}
or, since $\cD(\mu_\scS)\,\cI(\mu_\scS)$ is independent of $\mu_\scS$,
\begin{equation}
\label{eq:generatorofcI}
\mu_\scS \frac{d}{d\mu_\scS}\,\cI(\mu_\scS) = - \cS(\mu_\scS)\,\cI(\mu_\scS)
\;.
\end{equation}
This is a renormalization group equation for the infrared renormalization of $\cI(\mu_\scS)$. The operator $\cS(\mu)$ is the analogue of the anomalous dimension in the renormalization group associated with ultraviolet renormalization.

We will return to an examination of the structure of $\cI(\mu_\scS)$ in Sec.~\ref{sec:OperatorcI}.

%------------------------------------
\subsection{Parton shower from $\cD$}
\label{sec:showerfromcD}

The operator $\cD(\mu)$ is used to construct a parton shower according to the argument of Ref.~\cite{NSAllOrder}. We review this argument here. Consider the cross section $\sigma_\scJ$ for an infrared safe QCD observable\footnote{One could consider the production of particles that do not have strong interactions, but that is beyond the scope of this paper.}  $\cO_\LJ$ based on a statistical state $\isket{\rho}$:
\begin{equation}
\begin{split}
\label{eq:sigmaJstart}
\sigma_\scJ
={}& 
\sbra{1} \cO_\LJ \sket{\rho}
\;.
\end{split}
\end{equation}
We take $\isket{\rho}$ to be the statistical state for $e^+e^- \to \mathrm{hadrons}$ calculated at an arbitrarily high order of perturbation theory, without approximations. We also take the operators that we will use to be calculated at an arbitrarily high order of perturbation theory. We will then discuss approximations to this that account for the fact that calculations to arbitrarily high orders of perturbation theory are not available.

Now let $\mu_\scS$ be a hardness scale as described in the previous sections. In operators that depend on $\mu_\scS$, we set the renormalization scale to $\mur = \mu_\scS$. We will normally choose $\mu_\scS$ to be a hard scale $\muh$ close to $\sqrt{Q^2}$, but we can leave this choice adjustable. 

We insert $\cD(\mu_\scS)\,\cD^{-1}(\mu_\scS)$ into Eq.~(\ref{eq:sigmaJstart}), giving
\begin{equation}
\begin{split}
\sigma_\scJ
={}& 
\sbra{1} \cO_\LJ \cD(\mu_\scS)\,\cD^{-1}(\mu_\scS)\sket{\rho}
\;.
\end{split}
\end{equation}
The statistical state $\sket{\rho}$ contains infrared singularities corresponding to emissions of partons that are nearly collinear or zero momentum and poles corresponding to integrations over virtual loops. However, the operator $\cD^{-1}(\mu_\scS)$ removes these infrared poles and removes collinear and soft singularities below the scale $\mu_\scS$. Thus $\cD^{-1}(\mu_\scS)\isket{\rho}$ is infrared finite.

There are now two cases to consider for the operator $\cO_\LJ$. First, suppose that $\cO_\LJ$ is infrared safe at a scale $\mu_\scJ$ that is larger than $\mu_\scS$.\footnote{To be a little more precise, $\mu_\scJ$ should be enough larger than $\mu_\scS$ that powers of $\mu_\scS/\mu_\scJ$ can be neglected.} Then $\cO_\LJ$ commutes with $\cD(\mu_\scS)$, so that
\begin{equation}
\begin{split}
\sigma_\scJ
={}& 
\sbra{1} \cD(\mu_\scS)\,\cO_\LJ \,\cD^{-1}(\mu_\scS)\sket{\rho}
\;.
\end{split}
\end{equation}
Now use $\isbra{1}\cD(\mu_\scS) = \isbra{1}$, which we take to hold at arbitrarily high perturbative order even though we do not know $\cD(\mu_\scS)$ beyond a few perturbative orders. This gives
\begin{equation}
\begin{split}
\sigma_\scJ
={}& 
\sbra{1} \cO_\LJ \cD^{-1}(\mu_\scS)\sket{\rho}
\;.
\end{split}
\end{equation}
With the choice $\mu_\scS = \muh$, this is a compact statement of the standard method for calculating a cross section in QCD at next-to-leading order or a higher perturbative order. The calculation is in $\cD^{-1}(\mu_\scS)\isket{\rho}$, which is derived from Feynman graphs with their infrared singularities subtracted. Of course, the perturbative expansion of $\cD^{-1}(\mu_\scS)\isket{\rho}$ is truncated at whatever order is available.

Consider now what is needed when the infrared safety scale $\mu_\scJ$ of $\cO_\LJ$ is  much smaller than $\sqrt{Q^2}$. Then we could choose a scale $\mu_\scS = \mu_\Lf$ that is smaller than $\mu_\scJ$ and use
\begin{equation}
\begin{split}
\label{eq:standardsigmaJmus}
\sigma_\scJ
={}& 
\sbra{1} \cO_\LJ \cD(\mu_\Lf)\,\cD^{-1}(\mu_\Lf)\sket{\rho}
\\
={}& 
\sbra{1} \cD(\mu_\Lf)\,\cO_\LJ \,\cD^{-1}(\mu_\Lf)\sket{\rho}
\\
={}& 
\sbra{1} \cO_\LJ \,\cD^{-1}(\mu_\Lf)\sket{\rho}
\;.
\end{split}
\end{equation}

This formulation has a problem: the perturbative expansion of $\cD^{-1}(\mu_\Lf)\isket{\rho}$ contains large logarithms of $\mu_\Lf/\sqrt{Q^2}$. We can calculate $\cD^{-1}(\mu_\Lf)$ only to low perturbative orders, so that the large logarithms hinder the usefulness of Eq.~(\ref{eq:standardsigmaJmus}). Instead, we can define
\begin{equation}
\begin{split}
\cU(\mu_\Lf,\mu_\scH) ={}&  \cD(\mu_\Lf)^{-1} \cD(\mu_\scH) 
\;.
\end{split}
\end{equation}
Then
\begin{equation}
\begin{split}
\label{eq:showersigma}
\sigma_\scJ
={}& 
\sbra{1} \cO_\LJ\, \cU(\mu_\Lf,\mu_\scH)
\cD^{-1}(\mu_\scH)
\sket{\rho}
\;.
\end{split}
\end{equation}
Here it is sensible to truncate the perturbative expansion of $\cD^{-1}(\mu_\scH)\sket{\rho}$ at whatever order is available. Then, instead of expanding $\cU(\mu_\Lf,\mu_\scH)$ to a fixed order of perturbation theory, we write it as
\begin{equation}
\label{eq:cUexponential}
\cU(\mu_\Lf,\mu_\scH) = 
\mathbb{T}\exp\left(
\int_{\mu_\Lf}^{\mu_\scH}\!\frac{d\mus}{\mus}\,\cS(\mus)
\right)
\;,
\end{equation}
where $\cS(\mus)$ is the shower generator defined in Eq.~(\ref{eq:generatorofcD}). We can expand $\cS(\mus)$ in powers of $\as$:
\begin{equation}
\cS(\mus) = \sum_{n=1}^\infty 
\left[\frac{\as(\mur)}{2\pi}\right]^n
\cS^{(n)}(\mus,\mur)
\;.
\end{equation}
The coefficients $\cS^{(n)}(\mus,\mur)$ depend on the renormalization scale $\mur$, but in $\cS(\mus)$ this dependence is cancelled by the dependence of $\as(\mur)$ on $\mur$. Thus one can choose the value of $\mur$. In order to avoid large logarithms of $\mus/\mur$ in the coefficients, a sensible choice is $\mur = \mus$. Now if we truncate the perturbative expansion of $\cS(\mus)$ at a finite order, then the exponential in Eq.~(\ref{eq:cUexponential}) will give an approximation to $\cU(\mu_\Lf,\mu_\scH)$ that contains all orders of $\as$. This has at least the possibility of providing a useful summation of the terms with large logarithms.

With the formulation (\ref{eq:showersigma}), we start with $\cD^{-1}(\mu_\scH) \sket{\rho}$, which has had the infrared singularities that were present in $\isket{\rho}$ removed at a large scale $\muh$, so that this calculation does not generate large logarithms, Then we generate a parton shower with $\cU(\mu_\Lf,\mu_\scH)$, then measure the observable with $\isbra{1} \cO_\LJ$.

An alternative expression to Eq.~(\ref{eq:showersigma}) for a perturbative hard scattering followed by a parton shower starts with a hard scattering state $\tilde\cD(\mu_\scH)\isket{\rho}$, where $\tilde\cD(\mu_\scH)\isket{\rho}$ represents a different perturbative calculation with its infrared singularities removed. Then
\begin{equation}
\begin{split}
\label{eq:showersigmaALT}
\sigma_\scJ
={}& 
\sbra{1} \cO_\LJ\, \cU(\mu_\Lf,\mu_\scH)\,
\cD^{-1}(\mu_\scH)\tilde\cD(\mu_\scH)
\\&\times 
\tilde\cD^{-1}(\mu_\scH)
\sket{\rho}
\;.
\end{split}
\end{equation}
The operator $\cD^{-1}(\mu_\scH)\tilde\cD(\mu_\scH)$ matches the fixed order perturbative calculation to the parton shower according to the {\makeatletter\sc MC@NLO\makeatother} prescription \cite{MCatNLO}.

There is a complication in the argument for adjusting $\mur$. The Green functions represented in $\cS^{(n)}(\mus,\mur)$ depend on the gauge parameters $v^2$ and $\xi$, although we have generally not indicated this dependence. The gauge parameters are renormalized, so that they depend on $\mur$. That is, for general Green functions, we should use not only the running $\as(\mur)$ but also running gauge parameters $v^2(\mur)$ and $\xi(\mur)$. However, $\cD(\mus)$ is constructed to represent the infrared singularities of the cross section for $e^+ + e^- \to $ {\em hadrons}. The cross section, and thus its infrared behavior, is independent of $v^2$ and $\xi$. The construction of $\cD(\mus)$ involves approximations that can introduce $v^2(\mur)$ and $\xi(\mur)$ dependence to contributions to the cross section that are not infrared singular. As we will see in Sec.~\ref{sec:firstorder}, there is no $v^2$ or $\xi$ dependence in the first order shower as defined in this paper. However $v^2$ or $\xi$ could appear at higher orders, depending on how the higher order splitting operators are defined. The infrared nonsingular contributions can be adjusted to simplify the shower evolution, as we have already done for other issues. We conclude that when we adjust $\mur$, we can leave $v^2$ and $\xi$ at fixed values, for example $(v^2,\xi) = (4,1)$. This adjusts the behavior of the splitting operators away from the infrared singularities, but gives a physically equivalent parton shower. We could, alternatively, use running gauge parameters $v^2(\mur)$ and $\xi(\mur)$. We leave the investigation of these possibilities to future work.

%----------------------------------
\subsection{The operator {$\cI(\mu_\scS)$}}
\label{sec:OperatorcI}

In Sec.~\ref{sec:probabilityconservation}, we defined the operator $\cI(\mu_\scS)$ which we described as creating jets that are infrared renormalized at scale $\mu_\scS$. If we take the scale to be $\mu_\scS = \mu_\scH$, the state $\cI(\mu_\scH)\isket{\{p,f,c,c',s,s'\}_m}$ in Eq.~(\ref{eq:sigmaJresult}) is close to the starting basis state $\isket{\{p,f,c,c',s,s'\}_m}$, with perturbative corrections that are not enhanced by large logarithms. However, if $\mu_\scS$ is substantially smaller than $\mu_\scH$, the perturbative corrections are enhanced by large logarithms and can contain many more partons than $m$. In that case, we can write
\begin{equation}
\cI(\mu_\scS) = \cU(\mu_\scS,\mu_\scH)\,\cI(\mu_\scH)
\;.
\end{equation}
Now, $\cI(\mu_\scS)$ is expressed as an operator $\cI(\mu_\scH)$ with a well behaved perturbative expansion times a parton shower operator $\cU(\mu_\scS,\mu_\scH)$ that takes the state from the large scale $\mu_\scH$ to the smaller scale $\mu_\scS$. The shower then stops at scale $\mu_\scS$, so that $\mu_\scS$ serves as an infrared renormalization scale for the jets produced by $\cI(\mu_\scS)$.

%----------------------------------
\subsection{Infrared singularities}
\label{sec:IRsingularitiesforfactorization}

Consider the matrix element of $\sket{\rho}$ with a statistical space basis state, $\isbrax{\{\hat p,\hat f,\hat c,\hat c',\hat s,\hat s'\}_{\hat m}}\isket{\rho}$. This matrix element contains infrared poles and has logarithmic infrared singularities whenever two or more of the momenta $\hat p_i$ become collinear or some of these momenta become soft. How can we characterize these poles and singularities?

We can write the matrix element in the form
\begin{align}
\sbrax{\{\hat p, \hat f,\hat c,\hat c',\hat s,\hat s'\}_{\hat m}}\sket{\rho}
\hskip - 2.5 cm
\\ \nonumber
={}&
\sbra{\{\hat p,\hat f,\hat c,\hat c',\hat s,\hat s'\}_{\hat m}}
\cD(\mu_\scS)\,\cD^{-1}(\mu_\scS)\sket{\rho}
\\ \nonumber
={}&
\sbra{\{\hat p,\hat f,\hat c,\hat c',\hat s,\hat s'\}_{\hat m}}
\cD(\mu_\scS)\,\cU(\mu_\scS,\muh)
\cD^{-1}(\muh)
\sket{\rho}
\;.
\end{align}
Inserting a sum over intermediate states, this becomes
\begin{align}
\label{eq:rhosingularity}
\nonumber
\sbrax{\{\hat p,\hat f,\hat c,\hat c',\hat s,\hat s'\}_{\hat m}}\sket{\rho}
\hskip - 2 cm
\\={}&
\sum_m
\int\!d\{p,f,c,c',s,s'\}_m
\\& \nonumber \times
\sbra{\{\hat p, \hat f,\hat c,\hat c',\hat s,\hat s'\}_{\hat m}}
\cD(\mu_\scS)
\sket{\{p,f,c,c',s,s'\}_m}
\\ \nonumber &\times
\sbra{\{p,f,c,c',s,s'\}_m}
\cU(\mu_\scS,\muh)
\cD^{-1}(\muh)
\sket{\rho}
\;.
\end{align}

In this expression, $\cU(\mu_\scS,\muh)\cD^{-1}(\muh)\isket{\rho}$ equals $\cD^{-1}(\mu_\scS)\isket{\rho}$ when expanded to arbitrarily high perturbative order. However, if $\cD^{-1}(\mu_\scS)\isket{\rho}$ is known only up to order $\as^N$, then $\cD^{-1}(\mu_\scS)\isket{\rho}$ truncated at order $\as^N$ contains only up to $m = N+2$ partons. When $\mu_\scS$ is substantially smaller than $\muh$, we will be interested in many more partons than $N+2$. Thus we write $\cD^{-1}(\mu_\scS)\isket{\rho}$ in the form $\cU(\mu_\scS,\muh)\cD^{-1}(\muh)\isket{\rho}$. Then it is a good approximation to expand $\cD^{-1}(\muh)\isket{\rho}$ to order $N$ and to expand the exponent in $\cU(\mu_\scS,\muh)$ to order $N$. That is, we use perturbation theory matched to a parton shower as a good approximation to $\cD^{-1}(\mu_\scS)\isket{\rho}$.

Note that the matrix element in the last line of Eq.~(\ref{eq:rhosingularity}) is infrared finite. It has no infrared poles and its singularities as functions of the parton momenta $\{p\}_m$ are removed below scale $\mu_\scS$.

The remaining factor in Eq.~(\ref{eq:rhosingularity}) is the matrix element of $\cD(\mu_\scS)$. This matrix element contains the infrared singularities. 

Fig.~\ref{fig:d11exampleB} from Sec.~\ref{sec:cD} is a pictorial representation of Eq.~(\ref{eq:rhosingularity}). As we have seen in this paper, there is some freedom available in the definition of $\cD(\mu_\scS)$ since we can remove parts of $\cD(\mu_\scS)$ that are not infrared singular. We have done that in the previous sections. Modifying $\cD(\mu_\scS)$ then modifies the infrared finite matrix element in the last line of Eq.~(\ref{eq:rhosingularity}) without changing $\sket{\rho}$. We use Eq.~(\ref{eq:rhosingularity}) to define $\cD(\mu_\scS)$ at order $\as^N$, which allows us to define shower splitting functions at order $\as^N$. 

%--------------------------------------------------
\section{First order splitting functions}
\label{sec:firstorder}

In this section we provide the first order operators needed to build the splitting functions for a first order parton shower. These operators also provide building blocks for order $\as^2$ parton shower evolution. We also provide a brief review of the construction that we have used in this paper as it applies at order $\as$.

%-----------------------
\subsection{Introduction}
\label{sec:firstorderintro}

In Eq.~\eqref{eq:GisLWGammaH}, we defined the operator $\qG$ that represents the graphs that can contribute to the infrared singular part of the Feynman amplitude:
\begin{equation}
 \label{eq:full-G-operator1}
\qG = \qL\qW\qGamma_\scH
\;.
\end{equation}
The operator $\qL$ provides the LSZ factor for the outgoing partons. The operator $\qW$ provides real radiation with possible vertex and self-energy corrections. The operator $\qGamma_\scH$ represents the 1PI graphs involving the hard vertex that was introduced by the composite operator $\cH(x)$, but defined so that there is no contribution from the ultraviolet region of loop momenta $\ell$. The dependence of $\qGamma_\scH$ on the renormalization scale is also adjusted according to the anomalous dimensions of the quark and gluon fields in $\cH(x)$. In Sec.~\ref{sec:FeynmanAmplitude} we defined and discussed the perturbative expansion of the these operators.

We expand $\qG$ in perturbation theory. Working at lowest order, we will need the order $g_\Ls$ part of $\qW$: $\qGamma^{(1,0)}_l$, which describes the splitting of parton $l$ into two partons, Eq.~(\ref{eq:Gammal10}). We will also need the order $\as$ part of $\qL$, $\qL_l^{(1)}$, which describes a self energy graph on line $l$, Eq.~(\ref{eq:qLlpert}). Finally, we will need the order $\as$ part of $\qGamma_\scH$ from Eq.~(\ref{eq:qGammaH01}). This includes $\qE_{lk}^{(0,1)}$, which describes the exchange of a parton between partons $l$ and $k$, Eq.~(\ref{eq:qGammalk01def}). This is to be combined with the corresponding contribution from the operators $\qN_l^{(0,1)}$, Eq.~(\ref{eq:qKlk01}).

%-----------------------
\subsubsection*{Quantum evolution space}

We use the elementary operators in the operator $\qG$ to define an operator $\eG$ that acts on the quantum evolution space and represents the product of Feynman graphs for the ket amplitude, Feynman graphs for the conjugate bra amplitude, and an operator  $\eh$ that measures of the hardness $\mu$ of splittings. Following the notation that $\eG^{(n_\scR,n_\scV)}$ is the part of $\eG$ with $n_\scR$ real emissions and $n_\scV$ virtual loops, we have at order $\as$,
\begin{equation}
\label{eq:G-expantion-1st}
\eG = 1 + \aspi\left(\eG^{(1,0)} 
+  \eG^{(0,1)}\right) +\cdots
\;.
\end{equation}
The real emission part is
\begin{equation}
\label{eq:eG-real1}
\eG^{(1,0)} =  
\frac{1}{2} \sum_l \qGamma^{(1,0)}_l\otimes\qGamma^{(1,0)\dagger}_l
+ \sum_{k\neq l} \qGamma^{(1,0)}_l\otimes\qGamma^{(1,0)\dagger}_k
\;.
\end{equation}
The first term describes the splitting of parton $l$ both in the ket and bra amplitudes. This operator leads to collinear and soft singularities. The statistical factor 1/2 is needed because the two partons produced by the splitting are indistinguishable. The second term represents the graph in which parton $m+1$ is emitted from parton $l$ in the ket amplitude and by parton $k$ in the bra amplitude. This operator is singular only in the limit in which parton $m+1$ is soft, and only when this parton is a gluon. We can represent the full real emission operator by Feynman graphs:
%%%%%%%%%%%%%%%%%%%% FIGURE %%%%%%%%%%%%%%%%%%%%%%%%%%
% -------------------- Figure -----------------------------
\begin{equation}
  \begin{split}
    &\sbra{\{\bar{w}, \bar{w}'\}_{m+1}}
    \eG^{(1,0)}\sket{\{w,w'\}_m}
    \\
    &\quad
    = \frac{1}{2}\sum_{l=1}^m
    \begin{prdfig}{8e3a485530e80054cbae7a6c9fb1b5fc}{G10-topology1}
      \begin{tikzpicture}[baseline=(current bounding box.center)]
        \begin{feynman}[]
          \vertex[dot] (vl) at (0,0) {};
          \vertex[empty dot,label={[above] \small$w_l$}] (ol) 
          at ($(vl) + (180:1cm)$) {};
          \coordinate [label={[above] \small$\bar{w}_l$}] (l) 
          at ($(vl) + (45:1cm)$);
          \coordinate [label={[below] \small$\bar{w}_{m+1}$}] (m1) 
          at ($(vl) + (-45:1cm)$);  
          \vertex[dot] (vk) at ($(vl)+(2.5cm,0)$) {};
          \vertex[empty dot,label={[above] \small$w'_l$}] (ok)  
          at ($(vk) + (0:1cm)$) {};
          \coordinate [label={[above] \small$\bar{w}'_l$}] (k) 
          at ($(vk) + (135:1cm)$);
          \coordinate [label={[below] \small$\bar{w}'_{m+1}$}] (m1b)
           at ($(vk) + (-135:1cm)$);  
          \diagram*{
            (ol)--[](vl);
            (vl)--[](l);
            (vl)--[](m1);
            (ok)--[](vk);
            (vk)--[](k); 
            (vk)--[](m1b);
          };      
        \end{feynman}
      \end{tikzpicture}
    \end{prdfig}
    \\
    &\qquad
    + \sum_{l\neq k}^m
    \begin{prdfig}{6526ca995a2df9d5c06f1e0e9f1809f6}{G10-topology2}
      \begin{tikzpicture}[baseline=(current bounding box.center)]
        \begin{feynman}[]
          \vertex[empty dot,label={[above] $w_l$}] (ol) at (0,0) {};
          \vertex[dot] (vl) at ($(ol)+(1cm,0)$) {};
          \coordinate [label={[above] $\bar{w}_l$}] (l) at ($(vl) + (0:1cm)$);
          \coordinate [label={[below] $\bar{w}_{m+1}$}] (m1) 
          at ($(vl) + (-35:1cm)$);  
          \vertex[empty dot,label={[above] $w'_k$}] (ok)  
          at ($(ol)+(4.5cm,-1.5)$) {};
          \vertex[dot] (vk) at ($(ok)-(1cm,0)$) {};
          \coordinate [label={[below] $\bar{w}'_k$}] (k) 
          at ($(vk) + (-1cm,0cm)$);
          \coordinate [label={[above] $\bar{w}'_{m+1}$}] (m2) 
          at ($(vk) + (145:1cm)$);  
          \vertex[empty dot,label={[above] $w_k$}] (oi) at ($(ol)+(0,-1.5)$) {};
          \coordinate [label={[below] $\bar{w}_{k}$}] (i) at (oi-|m1);  
          \vertex[empty dot,label={[above] $w_l'$}] (l1) at (l-|ok) {};
          \coordinate [label={[above] $\bar{w}'_{l}$}] (l11) at (l-|k);  
         %%%%% 
          % 
          \diagram*{
            (ol)--[](vl);
            (vl)--[](l);
            (vl)--[](m1);
            (ok)--[](vk);
            (vk)--[](k); 
            (vk)--[](m2);
            (oi)--[](i);
            (l1)--[](l11);
          };      
        \end{feynman}
      \end{tikzpicture}
    \end{prdfig}
    \;.
  \end{split}
\end{equation}
Here the lines for spectator partons are not shown.

The operator $\eG^{(1,0)}$ can be written as
\begin{equation}
\label{eq:eG-real2}
\eG^{(1,0)} =  \sum_l \int_0^\infty \frac{d\mu}{\mu}\,
\bigg[ \eC_l^{(1,0)}(\mu)
+ \sum_{k\neq l} \eX_{lk}^{(1,0)}(\mu)\bigg]
\;.
\end{equation}
The definitions (\ref{eq:hatCl}) and (\ref{eq:eXlk}) of $\eC_l^{(1,0)}(\mu)$ and $\eX_{lk}^{(1,0)}(\mu)$ include an operator $\eh(A,B;\mu)$, Eq.~(\ref{eq:ehABdef}), that sets the scale $\mu$ of the splittings. However
\begin{equation}
\int_0^\infty \!\frac{d\mu}{\mu}\,\eh(A,B;\mu) = 1
\;.
\end{equation}
Thus the scale setting operator is not present in Eq.~(\ref{eq:eG-real1}). The scale definition is used later in Eq.~(\ref{eq:xAjetdecomposition}).

The part of $\eG$ that gives first order virtual contributions is
\begin{equation}
\label{eq:eG01oper}
\begin{split}
\eG^{(0,1)} ={}& \sum_l\left[\qL_l^{(1)}\otimes\bm{1} 
+\bm{1}\otimes\qL_l^{(1)}\right]
\\&
+ \sum_l \frac{1}{2} \sum_{k\ne l}
\left[\qE_{lk}^{(0,1)}\otimes\bm{1} 
+ \bm{1}\otimes \qE_{lk}^{(0,1)\,\dagger}\right]
\\&
+ \sum_l \left[\qN_l^{(0,1)}\otimes \bm{1}
+ \bm 1 \otimes \qN_l^{(0,1)\dagger}
\right]
\;.
\end{split}
\end{equation}
Here there is no splitting scale $\mu$ since there is no splitting. The first term is the contribution of the LSZ factors at one-loop level and the second term represents a single parton exchange between parton $l$ and parton $k$ in either the ket amplitude or the bra amplitude. The exchange operator $\bm E_{lk}^{(0,1)}$ was defined in Eq.~(\ref{eq:qGammalk01def}) and $\qN_l^{(0,1)}$ was defined in Eq.~(\ref{eq:qKlk01}).

%-----------------------
\subsubsection*{Extended statistical space}

These operators on the quantum evolution space give rise to operators in the extended statistical space, for which we need a momentum mapping operator. Thus, for instance, the operator $\eC_l^{(1,0)}(\mu)$ is extended to an operator $\xC_l^{(1,0)}(\mu)$. For the real emission operator,
\begin{equation}
\label{eq:xG-real2}
\xG^{(1,0)} =  
\sum_l \int_0^\infty \frac{d\mu}{\mu}\,\bigg[ \xC_l^{(1,0)}(\mu)
+ \sum_{k\neq l} \xX_{lk}^{(1,0)}(\mu)\bigg]
\;.
\end{equation}
The operator $\xC_l^{(1,0)}(\mu)$, Eq.~(\ref{eq:xC10}), comes with a collinear style mapping of the jet momenta, while the operator $\xX_{lk}^{(1,0)}(\mu)$, Eq.~(\ref{eq:hatXlk10}), has a soft style mapping. We define these mappings in Appendices~\ref{sec:CollinearMomentumMapping}~and~\ref{sec:SoftMomentumMapping}, respectively.

For the extension of the virtual operator $\eG^{(0,1)}$ to the extended statistical space, we supplement $\eG^{(0,1)}$ in Eq.~(\ref{eq:eG01oper}) by a  momentum mapping, which is simply the unit operator (as in Eq.~(\ref{eq:LSZcL1})):
\begin{equation}
\label{eq:xG1oper}
\xG^{(0,1)} = \eG^{(0,1)} \otimes \bm{1}_\Lp
\;.
\end{equation}

The next step is to simplify the operators in $\xG$ so as to retain parts that create infrared singularities but remove parts that are not infrared singular. We denote infrared singular parts by $\iIR{\cdots}$. According to Eq.~\eqref{eq:hatcAdef2}, the operator $\xG$ can be written as 
\begin{equation}
  \xG = \xA\xK\;,
\end{equation}
where $\xK$ is an infrared finite operator and $\xA$ retains the infrared singularities of $\xG$. Expanding this at first order, 
\begin{equation}
  \begin{split}
    \xA = {}& 1 + \aspi \left(\xG^{(1,0)} - \xK^{(1,0)} + \xG^{(0,1)}
     - \xK^{(0,1)}\right)
    \\
    &+ \cdots
    \;.
  \end{split}
\end{equation}

The part of the singular operator that represents first order real emissions is
\begin{equation}
\begin{split}
\label{eq:xA10-real}
\xA^{(1,0)} ={}& \xG^{(1,0)} - \xK^{(1,0)}
= \int_0^\infty \frac{d\mu}{\mu}\,\xS^{(1,0)}(\mu)
\;.
\end{split}
\end{equation}
Here we have introduced the operator
\begin{equation}
\begin{split}
\label{eq:xS10-real}
\xS^{(1,0)}(\mu) ={}&  
\sum_l \xC^{(1,0)}_l(\mu) + \sum_{\substack{k\neq l}} \xE^{(1,0)}_{lk}(\mu)\;,
\end{split}
\end{equation}
where, as in Sec.~\ref{sec:interferenceIR},
\begin{equation}
\begin{split}
\xE_{lk}^{(1,0)}(\mu) ={}&  \IR{\xX^{(1,0)}_{lk}(\mu)}
\;.
\end{split}
\end{equation}
From Eqs.~(\ref{eq:xG-real2}), (\ref{eq:xA10-real}), and (\ref{eq:xS10-real}), we obtain the infrared finite operator $\xK^{(1,0)}$:

\begin{equation}
  \xK^{(1,0)} = \int_0^\infty \frac{d\mu}{\mu}\,\omIR{\xX^{(1,0)}(\mu)}
\;.
\end{equation}

We will see in Sec.~\ref{sec:ProbabililtyConservation1st} that we do not need an explicit representation for $\xA^{(0,1)}$ in order to define the evolution of a first order shower. We can choose $\xK^{(0,1)} = 0$ so that $\xA^{(0,1)}$ is simply $\xG^{(0,1)}$ as in Eq.~(\ref{eq:xG1oper}). We leave other possible choices to future work at higher order, where $\xA^{(0,1)}$ appears explicitly.

%-----------------------
\subsubsection*{Introduction of the shower scale}

The next step is introduce the shower scale $\mus$ and to break up $\xA^{(1,0)}$ into the part that creates no jets that are resolvable at scale $\mus$, $\xA^{(1,0)}_0(\mus)$, and the part that creates at least one jet resolvable at scale $\mus$, $\xA^{(1,0)}_{\ge 1}(\mus)$, as in Eq.~(\ref{eq:hatcAjetdecomposition}):
\begin{equation}
\xA^{(1,0)} = \xA^{(1,0)}_0(\mus) + \xA^{(1,0)}_{\ge 1}(\mus)
\;.
\end{equation}
The jet measurement is based on the hardness scale of the splitting, so that
\begin{equation}
\begin{split}
\label{eq:xAjetdecomposition}
\xA^{(1,0)}_0(\mus) ={}& 
\int_0^{\mus} \frac{d\mu}{\mu}\, \xS^{(1,0)}(\mu)
\;,
\\
\xA^{(1,0)}_{\ge1}(\mus) ={}&  
\int_{\mus}^\infty \frac{d\mu}{\mu}\, \xS^{(1,0)}(\mu)
\;.
\end{split}
\end{equation}
%

%-----------------------
\subsubsection*{Evolution in the statistical space}

These considerations give us first order operators on the extended statistical space. The final step is to return to the statistical space, as described in Sec.~\ref{sec:HardSoftStatSpace}. Now the operator $\xA$ is translated into an operator $\cA$ that acts on the statistical space, with
\begin{equation}
\cA = \PS \xA \PH
\;.
\end{equation}
As in the extended statistical space, we divide $\cA$ into an operator $\cA_0(\mus)$ that creates no additional jets that are resolvable at scale $\mus$ and an operator $\cA_{\ge 1} (\mus)$ that creates at least one additional jet that is resolvable at scale $\mus$.

In the statistical space, we impose probability conservation in shower evolution by using an operator mapping $\iPop{\cdots}$ with the property
\begin{equation}
\sbra{1}\cA_0^{(n)}(\mu_\scS) = \sbra{1}\,\Pop{\cA_0^{(n)}(\mu_\scS)}
\;.
\end{equation}
This allows us to define the infrared singular operator $\cD(\mus)$ by
\begin{equation}
\cD(\mu_\scS) = \cA_0(\mu_\scS)\Pop{\cA_0(\mu_\scS)}^{-1}
\;,
\end{equation}
as in Sec.~\ref{sec:HardSoftStatSpace}. We will define the action of $\iPop{\cdots}$  at first order in Sec.~\ref{sec:ProbabililtyConservation1st} below.

Given $\cD(\mus)$, the splitting operator $\cS(\mu)$ that acts on the statistical space is defined by 
\begin{equation}
\begin{split}
\cS(\mu) ={}& \cD^{-1}(\mu)\,\mu\frac{d\cD(\mu)}{d\mu}
\;.
\end{split}
\end{equation}
The splitting operator has a perturbative expansion that starts with
\begin{equation}
\begin{split}
\cS(\mu) 
={}& \aspi\left(\cS^{(1,0)}(\mu) + \cS^{(0,1)}(\mu)\right) + \cO(\as^2)
\;.
\end{split}
\end{equation}
Then
\begin{equation}
\begin{split}
\cS^{(1,0)}(\mu) ={}& \mu\frac{d\cD^{(1,0)}(\mu)}{d\mu}
\;,
\\
\cS^{(0,1)}(\mu) ={}& \mu\frac{d\cD^{(0,1)}(\mu)}{d\mu}
\;.
\end{split}
\end{equation}
There are two contributions to $\cS^{(1,0)}(\mu)$, The first corresponds to the emission of a new parton $m+1$ from parton $l$ in the ket state and in the bra state. This has a collinear singularity and a soft singularity. The second corresponds to the emission of a new parton $m+1$ from parton $l$ in the ket state and from parton $k$ in the bra state. This has only a soft singularity. We describe these operators in the following subsections. 

There is also a part of $\cD$ that describes first order virtual graphs, $\cD^{(0,1)}(\mu_\scS)$. We will find in Sec.~\ref{sec:ProbabililtyConservation1st} that we can determine $\cD^{(0,1)}(\mu_\scS)$ from $\cD^{(1,0)}(\mu_\scS)$ by using the mapping $\iPop{\cdots}$.

%--------------------------------------------------
\subsection{Collinear splitting}
\label{sec:collinear1storder}

%%%%%%%%%%%%%%%%%%%% FIGURE %%%%%%%%%%%%%%%%%%%%%%%%%%
% -------------------- Figure -----------------------------
\begin{figure}[htb]
  \begin{equation*}
    \begin{split}
      \sbra{\{\bar w, \bar w'\}_{m+1}, \{\bar p\}_{m+1}} \xC_{l}^{(1,0)}(\mu)\sket{\{w,w'\}_m, \{p\}_m}
      \hskip -5cm &
      \\
      = {}& \sbra{\{\bar p\}_{m+1}}\cP^{(1)}_{\rm coll}(l)\sket{\{p\}_m}
      \\
      &\times
      2\mu^2\, \delta\big(\mu^2 - \max[h(q_l),h(q'_l)]\big)
      \\
      &\times
        \begin{prdfig}{1b8b6436ea7be37d7e7315340e72823f}{G10-topology1-again}
          \begin{tikzpicture}[baseline=(current bounding box.center)]
            \begin{feynman}[]
              \vertex[dot] (vl) at (0,0) {};
              \vertex[empty dot,label={[above] \small$w_l$}] (ol)  at ($(vl) + (180:1cm)$) {};
              \coordinate [label={[above] \small$\bar{w}_l$}] (l) at ($(vl) + (45:1cm)$);
              \coordinate [label={[below] \small$\bar{w}_{m+1}$}] (m1) at ($(vl) + (-45:1cm)$);  
              \vertex[dot] (vk) at ($(vl)+(2.5cm,0)$) {};
              \vertex[empty dot,label={[above] \small$w'_l$}] (ok) at ($(vk) + (0:1cm)$) {};
              \coordinate [label={[above] \small$\bar{w}'_l$}] (k) at ($(vk) + (135:1cm)$);
              \coordinate [label={[below] \small$\bar{w}'_{m+1}$}] (m1b) at ($(vk) + (-135:1cm)$);  
              \diagram*{
                (ol)--[](vl);
                (vl)--[](l);
                (vl)--[](m1);
                (ok)--[](vk);
                (vk)--[](k); 
                (vk)--[](m1b);
              };      
            \end{feynman}
          \end{tikzpicture}
        \end{prdfig}
        % 
      % \rBiigg[3.5cm]{]}
    \end{split}
  \end{equation*}
  \caption{\label{fig:selfenergy1}{\em First order, topology 1.} Collinear emission diagram, the non-interacting lines are omitted here.}
\end{figure}
% -------------------- Figure -----------------------------
%%%%%%%%%%%%%%%%%%% END FIGURE %%%%%%%%%%%%%%%%%%%%%%%%

For a first order collinear splitting, with the emission of parton $m+1$ from parton $l$ in the ket state and in the bra state, we have an operator ${\xC}^{(1,0)}_l(\mu)$ that acts on the extended statistical space. This is illustrated in Fig.~\ref{fig:selfenergy1}. In the statistical space, this becomes an operator $\cC^{(1,0)}_l(\mu)$ defined by
\begin{equation}
\begin{split}
\label{eq:xCl10}
\cC_l^{(1,0)}(\mu)
={}& 
\mathbb{P}_\scS\,
{\xC}^{(1,0)}_l(\mu)\, \mathbb{P}_\scH
\;.
\end{split}
\end{equation}
The operator $\hat{\cC}^{(1,0)}_l(\mu)$ is defined in Eq.~(\ref{eq:xC10}). To $\hat{\cC}^{(1,0)}_l(\mu)$, we apply the mapping $\mathbb{P}_\scH$ defined in Sec.~\ref{sec:MappingH} and  the mapping $\mathbb{P}_\scS$ defined in Sec.~\ref{sec:MappingS}. When we apply these definitions to a first order splitting using straightforward algebra, we obtain quite simple results. The result for $\cC_l^{(1,0)}(\mu)$ depends on the flavor choice and can be written in a rather compact form as
\begin{equation}
  \begin{split}
    \label{eq:cCl10qqg}
    \cC_l^{(1,0)}{}&(\mu)
    \sket{\{p,f,c,c',s,s'\}_m}
    \\
    ={}& 
    4\pi^2\!\!\!
    \sum_{\{\hat f, \hat c, \hat c', \hat s, \hat s'\}_{m\!+\!1}}
    \prod_{i=1}^m \delta_{\hat f_i, f_i}
    \int\!d\zeta_\scC\,
    \\
    &\times
    \theta(f_l = \hat{f}_l + \hat{f}_{m\!+\!1})
    \\
    &\times
    \sket{\{\hat p,\hat f, \hat c, \hat c', \hat s, \hat s'\}_{m\!+\!1}}\,
    \\
    &\times
    2\mu^2\, \delta\big(\mu^2 - h(\hat p_l + \hat p_{m\!+\!1})\big)
    \\
    &\times
    \prod_{\substack{i = 1 \\ i\ne l}}^m  \delta_{\hat s_i, s_i} \delta_{\hat s'_i, s'_i}\,
    \frac{1}{[(\hat p_l + \hat p_{m\!+\!1})^2]^2}
    \\
    &\times 
    \dualL\bra{\{\hat c\}_{m\!+\!1}} t_l^\dagger(f_l \to \hat{f}_l + \hat{f}_{m\!+\!1})\ket{\{c\}_{m}}\
    \\
    &\times 
    \bra{\{c'\}_{m}}t_l(f_l \to \hat{f}_l + \hat{f}_{m\!+\!1})\ket{\{\hat c'\}_{m\!+\!1}}\dualR
    \\
    &\times
    \bra{\hat{s}_l,\hat{s}_{m\!+\!1}}
    \bm{c}(p_l, f_l; \hat{p}_l,\hat{f}_l;\hat{p}_{m\!+\!1},\hat{f}_{m\!+\!1})
    \ket{s_l}
    \\
    &\times
    \bra{s'_l}\bm{c}^\dagger(p_l, f_l; \hat{p}_l,\hat{f}_l;\hat{p}_{m\!+\!1},\hat{f}_{m\!+\!1})
    \ket{\hat{s}'_l, \hat{s}'_{m\!+\!1}}
    \;.
  \end{split}
\end{equation}
We now turn to an explanation of the parts of Eq.~(\ref{eq:cCl10qqg}), including the operator $t_l^\dagger$, which gives the color dependence of $\cC_l^{(1,0)}(\mu)$, and the operator $\bm c$, which gives its spin dependence

The spectator partons retain their flavors. For the flavors of parton $l$ and its daughter partons, $l$ and $m+1$, we require flavor conservation: $f_l = \hat f_l+\hat f_{m\!+\!1}$ with the evident QCD definition of adding flavors. For each type of splitting except for $\Lg \to \Lg + \Lg$, there are two equivalent contributions related by interchanging the daughter parton quantum numbers. We give results for one of these choices. The other choice is obtained by the interchange $l \leftrightarrow m+1$.

There is an integration over the splitting variables $\zeta_\scC$ as given in general by Eq.~(\ref{eq:dzetaproperty}) and for the collinear splitting by Eq.~(\ref{eq:dzetacoll1}). The splitting scale $\mu$ is set according to $\mu^2 = h(\hat p_l + \hat p_{m\!+\!1})$ using the hardness function $h$ from Eq.~(\ref{eq:q2-measure}) or  Eq.~(\ref{eq:Lambdasqcoll}). The delta function setting $\mu$ restricts the integration over $\zeta_\scC$.

The momenta $\{\hat p\}_{m\!+\!1}$ of the partons after the splitting are given as functions of $\zeta_\scC$ and the momenta $\{p\}_{m}$ before the splitting using the momentum mapping function  given in general in Eq.~(\ref{eq:Rmap}) and for the collinear splitting in Eq.~(\ref{eq:Rcoll}). 

The momenta of the spectator partons are modified slightly by the momentum mapping. However, because of the mapping of spectator vector/spinor indices defined in Sec.~\ref{sec:SpectatorEvolution}, the spins of the spectator partons are unchanged by the interaction, both for the ket partons and for the bra partons. 

There is a new color state, obtained by inserting the color matrix for the given splitting onto line $l$ in the ket color state and in the bra final state, as defined in Eqs.~(\ref{eq:colorket}) and (\ref{eq:colorbra}) below. The result
depends on the color basis used. For the trace basis, the result is described in Sec.~7.2 of Ref.~\cite{NSI}. We can describe how the color factors come about using the definitions given after Eq.~(\ref{eq:xCl10}). In any color
basis, the color factors are  
\begin{equation}
  \label{eq:colorket}
  \begin{aligned}[c]
    \dualL\bra{\{\hat c\}_{m\!+\!1}} t_l^\dagger(f_l \to \hat{f}_l + \hat{f}_{m\!+\!1})\ket{\{c\}_{m}}
    \hskip - 4.3 cm &
    \\
    = {}&
    \sum_{\{\hat a\}_{m\!+\!1}}
    \sum_{\{a\}_m}
    \dualL\brax{\{\hat c\}_{m\!+\!1}}\ket{\{\hat a\}_{m\!+\!1}}
    \brax{\{a\}_{m}}\ket{\{c\}_{m}}
    \\
    &\quad\times 
    \bra{\hat{a}_{l}, \hat{a}_{m\!+\!1}}\bm{t}_l^\dagger(f_l; \hat{f}_l, \hat{f}_{m\!+\!1})\ket{a_l}
 %   \\
%    &\quad\times 
    \prod_{\substack{i = 1 \\ i\ne l}}^m \delta_{\hat a_i,a_i}
    \;,
  \end{aligned}
\end{equation}
where the single parton color operator is defined by its non-zero matrix elements,
\begin{equation}
  \label{eq:t-colorket}
  \begin{split}
    &\bra{\hat{a}_{l}, \hat{a}_{m\!+\!1}}\bm{t}_l^\dagger(q; q, \Lg)\ket{a_l} = t^{\hat a_{m\!+\!1}}_{\hat a_l,a_l}\;,
    \\
    &\bra{\hat{a}_{l}, \hat{a}_{m\!+\!1}}\bm{t}_l^\dagger(\bar q; \bar q, \Lg)\ket{a_l} = -t^{\hat a_{m\!+\!1}}_{a_l\hat a_l}\;,
    \\
    &\bra{\hat{a}_{l}, \hat{a}_{m\!+\!1}}\bm{t}_l^\dagger(\Lg; \Lg, \Lg)\ket{a_l} = \mi f_{a_l,\hat a_l,\hat a_{m\!+\!1}}\;,
    \\
    &\bra{\hat{a}_{l}, \hat{a}_{m\!+\!1}}\bm{t}_l^\dagger(\Lg; q, \bar q)\ket{a_l} = t^{a_l}_{\hat a_l,\hat a_{m\!+\!1}}\;.
  \end{split}
\end{equation}
Similarly for the bra amplitudes we have 
\begin{equation}
  \label{eq:colorbra}
  \begin{aligned}[c]
    \bra{\{c'\}_{m}}t_l(f_l \to \hat{f}_l + \hat{f}_{m\!+\!1})\ket{\{\hat c'\}_{m\!+\!1}}\dualR
    \hskip - 4.3 cm &
    \\
    = {}&
    \sum_{\{\hat a'\}_{m\!+\!1}}
    \sum_{\{a'\}_m}
    \brax{\{c'\}_{m}}\ket{\{a'\}_{m}}
    \brax{\{\hat a'\}_{m\!+\!1}}\ket{\{\hat c'\}_{m\!+\!1}}\dualR
    \\
    &\quad\times
    \bra{a'_l}\bm{t}_l(f_l; \hat{f}_l, \hat{f}_{m\!+\!1})\ket{\hat{a}'_{l}, \hat{a}'_{m\!+\!1}}
%    \\
 %   &\quad\times
    \prod_{\substack{i = 1 \\ i\ne l}}^m \delta_{\hat a'_i,a'_i}
    \;,
  \end{aligned}
\end{equation}
where the single parton color operators are 
\begin{equation}
  \label{eq:t-colorbra}
  \begin{split} %(f_l \to \hat{f}_l + \hat{f}_{m\!+\!1})
    &\bra{a'_l}\bm{t}_l(q; q, \Lg)\ket{\hat{a}'_{l}, \hat{a}'_{m\!+\!1}} = t^{\hat a'_{m\!+\!1}}_{a'_l, \hat a'_l}\;,
    \\
    &\bra{a'_l}\bm{t}_l(\bar q; \bar q, \Lg)\ket{\hat{a}'_{l}, \hat{a}'_{m\!+\!1}}  = -t^{\hat a_{m\!+\!1}}_{a'_l\hat a'_l}\;,
    \\
    &\bra{a'_l}\bm{t}_l(\Lg; \Lg, \Lg)\ket{\hat{a}'_{l}, \hat{a}'_{m\!+\!1}}  = \mi f_{a'_l,\hat a'_l,\hat a'_{m\!+\!1}}\;,
    \\
    &\bra{a'_l}\bm{t}_l(\Lg; q, \bar q)\ket{\hat{a}'_{l}, \hat{a}'_{m\!+\!1}}  = t^{a'_l}_{\hat a'_l,\hat a'_{m\!+\!1}}\;.
  \end{split}
\end{equation}

For the factor that describes the parton spins, we define an operator, $\bm c(p_l, f_l; \hat{p}_l,\hat{f}_l;\hat{p}_{m\!+\!1},\hat{f}_{m\!+\!1})$. For $q\to q+\Lg$ splitting, where $q$ is a quark flavor and $\hat{f}_{m\!+\!1} = \Lg$, we start with
\begin{equation}
  \begin{split}
    &\bra{\hat{s}_l,\hat{s}_{m\!+\!1}}\bm{c}(p_l,q; \hat p_l,q;\hat{p}_{m\!+\!1},\Lg)\ket{s_l}
    \\
    &\qquad
    =\overline U(\hat p_l, \hat s_l)\,
    \s{\varepsilon}(\hat p_{m\!+\!1}, \hat s_{m\!+\!1})^*\,
    \frac{\s{q}_l\s{n}U(p_l, s_l)}{2p_l\!\cdot\!n}
    \;,
  \end{split}
\end{equation}
with $q_l = \hat p_l + \hat p_{m\!+\!1}$. We can simplify this expression. The collinear momentum mapping in Appendix \ref{sec:CollinearMomentumMapping} gives us 
\begin{equation}
  \begin{split}
    \label{eq:qlcollinearbis}
    q_l^\mu ={}& \left[1 - \frac{q_l^2}{(2 p_l\cdot n)^2}\right] p_l^\mu
    + \frac{q_l^2}{2 p_l\cdot n}\, n^\mu
    \,.
  \end{split}
\end{equation}
(In Appendix \ref{sec:CollinearMomentumMapping}, $q_l^2/(2 p_l\cdot n)^2$ is abbreviated as $a_l y$.) The factors in Eq.~(\ref{eq:qlcollinearbis}) follow from the choice that $q_l - p_l$ is lightlike. Using $\s{p}_l U(p_l, s_l) = 0$, $\{\s{p}_l,\s{n}\} = 2 p_l\cdot n$, and $\s{n}\s{n} = 1$, we have 
\begin{equation}
  \begin{split}
    \s{q_l}\s{n}\,U(p_l, s_l) ={}& 
    \left[1 - \frac{q_l^2}{(2 p_l\cdot n)^2}\right] 2p_l\cdot n\, U(p_l, s_l)
    \\&
    + \frac{q_l^2}{2 p_l\cdot n}\,U(p_l, s_l)
    \\
    ={}& 2p_l\cdot n\, U(p_l, s_l)
    \,.
  \end{split}
\end{equation}
This gives
\begin{equation}
  \label{eq:cop-qqg}
  \begin{split}
    &\bra{\hat{s}_l,\hat{s}_{m\!+\!1}}\bm{c}(p_l,q; \hat p_l,q;\hat{p}_{m\!+\!1},\Lg)\ket{s_l}
    \\
    &\qquad
    = \overline U(\hat p_l, \hat s_l)\,\s{\varepsilon}(\hat p_{m\!+\!1}, \hat s_{m\!+\!1})^*U(p_l, s_l)
    \,.
  \end{split}
\end{equation}

For the case $\bar{q}\to\bar{q}+\Lg$ with $\hat{f}_{m\!+\!1} = \Lg$, an analogous derivation gives
\begin{equation}
\label{eq:cop-qbarqbarg}
  \begin{split}
    &\bra{\hat{s}_l,\hat{s}_{m\!+\!1}}\bm{c}(p_l,\bar q; \hat p_l,\bar q;\hat{p}_{m\!+\!1},\Lg)\ket{s_l}
    \\
    &\qquad\qquad
    = \overline V(p_l, s_l)\,\s{\varepsilon}(\hat p_{m\!+\!1}, \hat s_{m\!+\!1})^*V(\hat p_l, \hat s_l)
    \,.
  \end{split}
\end{equation}

For the case $\Lg\to q+\bar{q}$ with $\hat{f}_{m\!+\!1} = \bar{q}$, we start with 
\begin{equation}
\label{eq:cop-gqqbar}
  \begin{split}
    &\bra{\hat{s}_l,\hat{s}_{m\!+\!1}}\bm{c}(p_l,\Lg; \hat p_l,q;\hat{p}_{m\!+\!1},\bar{q})\ket{s_l}
    \\
    &\quad
    =  \overline U(\hat p_l, \hat s_l)\,\gamma^\mu V(\hat p_{m\!+\!1}, \hat s_{m\!+\!1})
    N_{\mu \alpha}(q_l)\,\varepsilon^\alpha(p_l,s_l)
    \,,
  \end{split}
\end{equation}
where $N_{\mu \alpha}(q_l)$ is the numerator of the gluon propagator, Eq.~(\ref{eq:gluonnumerator}). We note from Eq.~\eqref{eq:qlcollinearbis} that $q_l$ is a liner combination of $p_l$ and $n$. In addition, the vector $\tilde q$ in Eq.~(\ref{eq:gluonnumerator}) is also a liner combination of $p_l$ and $n$. Since $N_{\mu \alpha}(q_l)$ is contracted with $\varepsilon^\alpha(p_l,s_l)$, which is orthogonal to $p_l$ and $n$, we find that
\begin{equation}
\label{eq:Nonepsilon}
  N_{\mu \alpha}(q_l)\varepsilon^\alpha(p_l,s_l) = -\varepsilon_\mu(p_l,s_l)
\;.
\end{equation}
With this we have
\begin{equation}
  \begin{split}
    &\bra{\hat{s}_l,\hat{s}_{m\!+\!1}}
    \bm{c}(p_l,\Lg; \hat p_l,q; \hat{p}_{m\!+\!1},\bar{q})\ket{s_l}
    \\
    &\qquad\qquad
    = -\overline U(\hat p_l, \hat s_l)\,\s{\varepsilon}(p_l,s_l) V(\hat p_{m\!+\!1}, \hat s_{m\!+\!1})
    \,.
  \end{split}
\end{equation}

For a $\Lg\to\Lg+\Lg$ splitting, we start with
\begin{equation}
 \label{eq:cop-ggg}
  \begin{split}
    &\bra{\hat{s}_l,\hat{s}_{m\!+\!1}}
    \bm{c}(p_l,\Lg; \hat p_l,\Lg;\hat{p}_{m\!+\!1},\Lg)\ket{s_l}
    \\
    &\quad
    = 
    \varepsilon_{\alpha}(\hat p_l, \hat s_l)^*\,
    \varepsilon_{\beta}(\hat p_{m\!+\!1}, \hat s_{m\!+\!1})^*\,
    N_{\mu \gamma}(q_l)\varepsilon^\gamma(p_l,s_l)
    \\
    &\qquad
    \times
    \Big[
      g^{\alpha \beta}\,q_l^\mu
      + g^{\mu\beta}\,(\hat p_l^\alpha -2 \hat p_{m\!+\!1}^\alpha)
      \\&\qquad\qquad
      + g^{\mu \alpha }\,(-2\hat p_l^\beta + \hat p_{m\!+\!1}^\beta)
    \Big]
    \,.
  \end{split}
\end{equation}
Here we have included both parts of the $\Lg \to \Lg + \Lg$ splitting function in Eq.~(\ref{eq:Gammal10}) and have used $q_l^\mu = \hat p_l^\mu + \hat p_{m\!+\!1}^\mu$. Now we can use Eq.~(\ref{eq:Nonepsilon}) to replace $N_{\mu \alpha}(q_l)\varepsilon^\alpha(p_l,s_l)$ by $-\varepsilon_\mu(p_l,s_l)$. Then $q_l\cdot \varepsilon(p_l,s_l) = 0$ because, according to Eq.~(\ref{eq:qlcollinearbis}), $q_l$ is a linear combination of $p_l$ and $n$, which are both orthogonal to $\varepsilon(p_l,s_l)$. In addition, $\hat p_l\cdot \varepsilon(\hat p_l, \hat s_l)^* = 0$ and $\hat p_{m\!+\!1}\cdot \varepsilon(\hat p_{m\!+\!1}, \hat s_{m\!+\!1})^* = 0$. This leaves
\begin{equation}
  \begin{split}
    &\bra{\hat{s}_l,\hat{s}_{m\!+\!1}}
    \bm{c}(p_l,\Lg; \hat p_l,\Lg;\hat{p}_{m\!+\!1},\Lg)\ket{s_l}
    \\
    &\quad
    =
    -2\, \varepsilon_{\alpha}(\hat p_l, \hat s_l)^*\,
    \varepsilon_{\beta}(\hat p_{m\!+\!1}, \hat s_{m\!+\!1})^*\,
    \varepsilon_\mu(p_l,s_l)
    \\
    &\qquad
    \times
    \Big[
    g^{\mu\beta}\,\hat{p}_{m\!+\!1}^{\alpha}
    +g^{\mu\alpha}\,\hat{p}_l^{\beta}
    \Big]
    \,.
  \end{split}
\end{equation}

The bra amplitudes $\ibra{s'_l}\bm{c}^\dagger\iket{\hat{s}'_l,\hat{s}'_{m\!+\!1}}$ can be derived with similar reasoning. The results are listed in Appendix \ref{sec:collinearbra}.

%--------------------------------------------------
\subsection{Splitting with interference}
\label{sec:interference1storder}

%%%%%%%%%%%%%%%%%%%% FIGURE %%%%%%%%%%%%%%%%%%%%%%%%%%
% -------------------- Figure -----------------------------
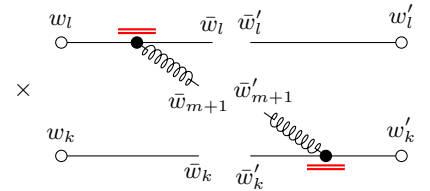
\begin{figure}[htb]
  \begin{equation*}
    \begin{split}
      \sbra{\{\bar w, \bar w'\}_{m+1}, \{\bar p\}_{m+1}} 
      \xE_{lk}^{(1,0)}(\mu)\sket{\{w,w'\}_m, \{p\}_m}
      \hskip -5cm &
      \\
      = {}& \sbra{\{\bar p\}_{m+1}}\cP^{(1)}_{\rm soft}\sket{\{p\}_m}
      \\
      &\times
      2\mu^2\, \delta\big(\mu^2 - \max[h(q_l),h(q'_k)]\big)
      \\
      &\times
        \begin{prdfig}{8b4e8579b1cadba42dd96c4a370e2ae7}{G10-topology2-again}
          \begin{tikzpicture}[baseline=(current bounding box.center)]
            \begin{feynman}[]
              \vertex[empty dot,label={[above] $w_l$}] (ol) at (0,0) {};
              \vertex[dot] (vl) at ($(ol)+(1cm,0)$) {};
              \coordinate [label={[above] $\bar{w}_l$}] (l) at ($(vl) + (0:1cm)$);
              \coordinate [label={[below] $\bar{w}_{m+1}$}] (m1)  
              at ($(vl) + (-35:1cm)$);  
              \vertex[empty dot,label={[above] $w'_k$}] (ok)  
              at ($(ol)+(4.5cm,-1.5)$) {};
              \vertex[dot] (vk) at ($(ok)-(1cm,0)$) {};
              \coordinate [label={[below] $\bar{w}'_k$}] (k) 
              at ($(vk) + (-1cm,0cm)$);
              \coordinate [label={[above] $\bar{w}'_{m+1}$}] (m2) 
              at ($(vk) + (145:1cm)$);  
              \vertex[empty dot,label={[above] $w_k$}] (oi) at ($(ol)+(0,-1.5)$) {};
              \coordinate [label={[below] $\bar{w}_{k}$}] (i) at (oi-|m1);  
              \vertex[empty dot,label={[above] $w_l'$}] (l1) at (l-|ok) {};
              \coordinate [label={[above] $\bar{w}'_{l}$}] (l11) at (l-|k);  
              %%%%%
              \draw[double, red, thick] ($(vl)+(-0.25,+0.15)$) -- 
              ($(vl)+(0.25,+0.15)$);
              \draw[double, red, thick] ($(vk)+(-0.25,-0.15)$) -- 
              ($(vk)+(0.25,-0.15)$);
              \diagram*{
                (ol)--[](vl);
                (vl)--[](l);
                (vl)--[gluon](m1);
                (ok)--[](vk);
                (vk)--[](k); 
                (vk)--[gluon](m2);
                (oi)--[](i);
                (l1)--[](l11);
              };      
            \end{feynman}
          \end{tikzpicture}
        \end{prdfig}
        % 
      % \rBiigg[3.5cm]{]}
    \end{split}
  \end{equation*}
  \caption{\label{fig:interference1}{\em First order, topology 2}: interference diagram. Parton $m\!+\!1$ must be a soft gluon, as in Sec.~\ref{sec:FeynmanAmplitude}.}
\end{figure}
% -------------------- Figure -----------------------------
%%%%%%%%%%%%%%%%%%% END FIGURE %%%%%%%%%%%%%%%%%%%%%%%%

We can now use $\xE_{lk}^{(1,0)}(\mu)$ to define
\begin{equation}
\begin{split}
\label{eq:xElk0}
\cE_{lk}^{(1,0)}(\mu)
={}& 
\mathbb{P}_\scS\,
\xE^{(1,0)}_{lk}(\mu)\, \mathbb{P}_\scH
\;,
\end{split}
\end{equation}
which is illustrated in Fig.~\ref{fig:interference1}. The $\xE^{(1,0)}_{lk}(\mu)$ operator is defined in Eq.~\eqref{eq:xElk10def} as the infrared approximation of the full interference operator $\xX_{lk}(\mu)$. Applying the mapping operators $\mathbb{P}_\scS$ and $\mathbb{P}_\scH$ and making straightforward algebraic
manipulations as in the collinear case, we find that 
\begin{equation}
  \label{eq:cElk10flfk}
  \begin{aligned}[c]
    \cE_{lk}^{(1,0)}(\mu)
    \sket{\{p,f,c,c',s,s'\}_m}
    \hskip - 3.2 cm {}&
    \\
    ={}& 8\pi^2\!\!
    \sum_{\{\hat f, \hat c, \hat c', \hat s, \hat s'\}_{m\!+\!1}}
    \!\!\delta_{\hat f_{m\!+\!1},\Lg}\,
    \prod_{i=1}^m 
    \delta_{\hat f_i, f_i}
    \int\!d\zeta_\scS\,
    \\
    &
    \times \sket{\{\hat p,\hat f, \hat c, \hat c', \hat s, \hat s'\}_{m\!+\!1}}\,
    \\
    &
    \times 2\mu^2\, \delta\big(\mu^2 - \max[h(q_l),h(q_k)]\big)
    \\
    &
    \times
    \prod_{\substack{i = 1 \\ i\ne l}}^m \delta_{\hat s_i, s_i}
    \prod_{\substack{j = 1 \\ j\ne k}}^m \delta_{\hat s'_j, s'_j}\,
   \\
    &
    \times \dualL\bra{\{\hat c\}_{m\!+\!1}} t^\dagger(f_l \to f_l + \Lg)\ket{\{c\}_{m}}
    \\
    &\times 
    \bra{\{c'\}_{m}}t(f_k \to f_k + \Lg)\ket{\{\hat c'\}_{m\!+\!1}}\dualR
    \\
    &\times
    \frac{\hat p_l\cdot\varepsilon(\hat p_{m\!+\!1}, \hat s_{m\!+\!1})^*}{\hat p_l \cdot \hat p_{m\!+\!1}}\,\delta_{\hat s_l s_l}
    \\
    &
    \times
    \frac{\hat p_k\cdot\varepsilon(\hat p_{m\!+\!1}, \hat s_{m\!+\!1})}{\hat p_k \cdot \hat p_{m\!+\!1}}\,\delta_{\hat s'_k s'_k}
    %
    % \frac{1}{2\hat p_l \cdot \hat p_{m\!+\!1}\, 2\hat p_k \cdot \hat p_{m\!+\!1} }
    % \overline U(\hat p_l, \hat s_l)
    % \s{\varepsilon}(\hat p_{m\!+\!1}, \hat s_{m\!+\!1})^*
    % \frac{\s{\hat p}_l\s{n}}{2 \hat p_l\cdot n}\,
    % U(\hat p_l, s_l)
    % \\
    % &
    % \times
    % \overline U(\hat p_k, s'_k)\,
    % \frac{\s{n}\s{\hat p}_k}{2 \hat p_k\cdot n}\,
    % \s{\varepsilon} (\hat p_{m\!+\!1}, \hat s'_{m\!+\!1})\,
    % U(\hat p_k, \hat s'_k)
    % 
    \;,
  \end{aligned}
\end{equation}
with $q_l = \hat p_l + \hat p_{m\!+\!1}$ and $q_k = \hat p_k + \hat p_{m\!+\!1}$.

The result in Eq.~(\ref{eq:cElk10flfk}) is similar to  that in Eq.~(\ref{eq:cCl10qqg}). Again, the spectator partons retain their flavors. Partons $l$ and $k$ also retain their flavors. The emitted parton carries the label $m+1$ and is a gluon. 

There is an integration over the splitting variables $\zeta_\scS$ for the soft momentum mapping as described in Sec.~\ref{sec:SoftMomentumMapping}. This is simply an integration over $\hat p_{m\!+\!1}$  The splitting scale $\mu$ is set according to $\mu^2 = \max[h(\hat p_l + \hat p_{m\!+\!1}),h(\hat p_k + \hat p_{m\!+\!1})]$. The delta function setting $\mu$ then restricts the integration over $\zeta_\scS$. 

The momenta $\{\hat p\}_{m\!+\!1}$ of the partons after the splitting are given by $\zeta_\scS$ and the momenta $\{p\}_{m}$ before the splitting by the momentum mapping function given for the soft splitting in Eq.~(\ref{eq:pitohatpisoft}). 

As in the collinear splitting, the momenta of the spectator partons are modified slightly by the momentum mapping. However, the spins of the spectator partons in the ket amplitude are unchanged, $\hat s_i = s_i$ for $i \ne l$ and the spins of the spectator partons in the bra amplitude are unchanged, $\hat s_j = s_j$ for $j \ne k$. 

As for the collinear splitting, there is a new color state, obtained by inserting the color matrix for the splitting $f_l \to f_l + \Lg$ into the ket color state and inserting the color matrix for the splitting  $f_k \to f_k + \Lg$ into the bra color state. 

In the final two lines of Eq.~(\ref{eq:cElk10flfk}), there is the spin dependent part of the ket amplitude and the bra amplitude. We have extracted the infrared singular part, $\cE_{lk}^{(1,0)}(\mu)$, of $\cX_{lk}^{(1,0)}(\mu)$, so the splitting amplitudes are given by the eikonal approximation.

%--------------------------------------------------
\subsection{Implementing probability conservation}
\label{sec:ProbabililtyConservation1st}

We have seen how to construct the splitting operator $\cS^{(1,0)}(\mu)$ that corresponds to one real emission. We now need the operator $\cS^{(0,1)}(\mu)$ that corresponds to first order virtual graphs. As explained in Sec.~\ref{sec:probabilityconservation}, if we start with $\cA_0(\mu_\scS) = \PS \xA_0(\mu_\scS) \PH$, then $\sbra{1} \cA_0(\mu_\scS)$ will be free of infrared singularities, but will not equal 1 exactly. However, we can adjust the definition of $\cD(\mu_\scS)$ so that we obtain an infrared singular operator with exact probability conservation,
\begin{equation}
\sbra{1} \cD(\mu_\scS) = 1
\;.
\end{equation}

To accomplish this, we define a linear mapping, $\cF \to \iPop{\cF}$, of operators on the statistical space \cite{NSThrustSum}.  The operator $\iPop{\cF}$ leaves the number $m$ of partons and $\{p,f\}_m$ unchanged, but can be an operator on the color and spin variables. The operator $\iPop{\cF}$ gives the same inclusive probability as $\cF$:
\begin{equation}
  \label{eq:Popdefbis}
  \sbra{1}\Pop{\cF} = \sbra{1}\cF \;.
\end{equation}
In this section, we will see how to define $\iPop{\cF}$ for the operators $\cF$ that occur at order $\as$.

Following Sec.~\ref{sec:probabilityconservation}, we define $\cD(\mu_\scS)$ according to Eq.~(\ref{eq:cDfromcA0}):
\begin{equation}
\label{eq:cDfromcA0bis}
\cD(\mu_\scS) = \cA_0(\mu_\scS)\Pop{\cA_0(\mu_\scS)}^{-1}
\;.
\end{equation}
\begin{widetext} %------------------------------
Up to first order, this is
\begin{equation}
\label{eq:cDfirstorder}
  \begin{aligned}[c]
    \cD(\mu_\scS) ={}& 
    \Big\{1 + \aspi\left(\cA_0^{(1,0)}(\mu_\scS)
    + \cA_0^{(0,1)}\right)+\cdots\Big\}
    \Big\{1 + \aspi\!\left(\Pop{\cA_0^{(1,0)}(\mu_\scS)} 
    + \Pop{\cA_0^{(0,1)}}\right)\! +\cdots \Big\}^{-1}
    \\
    ={}&
    1 + \aspi
    \Big(\cA_0^{(1,0)}(\mu_\scS) - \Pop{\cA_0^{(1,0)}(\mu_\scS)}
    + \cA_0^{(0,1)} -\Pop{\cA_0^{(0,1)}}\Big)
    +\cdots
    \;.
  \end{aligned}
\end{equation}
Here we have used the fact that $\cA_0^{(0,1)}$ does not depend on $\mu_\scS$, although it does depend on $\mur$.

To proceed, we need to define $\Pop{\cF}$ for the operators for the operators $\cF$ that are part of $\cA_0(\mu_\scS)$ at order $\as$ like, for instance, $\xE_{lk}^{(1,0)}(\mu)$. There is some freedom available for this definition. We want the operator $\Pop{\cF}$ to leave momenta and flavors unchanged. However it can be an operator on the color and spin space. A quite general operator with this structure takes the form
\begin{equation}
\begin{split}
\Pop{\cF} \sket{\{p,f,c,c',s,s'\}_{m}} ={}& 
\sum_{\{\hat c,\hat c',\hat s,\hat s'\}_{m}}
\sket{\{p,f,\hat c,\hat c',\hat s,\hat s'\}_{m}}
\\&\times
\sum_i
\dualL\bra{\{\hat c,\hat s\}_{m}}\bm{F}_i^\mathrm{ket}(\{p,f\}_m)\ket{\{c,s\}_{m}}
\bra{\{c',s'\}_{m}}\bm{F}_i^\mathrm{bra}(\{p,f\}_m)\ket{\{\hat c',\hat s'\}_{m}}\dualR
\;,
\end{split}
\end{equation}
with operators $\bm{F}_i^\mathrm{ket}$ and $\bm{F}_i^\mathrm{bra}$ to be defined.
The operators $\bm F^\mathrm{ket}_i$ act on the ket color and spin space and the operators $\bm F^\mathrm{bra}_i$ act on the bra color and spin space. Each of these operators depends on the parton momenta and flavors, $\{p,f\}_m$.  The inclusive probability that corresponds to $\Pop{\cF}$ is then, using Eq.~(\ref{eq:bra1}), 
\begin{equation}
\label{eq:1PopcFstart}
\sbra{1} \Pop{\cF}\sket{\{p,f,c,c',s,s'\}_{m}}
    = \sum_i
    \bra{\{c',s'\}_{m}}
    \bm F^\mathrm{bra}_i (\{p,f\}_m)
    \bm F^\mathrm{ket}_i (\{p,f\}_m)
    \ket{\{c,s\}_{m}}
    \;.
\end{equation}

We see that only the product $\bm F^\mathrm{bra}_i \bm F^\mathrm{ket}_i$ matters for the inclusive probability. Thus we could put the entire color$\times$spin dependence into a ket operator or into a bra operator or perhaps something else. At first order, we choose two contributions with $\bm F^\mathrm{ket}_1 (\{p,f\}_m) = \bm F^\mathrm{bra}_2 (\{p,f\}_m)=\bm F(\{p,f\}_m)$ and $\bm F^\mathrm{ket}_2 (\{p,f\}_m) = \bm F^\mathrm{bra}_1 (\{p,f\}_m)= (1/2)\, \bm 1$:
\begin{equation}
\begin{split}
\label{eq:PopcFstructure1}
\Pop{\cF} \sket{\{p,f,c,c',s,s'\}_{m}} ={}& 
\sum_{\{\hat c,\hat c',\hat s,\hat s'\}_{m}}
\sket{\{p,f,\hat c,\hat c',\hat s,\hat s'\}_{m}}
\\&\times
\Big\{ \frac{1}{2}\,
\dualL\bra{\{\hat c,\hat s\}_{m}}\bm{F}(\{p,f\}_m)\ket{\{c,s\}_{m}}
\bra{\{c',s'\}_{m}}\bm{1}\ket{\{\hat c',\hat s'\}_{m}}\dualR
\\&\quad +
\frac{1}{2}\,
\dualL\bra{\{\hat c,\hat s\}_{m}}\bm{1}\ket{\{c,s\}_{m}}
\bra{\{c',s'\}_{m}}\bm{F}(\{p,f\}_m)\ket{\{\hat c',\hat s'\}_{m}}\dualR
\Big\}
\;.
\end{split}
\end{equation}
That is
\begin{equation}
\begin{split}
\label{eq:PopcFstructure2}
\Pop{\cF} \sket{\{p,f,c,c',s,s'\}_{m}} ={}& 
\frac{1}{2}
\sum_{\{\hat c, \hat s\}_{m}}
\sket{\{p,f,\hat c,  c',\hat s,  s'\}_{m}}\,
\dualL\bra{\{\hat c,\hat s\}_{m}}\bm{F}(\{p,f\}_m)\ket{\{c,s\}_{m}}
\\& + \frac{1}{2}
\sum_{\{\hat c',\hat s'\}_{m}}
\sket{\{p,f, c,\hat c', s,\hat s'\}_{m}}\,
\bra{\{c',s'\}_{m}}\bm{F}(\{p,f\}_m)\ket{\{\hat c',\hat s'\}_{m}}\dualR
\;.
\end{split}
\end{equation}
We use Eq.~(\ref{eq:colorcompleteness}) to insert unit operators into Eq.~(\ref{eq:PopcFstructure2}), giving
\begin{equation}
\begin{split}
\label{eq:PopcFstructure3}
\Pop{\cF} \sket{\{p,f,c,c',s,s'\}_{m}} ={}& 
\frac{1}{2}
\sum_{\{\hat c, \tilde c, \hat s\}_{m}}
\sket{\{p,f,\hat c,  c',\hat s,  s'\}_{m}}\,
\dualL\brax{\{\hat c\}_{m}}\ket{\{\tilde c\}_{m}}\dualR\,  
\bra{\{\tilde c,\hat s\}_{m}}\bm{F}(\{p,f\}_m)\ket{\{c,s\}_{m}}
\\& + \frac{1}{2}
\sum_{\{\hat c',\tilde c',\hat s'\}_{m}}
\sket{\{p,f, c,\hat c', s,\hat s'\}_{m}}
\bra{\{c',s'\}_{m}}\bm{F}(\{p,f\}_m)\ket{\{\tilde c',\hat s'\}_{m}}\,
\dualL\brax{\{\tilde c'\}_{m}}\ket{\{\hat c'\}_{m}}\dualR\,  
.
\end{split}
\end{equation}

With the choice (\ref{eq:PopcFstructure1}) for $\Pop{\cF}$, Eq.~(\ref{eq:1PopcFstart}) becomes
\begin{equation}
\sbra{1}\Pop{\cF} \sket{\{p,f,c,c',s,s'\}_{m}} = \bra{\{c', s'\}_{m}}\bm{F}(\{p,f\}_m)\ket{\{c,s\}_{m}}
\;.
\end{equation}
We want $\sbra{1}\Pop{\cF} \sket{\{p,f,c,c',s,s'\}_{m}} = \sbra{1}\cF \sket{\{p,f,c,c',s,s'\}_{m}}$, so we define the matrix element of $\bm{F}$ by
\begin{equation}
\bra{\{c', s'\}_{m}}\bm{F}(\{p,f\}_m)\ket{\{c,s\}_{m}}
 = \sbra{1}\cF\sket{\{p,f,c,c',s,s'\}_{m}}
\;.
\end{equation}
Then Eq.~(\ref{eq:PopcFstructure3}) gives us the definition of $\Pop{\cF}$:
\begin{equation}
\begin{split}
\label{eq:PopcFstructure4}
\Pop{\cF} \sket{\{p,f,c,c',s,s'\}_{m}} ={}& 
\frac{1}{2}
\sum_{\{\hat c, \tilde c, \hat s\}_{m}}
\sket{\{p,f,\hat c,  c',\hat s,  s'\}_{m}}\,
\sbra{1}\cF\sket{\{p,f,c,\tilde c,s,\hat s\}_{m}}\,
\dualL\brax{\{\hat c\}_{m}}\ket{\{\tilde c\}_{m}}\dualR
\\& + \frac{1}{2}
\sum_{\{\hat c',\tilde c',\hat s'\}_{m}}
\sket{\{p,f, c,\hat c', s,\hat s'\}_{m}}\,
\sbra{1}\cF\sket{\{p,f,\tilde c',c',\hat s',s'\}_{m}}\,
\dualL\brax{\{\tilde c'\}_{m}}\ket{\{\hat c'\}_{m}}\dualR\,  
\;.
\end{split}
\end{equation}
\end{widetext}%-------------------------------

We will use this canonical form for $\Pop{\cA_0^{(1,0)}}$. For $\Pop{\cA_0^{(0,1)}}$, we simply take $\cA_0^{(0,1)} = \cG^{(0,1)}$ and obtain $\cG^{(0,1)}$ from $\xG^{(0,1)}$ in Eq.~(\ref{eq:eG01oper}). With this definition, $\cA_0^{(0,1)}$ is already in the canonical form, so that 
\begin{equation}
\Pop{\cA_0^{(0,1)}} = \cA_0^{(0,1)}
\;.
\end{equation}
This gives us a contribution to $\cD$ in Eq.~(\ref{eq:cDfirstorder}), $\cA_0^{(0,1)} - \Pop{\cA_0^{(0,1)}} = 0$. This leaves us with
\begin{align}
\cD(\mu_\scS) ={}& 
1 + \frac{\as}{2\pi}
\Big(
\cD^{(1,0)}(\mu_\scS)
+ \cD^{(0,1)}(\mu_\scS)
\Big)
+\cdots
\;,
\end{align}
where\footnote{Note that $-\Pop{\cA^{(1,0)}(\mu_\scS)}$ is first order in $\as$ and does not create any new partons, so it counts as part of $\cD^{(0,1)}(\mu_\scS)$. In fact, it is {\em all} of $\cD^{(0,1)}(\mu_\scS)$.}
\begin{equation}
\begin{split}
\cD^{(1,0)}(\mu_\scS) ={}& \cA_0^{(1,0)}(\mu_\scS)
\;,
\\
\cD^{(0,1)}(\mu_\scS) ={}& -\Pop{\cA_0^{(1,0)}(\mu_\scS)}
\;.
\end{split}
\end{equation}

The first order splitting operators are then
\begin{equation}
\begin{split}
\cS^{(1,0)}(\mu_\scS) ={}& 
\mu_\scS\,\frac{d \cD^{(1,0)}(\mu_\scS)}{d \mu_\scS}
\;,
\\
\cS^{(0,1)}(\mu_\scS) ={}& 
\mu_\scS\,\frac{d \cD^{(0,1)}(\mu_\scS)}{d \mu_\scS}
\;.
\end{split}
\end{equation}
That is
\begin{equation}
\begin{split}
\cS^{(1,0)}(\mu_\scS) ={}& 
\mu_\scS\,\frac{d \cA_0^{(1,0)}(\mu_\scS)}{d \mu_\scS}
\;,
\\
\cS^{(0,1)}(\mu_\scS) ={}& 
-\Pop{\cS^{(1,0)}(\mu_\scS)}
\;.
\end{split}
\end{equation}

This gives us the splitting operators
\begin{equation}
\begin{split}
\label{eq:S10expansion}
\cS^{(1,0)}(\mu) ={}& 
\sum_l \cC^{(1,0)}_l(\mu) 
+ \sum_{\substack{l,k\\ k \ne l}} \cE^{(1,0)}_{lk}(\mu)
\end{split}
\end{equation}
and
\begin{equation}
\label{eq:cS01fromcS10parts}
\cS^{(0,1)}(\mu) = 
- \sum_l \Pop{\cC^{(1,0)}_l(\mu)}
- \sum_{\substack{l,k\\ k \ne l}} \Pop{\cE^{(1,0)}_{lk}(\mu)}
\;.
\end{equation}
We are now ready to examine the operators that contribute to $\cS^{(0,1)}(\mu)$.

We begin with $\cE^{(1,0)}_{lk}(\mu)$. In order to be able to apply the $\iPop{\cdots}$ operator we need the inclusive probability associated with $\cE^{(1,0)}_{lk}(\mu)$. Using Eq.~(\ref{eq:cElk10flfk}) for $\cE^{(1,0)}_{lk}(\mu)$, we find
\begin{equation}
  \begin{aligned}[c]
    \label{eq:cElk-total}
    \sbra{1}\cE^{(1,0)}_{lk}(\mu)&\sket{\{p,f,c,c',s,s'\}_m}
    \\
    ={}&E(\mu, p_l, p_k)\,\brax{\{s'\}_{m}}\ket{\{s\}_m}
    \\
    &\times
    \bra{\{c'\}_m} \bm{T}_k\cdot \bm{T}_l\ket{\{c\}_m}
    \;,
  \end{aligned}
\end{equation}
where
\begin{equation}
  \label{eq:Eikonal-func}
  \begin{split}
    E(\mu, p_l, p_k)
    ={}&8\pi^2\int\!d\zeta_\scS\,\frac{\hat p_l \cdot P_\perp(\hat p_{m\!+\!1})\cdot\hat p_k}
    {\hat p_l\cdot \hat p_{m\!+\!1}\ \hat p_k \cdot \hat p_{m\!+\!1}}\,
    \\
    &
    \times
    2\mu^2\,\delta\!\left(\mu^2  - \max\left[h(q_l), h(q_k)\right]\right)
    \;,
  \end{split}
\end{equation}
with $q_l = \hat p_l+\hat p_{m\!+\!1}$, $q_k = \hat p_k+\hat p_{m\!+\!1}$, and
\begin{equation}
  \label{eq:gluonspinsum}
  \begin{split}
    P_\perp^{\mu\nu}(p) ={}& \sum_{s} \varepsilon^\mu(p, s)^* \varepsilon^\nu (p,  s)
    \\
    ={}& - g^{\mu\nu} + \frac{p^\mu n^\nu + n^\mu p^\nu}{ p\cdot n} - \frac{p^\mu p^\nu}{(p\cdot n)^2}
    \;.
  \end{split}
\end{equation}
The integrand in Eq.~(\ref{eq:Eikonal-func}) depends on the momenta $\hat p_l$, $\hat p_k$ and $\hat p_{m\!+\!1}$, which are related to the initial momenta $p_l$ and $p_k$ and to the splitting variables $\zeta_\scS$ via the soft momentum mapping discussed in Appendix~\ref{sec:SoftMomentumMapping}. We note that $E(\mu, p_l, p_k)$ contains the function defining the hardness scale $\mu$ together with the amplitudes for emitting the soft gluon from parton $l$ in the ket state and from parton $k$ in the bra state, evaluated in the eikonal approximation.

We introduce a shorthand notation $\bm T_k \cdot \bm T_l$ for the color operator describing emitting a gluon from line $l$ and then absorbing this gluon on line $k$, summed over the color indices $a$ of the gluon:
\begin{equation}
  \begin{split}
    \sbra{1_\Lc}t_l^\dagger(&f_l \to f_l + \Lg) \otimes 
    t_k(f_k \to f_k + \Lg)\sket{\{c,c'\}_m}
    \\
    = {}&\sum_{\{\hat c, \hat c'\}_{m+1}} 
    \brax{\{\hat c'\}_{m+1}}\ket{\{\hat c\}_{m+1}}
    \\
    &\qquad\times
    \bra{\{c'\}_m}t_k(f_k \to f_k + \Lg)\ket{\{\hat c'\}_{m+1}}\dualR
    \\
    &\qquad\times
    \dualL\bra{\{\hat c\}_{m+1}}t_l^\dagger(f_l\! \to f_l + \Lg)\ket{\{c\}_m}
     \\
    ={}&
    \bra{\{c'\}_m}t_k(f_k\!\to\! f_k \!+\! \Lg)\,
    t_l^\dagger(f_l\! \to \!f_l\! +\! \Lg)\ket{\{c\}_m}
    \\
     ={}& \bra{\{c'\}_m} \bm{T}_k\cdot \bm{T}_l\ket{\{c\}_m}
    \;.
  \end{split}
\end{equation}
Here we have used the completeness relation of the color basis vectors.

Now we can construct the operator $\Pop{\cE_{lk}(\mu)}$ according to the definition (\ref{eq:PopcFstructure4}):
\begin{equation}
  \begin{aligned}[c]
    \label{eq:cElkendP}
    \Pop{\cE^{(1,0)}_{lk}(\mu)}\sket{\{p,f,c,c',s,s'\}_m}
    \hskip - 4.0 cm &
    \\
    ={}&E(\mu, p_l, p_k)\,\frac12 \sum_{\{\tilde c, \tilde c'\}_{m}}\sket{\{p,f,\tilde c, \tilde c', s, s'\}_m}
    \\
    &\times
    \Big\{
    \dualL\bra{\{\tilde c\}_m} \bm{T}_k\cdot \bm{T}_l\ket{\{c\}_{m}}
    \brax{\{c'\}_m}\ket{\{\tilde c'\}_{m}}\dualR   
    \\
    &\quad
    +
    \dualL\brax{\{\tilde c\}_m}\ket{\{c\}_{m}}
    \bra{\{c'\}_m} \bm{T}_k\cdot \bm{T}_l\ket{\{\tilde c'\}_{m}}\dualR
    \Big\}
    \;.
  \end{aligned}
\end{equation}
Note that in the first term, the matrix element $\ibrax{\{c'\}_m}\iket{\{\tilde c'\}_{m}}\dualR$ sets $\{\tilde c'\}_{m} = \{c'\}_{m}$, while in the second term, the matrix element ${}_\scD\!\ibrax{\{\tilde c\}_m}\iket{\{c\}_{m}}$ sets $\{\tilde c\}_{m} = \{c\}_{m}$. We see that the state obtained by applying $\Pop{\cE_{lk}(\mu)}$ has the same $m$ partons with momenta, flavors and spins, $\{p,f,s,s'\}_m$, as in the original state. However, the color state of the partons is changed.

We now consider $\Pop{\cC^{(1,0)}_{l}(\mu)}$, beginning with Eq.~(\ref{eq:cCl10qqg}) for $\cC^{(1,0)}_{l}(\mu)$. First we calculate the inclusive probability associated with $\cC^{(1,0)}_{l}(\mu)$:
\begin{equation}
  \label{eq:cCl-inclusive}
  \begin{aligned}[c]
    \sbra{1}\cC^{(1,0)}_{l}(\mu)\sket{\{p,f,c,c',s,s'\}_m}
    \hskip -4 cm &
    \\
    ={}& \sum_{\substack{f_1,f_2\\ f_l=f_1+f_2}}\bra{s'_l}\bm{P}_{f_1, f_2}(p_l, f_l)\ket{s_l}
    \prod_{\substack{i=1\\i\ne l}}^m
    \delta_{s_i, s'_i}\,
    \\
    &
    \times
    \frac{\bra{\{c'\}_{m}}t_l(f_l \to f_1 + f_2)\,t_l^\dagger(f_l \to f_1 + f_2)\ket{\{c\}_{m}}}{C(f_1,f_2)}
    \;.
  \end{aligned}
\end{equation}
The spins of the spectator partons are unchanged. We address the operator $\bm{P}_{f_1, f_2}(p_l, f_l)$ that acts on the spin space for parton $l$ below. First, we examine the color operator.
 
The color factor in Eq.~(\ref{eq:cCl-inclusive}) can be simplified by using the definition of the color operator from Eqs.~(\ref{eq:colorket}), (\ref{eq:t-colorket}), (\ref{eq:colorbra}), and (\ref{eq:t-colorbra}): 
\begin{equation}
  \begin{aligned}[c]
    \bra{\{c'\}_{m}}t_l(f_l \to  f_1+f_2 )\,t^\dagger_l(f_l \to  f_1+f_2 )\ket{\{c\}_{m}}
    \hskip - 6.4 cm &
    \\
    = {}&
    \sum_{\{a,a'\}_m}
    \brax{\{c'\}_{m}}\ket{\{a'\}_{m}}\brax{\{a\}_{m}}\ket{\{c\}_{m}}
    \\
    &\quad\times 
    \bra{a'_{l}}\bm{t}_l(f_l; f_1, f_2)\bm{t}_l^\dagger(f_l; f_1, f_2)\ket{a_l}
    \\
    &\quad\times 
    \prod_{\substack{i = 1 \\ i\ne l}}^m \delta_{a'_i,a_i}
    \\
    ={}& C(f_1,f_2)\,\brax{\{c'\}_{m}}\ket{\{c\}_{m}}
    \;,
  \end{aligned}
\end{equation}
where the flavor dependent color constant $C(f_1,f_2)$ is 
\begin{equation}
  \begin{split}
    &C(f_1,f_2) = 
    \begin{cases}
      C_\LF\quad & f_1 \neq \Lg\; \&\; f_{2} = \Lg\;,
      \\
      C_\LF\quad & f_1 = \Lg\; \&\; f_{2} \neq \Lg\;,
      \\
      C_\LA\quad & f_1 = f_2 = \Lg\;,
      \\
      T_\LR\quad &  \hat f_1= -f_2 \neq \Lg\;.
    \end{cases}
  \end{split}
\end{equation}
This is the usual color structure for a collinear emission.

Now, we examine the operator $\bm{P}_{f_1, f_2}(p_l, f_l)$, which is defined by
\begin{equation}
  \label{eq:sPs-splitkern}
  \begin{split}
    \bra{s'}\bm{P}_{f_1, f_2}(\mu, p_l, f_l)\ket{s} 
    \hskip -2cm &
    \\
    = {}& 8\pi^2 C(f_1,f_2) \int d\zeta_\scC \,
    %\\&\times 
    \big[(\hat p_l + \hat p_{m\!+\!1})^2\big]^{-1}
    \\
    &\times
    \bra{s'_l}\bm{c}^\dagger(p_l,f_l; \hat p_l, \hat f_l;\hat{p}_{m\!+\!1},\hat{f}_{m\!+\!1})
    \\&\qquad \times
      \bm{c}(p_l,f_l; \hat p_l, \hat f_l;\hat{p}_{m\!+\!1},\hat{f}_{m\!+\!1})\ket{s_l}
    \;.
  \end{split}
\end{equation}
There is an integration over the collinear splitting variables, $\zeta_\scC$. The  momenta $\hat p_l$, $\hat p_{m\!+\!1}$ are related to the momentum of the emitter parton, $p_l$ and the splitting variables by the collinear momentum mapping  described in Appendix~\ref{sec:CollinearMomentumMapping}. The operators $\bm{c}$ depend on the parton flavors involved and are defined in Eqs.~(\ref{eq:cop-qqg}), (\ref{eq:cop-qbarqbarg}), (\ref{eq:cop-gqqbar}), and (\ref{eq:cop-ggg}) in Sec.~\ref{sec:collinear1storder}.   

The expression in Eq.~\eqref{eq:sPs-splitkern} has a simple spin dependence. After integrating over the splitting variables $\zeta_\scC$, the spin matrix is invariant under rotations about the $\vec{p}_l$ axis. For this reason, it is proportional to the unit matrix in the spin space and can be obtained by averaging over the spin variables. We define a splitting probability function summed over the daughter parton flavors as
\begin{equation}
  P(\mu, p_l, f_l) = \!\!\sum_{\substack{f_1,f_2\\ f_l=f_1+f_2}}\!\!
  \sum_{s_l}\frac{\bra{s_l}\bm{P}_{f_1, f_2}(\mu, p_l, f_l)\ket{s_l}}{n_\Ls(f_l)}
  \,,
\end{equation}
where $n_\Ls(f)$ is the number of the spin states: for fermions $n_\Ls(q)=n_\Ls(\bar q)=2$ and for gluons $n_\Ls(\Lg) = d-2$.

With this, the inclusive probability associated with $\cC^{(1,0)}_{l}(\mu)$ is
\begin{equation}
  \label{eq:cCl-inclusive1}
  \begin{aligned}[c]
    \sbra{1}\cC^{(1,0)}_{l}(\mu)\sket{\{p,f,c,c',s,s'\}_m}
    \hskip -3.5 cm &
    \\
    ={}& P(\mu, p_l, f_l)\, \brax{\{c',s'\}_{m}}\ket{\{c,s\}_{m}}
    \;.
  \end{aligned}
\end{equation}
Given this result for $\sbra{1}\cC^{(1,0)}_{l}(\mu)$, we can use Eq.~\eqref{eq:PopcFstructure4} to obtain $\Pop{\cC^{(1,0)}_{l}(\mu)}$:
\begin{equation}
  \label{eq:cClendP}
  \begin{aligned}[c]
    \Pop{\cC^{(1,0)}_{l}(\mu)}\sket{\{p,f,c,c',s,s'\}_m}
    \hskip -3.5 cm &
    \\
    ={}& P(\mu, p_l, f_l)\, \sket{\{p,f,c,c',s,s'\}_m}
    \;.
  \end{aligned}
\end{equation}
This operator leaves the flavors, colors, and spins of the spectator partons and also the active parton $l$ unchanged. It does depend on the flavor $f_l$ of parton $l$.

%-------------------------------------------------------------
%-------------------------------------------------------------
%---------------------------------
\section{Conclusions and outlook}
\label{sec:conclusions}

Our aim in this paper has been to provide a formalism in which one can define the evolution operator $\cS(\mu)$ for a parton shower for $e^+e^-$ annihilation at second order in $\as$ or beyond. 

To construct $\cS(\mu)$ at order $\as^2$, we need the infrared singular operator $\cD(\mu)$ \cite{NSAllOrder} at order $\as^2$. Since $\cD^{-1}(\mu)$ provides the subtraction of infrared singularities from a perturbative calculation of a cross section, constructing $\cD(\mu)$ at order $\as^2$ is similar in complexity to constructing a fixed order cross section at order $\as^2$ beyond the Born level. This has proven to be a problem of some difficulty. Thus the formalism that we have presented is necessarily rather complex. It is perhaps useful here to review a few of the most important features of this formalism. 

For a state of many partons, we define a scale $\mu_\scS$ that measures how well separated the parton momenta are. That is, $\mu_\scS$ is small when any two partons are nearly collinear or when any parton momentum is very soft. We regard parton shower evolution as a renormalization group evolution that tells how a vector $\isket{\rho(\mu_\scS)}$ that represents the state of the system evolves as the scale $\mu_\scS$ of the partons decreases from a very large value associated with the hard process. This evolution is controlled by the infrared singularities of QCD. 

The formalism defines a {\em quantum shower} evolution in the sense that $\isket{\rho(\mu_\scS)}$ is a vector in a space, the {\em statistical space}, described by momenta and flavors for the partons and by colors and spins for the partons in the ket amplitude and separate colors and spins for the partons in the conjugate bra amplitude. In this way, the shower evolution can match the infrared singularities of Feynman graphs for the amplitude and for the conjugate amplitude, including quantum interference. 

It is usual to define the amplitudes needed for shower evolution using Feynman gauge. However, this leads to complications when one tries to go beyond leading order for the splitting operators. Instead, we use interpolating gauge \cite{Doust, BaulieuZwanziger, Gauge}, which simplifies the infrared singularities of QCD. Indeed, the use of interpolating gauge simplifies the treatment of infrared singularities from interference graphs even at first order. 

To go beyond first order for the evolution operators, we describe Feynman diagrams for the ket and bra amplitudes in a precise way that allows for the use of linear algebra to organize the factors in the diagrams. For this purpose, we introduce a bigger space than the statistical space, the {\em extended statistical space}. In the extended statistical space, parton momenta are generally off shell. Furthermore, a contribution from a Feynman graph for the ket amplitude can multiply a contribution to the conjugate bra amplitude from a different Feynman graph. Thus evolution in the extended statistical space includes quantum interference in momentum and flavor as well as color and spin.

The extended statistical space also includes information that allows a momentum mapping from $m$ on-shell momenta to $\hat m$ on-shell momenta after splittings while preserving exact momentum conservation.

Based on what we anticipate is the general structure of the operators $\xG$ that describe graphs in the extended statistical space, we find that the graphs factor into a hard factor with scales greater than a separation scale $\mu_\scS$ and an infrared singular operator $\xA_0(\mu_\scS)$ with scales smaller than $\mu_\scS$.

Once we have $\xA_0(\mu_\scS)$ at a certain perturbative order, we map it into an operator $\cD(\mu_\scS)$ that operates on the statistical space. Then the evolution operator in the statistical space is $\cS(\mu_\scS)$ defined by $\mu_\scS\,d\cD(\mu_\scS)/d\mu_\scS = \cD(\mu_\scS)\,\cS(\mu_\scS)$, evaluated at the available perturbative order. The definition is arranged so that the splitting operator $\cS(\mu_\scS)$ conserves probability: the total probability for partons to split into more partons plus the probability for the partons not to split equals 1.

We have seen how this formalism works to define the evolution operators $\cS^{(1,0)}(\mu_\scS)$ for first order parton emission and $\cS^{(0,1)}(\mu_\scS)$ for first order virtual graphs. With probability conservation, $\cS^{(0,1)}(\mu_\scS)$ is defined by integrating $\cS^{(1,0)}(\mu_\scS)$ over the splitting variables.

We leave for future work the operators $\cS^{(2,0)}(\mu_\scS)$ for two real parton emissions, $\cS^{(1,1)}(\mu_\scS)$ for one real parton emission and a first order virtual graph, and $\cS^{(0,2)}(\mu_\scS)$ for second order virtual graphs.

Knowing the evolution operator $\cS(\mu_\scS)$ gives us an evolution equation in the statistical space. This does not give us a practical parton shower event generator, any more than knowing the Schr\"odinger equation for molecules made from nuclei and electrons gives us directly a practical method for calculating the energy levels of these molecules. Rather, once we have the evolution equation with an evolution operator $\cS(\mu_\scS)$, we need to devise an approximate evolution operator $\cS_\mathrm{approx}(\mu_\scS)$ that is simple enough that it can be implemented in a parton shower event generator. One can then define
\begin{equation}
\Delta \cS(\mu_\scS) = \cS(\mu) - \cS_\mathrm{approx}(\mu_\scS)
\;.
\end{equation}
Then one can check whether including just a few powers of $\Delta \cS(\mu_\scS)$ in the shower makes a substantial difference in the results for an observable of interest. 

This has been the strategy in the first order parton shower event generator \textsc{Deductor} \cite{Deductor}. Here spin is simply neglected by averaging over spins, but some approximation is needed for color. The operator $\cS_\mathrm{approx}(\mu_\scS)$ is obtained by making an approximation denoted as the LC+ approximation \cite{NScolor}. This is plausibly a very good approximation because it is exact for collinear emissions and for collinear$\times$soft emissions. Note that the statement that the LC+ approximation is exact for collinear emissions and for collinear$\times$soft emissions depends on knowing the exact $\cS(\mu_\scS)$. We can then test the numerical effect of including a few powers of $\Delta \cS(\mu_\scS)$ in the shower \cite{NSNewColor}.

For the first order splitting operator $\cS^{(1)}(\mu_\scS)$ constructed in this paper, we would anticipate using $\cS^{(1)}_\mathrm{approx}(\mu_\scS)$ based on the LC+ approximation. We leave finding an approximation for second order for future work. One possibility is to define $\cS^{(2)}_\mathrm{approx}(\mu_\scS)$ to be zero, so that the entire $\cS^{(2)}(\mu_\scS)$ is treated as part of $\Delta \cS(\mu_\scS)$. One could then test the effect of including a few second order splittings in the shower.

%-------------------------------------------------
\acknowledgments{ 
This work was supported in part by the United States Department of Energy under grant DE-SC0011640. One of us (DES) thanks the theory group of DESY Laboratory, Hamburg for its hospitality during some of the work on this paper and also thanks the Mainz Institute for Theoretical Physics for its hospitality during some of this work.
}
%----------------------------------------------------------------------------

%=======================================================================
\appendix
%----------------------
%---------------------------------
%-----------------------------------
\section{Collinear momentum mapping}
\label{sec:CollinearMomentumMapping}

In order to describe a splitting of parton $l$ in the ket state and the same parton in the bra state into two partons, we need a momentum mapping operator $\bm R_{\mathrm{coll}}^{(1)}(l)$ that maps the momenta $\{p\}_m$ to new momenta $\{\hat p\}_{m\!+\!1}$, which are functions of the $m$ initial momenta $p_i$ and splitting variables $\zeta_\scC$, as in Eq.~(\ref{eq:bmRdef}). We will also need a momentum mapping $\bm R_{\mathrm{coll}}^{(2)}(l)$ for the splitting of one parton into three partons.

Consider first the collinear splitting of Fig.~\ref{fig:selfenergy1}. We start with the momentum $p_l$ of the parton that splits, with $p_l^2 = 0$. We use an auxiliary lightlike vector $n_l$ in the plane of $p_l$ and the total momentum $Q$:
\begin{equation}
n_l = \frac{2 p_l\cdot Q\,Q - Q^2\,p_l}{Q^2}
\;.
\end{equation}
Then
\begin{equation}
\label{eq:Qfromalnl}
Q = a_l(p_l + n_l)
\;,
\end{equation}
where
\begin{equation}
a_l = \frac{Q^2}{2 p_l\cdot Q}
\;.
\end{equation}
Parton $l$ splits into partons $l$ and $m\!+\!1$ with momenta parameterized by splitting parameters $z$ and $k_\perp$, where the vector $k_\perp$ is orthogonal to both $p_l$ and $n_l$:
\begin{equation}
\begin{split}
\label{eq:collinearsplitting1}
\hat p_l ={}& z p_l + \frac{\bm k^2}{2 z\, p_l\cdot n_l}\, n_l
+ k_\perp
\;,
\\
\hat p_{m\!+\!1} ={}& (1-z) p_l + \frac{\bm k^2}{2 (1-z)\, p_l\cdot n_l}\, n_l
- k_\perp
\;.
\end{split}
\end{equation}
Considering $k_\perp$ to be a vector $\bm k$ in $2 - 2 \epsilon$ dimensions, we have $k_\perp^2 = - \bm k^2$. We define
\begin{equation}
P = \hat p_l + \hat p_{m\!+\!1}
\;.
\end{equation}
Then $P^2 > 0$. It is useful to define a dimensionless measure $y$ of the virtuality in the splitting by
\begin{equation}
y = \frac{P^2}{2 p_l\cdot Q}
= \frac{P^2}{a_l 2 p_l\cdot n_l}
\;.
\end{equation}
Using $y$, we can write $P$ as
\begin{equation}
P = p_l + a_l y\, n_l
\;.
\end{equation}
The virtuality measure $y$ is related to the splitting variables by
\begin{equation}
a_l y = \frac{\bm k^2}{z(1-z)2 p_l \cdot n_l}
\;.
\end{equation}

Denote the momenta of the remaining partons after the splitting by $\hat p_i$. For a splitting with $y = 0$, we have $P = p_l$. Then we can let $\hat p_i = p_i$ for with $1 \le i \le m$ and $i \ne l$. However, when $y > 0$ we need to map $p_i$ for the remaining partons to new momenta $\hat p_i$ with $\hat p_i^2 = 0$ and
\begin{equation}
\label{eq:momentumconservation}
\sum_{\substack{i = 1\\i\ne l}}^m \hat p_i + \hat p_l + \hat p_{m\!+\!1}
= \sum_{\substack{i = 1\\i\ne l}}^m p_i + p_l
\;.
\end{equation}
This can be accomplished quite simply by letting $\hat p_i$ be related to $p_i$ by a scaling times a Lorentz transformation\footnote{In a first order dipole shower, one sometimes takes the momentum $P - p_l$ from a single dipole partner parton. Here, as in \textsc{Deductor}, we take the needed momentum from {\em all} of the remaining partons, so that the fractional change in the momentum of any one parton is small.}
\begin{equation}
\label{eq:pitohatpi}
\hat p_i^\mu =\sqrt{1-y}\, \Lambda^\mu_\nu p_i^\nu
\;.
\end{equation}
The Lorentz transformation is a boost in the $p_l$-$n_l$ plane. The transformation is most simply stated in component form. Let
\begin{equation}
p_i = \frac{p_i\cdot n_l}{p_l\cdot n_l}\, p_l
+ \frac{p_i\cdot p_l}{p_l\cdot n_l}\, n_l
+ p_{i\perp}
\;.
\end{equation}
with $p_{i\perp}\cdot p_l = p_{i\perp}\cdot n_l = 0$. Then define
\begin{equation}
\hat p_i = \frac{p_i\cdot n_l}{p_l\cdot n_l}\, p_l
+ (1-y) \frac{p_i\cdot p_l}{p_l\cdot n_l}\, n_l
+ \sqrt{1-y}\,p_{i\perp}
\;.
\end{equation}
We leave the component of $p_i$ along $p_l$ unchanged but scale the component along $n_l$ and also the transverse components so as to leave $p_i^2$ unchanged. Using
\begin{equation}
\sum_{\substack{i = 1\\i\ne l}}^m p_i = Q - p_l = (a_l - 1) p_l + a_l n_l
\;,
\end{equation}
we recover the momentum conservation equation (\ref{eq:momentumconservation}):
\begin{equation}
\sum_{\substack{i = 1\\i\ne l}}^m (\hat p_i - p_i)
= - a_l y\, n_l
= - (P - p_l)
\;.
\end{equation}

The splitting variables for the collinear splitting are $\zeta_\scC = \{z, \bm k\}$. We define a measure for integration over the splitting variables by
\begin{equation}
\label{eq:dzetacoll1}
d\zeta_\scC = J_\scC \frac{1}{(2\pi)^{3-2\epsilon}}
\frac{dz}{2 z (1-z)}\,d^{2 - 2\epsilon}d\bm k
\;.
\end{equation}
Here $J_\scC$ is an additional Jacobian factor to be specified. 

We integrate over the momenta before the splitting using the measure $\{d p\}_m$ in Eq.~(\ref{eq:dp}). We integrate over the momenta after the splitting using the measure $\{d \hat p\}_{m\!+\!1}$ for $m+1$ partons. This should match the measure of integrating over the momenta before the splitting together with integrating over the splitting variables, as in Eq.~(4.37) of Ref.~\cite{NSI} and Eq.~(\ref{eq:dzetaproperty}):
\begin{equation}
\{d \hat p\}_{m\!+\!1} = \{d p\}_m\,d\zeta_\scC
\;.
\end{equation}
Enforcing this matching gives us, after some calculation,
\begin{equation}
\label{eq:Jpresult}
J_\scC = (1-a_l y)^{2 - 2\epsilon}
(1-y)^{m-3 - (m-2)\epsilon}
\;.
\end{equation}
Of course, we could choose any functions of $z$ and $\bm k$ as alternative splitting variables $\zeta_\scC$ \cite{NSI}. Then we would use an alternative Jacobian $J_c$ obtained by multiplying $J_c$ given in Eq.~(\ref{eq:Jpresult}) by the Jacobian for the change to the new splitting variables.

The action of the momentum mapping operator $\bm R_{\mathrm{coll}}^{(1)}(l)$ in Eq.~(\ref{eq:bmRdef}) for the first order collinear splitting can now be specified as
\begin{equation}
\label{eq:Rcoll}
\bm R_{\mathrm{coll}}^{(1)}(l) \sket{\{p\}_m} = \int\!d\zeta_\scC\ \sket{\{\hat p\}_{m\!+\!1}}
\;,
\end{equation}
where $\{\hat p\}_{m\!+\!1}$ is given in Eqs.~(\ref{eq:collinearsplitting1}) and (\ref{eq:pitohatpi}) and $d\zeta_\scC$ is given in Eq.~(\ref{eq:dzetacoll1}).

We will also need a momentum  mapping for a $1 \to 3$ parton splitting. We can follow rather closely the construction above for a $1 \to 2$ splitting. We use splitting variables $\zeta_{\Lc,3} = \{z_1,z_2,\bm k_1, \bm k_2\}$ with
\begin{equation}
\begin{split}
\label{eq:coll3zeta}
\hat p_{m\!+\!1} ={}& z_1 p_l 
+ \frac{\bm k_1^2}{2 z_l\, p_l\cdot n_l}\, n_l
+ k_{1,\perp}
\;,
\\
\hat p_{m\!+\!2} ={}& z_2 p_l 
+ \frac{\bm k_2^2}{2 z_2\, p_l\cdot n_l}\, n_l
+ k_{2,\perp}
\;,
\\
\hat p_{l} ={}& z_l p_l 
+ \frac{\bm k_l^2}{2 z_l\, p_l\cdot n_l}\, n_l
+ k_{l,\perp}
\;,
\end{split}
\end{equation}
where
\begin{equation}
\begin{split}
\label{eq:coll3zetapart2}
z_l ={}& 1 - z_1 - z_2
\;,
\\
k_{l,\perp} ={}& -k_{1,\perp} - k_{2,\perp}
\;.
\end{split}
\end{equation}
The remaining momenta need to be defined so that momentum conservation is maintained:
\begin{equation}
\sum_{i = 1}^{m\!+\!2} \hat p_i
= \sum_{i=1}^m p_i
\;.
\end{equation}
This can be achieved with the definition of $\hat p_i$ with $i \le m, i \ne l$,
\begin{equation}
\label{eq:hatpi1to3}
\hat p_i^\mu = \sqrt{1 - y_3}\, \Lambda^\mu_\nu p_i^\nu
\;,
\end{equation}
where
\begin{equation}
y_3 = \frac{(\hat p_l + \hat p_{m\!+\!1} + \hat p_{m\!+\!2})^2}
{2 p_l \cdot Q}
\end{equation}
and $\Lambda^\mu_\nu$ is the Lorentz transformation defined by
\begin{equation}
Q^\mu - \hat p_l^\mu - \hat p_{m\!+\!1}^\mu - \hat p_{m\!+\!2}^\mu 
= \sqrt{1 - y_3}\, \Lambda^\mu_\nu (Q^\nu - p_l^\nu)
\;.
\end{equation}

This Lorentz transformation is \cite{NSI}
\begin{equation}
\begin{split}
\label{eq:StdBoost}
\Lambda^\mu_{\,\nu} ={}& g^\mu_{\,\nu}
- \frac{2(\hat K + K)^\mu (\hat K + K)_\nu}{(\hat K + K)^2}
+ \frac{2 \hat K^\mu K_\nu}{K^2}
\;,
\end{split}
\end{equation}
where
\begin{equation}
\begin{split}
K^\mu ={}& \sqrt{1 - y_3}\, (Q^\nu - p_l^\nu)
\;,
\\
\hat K^\mu ={}&
Q^\mu - \hat p_l^\mu - \hat p_{m\!+\!1}^\mu - \hat p_{m\!+\!2}^\mu 
\;.
\end{split}
\end{equation}

To integrate over the splitting variables, we define a measure $d\zeta_{\Lc,3}$ such that $\{d \hat p\}_{m\!+\!2} = \{d p\}_m\,d\zeta_{\Lc,3}$:
\begin{equation}
\label{eq:dzetacoll2}
d\zeta_{\Lc,3} =  \frac{J_\scC}{(2\pi)^{6-4\epsilon}}
\frac{dz_1\,dz_2}{4 z_1 z_2 (1-z_1- z_2)}\,d^{2 - 2\epsilon}\bm k_1
\,d^{2 - 2\epsilon}\bm k_2
\;.
\end{equation}
We find that the Jacobian here is $J_\scC$ as given in Eq.~(\ref{eq:Jpresult}).

We can now define a momentum mapping operator $\bm R^{(2)}_{\mathrm{coll}}(l)$ for the second order collinear splitting as
\begin{equation}
\label{eq:Rcoll2}
\bm R^{(2)}_{\mathrm{coll}}(l) \sket{\{p\}_m} 
= \int\!d\zeta_{\Lc,3}\ \sket{\{\hat p\}_{m\!+\!2}}
\;,
\end{equation}
where $\{\hat p\}_{m\!+\!2}$ is given in Eqs.~(\ref{eq:coll3zeta}), (\ref{eq:coll3zetapart2}) and (\ref{eq:hatpi1to3}) and $d\zeta_{\Lc,3}$ is given in Eq.~(\ref{eq:dzetacoll2}).

%---------------------------------
\section{Soft momentum mapping}
\label{sec:SoftMomentumMapping}

Consider the interference diagram of Fig.~\ref{fig:interference1}. The momentum mapping needed here is simple because there is no collinear singularity, only a soft singularity when $\hat p_{m\!+\!1} \to 0$. We can use the components of $\hat p_{m\!+\!1}$ as the splitting variables $\zeta_\scS$ for the soft splitting. 

In fact, we can easily consider a slightly more general case, in which several soft partons are emitted. Let the momenta of the soft partons be $\hat p_{m\!+\!1},\dots,\hat p_{m\!+\!n}$. Then we can write the measure for integrating over the splitting variables as
\begin{equation}
\label{eq:dzetasoft1}
d\zeta_\scS = \prod_{J=1}^n \left\{\frac{d^{4-2\epsilon}p_{m\!+\!J}}{(2\pi)^{4-2\epsilon}}
(2\pi)\,\delta_+(p_{m\!+\!J}^2) \right\} J_\scS
\;,
\end{equation}
with a Jacobian factor $J_\scS$ to be specified. Define
\begin{equation}
q = \sum_{J=1}^n p_{m\!+\!J}
\;.
\end{equation}
We need to transform the remaining momenta so as to maintain momentum conservation
\begin{equation}
\sum_{i = 1}^m \hat p_i + q
= \sum_{i = 1}^m p_i
\;.
\end{equation}
This can be accomplished quite simply by letting $\hat p_i$ for $i \le m$ be related to $p_i$ by a scaling times a Lorentz transformation in the $Q,q$ plane:
\begin{equation}
\label{eq:pitohatpisoft}
\hat p_i^\mu =\sqrt{1-\alpha}\, \Lambda^\mu_{\,\nu} p_i^\nu
\;.
\end{equation}
Then momentum conservation yields
\begin{equation}
Q^\mu - q^\mu = \sqrt{1-\alpha}\, \Lambda^\mu_{\,\nu} Q^\nu
\;.
\end{equation}
Since the Lorentz transformation preserves squares of vectors, this requires
\begin{equation}
(Q-q)^2 = (1-\alpha) Q^2
\;.
\end{equation}
This gives
\begin{equation}
\label{eq:alphasoftdefA}
\alpha =\frac{2 Q\cdot q - q^2}{Q^2}
\;.
\end{equation}
Then the Lorentz transformation is given by Eq.~(\ref{eq:StdBoost}), with
\begin{equation}
\begin{split}
K^\mu ={}& \sqrt{1-\alpha}\,Q^\mu
\;,
\\
\hat K^\mu ={}& Q^\mu - q^\mu
\;.
\end{split}
\end{equation}

We require
\begin{equation}
\{d \hat p\}_{m\!+\!n} = \{d p\}_m\,d\zeta_\scS
\;.
\end{equation}
A simple calculation shows that this relation holds if
\begin{equation}
J_\scS = (1-\alpha)^{m-2 - (m-1)\epsilon}
\;.
\end{equation}

At first order, $q = \hat p_{m\!+\!1}$ is lightlike. Then the mapping is quite simple. The value of $\alpha$ is
\begin{equation}
\label{eq:alphasoftdef}
\alpha =\frac{2 Q\cdot \hat p_{m\!+\!1}}{Q^2}
\;.
\end{equation}
Define a vector $u$ by
\begin{equation}
\label{eq:alphaQ}
\alpha\,Q = u + \hat p_{m\!+\!1}
\;.
\end{equation}
Then using Eq.~(\ref{eq:alphasoftdef}), we see that $u^2 = 0$. The Lorentz transformation is defined by
\begin{equation}
Q^\mu - \hat p_{m\!+\!1}^\mu 
= \sqrt{1-\alpha}\, \Lambda^\mu_\nu Q^\nu
\;.
\end{equation}
Multiplying by $\alpha$ and using Eq.~(\ref{eq:alphaQ}), we have
\begin{equation}
u^\mu + (1-\alpha) \hat p_{m\!+\!1}^\mu 
= \sqrt{1-\alpha}\, \Lambda^\mu_\nu 
(u^\nu + \hat p_{m\!+\!1}^\mu)
\;.
\end{equation}
If we call the boost angle $\omega$, this is
\begin{equation}
u^\mu + (1-\alpha) \hat p_{m\!+\!1}^\mu 
= \sqrt{1-\alpha}\, 
(e^{-\omega} u^\mu + e^\omega \hat p_{m\!+\!1}^\mu)
\;.
\end{equation}
We thus identify the boost factor as
\begin{equation}
e^\omega = \sqrt{1-\alpha}
\;.
\end{equation}
We see that as long as the emitted gluon is soft, the parton momenta $p_i$ for $1 \le i \le m$ undergo a small scaling and a small boost in the $Q$-$\hat p_{m\!+\!1}$ plane.

The action of the momentum mapping operator $\bm R_{\mathrm{soft}}^{(n)}$ in Eq.~(\ref{eq:bmRdef}) for $n$ soft parton emissions can now be specified as
\begin{equation}
\label{eq:Rsoftn}
\bm R_{\mathrm{soft}}^{(n)} \sket{\{p\}_m} = 
\int\!d\zeta_\scS\ \sket{\{\hat p\}_{m\!+\!n}}
\;,
\end{equation}
where $\{\hat p\}_{m\!+\!n}$ is given by Eq.~(\ref{eq:pitohatpisoft}) and the choice of the components of $\hat p_{m\!+\!J}$ as the splitting variables $\zeta_\scS$ and where $d\zeta_\scS$ is given in Eq.~(\ref{eq:dzetasoft1}). 

%---------------------------------
\section{Momentum mapping operator}
\label{sec:MomentumMappingOperator} 

We have specified momentum mapping functions $R$ corresponding to graphs $G$ that map parton momenta $\{p\}_m$ for $m$ partons together with splitting parameters $\zeta_{G}$ into parton momenta $\{\hat p\}_{\hat m}$ for more partons,
\begin{equation}
\label{eq:RmapAppendix}
\{\hat p\}_{\hat m} = R(G;\zeta_{G},\{p\}_m)
\;.
\end{equation}
In this paper $\hat m = m+1$ or $\hat m = m+2$. 

One can choose the splitting parameters to be
\begin{equation}
\zeta_G = \{p_{m\!+\!1},\dots,p_{\hat m}\}
\;,
\end{equation}
as in Appendix \ref{sec:SoftMomentumMapping}, although this is sometimes not the most useful choice, as in Appendix \ref{sec:CollinearMomentumMapping}. To integrate over the splitting parameters we use a measure $d\zeta_{G}$ that includes an appropriate Jacobian such that
\begin{equation}
\label{eq:dzetapropertyAppendix}
d\{\hat p\}_{\hat m} = d\zeta_{G}\ d\{p\}_{m}
\;.
\end{equation}
The needed functions $R$ and the measure $d\zeta$ were defined in Appendices \ref{sec:CollinearMomentumMapping} and \ref{sec:SoftMomentumMapping}.

In the extended statistical space, we define a space that specifies just on-shell parton momenta, considered as jet momenta, with basis vectors $\sket{\{p\}_m}$. We define the completeness relation for these basis vectors to be
\begin{equation}
\label{eq:completenesspAppendix}
1 = \sum_{m=1}^\infty \frac{1}{m!}
\int\! d\{p\}_m\
\sket{\{p\}_m}\sbra{\{p\}_m}
\;.
\end{equation}

We have defined an operator $\bm R(G)$ that maps a basis vector in the $m$ parton space to the a linear combination of vectors in the $\hat m$ parton space.\footnote{The notation $\cP(G)$ was used for an operator similar to $\bm R(G)$ for first order splittings in Ref.~\cite{NSI}. The operator $\cP(G)$ included flavor. If we simply let all partons be gluons so that flavor plays no role, then the definition was $\cP\sket{\{p\}_m} = [1/(m+1)] \bm R\sket{\{p\}_m}$.} This operator was defined briefly in Sec.~\ref{sec:ExtendedStatisticalSpace} and in Appendices \ref{sec:CollinearMomentumMapping} and \ref{sec:SoftMomentumMapping}. In this appendix, we provide a justification for the definition, paying particular attention to the choice of normalization.

Let $\sket{\rho}$ be a vector in the $m$ parton space. Define
\begin{equation}
\rho(\{p\}_m) = \sbrax{\{p\}_m}\sket{\rho}
\;.
\end{equation}
The total probability associated with $\sket{\rho}$ is $\isbrax{1_\Lp}\isket{\rho}$ where $\isbrax{1_\Lp}\isket{\{p\}_m}$ = 1, as in Eq.~(\ref{eq:bra1}), If we normalize the total probability to 1, we have
\begin{equation}
\begin{split}
\label{eq:normalizationm}
1 = {}&
\sbrax{1_\Lp}\sket{\rho} 
\\={}&  
\frac{1}{m!}\int\! d\{p\}_m\
\sbrax{1}\sket{\{p\}_m}\sbrax{\{p\}_m}\sket{\rho}
\\
={}& \int\! d\{p\}_m\ 
\frac{\rho(\{p\}_m)}{m!}
\;.
\end{split}
\end{equation}
This tells us that $\rho(\{p\}_m)/m!$ is to be interpreted as the probability density for parton 1 to have momentum $p_1$, parton 2 to have momentum $p_2$, \dots, and parton $m$ to have momentum $p_m$.

Given $\rho(\{p\}_m)$, we can define a symmetrized version by summing over permutations $\pi$ of the indices:
\begin{equation}
\label{eq:rhosymmetrized}
\rho_\scS(\{p\}_m) = \frac{1}{m!}\sum_{\pi \in S_m} \rho(\{p_\pi\}_m)
\;.
\end{equation}
Since the partons are identical particles, the symmetric version is the one that has a physical meaning independent of labeling. Its normalization is
\begin{equation}
\begin{split}
\frac{1}{m!}\int\!d\{p\}_m&\ \rho_\scS(\{p\}_m) 
\\={}& 
\frac{1}{m!}\int\!d\{p\}_m\ \frac{1}{m!}\sum_{\pi \in S_m} 
\rho(\{p_\pi\}_m) 
\\
={}& \frac{1}{m!}\sum_{\pi \in S_m}\int\!d\{p\}_m\ 
\frac{\rho(\{p_\pi\}_m)}{m!}
\\
={}& \frac{1}{m!}\sum_{\pi \in S_m}\int\!d\{p_\pi\}_m\ 
\frac{\rho(\{p_\pi\}_m)}{m!}
\\
={}& \frac{1}{m!}\sum_{\pi \in S_m}\int\!d\{p\}_m\ 
\frac{\rho(\{p\}_m)}{m!}
\\
={}& \frac{1}{m!}\sum_{\pi \in S_m}1
\\
={}& 1
\;.
\end{split}
\end{equation}
This is the expected normalization condition, interpreted as meaning that $\int\!d\{p\}_m$ counts the same physical state $m!$ times.

We now need an operator that represents the momentum mapping. As in Eq.~(\ref{eq:bmRdef}), we define
\begin{equation}
\label{eq:bmRdefbis}
\bm R(G) \sket{\{p\}_m} = 
\int\!d\zeta_G\ \sket{R(G;\zeta_G,\{ p\}_{m})}
\;.
\end{equation}
This clearly maps the momenta into the right place according to the given momentum mapping function $R(G;\zeta_G,\{ p\}_{m})$. However, it is not immediately clear if the normalization is right so as to conserve probability. We now investigate this question.

Let $\isket{\hat\rho}$ be the state vector that we get by applying $\bm R(G)$ to a vector $\isket{\rho}$ with $m$ partons:
\begin{equation}
\sket{\hat\rho} = \bm R(G)\sket{\rho}
\;.
\end{equation}
Note that only $\bm R(G)$, with no Feynman graph factors, is applied to $\sket{\rho}$ here. Then, using $\{\hat p\}_{\hat m} = R(G;\zeta_G,\{ p\}_{m})$,
\begin{equation}
\begin{split}
\sket{\hat\rho} ={}& 
\frac{1}{m!}\int\!d\{p\}_m\ \bm R(G)
\sket{\{p\}_m} \sbrax{\{p\}_m}\sket{\rho}
\\
={}& \frac{1}{m!}\int\!d\{p\}_m
\int\!d\zeta_G\ \sket{\{\hat p\}_{\hat m}}
\sbrax{\{p\}_m}\sket{\rho}
\\
={}& \frac{1}{m!}\int\!d\{\hat p\}_{\hat m}\
\sket{\{\hat p\}_{\hat m}}
\sbrax{\{p\}_m}\sket{\rho}
\\
={}& \int\!d\{\hat p\}_{\hat m}\
\sket{\{\hat p\}_{\hat m}}
\frac{\rho(\{p\}_m)}{m!}
\;.
\end{split}
\end{equation}
We also have
\begin{equation}
\begin{split}
\sket{\hat\rho}
={}& \frac{1}{\hat m!}\int\!d\{\hat p\}_{\hat m}\
\sket{\{\hat p\}_{\hat m}}\,
\hat\rho(\{\hat p\}_{\hat m})
\;.
\end{split}
\end{equation}
Comparing these, we have
\begin{equation}
\frac{\hat\rho(\{\hat p\}_{\hat m})}{\hat m!} = \frac{\rho(\{p\}_m)}{m!}
\;.
\end{equation}
We see that the normalization according to Eq.~(\ref{eq:normalizationm}) and its analogue for $\{\hat p\}_{\hat m}$ is preserved.

%---------------------------------
\section{Hardness inequality}
\label{sec:hardnessinequality}

In this section, we provide a proof that the inequalities (\ref{eq:Hinequality}) hold for the hardness function $h(q)$ defined  in Eq.~(\ref{eq:Lambdasqcoll}).

Consider a splitting of parton $l$ to partons $i$ and $j$ for partons with $q_l = \bar q_i + \bar q_j$ and with $\bar q_i^2 \ge 0$, $\bar q_i \cdot Q > 0$ and $\bar q_j^2 \ge 0$, $\bar q_j^2 \cdot Q > 0$. We pick a lightlike vector $n$. Then we choose a reference frame in which $\vec Q = 0$ and $\vec n$ lies along the $-z$ axis. In this frame, $(n^+,n^-, \bm n_\perp) = (0,n^-,\bm 0)$. We let the vectors have components in this reference frame
\begin{equation}
\begin{split}
\label{eq:p012components}
q_l ={}& \left(P^+,\,\frac{(\bar {\bm q}_i + \bar {\bm q}_j)^2 + q_l^2}
{2 P^+},\,\bar {\bm q}_i + \bar {\bm q}_j\right)
\;,
\\
\bar q_i ={}& \left(z P^+,\,\frac{\bar {\bm q}_i^2  + \bar q_i^2}
{2z P^+},\,
\bar {\bm q}_i \right)
\;,
\\
\bar q_j ={}& \left((1-z) P^+,\,\frac{\bar {\bm q}_j^2 + \bar q_j^2}
{2(1-z)P^+},\, 
\bar {\bm q}_j \right)
\;.
\end{split}
\end{equation}
We then set $q_l^- = \bar q_i^- + \bar q_j^-$ and solve for $q_l^2$, giving
\begin{equation}
\label{eq:LambdaEquality0}
q_l^2 
=
\frac{((1-z)\bar {\bm q}_i - z \bar {\bm q}_j)^2}
{z(1-z)}
+\frac{\bar q_i^2}{z}
+\frac{\bar q_j^2}{(1-z)}
\;.
\end{equation}
The definitions of $z$ and $1-z$ are
\begin{equation}
\begin{split}
z ={}& \bar q_i\cdot n/q_l\cdot n
\;,
\\
(1-z) ={}& \bar q_j \cdot n/q_l\cdot n
\;.
\end{split}
\end{equation}
This gives \cite{ShowerTime}
\begin{equation}
\begin{split}
\label{eq:LambdaEquality}
\frac{q_l^2}{2 q_l\cdot n}\,Q^2
={}&
\frac{((1-z)\bar {\bm q}_i - z \bar {\bm q}_j)^2}
{z(1-z)2 q_l\cdot n}\,Q^2
\\&
+\frac{\bar q_i^2}{2 \bar q_i \cdot n}\,Q^2
+\frac{\bar q_j^2}{2 \bar q_j \cdot n}\,Q^2
\;.
\end{split}
\end{equation}

We can use this to derive the inequality that we need. Since the terms on the right-hand side of Eq.~(\ref{eq:LambdaEquality}) are all nonnegative, we have
\begin{equation}
\frac{q_l^2}{2 q_l\cdot n}\,Q^2
>
\frac{\bar q_i^2}{2 \bar q_i \cdot n}\,Q^2
\;.
\end{equation}
This implies
\begin{equation}
2 \bar q_i \cdot n\, q_l^2 \,Q^2
>
{2 q_l\cdot n}\,\bar q_i^2\,Q^2
\;.
\end{equation}
Now let $n_\scC$ be another lightlike vector chosen so that $\vec n_\scC$ lies along the $+z$ axis in the reference frame used above. Then  $(n_\scC^+,n_\scC^-, \bm n_\scC) = (n_\scC^+,0,\bm 0)$. Then the same derivation with the roles of $+$ and $-$ components interchanged gives
\begin{equation}
2 \bar q_i \cdot n_\scC\, q_l^2 \,Q^2
>
2 q_l\cdot n_\scC\,\bar q_i^2\,Q^2
\;.
\end{equation}
Choose $n_\scC^+ = Q^+$ and, for the previous lightlike vector $n$, $n^- = Q^-$. Then
\begin{equation}
n + n_\scC = Q
\;.
\end{equation}
Then adding the two inequalities gives
\begin{equation}
2 \bar q_i \cdot Q\, q_l^2 \,Q^2
>
2 q_l\cdot Q\,\bar q_i^2\,Q^2
\;.
\end{equation}
That is
\begin{equation}
\label{eq:LambdaInequality1}
\frac{q_l^2}{2 q_l\cdot Q}\,Q^2
>
\frac{\bar q_i^2}{2 \bar q_i \cdot Q}\,Q^2
\;.
\end{equation}
We also derive
\begin{equation}
\label{eq:LambdaInequality2}
\frac{q_l^2}{2 q_l\cdot Q}\,Q^2
>
\frac{\bar q_j^2}{2 \bar q_j \cdot Q}\,Q^2
\;.
\end{equation}
These are the inequalities (\ref{eq:Hinequality}).

%-------------------
\section{Conjugate amplitudes for collinear splittings}
\label{sec:collinearbra}

In this appendix, we list the bra amplitudes $\bra{s'_l}\bm{c}^\dagger\ket{\hat{s}'_l,\hat{s}'_{m\!+\!1}}$ for the spin dependence in collinear splittings. The derivations are similar to those in Sec.~\ref{sec:collinear1storder} for the ket amplitudes.

For a $q \to q + \Lg$ splitting, we find
\begin{equation}
  \begin{split}
    &\bra{s'_l}\bm{c}^\dagger(p_l,q;\hat p_l,q;\hat{p}_{m\!+\!1},\Lg)
    \ket{\hat{s}'_l,\hat{s}'_{m\!+\!1}}
    \\
    &\qquad
    = \overline U(p_l, s'_l)\,
    \s{\varepsilon}(\hat p_{m\!+\!1}, \hat s'_{m\!+\!1})\,
    U(\hat p_l, \hat s'_l)
    \,.
  \end{split}
\end{equation}
For a $\bar q \to \bar q + \Lg$ splitting, we find
\begin{equation}
  \begin{split}
    &\bra{s'_l}\bm{c}^\dagger(p_l,\bar q; \hat p_l,\bar q;\hat{p}_{m\!+\!1},\Lg)
    \ket{\hat{s}'_l,\hat{s}'_{m\!+\!1}}
    \\
    &\qquad
    = \overline V(\hat p_l, \hat s'_l)\,\s{\varepsilon}(\hat p_{m\!+\!1}, \hat s'_{m\!+\!1})V(p_l,  s'_l)
    \,.
  \end{split}
\end{equation}
For a $\Lg \to q + \bar q$ splitting, we find
\begin{equation}
  \begin{split}
    &\bra{s'_l}\bm{c}^\dagger(p_l,\Lg; \hat p_l, q;\hat{p}_{m\!+\!1},\bar q)
    \ket{\hat{s}'_l,\hat{s}'_{m\!+\!1}}
    \\
    &\qquad
    =  -\overline V(\hat p_{m\!+\!1}, \hat s'_{m\!+\!1})\,\s{\varepsilon}(p_l,s'_l)^* U(\hat p_l, \hat s'_l)
    \,.
  \end{split}
\end{equation}
For a $\Lg \to \Lg + \Lg$ splitting, we find
\begin{equation}
  \begin{split}
    &\bra{s'_l}\bm{c}^\dagger(p_l,\Lg;\hat p_l, \Lg;\hat{p}_{m\!+\!1},\Lg)
    \ket{\hat{s}'_l,\hat{s}'_{m\!+\!1}}
    \\
    &\qquad
    =  -2\, \varepsilon_{\alpha}(\hat p_l, \hat s'_l)\,
    \varepsilon_{\beta}(\hat p_{m\!+\!1}, \hat s'_{m\!+\!1})\,
    \varepsilon_\mu(p_l,s'_l)^*
    \\
    &\qquad
    \times
    \Big[
    g^{\mu\beta}\,\hat{p}_{m\!+\!1}^{\alpha}
    +g^{\mu\alpha}\,\hat{p}_l^{\beta}
    \Big]
    \,.
  \end{split}
\end{equation}
%

%%%%%%%%%%%%%%%%%%%%%%%%%%%%%%%%%%%%%%%%%%%%%%%%%%%%%%%%%%%%%%%%%%%%%%%%%%%%%
%%%%%%%%%%%%%%%%%%%%%%%%%%%%%%%%%%%%%%%%%%%%%%%%%%%%%%%%%%%%%%%%%%%%%%%%%%%%%

\end{document}